\pdfoutput=1

\documentclass[twoside,11pt,openright]{report}

\usepackage[latin1]{inputenc}
\usepackage[american]{babel}
\usepackage{a4}
\usepackage{latexsym}
\usepackage{amssymb}
\usepackage{amsmath}
\usepackage{graphics}
\usepackage{graphicx}
\usepackage{calc}
\usepackage{cite}
\usepackage{multirow}
\usepackage{url}
\usepackage{subfigure}
\usepackage{snapshot}

\makeatletter
\def\@makechapterhead#1{%
  \vspace*{50\p@}%
  {\parindent \z@ \raggedright \normalfont
    \ifnum \c@secnumdepth >\m@ne
    \Huge\bfseries \@chapapp\space \thechapter
    \par\nobreak
    \vskip 20\p@
    \fi
    \interlinepenalty\@M
    \LARGE \normalfont \bfseries #1\par\nobreak
    \vskip 40\p@
    }}
\makeatother

\newenvironment{publist}[1]%
{\begin{list}{}{%
    \settowidth{\labelwidth}{#1\hspace{.1em}}%
    \setlength{\leftmargin}{\labelwidth+\labelsep}}}%
{\end{list}}

\newenvironment{myabstract}
{\small\begin{center}\bf\abstractname\vspace{-0.5em}\end{center}\quotation}
{\endquotation}

\newcommand{\mytitle}[3]{
  {\renewcommand{\thefootnote}{\fnsymbol{footnote}}
    \vspace*{2\baselineskip}
    \begin{center}
      {\LARGE #1}\par\vskip2em
      {\large\lineskip.5em\begin{tabular}[t]{c}#2\end{tabular}}
    \end{center}
    \vskip1.5em #3}}

\newcommand{\mycitation}[2]{
  {\begin{flushright}
      #1 \\ 
      --- \textit{#2}
    \end{flushright}}}

\newcommand{\clearemptydoublepage}{\newpage{\pagestyle{empty}\cleardoublepage}}

\begin{document}

\pagestyle{empty} 
\pagenumbering{roman} 
\setcounter{secnumdepth}{-1}
\vspace*{\fill}
\begin{flushright}
  {\Huge\sf Indoor Positioning with\\ Radio Location Fingerprinting}\\[3ex]
  {\huge\sf Mikkel Baun Kjærgaard} 
\end{flushright}
\noindent\rule{\linewidth}{1mm}\\[-.5ex]
\noindent\rule{\linewidth}{2.5mm}
\vfill
\begin{center}
  {\huge\sf PhD Dissertation}\\[\fill]
  \includegraphics{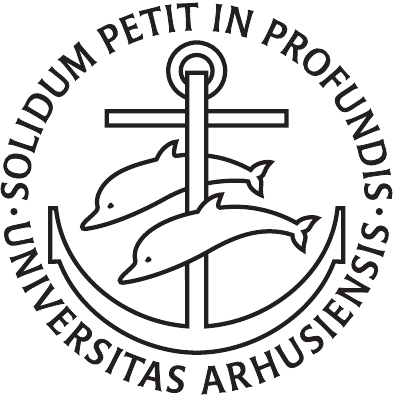}\\[\fill]
  {\sf Department of Computer Science\\University of Aarhus\\Denmark}
\end{center}
\vspace*{\fill}
\cleardoublepage
\begin{center}
  \vspace*{\stretch{1}}
  {\huge Indoor Positioning with Radio Location Fingerprinting}\\[\fill]
  A Dissertation\\
  Presented to the Faculty of Science\\
  of the University of Aarhus\\
  in Partial Fulfilment of the Requirements for the\\
  PhD Degree\\[\stretch{2}]
  by\\
  Mikkel Baun Kjærgaard\\
  \makeatletter\@date\makeatother
\end{center}
\vspace*{\stretch{1}}


\clearemptydoublepage
\pagestyle{plain}
\chapter*{{\Huge Abstract}}
\addcontentsline{toc}{chapter}{Abstract}
An increasingly important requirement for many novel applications is sensing the positions of people, equipment, animals, etc. GPS technology has proven itself as a successfull technology for positioning in outdoor environments but indoor no technology has yet gained a similar wide-scale adoption. A promising indoor positioning technique is radio-based location fingerprinting, having the major advantage of exploiting already existing radio infrastructures, like IEEE 802.11 or GSM, which avoids extra deployment costs and effort. The research goal of this thesis is to address the limitations of current indoor location fingerprinting systems. 

In particular the aim is to advance location fingerprinting techniques for the challenges of handling heterogeneous clients, scalability to many clients, and interference between communication and positioning. The wireless clients used for location fingerprinting are heterogeneous even when only considering clients for the same technology. The heterogeneity is due to different radios, antennas, and firmwares causing measurements for location fingerprinting not to be directly comparable among clients. Heterogeneity is a challenge for location fingerprinting because it severely decreases the precision of location fingerprinting. To support many clients location fingerprinting has to address how to scale estimate calculation, measurement distribution, and distribution of position estimates. This is a challenge because of the number of calculations involved and the frequency of measurements and position updates. Positioning using location fingerprinting requires the measurement of, for instance, signal strength for nearby base stations. However, many wireless communication technologies block communication while collecting such measurements. This interference is a challenge because it is not desirable that positioning disables communication. 

In summary, this thesis contributes to methods, protocols, and techniques of location fingerprinting for addressing these challenges. An additional goal is to improve the conceptual foundation of location fingerprinting. A better foundation will aid system developers and researchers to better survey, compare, and design location fingerprinting systems.


\clearemptydoublepage
\chapter*{{\Huge Acknowledgements}}
\addcontentsline{toc}{chapter}{Acknowledgements}

{
\setlength{\parskip}{0.2cm}
There are many people who I would like to thank for their encouragement and support in making my period of study a pleasant time. Here I can only mention a few of them.

I would like to thank my supervisor Klaus Marius Hansen for his valuable guidance during the last four years. I would also like to thank my second 
supervisor Søren Christensen for his guidance. During my Ph.D studies I have greatly benefitted from working together with Lisa Wells, Doina Bucur, and Carsten Valdemar Munk and I would like to thank them for their invaluable help and support. I would also like to thank Jonathan Bunde-Pedersen and Martin Mogensen for being great fellow students during the last eight years and for all the good discussions about doing research and life as a Ph.D student.

Furthermore, I would like to thank the members of the Mobile and Distributed Systems group for hosting my stay at the Ludwig-Maximilian-University Munich and for a lot of inspiring work and discussions while I was there. I would also like to thank Thomas King for the great collaboration during the past year and for his fruitful visit to Aarhus.

I would also like to acknowledge the financial support from the software part of the ISIS Katrinebjerg Competence Center and Kirk Telecom. Furthermore, I would like to thank the people working at Kirk Telecom for a good working relationship and for being a source of inspiration for my research.

But doing a Ph.D would not have made much fun without the support, love and joy from Sebastian, Mathilde and Mia and the rest of my family.

}
\vspace{2ex}
\begin{flushright}
  \emph{Mikkel Baun Kjærgaard,}\\
  \emph{Århus, \today.}
\end{flushright}


\clearemptydoublepage
\chapter*{{\Huge Structure of the Thesis}}
\addcontentsline{toc}{chapter}{Structure of the Thesis}

Part I of my PhD thesis entitled "Indoor Positioning with Radio Location Fingerprinting" gives an overview of my work. It summarizes my research and relates this to relevant literature and research. The text assumes a basic knowledge of statistics, and methods for machine learning and estimation.

This part is structured as follows:

\begin{description}
	\item[Chapter 1: Introduction and Motivation] motivates the need for indoor positioning and introduces location fingerprinting as a solution for this problem. Furthermore it discusses the research objectives and approach of the thesis and describes the empirical background of the thesis.
	\item[Chapter 2: Background] provides an overview of techniques for indoor positioning and describes the details and limitations of signal strength measurement using IEEE 802.11.
	\item[Chapter 3: A Conceptual Foundation for Location Fingerprinting] motivates the need for a better conceptual foundation for location fingerprinting. The chapter then discusses the thesis' contribution to this problem in the form of a taxonomy for location fingerprinting.
	\item[Chapter 4: Handling Heterogeneous Clients] motivates the problem of handling heterogeneous clients and discusses the thesis' contributions to this problem in the form of several methods for handling heterogeneity.
	\item[Chapter 5: Scalability to Many Clients] introduces the problem of scalability to many clients and discusses the thesis' contributions for this problem in the form of methods and protocols for improving the efficiency of location fingerprinting. 
	\item[Chapter 6: Interference between Communication and Positioning] introduces the problem of interference between communication and positioning and discusses the thesis' contributions to this problem in the form of methods to minimize such interference.
	\item[Chapter 7: Conclusions and Future Work] summarizes the main contributions of the thesis and discusses directions of future work.
\end{description}

Part II consists of six published papers. References to these papers are marked with
square brackets, i.e., "[. . .]" in Part I of the thesis.

\begin{description}
	\item[Paper 1:] \textbf{A Taxonomy for Radio Location Fingerprinting} presents a taxonomy for improving the conceptual foundation of location fingerprinting. The taxonomy consists of eleven main taxons and 88 subtaxons that in more detail classifies location fingerprinting systems. The taxonomy has been constructed based on a literature study of 51 papers and articles. The 51 papers and articles propose 30 different systems which have been analyzed, and methods and techniques have been grouped to form taxons for the taxonomy.
	
	\small M.\ B.\ Kjærgaard. A Taxonomy for Radio Location Fingerprinting. In \emph{Proceedings of the Third International Symposium on Location and Context Awareness}, pages~139--156, Springer, 2007. Acceptence rate 31\% (17/55).
	\normalsize
	
	\item[Paper 2:] \textbf{Automatic Mitigation of Sensor Variations for Signal Strength Based Location Systems} presents methods for classifying a client's measurement quality. Quality is classified in terms of if a client is caching, if it has a low measurement frequency, or if it provides measurements that do not correspond to signal strength measurements. Furthermore the paper proposes an automatic linear-mapping method for handling signal-strength differences. The method uses a linear mapping to transform one client's measurements to match another client's measurements. The method is automatic, but requires a learning period to find the parameters for the linear mapping.
	
	\small M.\ B.\ Kjærgaard. Automatic Mitigation of Sensor Variations for Signal Strength Based Location Systems. In \emph{Proceedings of the Second International Workshop on Location and Context Awareness}, pages~30--47, Springer, 2006. Acceptence rate 24\% (18/74).
	\normalsize
	
	\item[Paper 3:] \textbf{Hyperbolic Location Fingerprinting: A Calibration-Free Solution for Handling Differences in Signal Strength} presents a method named hyperbolic location fingerprinting for handling signal-strength differences. The key idea behind hyperbolic location fingerprinting is that fingerprints are recorded as signal-strength ratios between pairs of base stations instead of as absolute signal strength. The advantage of hyperbolic location fingerprinting is that it can resolve signal-strength differences \textit{without} requiring any extra calibration. Furthermore the paper proposes a method in the form of a filter to handle sensitivity differences among clients.
	
	\small M.\ B.\ Kjærgaard and C.\ V.\ Munk. Hyperbolic Location Fingerprinting: A Calibration-Free Solution for Handling Differences in Signal Strength. In \emph{Proceedings of the Sixth Annual IEEE International Conference on Pervasive Computing and Communications}, pages~110--116, IEEE, 2008. Acceptence rate 16\% (25/160).
		\normalsize
	
	\item[Paper 4:] \textbf{Zone-based RSS Reporting for Location Fingerprinting} presents an efficient zone-based signal strength protocol for terminal-assisted location fingerprinting. The protocol works as follows: a location server dynamically configures a client with update zones defined in terms of signal strength patterns. Only when the client detects a match between its current measurements and these patterns, that is, when it enters or leaves the zone, it notifies the server about the fact. The associated challenge is the adequate definition of signal strength patterns for which the paper proposes several methods.
	
	\small M.\ B.\ Kjærgaard, G. Treu, and C. Linnhoff-Popien. Zone-based RSS Reporting
for Location Fingerprinting. In \emph{Proceedings of the 5th International Conference on Pervasive Computing}, pages~316--333, Springer, 2007. Acceptance rate 16\% (21/132).
	\normalsize
	
	\item[Paper 5:] \textbf{Efficient Indoor Proximity and Separation Detection for Location Fingerprinting} presents an efficient method for walking-distance-based proximity and separation detection for location fingerprinting. The method uses a detection strategy that dynamically assigns clients' update zones in order to correlate the positions of multiple clients. In indoor environments such update zones can be effectively realized with the zone-based signal strength protocol together with a novel semantic for indoor distances.
		
	\small M.\ B.\ Kjærgaard, G. Treu, P. Ruppel and A. Küpper. Efficient Indoor Proximity and Separation Detection for Location Fingerprinting. In \emph{Proceedings of the First International Conference on MOBILe Wireless MiddleWARE, Operating Systems, and Applications}, pages~1--8, ACM, 2008. Invited Paper.
	\normalsize
	
	\item[Paper 6:] \textbf{ComPoScan: Adaptive Scanning for Efficient Concurrent Communications and Positioning with 802.11} presents a solution to address interference between communication and positioning. The solution, named ComPoScan, is based on movement detection to switch between light-weight monitor sniffing and invasive active scanning. Only in the case that the system detects movement of the user active scans are performed to provide the positioning system with the signal strength measurements it needs. If the system detects that the user is standing still it switches to monitor sniffing to allow communications to be uninterrupted.
	
	\small T.\ King and M.\ B.\ Kjærgaard. ComPoScan: Adaptive Scanning for Efficient Concurrent Communications and Positioning with 802.11. In \emph{Proceedings of the 6th ACM International Conference on Mobile Systems, Applications, and Services}, ACM, 2008. Acceptance rate 17\% (22/132).
	\normalsize
	
\end{description}

Other publications not included in the thesis.
\begin{description}
	\item[Paper 7:] Mikkel Baun Kjærgaard. Cleaning and Processing RSS Measurements for Location Fingerprinting.
In \textit{Proceedings of the Third International Conference on Autonomic and Autonomous
Systems (ICAS 2007)}. IEEE, 2007. Acceptence rate 27\% (56/207).

	\item[Paper 8:] Mikkel Baun Kjærgaard. Cyclic Processing for Context Fusion. In \textit{Adjunct Proceedings of the
Fifth International Conference on Pervasive Computing (Pervasive 2007)}. OCG, 2007. Acceptence rate 48\% (14/29).

	\item[Paper 9:] Mikkel Baun Kjærgaard and Jonathan Bunde-Pedersen. Towards a Formal Model of Context
Awareness. In \textit{Proceedings of the First International Workshop on Combining Theory and
Systems Building in Pervasive Computing (CTSB 2006)}, 2006.

	\item[Paper 10:] Mikkel Baun Kjærgaard. An API for Integrating Spatial Context Models with Spatial Reasoning
Algorithms. In \textit{Proceedings of the 3rd Workshop on Context Modeling and Reasoning (CoMoRea
2006)}. IEEE, 2006.	

  \item[Paper 11:] Kåre J. Kristoffersen, Mikkel Baun Kjærgaard, Jianjun Chen, Jim Sheridan, René Rønning,
and John Aa. Sørensen. Extending Wireless Broadband Network Architectures with Home
Gateways, Localization, and Physical Environment Surveillance. In \textit{Proceedings of the Second
International Conference on Next Generation Broadband, Content and User Perspectives (CICT
2005)}. CICT, 2005.

	\item[Paper 12:] Mikkel Baun Kjærgaard. On Abstraction Levels For Software Architecture Viewpoints. In
\textit{Procedings of the 17th International Conference on Software Engineering and Knowledge Engineering
(SEKE 2005)}. Knowledge Systems Institute, 2005. Acceptence rate 60\% (134/225)
\end{description}


\clearemptydoublepage
\tableofcontents
\clearemptydoublepage
\pagenumbering{arabic}
\setcounter{secnumdepth}{2}

\pagestyle{myheadings}
\renewcommand{\chaptermark}[1]{\markboth{\textit{\chaptername\
      \thechapter. #1}}{}} 
\renewcommand{\sectionmark}[1]{\markright{\textit{\thesection. #1}}}


\clearemptydoublepage
\part{Overview}

\clearemptydoublepage
\chapter{Introduction and Motivation}
\label{chap:intro}
\mycitation
{\textbf{position} \textit{(noun)} the place where somebody or something is situated.}
{Oxford Advanced Learner's Dictionary}
An increasingly important requirement for many novel applications is sensing the positions of people, equipment, animals, etc. This requirement is fundamental for novel applications within research areas such as pervasive computing, context-aware computing, sensor networks, and location-based services. Applications such as using the positions of people to support awareness among hospital staff \cite{Bardram2004}, using the positions of cars and trucks in fleet management systems, using the positions of equipment to optimize use, and using the positions of cows for smart farming \cite{Kuep05}.

How positions can be determined depend on what position sensors can be introduced or might already be available. A person might already carry possible position sensors around with them in their daily life such as mobile phones, cordless phones, laptops, PDAs or a \emph{Global Positioning System (GPS)} receiver. In other cases a position sensor might be attached to an animal or some equipment like a \emph{Radio-Frequency IDentification (RFID)} tag, an ultrasound tag, or an ultra-wide band tag.

A fundamental challenge when estimating the positions of sensors is the impact of the environment. One can here distinguish between outdoor and indoor environments. Outdoor environments cover huge areas and signals are impacted by a moderate number of obstructions. Indoor environments cover only moderate areas but signals are impacted by a large number of obstructions. Therefore each environment has its main challenge: outdoor is challenging because of the huge coverage and indoor is challenging because of the high number of obstructions. So far, there is no single positioning technology that supports both environments in an acceptable quality. GPS technology has proven itself as a successfull technology for outdoor environments but indoor no technology has yet gained a similar wide-scale adoption.

In the mentioned application areas, positioning of single sensors is not enough. Positioning technologies should support the positioning of a large number of sensors. Applications also require more information than just positions. They have to observe relationships such as line-of-sight distance or walking distance, between sensors or between sensors and static points of interests. This requires that positioning technologies support the distribution and comparison of position information to observe such relationships.

\section{Location Fingerprinting}
\begin{figure}[h]
	\centering
		\includegraphics[viewport=150 210 420 450,width=0.55\textwidth,clip]{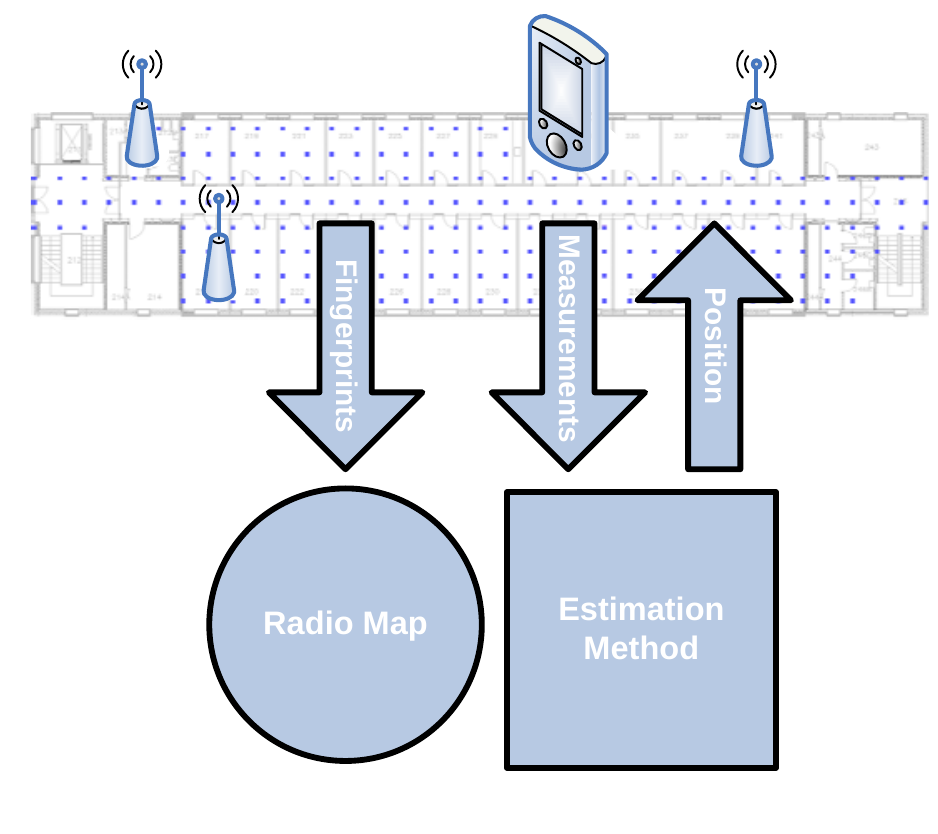}
	\caption{Location Fingerprinting.}
	\label{intro:fig:LF}
\end{figure}

A promising indoor positioning technique is \emph{Location Fingerprinting (LF)}, having the major advantage of exploiting already existing radio infrastructures, like IEEE 802.11 or GSM, which avoids extra deployment costs and effort. LF uses a radio map of pre-recorded measurements from different locations, denoted as \emph{fingerprints}, which is illustrated as small squares in Figure \ref{intro:fig:LF}. The most common type of measurements used for LF is the strength of radio signals. Later, a sensor's position is calculated using an estimation method by comparing current measurements with the pre-recorded radio map. When LF is used in connection with radio infrastructures, like IEEE 802.11 or GSM, mobile phones, laptops or PDAs already carried by persons can be used as position sensors. However, it is also possible to embed an IEEE 802.11 or GSM radio in a tag, for instance, for animal or equipment tracking. In the remaining parts of this thesis a radio-based LF position sensor will be denoted as a wireless client.

\subsection{Overview}
This section gives an introduction to existing LF systems to discuss the systems' precision, support for privacy, and need for calibration in terms of fingerprint collection. In this section LF systems will be classified with respect to the three properties; \emph{scale}: the size of a system's deployment area, \emph{roles}: the division of responsibilities between wireless clients, base stations, and servers, and \emph{collector}: who or what collects fingerprints. These three properties are important factors when considering systems' precision, support for privacy, and need for calibration. In Chapter \ref{chap:rlf} a detailed taxonomy for LF is presented that covers other relevant properties.

\textit{Scale} describes a system's targeted size of deployment. Scale is important because size of deployment impacts how fingerprints can be collected and some systems are limited in scale because of specific assumptions. Scale can be classified as \emph{building}, \emph{campus}, or \emph{city}. Many LF systems have been proposed for a \emph{building} scale of deployment \cite{Bahl2000,Prasithsangaree2002,Battiti2002,Roos2002a}. Some systems are limited to this scale because they assume knowledge about the physical layout of buildings \cite{Krumm2004,Castro2001,Ladd2002,Haeberlen2004}; others because they assume the installation of a special infrastructure \cite{Bahl2000b,Krishnan2004}. \emph{Campus}-wide systems \cite{Bhasker2004} scale by proposing more practical schemes for fingerprint collection. \emph{City}-wide systems \cite{Laitinen2001,Roos2002b,LaMarca2005} scale even further by not assuming that a system is deployed by or for a single organization. City wide systems could scale to any area size that is covered by base stations.

\begin{figure}[h]
	\centering
		\includegraphics[viewport=20 200 630 570,width=0.85\textwidth,clip]{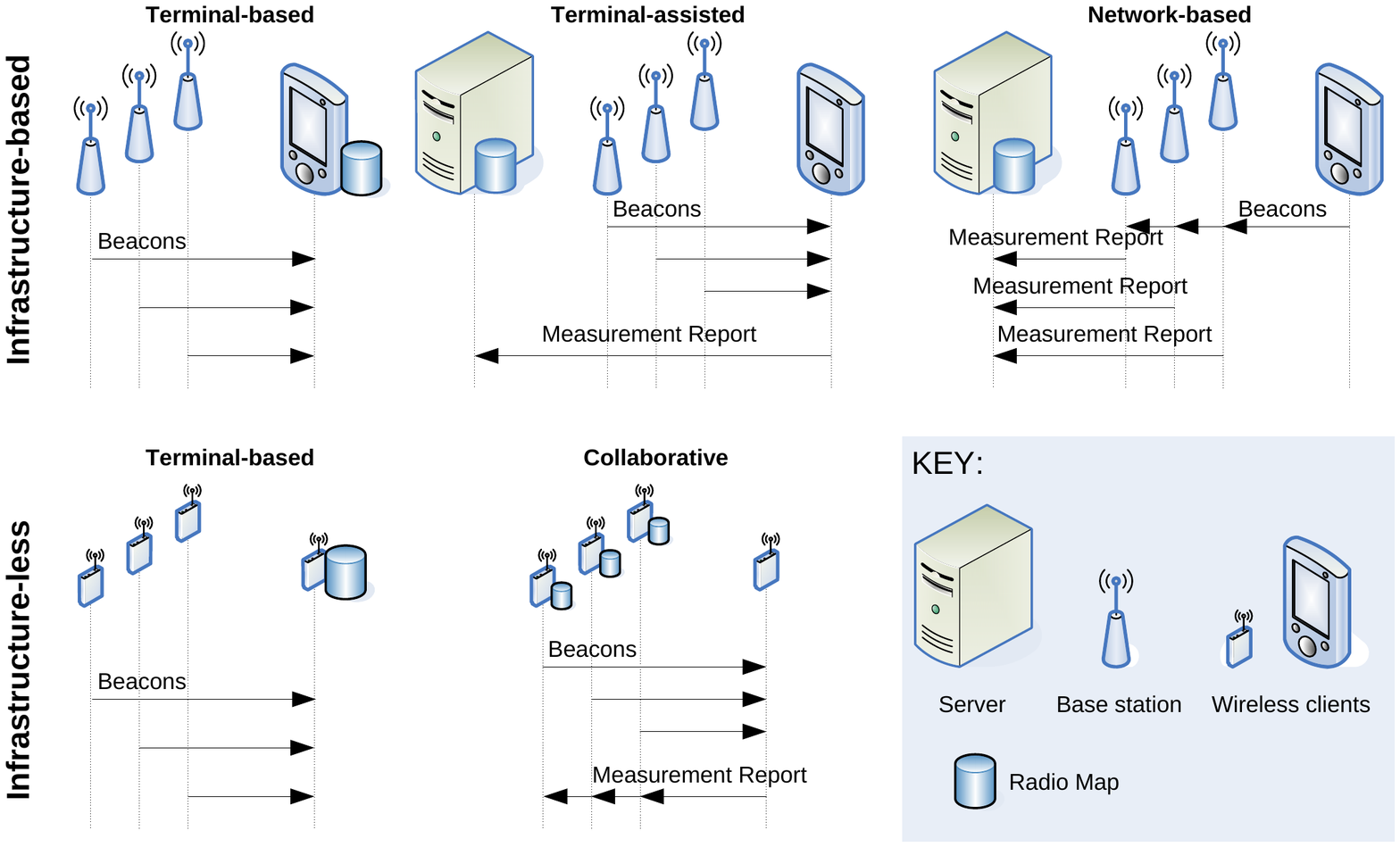}
	\caption{Different assignments of responsabilities to wireless clients, base stations, and servers.}
	\label{intro:fig:LFRoles}
\end{figure}

\textit{Roles} denotes the division of responsibilities between wireless clients, base stations, and servers. How roles are assigned impact both how systems are realized, but also important non-functional properties like privacy and scalability. The two main categories for roles are \textit{infrastructure-based} and \textit{infrastructure-less}. Infrastructure-based systems depend on a pre-installed powered infrastructure of base stations. Infrastructure-less systems consist of ad-hoc-installed battery-powered wireless clients where some of them act as "base stations". Infrastructure-based systems can according to K{\"u}pper \cite{Kuep05} be further divided into \emph{terminal-based}, \emph{terminal-assisted}, and \emph{network-based} systems. The infrastructure-less systems are divided into \emph{terminal-based} and \emph{collaborative} systems. The different types of systems differ in who transmits wireless packages, denoted as beacons, for other to measure and who makes measurements from the beacons. Furthermore they differ in who stores the radio map and runs LF estimation, as illustrated in Figure \ref{intro:fig:LFRoles}. Most LF systems have been built as infrastructure-based and terminal-based \cite{Youssef2005b,Prasithsangaree2002,LaMarca2005}, which is attractive because this setup supports privacy because the wireless clients do not transmit any beacons or measurement reports that reveal their existence. Terminal-assisted \cite{Castro2001,Bhasker2004} and network-based systems \cite{Bahl2000,Krishnan2004} have also been built offering good support for resource-weak wireless clients. Infrastructure-less LF-systems have to be optimized for the resource-weak wireless clients, which is addressed by the collaborative setup \cite{Lorincz2005,Lorincz2006}.

\textit{Collector} describes who or what collects fingerprints. There are three categories: \emph{user}, \emph{administrator}, and \emph{system}. A user is a person who is either tracked by or uses information from a LF system \cite{Bhasker2004,LaMarca2005}. An administrator is a person who manages a LF system \cite{Bahl2000,Haeberlen2004,Seshadri2005} and a system is a specially-installed infrastructure for collecting fingerprints \cite{Krishnan2004}.

Previous litterature on LF has proposed systems with different choices for the properties of \emph{scale}, \emph{roles}, and \emph{collector}. The implications of different combinations will be discussed in the following focusing on precision, support for privacy, and need for calibration. Table \ref{tab:systems} lists four examples of LF systems: RADAR, LEASE, Place Lab, and Active Campus. Each entry in the list describes a system's scale, division of roles, and type of collector together with the precision at median accuracy as reported by papers for the specific system.

\begin{table}[h]
\small
\centering
\begin{tabular}{p{2.7cm}|p{1.3cm}|p{1.8cm}|p{2.2cm}|p{3.2cm}}
 & \textit{Scale} & \textit{Roles} & \textit{Collector} & \textit{Precision}\\
 \hline
 RADAR \cite{Bahl2000} & Building & Network & Administrator & 2.75 meter\\
 \hline
 LEASE \cite{Krishnan2004} & Building & Network & System & 2.1 meter\\
 \hline
 ActiveCampus \cite{Bhasker2004}& Campus & Terminal-Assisted & Users & Room recognition with 90\% accuracy\\
 \hline
 Place Lab \cite{LaMarca2005}& City & Terminal & Users & Urban: 21.8 meter\\
           & & & & Residential: 13.4 meter\\
           & & & & Suburban: 31.3 meter\\
\end{tabular}
\caption{The accuracy of LF systems with different scales, division of roles and collectors}
\label{tab:systems}
\end{table}

The precision of LF systems depends on numerous factors. The impact of a system's scale on the precision can mainly be attributed to how the scale imply coverage over indoor and outdoor areas. Indoor areas generally have a high LF precision because the high number of obstructions makes fingerprints more distinctive and thereby easier for a LF system to recognize. Indoor areas also tend to be smaller which makes it practical to increase precision by collecting a more dense set of fingerprints. Furthermore indoor areas are normally covered with a more dense set of access points which also increase precision. For a more detailed analysis of the factors of fingerprint and access point denseness we refer to the study by King et al. \cite{KingLoca2007}. However, precision also depends on other factors such as people present, building materials and building structure. Compared to indoor areas, outdoor areas tend to have a lower LF precision because of fewer obstructions and a lower number of access points.

These factor's impact on the LF precision can be noticed from Table \ref{tab:systems}. The two building scale systems have the highest precision with a median accuracy of approximately two meters. The listed result for Active Campus only covers indoor areas and can as such only be considered as a building-scale evaluation of a campus-scale system. The result is not reported in meters but with a precision of distinctive rooms for which the system has a recognition accuracy of 90\%. The city scale system PlaceLab has the lowest LF precision with a median precision between 13.4 to 31.3 meters. The precision is best in urban and residential areas which have the highest number of access points and is lower in suburban areas with fewer access points.

Fingerprint collection is above classified into user, administrator, and system. That a user can collect fingerprints makes it easy for people to increase coverage of a system to new areas or for them to re-calibrate the system. The need for re-calibration can, for instance, be due to outdated fingerprints because of building changes or movement of base stations. However, the drawback is how to maintain the validity of user-reported data as discussed by Bhasker et al. \cite{Bhasker2004}. The administrator solution solves the validity problem but adds a second step to the process of updating fingerprints. The system approach makes it easy to update fingerprints but requires a specially installed infrastructure. Therefore each of the collection methods has it benefits and drawbacks. The in Table \ref{tab:systems} listed systems have been based on different methods. One trend that can be noticed from the list is that the campus and city systems apply user-based fingerprinting to scale beyond building scale systems.

An important aspect of any positioning technology is the support for privacy. Privacy is the property that a position sensor does not reveal its existence and thereby its position to others. Privacy was briefly mentioned above when discussing the division of roles which has a major impact on LF systems support for privacy. The reason is that if a wireless client has to sent out beacons to position it-self it reveals both its existence and makes it possible for others to estimate the client's position. Therefore it is only terminal-based LF systems that are able to hide their existence from others and there-by support full control over privacy. For IEEE 802.11 technical details do complicate the control of privacy a bit more which will be discussed in Section \ref{sec:ieee80211scanning}. However, for many novel applications to work wireless clients have to share their positions with others. One example of such an application is the ActiveCampus \cite{Griswold2004} system created to foster social-interactions in a campus setting. One of the services offered by this application provides users with a list of nearby buddies and shows maps overlaid with information about buddies, sites, and current activities. In such an application the privacy goal is not that sensors' positions are never revealed but only to trusted parties in user-desired time intervals and with user-desired precision. Mechanisms for privacy control, for instance, the ones proposed by Beresford et al. \cite{Beresford2003} can be built on top of LF systems to satisfy such needs.

\subsection{Challenges}
\begin{figure}[h]
	\centering
		\includegraphics[viewport=20 535 560 815,width=1.0\textwidth,clip]{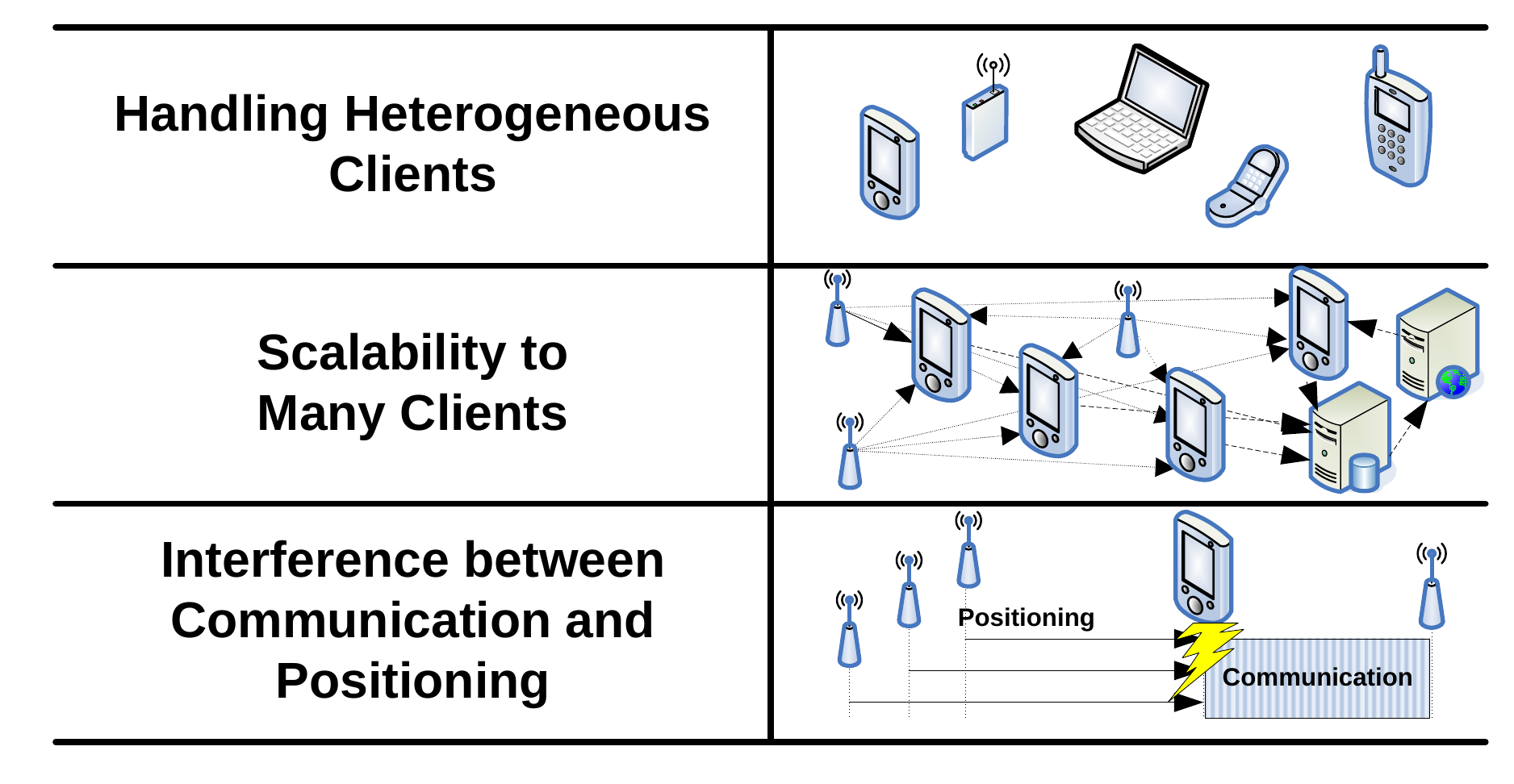}
	\caption{LF Challenges.}
	\label{fig:Challenges}
\end{figure}

The preceding sections introduced LF and discussed precision, support for privacy, and need for calibration. This section outlines the important LF challenges of heterogeneous clients, scalability to many clients, and interference between communication and positioning. These challenges are all illustrated in Figure \ref{fig:Challenges}.

\begin{description}
	\item[Handling Heterogenous Clients:] The wireless clients used for LF are heterogeneous even when only considering clients for the same technology. The heterogeneity is due to different radios, antennas, and firmwares causing measurements for LF not to be directly comparable among clients. For instance, signal strength measurements might be lower or higher at the same position or radio sensitivity, the limit for how weak signals a client can hear might also be different. Heterogeneity is a challenge for LF because it severely decreases the precision of LF.
	\item[Scalability to Many Clients:] To support many clients LF has to address how to scale estimate calculation, measurement distribution, and distribution of position estimates. To calculate estimates for a large number of clients is demanding due to the number of calculations involved. Furthermore if position estimates are not calculated on the measuring client measurements have to be distributed which is challenging due to the frequency of measurements. Finally, position estimates have to be distributed to interested parties, for instance, for observing various relationships. This distribution is also a challenge due to the amount of updates.
	\item[Interference between Communication and Positioning:] Positioning using LF requires the measurement of, for instance, signal strength for nearby base stations. However, many wireless communication technologies separate communication by dividing their frequency bands into separate channels. Base stations for a technology will normally only operate on one channel. Therefore to measure all nearby base stations clients have to scan all channels and therefore block communication by leaving the current communication channel. This is a challenge because it is not desirable that LF positioning when in use disables communication.
\end{description}

\section{Research Objectives}
\label{introduction:resobj}
\begin{figure}[h]
	\centering
		\includegraphics[viewport=0 80 830 530,width=1.0\textwidth,clip]{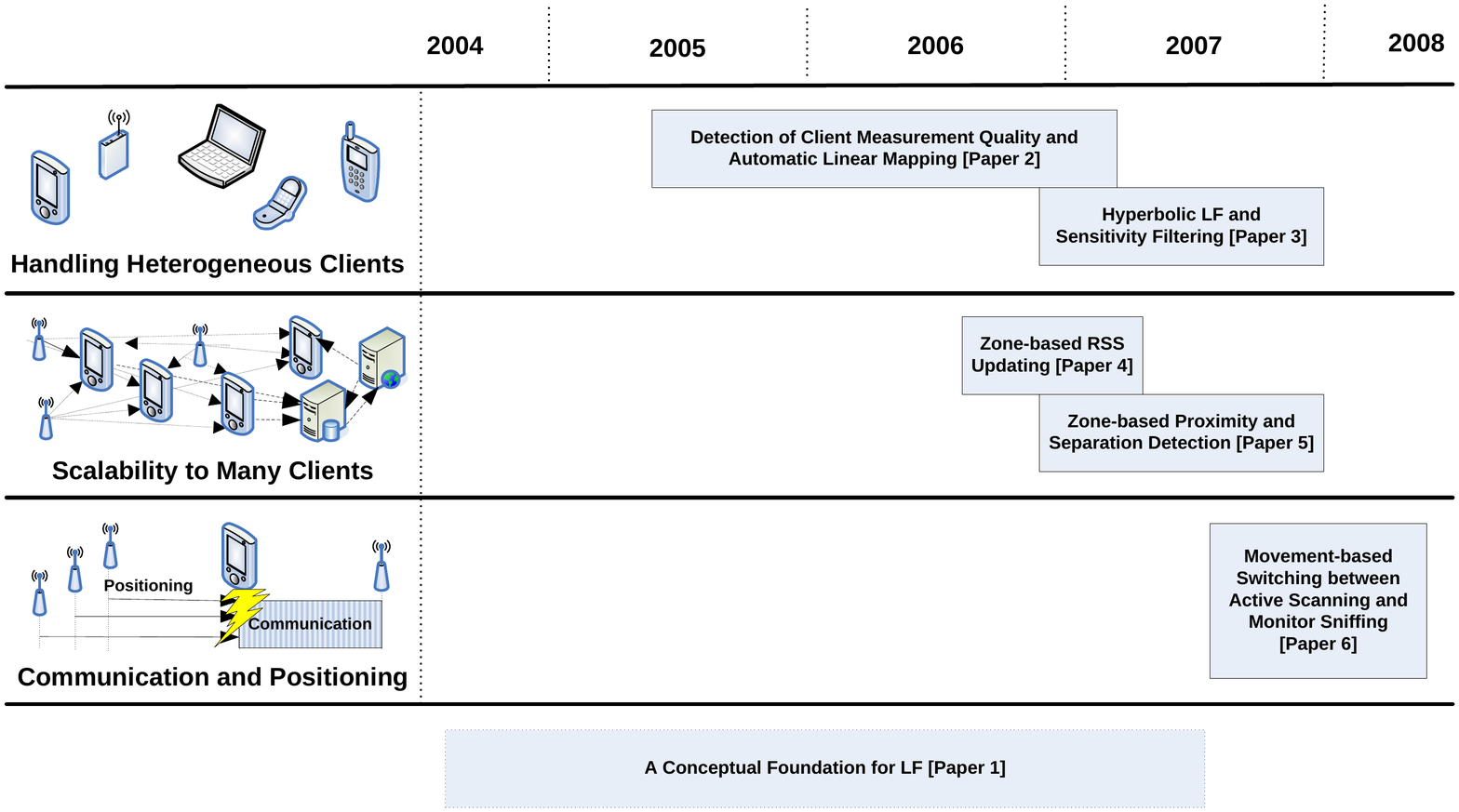}
	\caption{Time-based overview over challenges, papers, and techniques.}
	\label{fig:ChallengesHypo}
\end{figure}

The research objective of this thesis is to address the limitations of current indoor LF systems. In particular, the aim is to advance LF for the challenges of handling heterogeneous clients, scalability to many clients, and interference between communication and positioning. A set of techniques for these challenges will enable the use of LF with heterogeneous clients, with more clients, and with less interference all together enabling a more succesful use of LF. An additional goal is the improvement of the conceptual foundation of LF. A better foundation will aid LF system developers and researchers better survey, compare, and design LF systems. Figure \ref{fig:ChallengesHypo} gives a time-based overview over the work presented in the papers of this thesis for each of the three challenges. From the figure, it can also be seen how work on the different problems have progressed during the project period.

\section{Research Approach}
The research approach of this thesis is one of asking research questions, stating hypotheses, and providing evidence. One of the research questions is \textit{"how to address the challenge of handling heterogeneous clients"}. For this question several hypotheses were proposed, eventually four of these hypotheses were fruitful (all described in Chapter \ref{chap:hdlf}) and supporting evidence was assembled. All of the four hypotheses are constructive in the sense that they describe a solution for the research question. The use of such constructive hypotheses is a common element within computer science \cite{Zobel2004}.

The proposed hypotheses have been tested by assembling supporting evidence. Evidence has been provided by the use of controlled experiments which according to Zobel \cite{Zobel2004} is defined as "\textit{a full test of a hypothesis based on an implementation of the proposal and on real - or at least realistic - data"}. Two kinds of controlled experiments have been used: emulation and validation. Emulation is a full test of a hypothesis which is tested in an environment emulated by recorded real data. The purpose of emulation is testing and parameter optimization on a stable set of data. For evaluating the proposed techniques during the project period several data sets have been collected of signal strength measurements. Validation is a full test of a hypothesis as a deployed system with fixed parameters in a real setting. The purpose of validation is testing a system in a manner so no real-world effects are missed. During the project period several of the proposed techniques have been implemented and deployed for evaluation by validation. The methods have also been combined by, first, testing and optimizing parameters using emulation and then, later, real-world testing using validation.

\section{Empirical Background}
The empirical foundation of this thesis is the following three projects. The "Focus on the Future"-project targeted positioning in a DECT radio-infrastructure, the IEEE 802.11 LF-project has been a continuous effort to enable IEEE 802.11 positioning at The Department of Computer Science at the University of Aarhus, and the TraX-project targeted the creation of a novel platform for location-based applications.

\subsection{Focus on the Future}
The project ''Focus on the Future'' was a combined project between the University of Aarhus, ISIS Katrinebjerg Software, and an industrial partner KIRK which ran from 2004 to 2006. The company KIRK develops and sells products based on Digital Enhanced Cordless Telecommunications (DECT) technology. DECT is a digital radio access standard for cordless communication in residential, corporate, and public environments. Today DECT technology is used in many types of products where the most common product is cordless phones. A DECT infrastructure consists of a number of base stations. For small residential systems there might only be one base station but for corporate systems there might be hundreds. This infrastructure can then be utilized by DECT clients, for instance, in the form of phones delivering telephone services to users. If, however, these infrastructures were extended with positioning, it would open up the possibility to make new location-based applications on DECT clients.

\begin{figure}[h]
	\centering
	\includegraphics[width=0.8\textwidth]{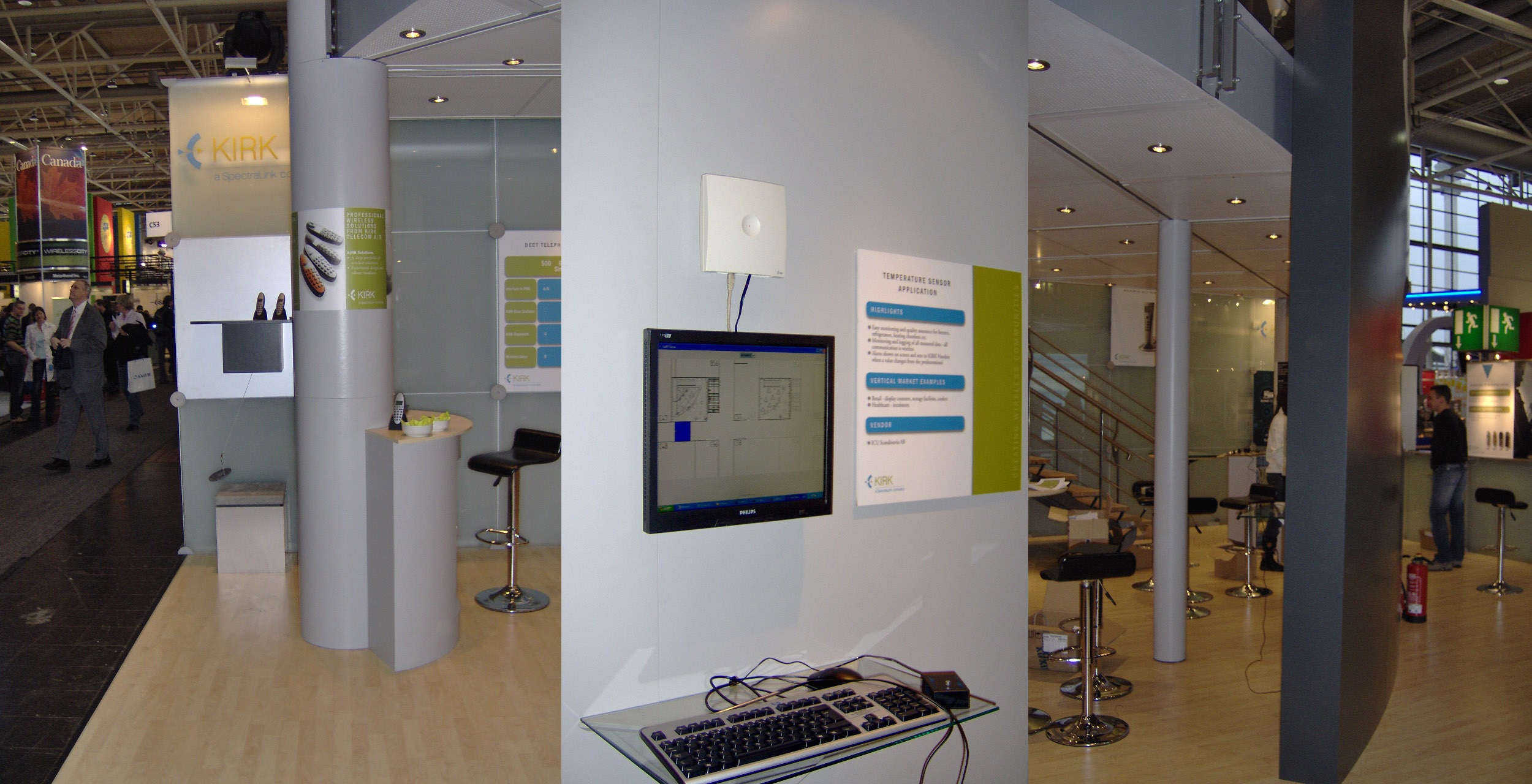}
	\caption{Prototype deployment at CEBIT 2006.}
	\label{fig:LocalizationCebitDeployment}
\end{figure}

During the project several prototypes were realized of positioning extensions to DECT infrastructures. The prototypes have been tested at eight sites including a deployment at KIRK's stand at CEBIT 2006 as shown in Figure \ref{fig:LocalizationCebitDeployment}. The test results for precision of indoor DECT LF were comparable to that of indoor IEEE 802.11 LF positioning which is consistent with the results for DECT reported by Rauh et al.\cite{Rauh2003} and Schwaighofer et al. \cite{Schwaighofer2004}. For the thesis this project has mainly served as inspiration for the research carried out in the context of the IEEE 802.11 LF-project.

\subsection{IEEE 802.11 Location Fingerprinting}
\begin{figure}[h]
	\centering
		\includegraphics[width=0.75\textwidth]{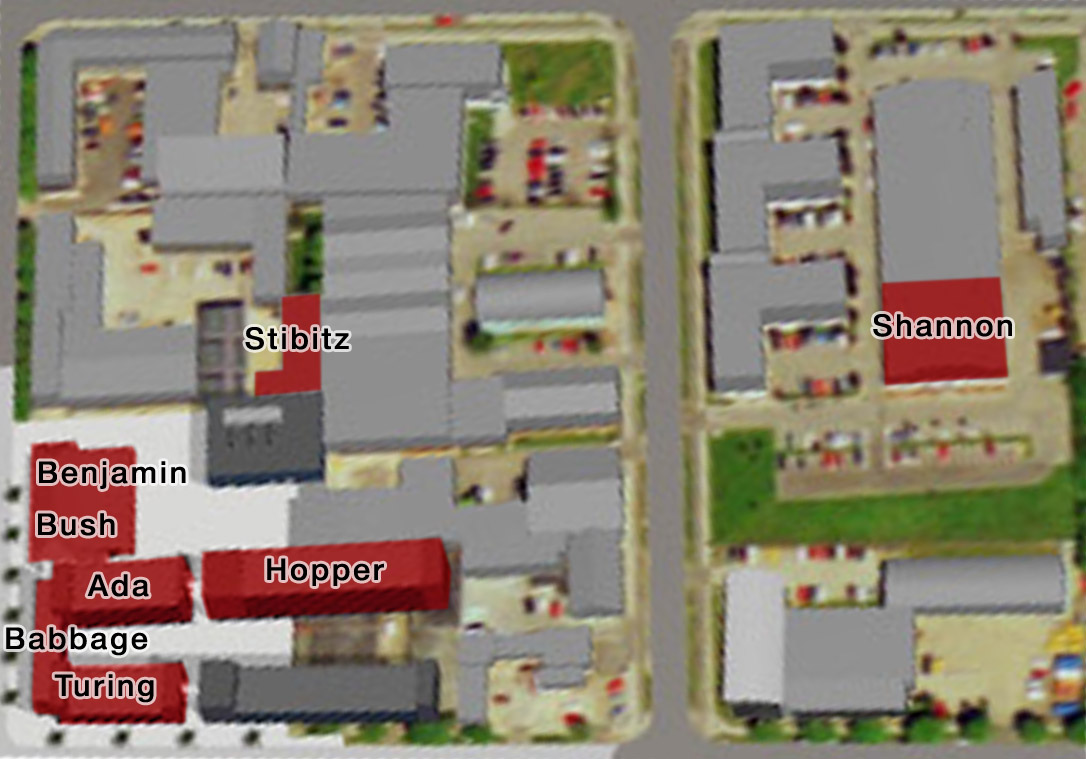}
	\caption{Test-site buildings highlighted in red.}
	\label{fig:Map}
\end{figure}

The empirical background of the thesis also includes a continuing effort to enable positioning on the IEEE 802.11 installations at the Department of Computer Science at the University of Aarhus from 2004 to 2008. These installations have been used for both emulation and validation. For emulation an extensive set of data has been collected totalling more than two million base station measurements during the project period. To use the data for hypotheses testing the data set consists of measurements collected with different properties, for instance, measurements collected with different types of clients. The IEEE 802.11 installations cover several buildings and eight of these have been used as test sites in the research as illustrated in Figure \ref{fig:Map}. The buildings also have different properties in terms of age, building materials, size of rooms which supports the correctness of emulation and validation results with respect to other buildings. The buildings used have the following properties:

\begin{description}
	\item[Turing, Ada, Hopper:] Newer office buildings.
	\item[Babbage:] New building consisting of one large atrium.
	\item[Bush, Stibitz, Shannon:] Older warehouse buildings refitted to lecture halls.
	\item[Benjamin:] Old warehouse building refitted to one large lecture hall.
\end{description}

During the project several LF system prototypes have been realized including several map-based GUI interfaces for easy visualization and fingerprint collection. The prototypes have also contributed to the development of a stream-based software architecture for LF systems and an indoor location modelling framework. The stream-based software architecture combines component and stream abstractions to provide flexible processing for LF systems as described in Kjærgaard \cite{KjaergaardIcas2007} and Kjærgaard \cite{KjaergaardPervAdj2007}. The indoor location modelling framework provides various facilities for handling location information such as model querying and storage, coordinate transformations, and calculation of various graph and geometric-based metrics. The framework is described in more detail in Kjærgaard \cite{KjaergaardComorea2006}.

\subsection{TraX}
The empirical background further includes the TraX (Tracking and X-change)-project. The author worked within the scope of the TraX project while visiting the mobile and distributed systems group at the Ludwig-Maximilian-University of Munich in the fall of 2006. The focus of the TraX-project was to create a platform for enabling proactive location-based applications. In contrast to conventional reactive applications, proactive applications are not initialized by the user. Rather, they are event-based, i.e., they are automatically triggered as soon as the user enters a predefined point of interest. In the context of the TraX-project new concepts and a platform were developed and evaluated for efficient support of proactive location-aware applications. The TraX-project and platform are described in more detail in Küpper et al. \cite{KupperCM2006}.

\section{Summary}
To sum up, this chapter motivated the need for and challenge of indoor positioning. A promising technique to address the indoor positioning problem is LF. Three important properties of LF systems are precision, calibration, and privacy and how LF systems are built and deployed impact these three properties. Three important research challenges of LF are how to handle heterogeneous clients, scalability to many clients, and the interference between communication and positioning. Furthermore this thesis also contributes to the conceptual foundation of LF. To address these three challenges the work presented in this thesis have used a research approach of putting forward research questions, stating hypotheses, and providing evidence. The empirical background of the work has been within the three projects of "Focus on the Future", IEEE 802.11 Location Fingerprinting, and TraX.


\clearemptydoublepage
\chapter{Background}
\label{chap:background}
\mycitation
{\textbf{background} \textit{(noun)} the circumstances or past events which help explain why something is how it is.}
{\textit{Oxford Advanced Learner's Dictionary}}
LF is not the only technique that can be applied to address the indoor positioning problem. Therefore this chapter will cover other techniques and discuss their relationship to LF. Furthermore one of the primary measurement types used for LF is signal strength measurements. Therefore this chapter also covers the details and limitations for the measurement of signal strength using IEEE 802.11.

\section{Indoor Positioning}
This section gives an overview over indoor positioning. Indoor positioning is a complex engineering problem that has been approached by many computing communities: networking, robotics, vision, and signal processing. The overview will be divided into a discussion of signals and methods. The signals are the physical phenomenons that are used to position sensors. Signals are sent between the position sensors to make distance-related measurements. Afterwards sensor positions are estimated from measurements by a positioning method.

\subsection{Signals}
Many types of physical signals can be used for positioning and therefore this section only discusses the most common signal types: radio, light, and sound. Radio and light signals are both electromagnetic waves which traditionally are classified by their wavelengths. The types of electromagnetic waves that are important for positioning are radio waves with wavelengths around $10^3$ meters, infrared light with wavelengths around $10^{-5}$ meters, and visible light with wavelengths around $0.5 \times 10^{-6}$ meters. An important property for positioning is the propagation speed of signals. In vacuum electromagnetic waves propagate at the speed of light but for other mediums the speed depends on the properties of the medium. 

Sound signals are waves of vibrational mechanical energy. Sound signals are traditionally classified by their frequency. Relevant for positioning are ultrasound waves with a frequency of more than 20.000 Hz and human-hearable acoustic sound waves with a frequency between 20 Hz and 20.000 Hz. Sound's propagation speed depends on the medium's properties, for instance, in air at sea level the speed is approximately 343 meter pr. second.

Given that a signal can be transmitted between position sensors, several types of distance-related measurements can be collected. If a signal's propagation speed is known one can estimate distance by measuring the time delay from sensor to sensor. This is know as Time-Of-Flight (TOF)\footnote{Also sometimes referred to as Time-Of-Arrival (TOA)} measurements. One can also measure the relative time delay by measurering a signal's arrival time at several sensors, something that is known as Time-Difference-Of-Arrival (TDOA). Distances can also be measured by comparing the strength of a signal when it was sent to when it was received. Another option is to measure the angle to a sensor by observing what angle a signal from this sensor arrives in which is known as Angle-Of-Arrival (AOA) measurements.  \cite{Kuep05}

\subsection{Methods}
\begin{figure}[h]
	\centering
		\includegraphics[viewport=100 170 550 480,width=0.8\textwidth,clip]{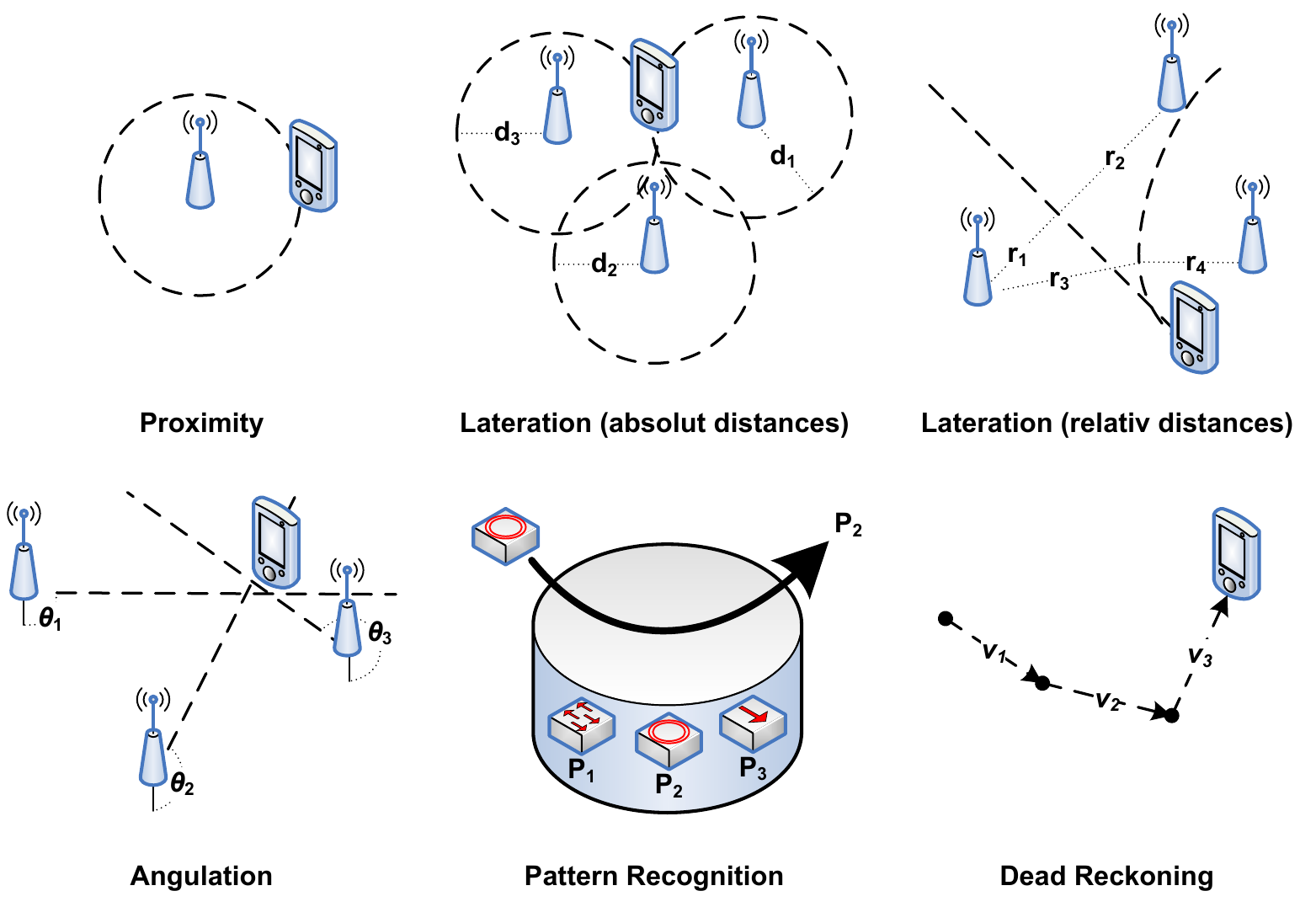}
	\caption{Methods.}
	\label{fig:indoormethods}
\end{figure}

There exist many different positioning methods that given suitable measurements can be used to estimate sensor positions. Each method has specific requirements as to what types of measurements are needed. This section covers the position methods of proximity, lateration, angulation, pattern recognition, and dead reckoning, all illustrated in Figure \ref{fig:indoormethods}. The methods can be applied alone but they can also be combined to build various kinds of hybrid systems. Another option is to apply the methods in parallel and then combine all the estimates into one final estimate.

\subsubsection{Proximity}
The proximity method estimates positions by logging when mobile sensors come into proximity of fixed sensors, as illustrated in Figure \ref{fig:indoormethods}. The position of mobile sensors is then estimated as the position of the fixed sensor which last logged it. That a target is in proximity can, for instance, be detected as the ability to transmit either radio or light signals between sensors. 

A system that uses the proximity method with infrared light is the Active Badge system \cite{Want1992a,Want1992b,Harter1994}. The Active Badge system is designed for position estimation with room-size precision. The system consists of people-worn tags\footnote{We consider badges as a special type of tags designed to be worn by the neck.} identifying themselves via infrared light to fixed sensors. A server is responsible for pulling sensors for tag sightings and a tag's position is then predicted as the position of the sensor which last sighted it. Another example based on radio signals is passive \emph{Radio-Frequency IDentification (RFID)} where a passive RFID tag's position is known when in proximity of a RFID scanner.

The proximity method has several advantages. First, it can be used with nearly all types of existing radio infrastructures. Second, because targets only have to emit an identification code they can be designed to be very low-cost as in the case of RFID. However, the method also has some disadvantages. First, precision is limited by the range of the sensors. Second, targets can only be positioned when in proximity. Third, the area where devices are in range is not static and can therefore take arbitrary shapes. This means that if a fixed sensor is installed in a room to log which sensors are in the room it is very likely that it will also log sensors in the adjacent hallway or miss sensors in the room.

\subsubsection{Lateration}
The lateration method estimates positions from distance-related measurements to fixed sensors with known positions. For lateration there exists a number of different schemes \cite{Kuep05} where the two main types are: lateration with absolute distances and lateration with relative distances, also illustrated in Figure \ref{fig:indoormethods}.

Lateration with absolute distances uses measurements that directly describe the distance between a mobile sensor and several fixed sensors. Each of the distances $d_1,d_2,d_3$ in Figure \ref{fig:indoormethods} form a circle of possible positions around the fixed sensors. The position estimate can then be found as the most likely position given a specific error criteria with respect to these circles. Lateration with relative distances uses measurements that describe the relation between the distances from a mobile sensor to fixed sensors. Given measurements $r_1,r_2,r_3,r_4$ that describe the relative distance between a mobile sensor and several fixed sensors. Each of the relations $r_1:r_2$ and $r_3:r_4$ in Figure \ref{fig:indoormethods} form a hyperbola of possible positions related to pairs of fixed sensors. The position estimate can then be found as the most likely position given a specific error criteria with respect to these hyperbolas.

A system that uses lateration with absolute distances is the Bat system \cite{Ward1997,Addlesee2001,Harle2005}. The Bat system is designed for positioning with centimetre precision. The system consists of people-worn tags emitting ultrasonic pulses when requested via a radio signal. The ultrasound is picked up by a set of ultrasound receivers installed at fixed positions in the ceiling and forwarded to a server for positioning. The system uses TOF measurements that are measured as the time difference between the sending of the radio signal request and the receiving of the responding ultrasonic pulse. This measurement method works because the time for the radio signal to propagate from sensor to tag takes a fraction of the time it takes the ultrasonic pulse to propagate from tag to sensor.

A system that uses lateration with relative distances is the system proposed by Yamasaki et al. \cite{Yamasaki2005}. The system is designed for positioning with meter precision. The system consists of extended IEEE 802.11 base stations with clocks synchronized down to nanoseconds. The system uses TDOA measurements that are measured as the differences in propagation time for base station pairs that receive a special location packet from a mobile sensor. Because the access points are time synchronized the differences can be computed by the difference in their own clock time. A server then estimates a position by finding a solution for the hyperbolas formed by the measurements.

The lateration method has several advantages. First, it be can be used for designing systems with high precision. Second, it enables systems with large coverage because positions can be found in all areas covered by sensors. However, the method also has some disadvantages. First, most systems require that special sensors are installed in the covered area. Second, the positions of the fixed sensors have to be established which is not an easy task in large and complex indoor environments. Third, many lateration systems depend on some form of time synchronization that often requires a direct cabling between the fixed sensors. Finally, the precision can be severely degraded by multipathed signals. Multipathed signals are signals that do not propagate by the direct path between two sensors. Such signals can impact measurements so sensors appear to be further away than they really are and thereby degrade the precision of the final position estimate.

\subsubsection{Angulation}
The angulation method estimates positions from angle measurements to fixed sensors with known locations. Each of the angle measurements $\theta_1,\theta_2,\theta_3$ in Figure \ref{fig:indoormethods} describes a line of possible positions through the positions of the fixed sensors. The position estimate can then be found as the most likely position given a specific error criteria with respect to these lines.

A system that uses angulation is the system of VHF Omnidirectional Ranging (VOR) base stations proposed by Niculescu et al. \cite{Niculescu2004}. The system is designed for positioning with meter precision. The system is based on extended 802.11 access points that can make AOA measurements. Given the AOA measurements for a number of fixed points the position of a target can be estimated.

The angulation method generally has the same advantages and disadvantages as the lateration method. However, the angulation method is even more sensitive to multipathed signals than lateration. The reason is that multipathed signals can come from the opposite direction than the signals which propagate by the direct path and thereby severely degrade the precision of the final position estimate.

\subsubsection{Pattern Recognition}
The pattern recognition method estimates positions by recognizing position-related patterns in measurements. Each pattern to be recognized has to be available in some encoding. The encoding should for each pattern contain a mapping from the pattern to a position, as illustrated in Figure \ref{fig:indoormethods}. The method can be applied with many types of measurements, for instance, vision systems recognizing patterns in video feeds from cameras or LF recognizing patterns in signal strength measurements.

A system that uses pattern recognition is the Cantag system \cite{Rice2006}. The Cantag system is designed for centimetre precision. The system uses video feeds from cameras to position physical markers represented as 2D barcodes. The recognition process uses video feeds from two cameras to recognize the information encoded in the barcode and from the barcode size and orientation estimate its position with respect to the cameras. 

Pattern recognition has several advantages: First, it can support tracking of non-tagged people or items. Second, it can be applied to many types of measurements. However, the method also has some disadvantages: First, the patterns have to be recorded / encoded for the method to work. Second, in the case of vision systems an infrastructure of cameras are needed and the cameras need direct line of sight to tracked objects.

\subsubsection{Dead Reckoning}
The dead reckoning method estimates positions by advancing previous estimates by known speed, elapsed time, and direction. Each vector $v_1,v_2,v_3$ in Figure \ref{fig:indoormethods} is a measurement of the movement since the previous position estimate. The position estimate can then be found by advancing the previous estimate by this vector.
 
A system that uses dead reckoning is the GETA sandals proposed by Yeh et al. \cite{Yeh2007}. The GETA sandals are designed for meter precision. The system uses force, ultrasonic, accelerometer, and orientation sensors to measure displacement vectors along a trail of footsteps. Each displacement vector is formed by drawing a line between each pair of footsteps. The system estimate positions by summing up the current and all previous displacement vectors.

The dead reckoning method has the advantage that it can be applied without an infrastructure in the coverage area. All needed sensors can be placed on the tracked person or equipment. However, the method also has some disadvantages: First, to compare dead reckoning positions among sensors starting positions have to be known in a relevant coordinate system. Second, position errors will increase over time because small errors in each estimate will quickly built up.

\subsubsection{Location Fingerprinting}
In this section LF was classified as an example of the method of pattern recognition. LF encondes patterns in a radio map based on fingerprints. The radio map contains a mapping for each encoded pattern to a position. With respect to the disadvantages of pattern recognition LF has the same disadvantage of need for calibration. However, radio-based LF systems avoid the need for a specially installed infrastructure by using already available infrastructures. Compared to other types of positioning radio-based LF is not able to provide the centimetre precision realized with some of the other methods. As mentioned earlier methods can also be combined. For instance, Niculescu et al. \cite{Niculescu2004} in an extended version of their VOR system combine angulation with LF thereby improving the overall precision of thier system.

\section{Measuring Signal Strength with IEEE 802.11}
\label{sec:ieee80211scanning}
IEEE 802.11 \cite{StandardIEEE802.11} is a wireless networking technology that today is widely used for wireless connectivity for mobile devices such as laptops, phones, PDA, etc. To connect a mobile device to a base station it first has to be discovered. The standard describes two client base-station discovery techniques, namely active scanning and passive scanning. As part of scanning signal strength measurements will be collected for the discovered base stations. Therefore such scanning techniques can collect signal-strength measurements at clients for LF. To collect signal strength measurements at base stations no standardized technique is available. Therefore base stations must measure signal strength of packets received from clients during normal operation.

IEEE 802.11 subdivides the used radio spectrum into a set of channels (13 in Europe for 802.11g). This is important for scanning because a wireless client can only listen to one channel at a given time. Therefore during scanning a wireless client has to tune to each channel, one after another to discover all base stations in communication range.

\begin{figure}[h]
	\centering
		\includegraphics[viewport=40 350 410 570,width=0.8\textwidth,clip]{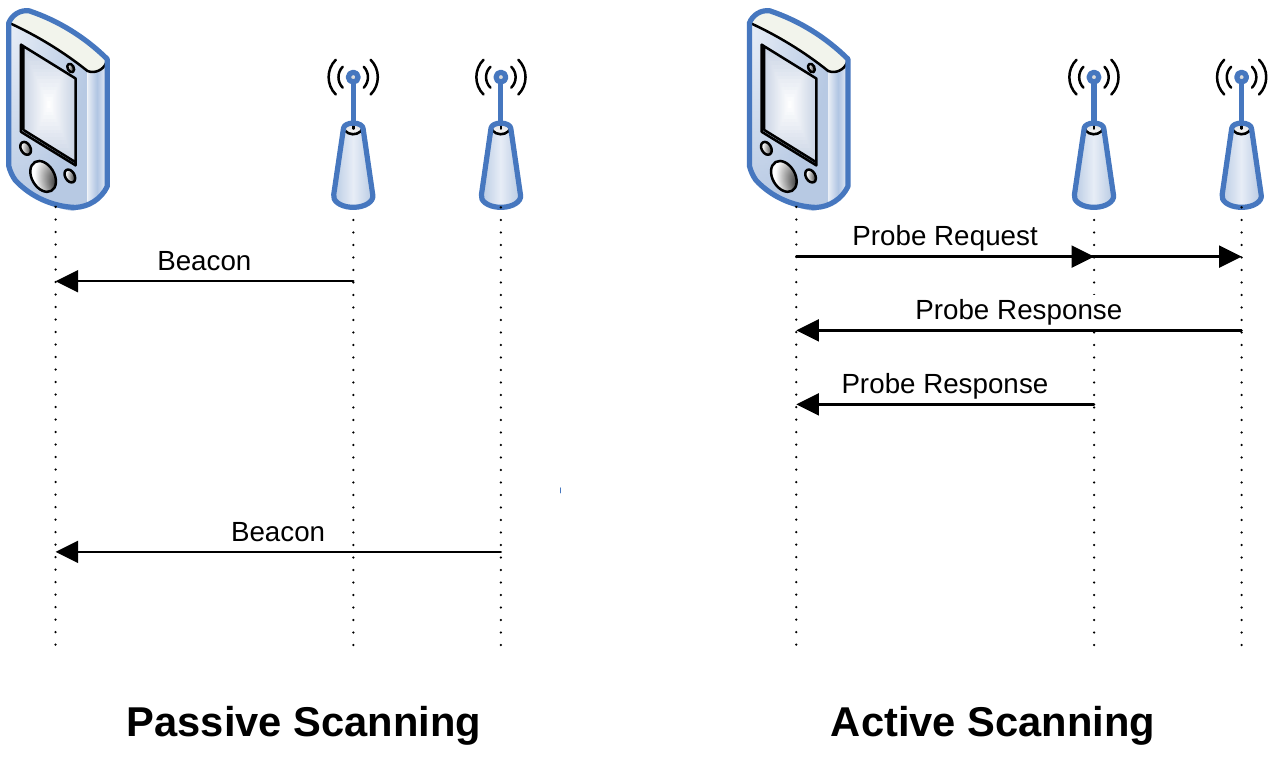}
	\caption{Passive and Active Scanning.}
	\label{fig:IEEE80211Scanning}
\end{figure}

\subsection{Passive Scanning}
Passive scanning is passive in the sense that it only requires the wireless client to listen. The technique works by listening for beacon frames on each channel, as illustrated in Figure \ref{fig:IEEE80211Scanning}. Beacon frames are sent out by IEEE 802.11 base stations on a regular basis to maintain the network. Beacon frames contain information about the network, for instance, the name of the network and supported data rates. Beacon frames are normally sent out every 100 milliseconds, however, this is a configurable value. Therefore passive scanning has to listen for at least 100 milliseconds on each channel to hear all base stations on a specific channel. This means that passive scanning takes at least 1.3 seconds not counting the small delay involved when changing channels as discussed by King et al. \cite{KingPercom2007}. 

Passive scanning has several advantages. First, because no communication is required the technique is light-weight in terms of power consumption. Second, it preserves the privacy of the client because the client's existence is not revealed. Therefore the wireless client can position it-self using LF but remains private as discussed by LaMarca et al. \cite{LaMarca2005}. The main disadvantage of this technique is that it takes over a second to perform each scan.

\subsection{Active Scanning}
Active scanning is active in the sense that it requires the wireless client to actively ask base stations to identify themselves to the wireless client. Active scanning works by on each channel the client sends a \emph{probe request} and listen for \emph{probe responses} from base stations as illustrated in Figure \ref{fig:IEEE80211Scanning}. When a base station receives a probe request it will as quickly as possible answer with a probe response. The probe response will contain information about the network, for instance, the name of the network and supported data rates. During an active scan the wireless client has to stay on each channel to send out the request and then wait for any responses. The time a wireless client waits for response is a configurable parameter. King et al. \cite{KingPercom2007} reports that at most 20 milliseconds are required for each channel. This means that in total a scan over all channels takes less than 260 milliseconds.

Active scanning has the advantage of requiring less than 260 milliseconds supporting a sampling frequency of nearly 4 Hz. The main disadvantage is that clients need to actively sent out requests which reveal both the existence of the client and consumes power.

The work presented in this thesis is based on measurements collected with active scanning. The reason for this is that active scanning supports the highest sampling frequency and that active scanning is better supported by clients. However, there exists other novel options such as \emph{monitor sniffing} which will be discussed in Chapter \ref{chap:lfcom}.

\section{Summary}
This chapter presented background material on signals and methods for indoor positioning where LF was classified as an example of pattern recognition. Furthermore the measurement of signal strength for IEEE 802.11 was discussed and it was argued for why mainly active scanning has been used to collect measurements with.


\clearemptydoublepage
\chapter{A Conceptual Foundation for Location Fingerprinting}
\label{chap:rlf}
\mycitation
{\textbf{conceptual} \textit{(formal)} related to or based on ideas.}
{\textit{Oxford Advanced Learner's Dictionary}}
This chapter discusses [Paper 1] (\emph{A Taxonomy for Radio Location Fingerprinting}). Section \ref{radio:intro} discusses the motivation behind the development of the taxonomy and introduces the taxonomy. Section \ref{radio:contrib} summarises the main contributions of the paper, and Section \ref{radio:relatedwork} discusses related work.

\section{Introduction}
\label{radio:intro}
Many types of LF systems have been proposed in the literature. When surveying LF systems one has to answer many questions. For instance: How do systems differ in scale; can they be deployed to cover a single building or an entire city? What signals are measured? What are the roles of the wireless clients, base stations, and servers in the estimation process? Which estimation method is used? How are fingerprints collected and used? These questions are not only important for researchers surveying LF but also developers of LF systems who have to understand the different possibilities. A taxonomy will aid LF system developers and researchers better survey, compare, and design LF systems. Being able to better survey and compare existing work also makes it possible to use a taxonomy as an aid when finding ideas for future research. This is especially important as LF research moves more and more from understanding basic mechanisms to optimizing existing methods for non-functional properties such as robustness and scalability. 

The proposed taxonomy for LF is built around eleven taxons listed with definitions in Table \ref{radio:tab:deftaxon}. Three of the taxons were already introduced in Chapter \ref{chap:intro}. The taxons were partly inspired by earlier work on taxonomies for position technologies in general and from a literature study of 51 papers and articles. The four taxons: \emph{scale}, \emph{output}, \emph{measurements}, and \emph{roles} describe general properties of LF systems. We mean by \textit{scale} the size of the deployment area and by \textit{output} the type of provided location information. \textit{Measurements} means the types of measured network characteristics and \textit{roles} means the division of responsibilities between wireless clients, base stations, and servers.

\emph{Estimation method} and \emph{radio map} describe the location estimation process. Estimation method denotes a method for predicting locations from a radio map and currently measured network characteristics and radio map a model of network characteristics in a deployment area. The division into estimation method and radio map is used in many papers about LF, for instance, Youssef et al. \cite{Youssef2005b}. However, some papers use a slightly different naming, for instance, Otsason et al. \cite{Otsason2005} use \emph{localization algorithm} and \emph{radio map}. 

\begin{figure}[h]
	\centering
		\includegraphics[viewport=60 310 770 570,width=0.85\textwidth,clip]{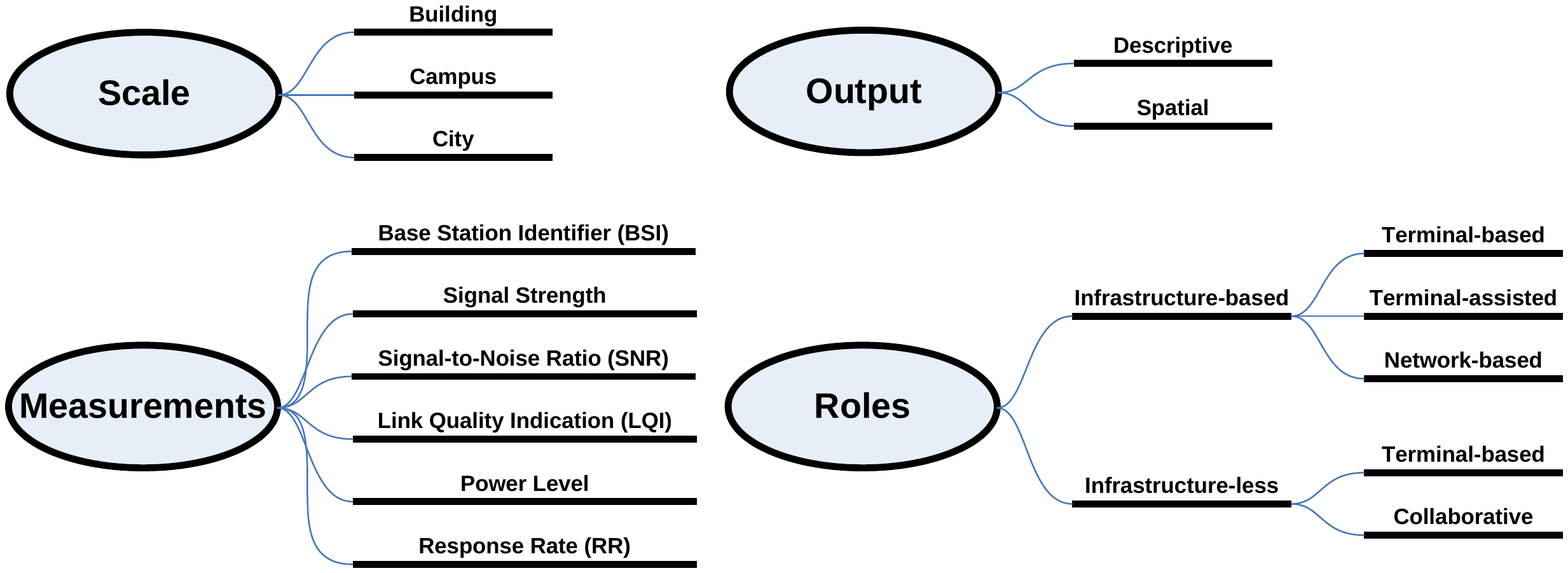}
	\caption{Scale, output, measurements and roles.}
	\label{radio:fig:LF}
\end{figure}

\begin{figure}[h]
	\centering
		\includegraphics[viewport=20 460 480 790,width=0.85\textwidth,clip]{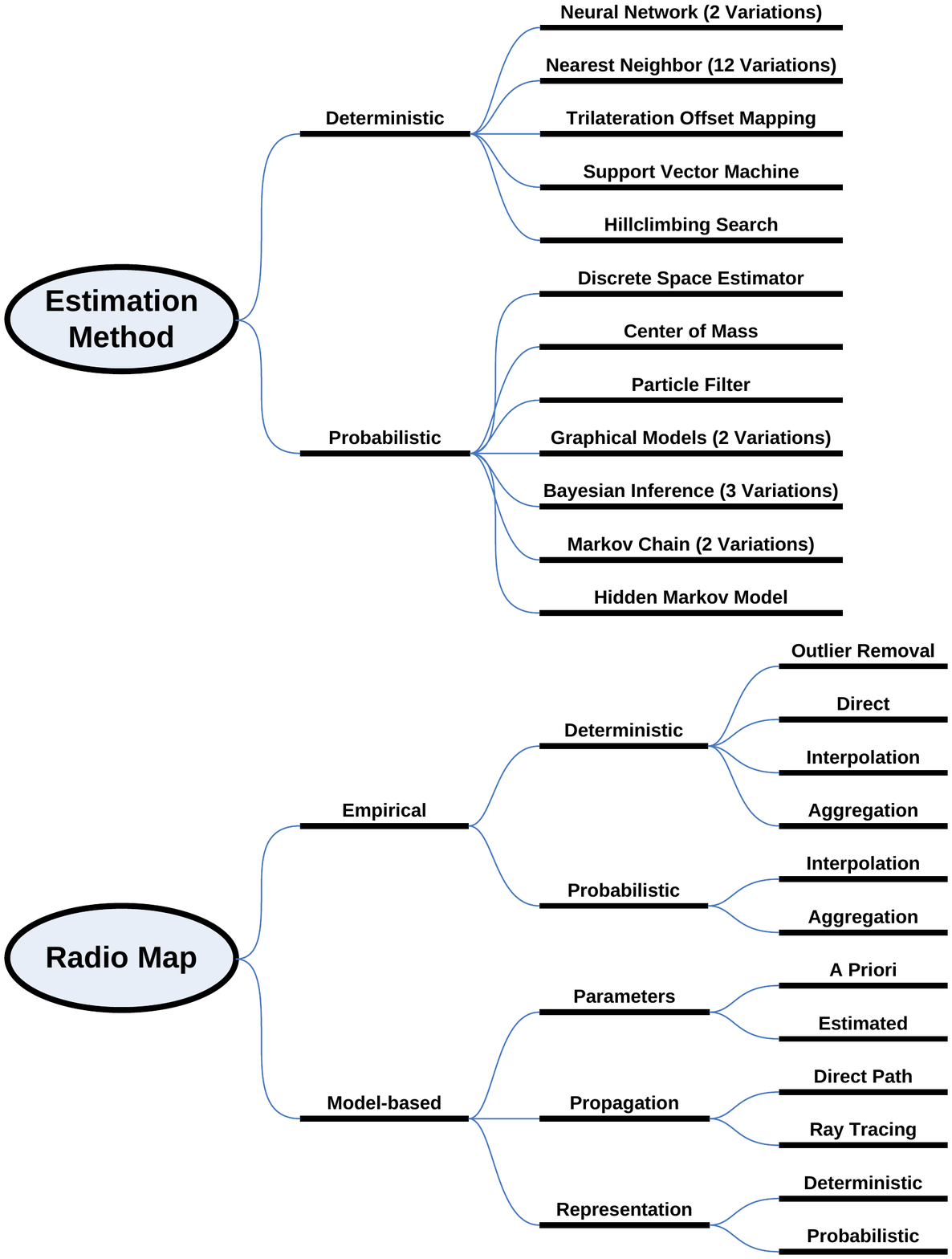}
	\caption{Estimation method}
	\label{radio:fig:LFEstimation}
\end{figure}

\begin{figure}[h]
	\centering
		\includegraphics[viewport=55 320 265 506,width=0.6\textwidth,clip]{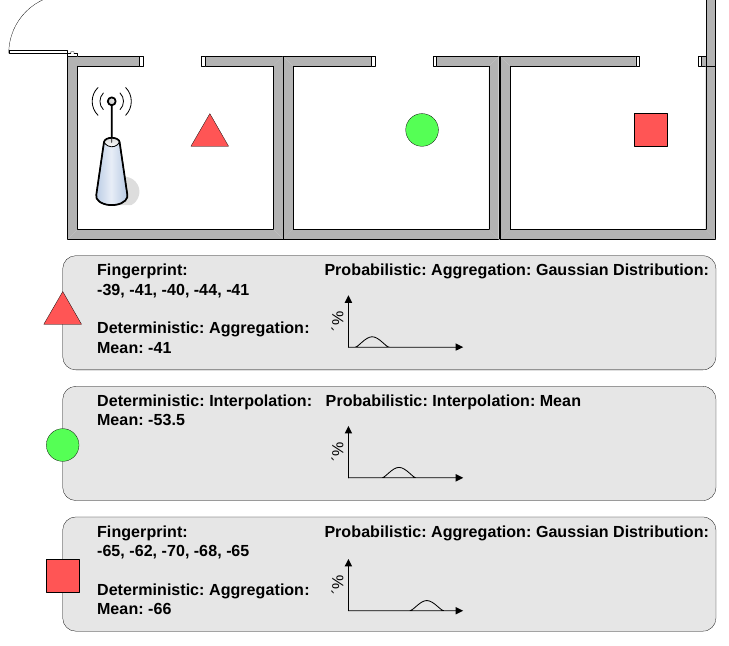}
	\caption{Deterministic and probabilistic aggregation and interpolation}
	\label{radio:fig:LFAggInt}
\end{figure}

\begin{figure}[h]
	\centering
		\includegraphics[viewport=20 365 550 810,width=0.85\textwidth,clip]{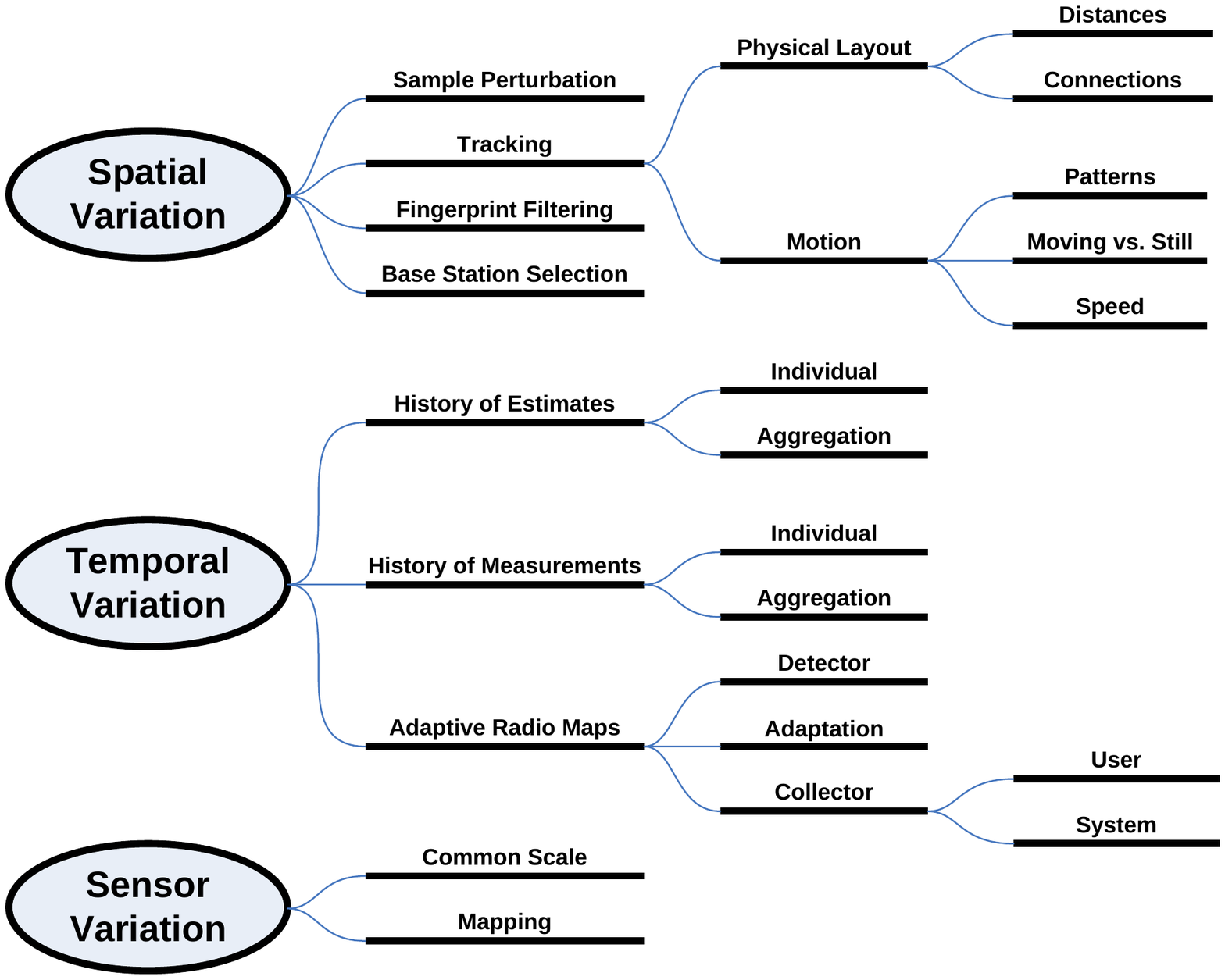}
	\caption{Spatial variations, temporal variations, and sensor variations.}
	\label{radio:fig:LFVariation}
\end{figure}

\begin{figure}[h]
	\centering
		\includegraphics[viewport=50 500 470 810,width=0.85\textwidth,clip]{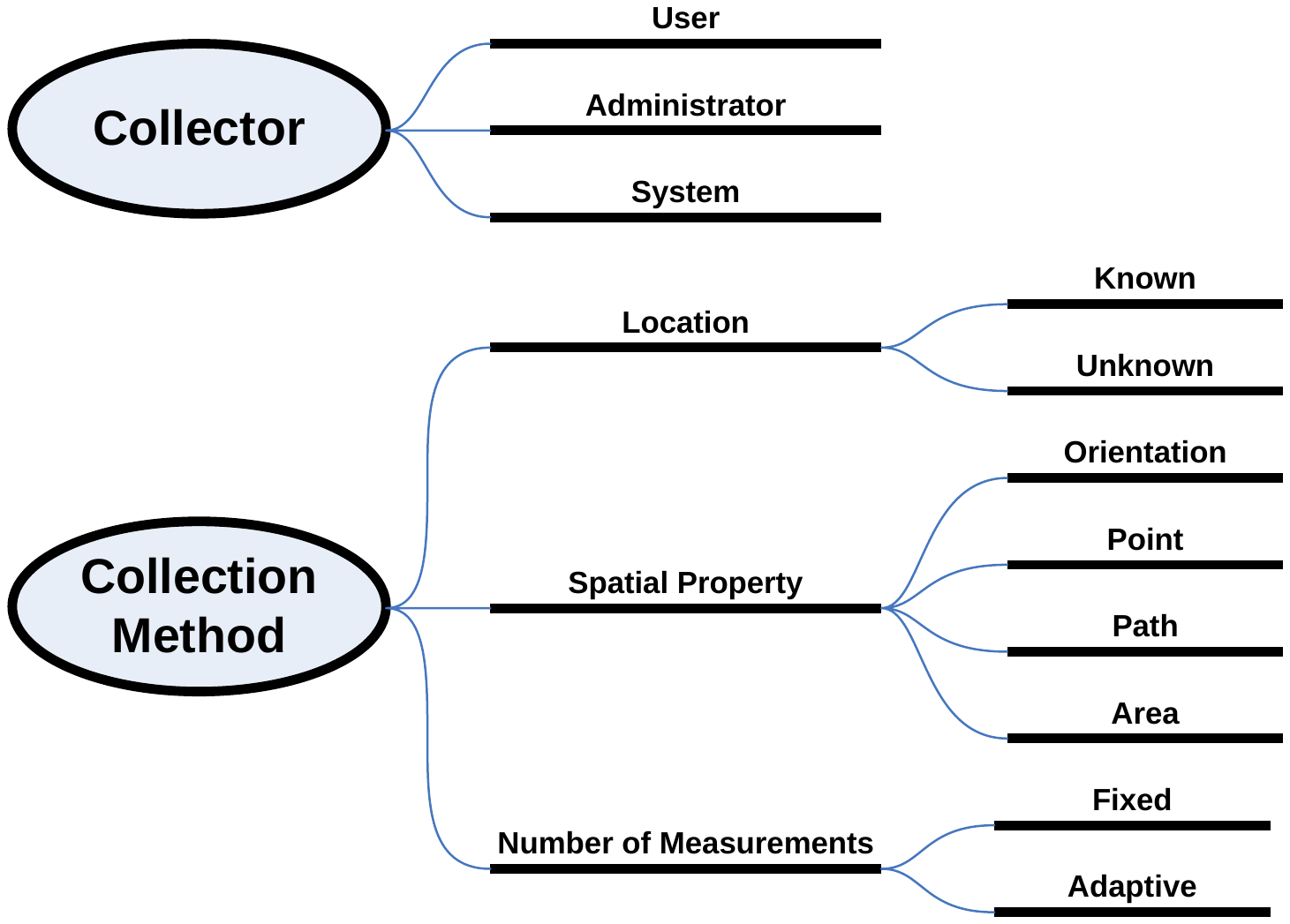}
	\caption{Collector and collection method.}
	\label{radio:fig:LFCollection}
\end{figure}

How changing network characteristics over space, time, and sensors can be handled is described by \emph{spatial, temporal, and sensor variations}. The spatial and temporal dimensions were introduced by Youssef et al. \cite{Youssef2005b}. The sensor dimension was introduced in [Paper 2]. The taxons \emph{collector} and \emph{collection method} describe how fingerprints are collected. These two taxons have been introduced to characterize the assumptions systems put on fingerprint collection. The proposed taxons and subtaxons are shown including subtaxons in Figure \ref{radio:fig:LF} to Figure \ref{radio:fig:LFCollection}.

\begin{table}[h]

\centering
\begin{tabular}{p{3.7cm}p{8.3cm}}
\textit{Taxon} & \textit{Definition} \\
\hline
\textbf{Scale} & Size of deployment area.\\ 
\textbf{Output} & Type of provided location information.\\
\textbf{Measurements} & Types of measured network characteristics.\\
\textbf{Roles} & Division of responsibilities between wireless clients, base stations, and servers.\\
\textbf{Estimation Method} & Method for predicting locations from a radio map and currently measured network characteristics.\\
\textbf{Radio Map} & Model of network characteristics in a deployment area.\\
\textbf{Spatial Variations} & Observed differences in network characteristics at different locations because of signal propagation characteristics.\\
\textbf{Temporal Variations} & Observed differences in network characteristics over time at a single location because of continuing changing signal propagation.\\
\textbf{Sensor Variations} & Observed differences in network characteristics between different types of wireless clients.\\
\textbf{Collector} & Who or what collects fingerprints.\\
\textbf{Collection Method} & Procedure used when collecting fingerprints.\\
\hline
 & \\
\end{tabular}
\caption{Taxon definitions}
\label{radio:tab:deftaxon}
\end{table}

\textit{Output} denotes the type of provided location information. The subtaxons for output are proposed to follow the notion introduced in K{\"u}pper \cite{Kuep05} of dividing location information into \emph{descriptive} and \emph{spatial} information. Descriptive locations are described by names, identifiers or numbers assigned to natural geographic or man-made objects\footnote{Some authors refer to this as symbolic locations}. Spatial locations are described by a set of coordinates stated with respect to a spatial reference system. Many LF systems output \emph{spatial} locations \cite{Bahl2000,Roos2002a,LaMarca2005,Smailagic2002} but systems have also been proposed that output \emph{descriptive} locations \cite{Castro2001,Haeberlen2004,Bhasker2004}. However, a location outputted as either of the two types can be mapped to the other type given a suitable location model.

\textit{Measurements} are the types of measured network characteristics. The following network characteristics have been used in existing systems: \emph{Base Station Identifiers (BSI)}, signal strength, \emph{Signal-to-Noise Ratio (SNR)}, \emph{Link Quality Indicator (LQI)}, \emph{power level}, and \emph{Response Rate (RR)}. BSI is a unique name assigned to a base station. Signal strength, SNR, and LQI are signal propagation metrics collected by radios for handling and optimizing communication. Scanning techniques for measuring signal strength were discussed in Chapter \ref{chap:background}. The power level is information from the signal sender about current sending power. The response rate is the frequency of received measurements over time from a specific base station. Many LF systems are based on BSI and signal strength \cite{Bahl2000,Roos2002a,Haeberlen2004,Smailagic2002}; other systems have used RR in addition to signal strength \cite{Krumm2004,Ladd2002,LaMarca2005}. BSI and SNR have also been used \cite{Castro2001} and the combination BSI, LQI, signal strength, and power level \cite{Lorincz2005,Lorincz2006}.

\label{radio:sec:estimation}

\begin{figure}[h]
	\centering
		\includegraphics[viewport=20 110 530 450,width=0.85\textwidth,clip]{paper1/Images/LFEstimation.pdf}
	\caption{Radio map}
	\label{radio:fig:LFRadioMap}
\end{figure}

A central part of a LF system is the \textit{estimation method} used for predicting locations from a radio map and currently measured network characteristics. It would, however, be very challenging to taxonomize all possible methods because nearly all methods developed for machine learning (see Witten et al. \cite{Witten2005} for a list of methods) or in the field of estimation (see Crassidis et al. \cite{Crassidis2004} for a list of methods) are applicable to the problem of LF estimation. Here we follow Krishnakumar et al. \cite{Krishnakumar2005} and divide methods only into deterministic and probabilistic methods. \emph{Deterministic methods} estimate location by considering measurements only by their value \cite{Bahl2000,Prasithsangaree2002,Laitinen2001,Smailagic2002}. \emph{Probabilistic methods} estimate location considering measurements as part of a random process \cite{Youssef2005b,Krumm2004,Castro2001,Haeberlen2004}. In Figure \ref{radio:fig:LFEstimation} examples of applied methods for LF are shown for each of the two categories, including number of identified varieties in our literature study\footnote{However, even this simple classification is fuzzy for instance when considering the machine learning technique of support vector machines (SVMs) as applied for LF \cite{Brunato2005}. Because SVMs are defined on a probabilistic foundation but when applied for LF, SVMs only consider the actual values of measurements.}. For example, the classical deterministic technique of Nearest Neighbor was identified during the literature study in twelve different variations. A comment is that many of the studied LF systems use more than one of the listed methods.

A \textit{radio map} provides a model of network characteristics in a deployment area. Radio maps can be constructed by methods which can be classified as either \emph{empirical} or \emph{model-based}. Empirical methods work with collected fingerprints to construct radio maps \cite{Youssef2005b,Bahl2000,Krumm2004,Haeberlen2004}. Model-based methods use a model parameterised for the LF-system-covered area to construct radio maps \cite{Bahl2000,Roos2002b,Wallbaum04,Ji2006}.

Empirical methods can be subdivided into \emph{deterministic} and \emph{probabilistic} methods in the same manner as estimation methods, depending on how they deal with fingerprint-collected measurements. Deterministic methods represent entries in a radio map as single values and probabilistic methods represent entries by probability distributions. Both of these can be further subcategorised into \emph{aggregation} and \emph{interpolation} methods. An aggregation method creates entries in a radio map by summarising fingerprint measurements from a single location \cite{Bahl2000,Roos2002a,Haeberlen2004,Berna2003}. Figure \ref{radio:fig:LFAggInt} illustrates two aggregation methods for five signal-strength measurements at two locations marked with a triangle and a square on the figure. The first aggregation method is a deterministic mean method which takes the five measurements and finds the mean and put this value as this location's entry in the radio map. The second aggregation method is a probabilistic Gaussian distribution method which takes the five measurements and fits them to a Gaussian distribution and puts the distribution as the location's entry in the radio map. An interpolation method generate entries in a radio map at unfingerprinted locations by interpolating from fingerprint measurements or radio map entries from nearby locations \cite{Krumm2004,Krishnan2004,LaMarca2005}. Figure \ref{radio:fig:LFAggInt} illustrates two interpolation methods at the location marked with a circle using the square-marked and triangle-marked locations as nearby locations. The first interpolation method is a deterministic mean interpolation which finds the mean of nearby radio-map entries and put this value as the entry in the radio map. The second interpolation method is a probabilistic mean method that finds the mean of nearby radio-map entries' Gaussian distributions and put the mean distribution as the entry in the radio map. Two other deterministic methods are \emph{outlier removal} filtering away outliers \cite{Saha2003} and \emph{direct} creating a radio map using a direct one-to-one mapping to measurements \cite{Otsason2005}.

Model-based methods can be categorized based on how \emph{parameters} for the model are specified, how signal \emph{propagation} is modelled, and what type of \emph{representation} is used by the generated radio map. Parameters can either be given \emph{a priori} \cite{Bahl2000} or they can be \emph{estimated} from a small set of parameter-estimation fingerprints \cite{Ji2006}. Propagation can either be modelled by only considering the \emph{direct path} between a location and a base station \cite{Bahl2000} or by considering multiple paths categorized as \emph{ray tracing} \cite{Ji2006}. The representation of the generated radio map can either be \emph{deterministic} (using single values) \cite{Bahl2000} or \emph{probabilistic} (using probability distributions) \cite{Madigan2005}.

\textit{Spatial variations} are the observed differences in network characteristics at different locations because of signal propagation characteristics. Because of how signals propagate, even small movements can create large variations in the measured network characteristics, for instance, because of multipathed signals. The main method for addressing spatial variations  is \emph{tracking}: the use of constraints to optimize sequential location estimates. Tracking can be based on motion in terms of target \emph{speed} \cite{LaMarca2005,Chai2005}, target being \emph{still versus moving} \cite{Krumm2004}, and knowledge about motion \emph{patterns} \cite{Chai2005}. Tracking can also be based on physical constraints such as how \emph{connections} exist between locations \cite{Castro2001} and the \emph{distance} between them \cite{Krumm2004,Bahl2000b}. Tracking using one or several of the listed constraints is implemented using an estimation method (such as the ones listed in Section \ref{radio:sec:estimation}) that is able to encode the constraints. Spatial variations can also be addressed by \textit{base station selection}, \textit{fingerprint filtering}, and \emph{sample perturbation}. Base station selection filters out measurements to base stations that are likely to decrease precision and accuracy \cite{Varshavsky2007,Kushki2007}. Fingerprint filtering limits the set of used fingerprints to only those that are likely to optimize precision and accuracy \cite{Kushki2007}. \emph{Sample perturbation} apply perturbation of measurements to mitigate spatial variations \cite{Youssef2005b}.

\textit{Temporal variations} are the observed differences in network characteristics over time at a single location because of continuing changing signal propagation. On a large-scale, temporal variations are the prolonged effects observed over larger periods of time such as day versus night. On a small-scale, temporal variations are the variations implied by quick transient effects, such as a person walking close to a client. Methods for handling temporal variations can be divided into methods that are based on a \emph{history of estimates}, a \emph{history of measurements}, or \emph{adaptive radio maps}. A history of either measurements or estimates here denotes a set of estimates or measurements inside a defined time window. The alternative to a history is to only use the most recent estimate or measurements. The history of either measurements or estimates can either be used as \emph{individual} \cite{Krumm2004,Haeberlen2004} measurements or estimates or, using some \emph{aggregation} \cite{Youssef2005b,Roos2002a}, can be combined to one measurement or estimate. The adaptive radio map method introduces the idea of handling temporal variations by making the radio map adapt to the current temporal variations \cite{Bahl2000b,Krishnan2004,Berna2003}. For this idea to work, some \emph{collector} has to make measurements that can be used by a \emph{detector} to control if some adaptation should be applied to the current radio map. The measurements can either be collected from the measurements a \emph{user} collects \cite{Berna2003} to run LF estimation on or it can be collected by some specially-installed \emph{system} infrastructure \cite{Bahl2000b,Krishnan2004}.  

\textit{Sensor variations} are the observed differences in network characteristics between different types of wireless clients also described as the problem of handling hetoregenous devices in Chapter \ref{chap:intro}. On a large-scale, variations can be observed between clients from different manufactures. On a small-scale, variations can be observed between different examples of similar clients. One method for addressing sensor varations is to define a \emph{common scale} and then, for each type of sensor, find out how this sensor's measurements can be converted to the common scale. A second approach is to use a single sensor to fingerprint with and then find a mapping from new sensors to the sensor that was used for fingerprinting \cite{KjaergaardLoca2006,Haeberlen2004}. The problem of handling heterogeneous clients is discussed in more detail in Chapter \ref{chap:hdlf}.

The fingerprints are collected following some \textit{collection method}. A collection method places assumptions on if fingerprints are collected on a \emph{location} that is either \emph{known} \cite{Otsason2005} or \emph{unknown} \cite{Madigan2005,Chai2005}. If fingerprints are collected to match a \emph{spatial property} such as: \emph{orientation} \cite{Bahl2000}, at a \emph{point} \cite{Krumm2004}, covering a \emph{path} \cite{LaMarca2005}, or covering an \emph{area} \cite{Haeberlen2004,Varshavsky2007}. If the collected \emph{number of measurements} for each fingerprint is \emph{fixed} \cite{Youssef2005b,Roos2002a} or determined based on some \emph{adaptive} strategy.

\subsection{Examples}
\begin{figure}[!]
	\centering
		\includegraphics[viewport=20 160 460 830,width=1.0\textwidth,clip]{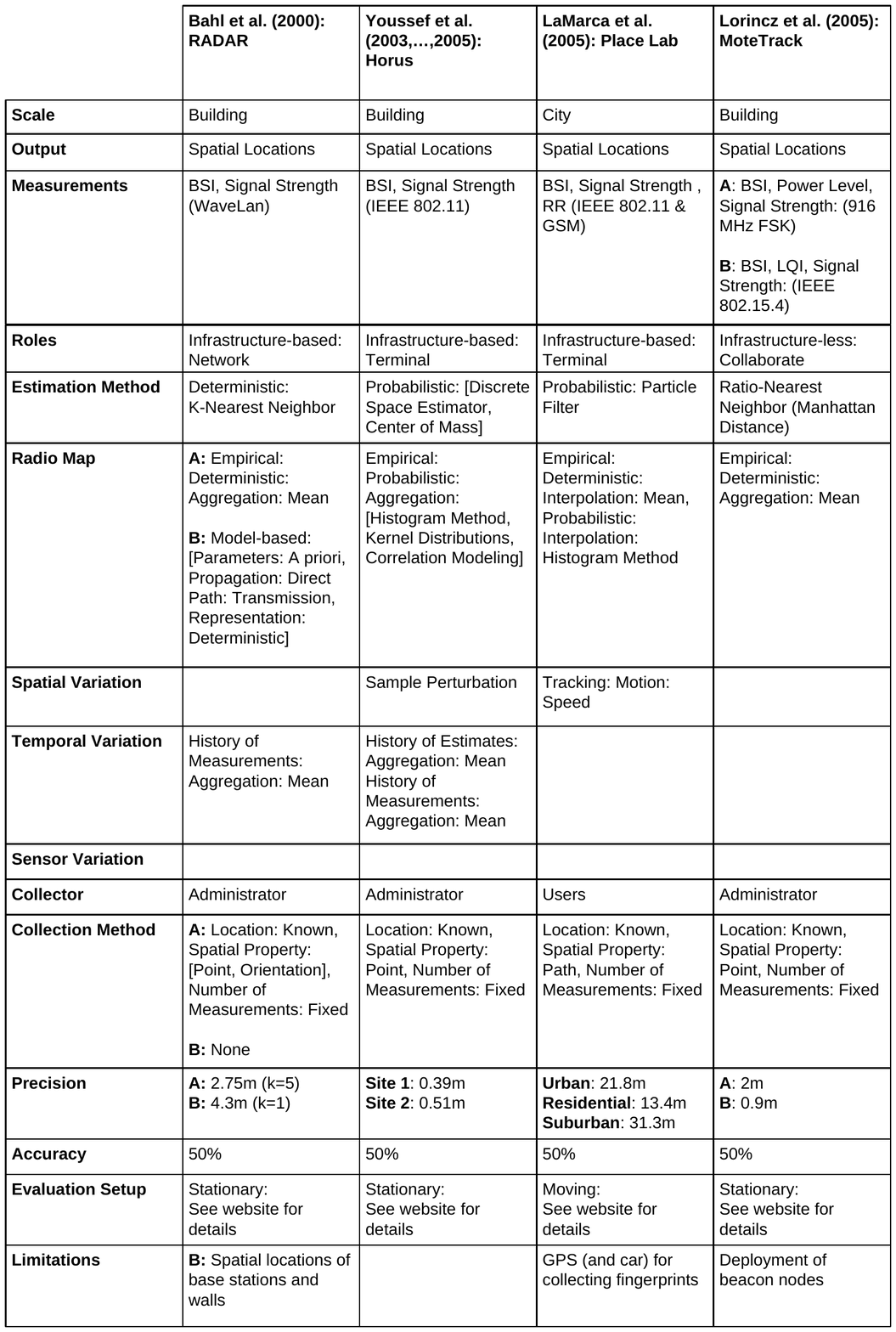}
	\caption{Analysis results for the four case studies.}
	\label{radio:fig:LFCases}
\end{figure}

To show the use of the proposed taxonomy, this section presents an analysis using the taxonomy of four LF systems. Figure \ref{radio:fig:LFCases} shows the analysis results in a compact form. The four systems have been selected to highlight different parts of the taxonomy. In addition to the eleven taxons, four extra categories describe the systems from an evaluation perspective; these are: \emph{accuracy, precision, evaluation setup and limitations}. The listed evaluation results have been taken from the original papers. Evaluation setup is grouped into \emph{stationary} (meaning that the authors' test data was collected while keeping a wireless client at a static position) or \emph{moving} (for which the wireless client was moved around mimicking normal use). 

The RADAR system proposed by Bahl et al. \cite{Bahl2000} is aimed at a building scale of deployment and provides spatial locations as output. The system measures BSI, and signal strength for the WaveLAN technology and roles are assigned as infrastructure-based: network. The estimation method is the deterministic k-nearest neighbor algorithm. They propose two setups, here named A and B. For A the radio map is constructed using deterministic aggregation using the mean from empirical-collected fingerprints. For B the radio map is deterministically constructed by a model which considers the direct path of transmission using a priori parameters. For A, an administrator will collect fingerprints at known locations standing at one point with different orientations collecting a fixed number of measurements and for B no fingerprints are collected. A limitation for setup B is that knowledge is needed of spatial locations of base stations and walls. 

The Horus system proposed by Youssef et al. \cite{Youssef2005b,Youssef2003a,Youssef2003b,Youssef2004a,Youssef2005a} also aims at a building scale of deployment and provide spatial locations as output. The system measures BSI, and signal strength for the IEEE 802.11 technology and the assigned roles match infrastructure-based: terminal. The estimation method is a combination of two probabilistic techniques: discrete space estimator and center of mass. The radio-map is built using probabilistic aggregation, either based on a histogram method or on a kernel distribution method; in addition, a method for correlation modelling is also applied. To handle spatial variations sample perturbation is applied and temporal variations are handled by both mean aggregating measurements and estimates. An administrator collects fingerprints at known locations standing at each point collecting a fixed number of measurements.

The Place Lab system proposed by LaMarca et al. \cite{LaMarca2005,Cheng2005,Hightower2004} aims at a city-wide deployment and provides spatial locations as output. The system measures BSI, signal strength, and RR for both IEEE 802.11 and GSM and the assigned roles match infrastructure-based: terminal. The most advanced of the system's estimation methods uses a particle filter. The radio map is built in two steps, first applying deterministic interpolation based on means and then probabilistic interpolation based on the histogram method. Spatial variations are addressed by tracking based on motion by speed constraints. The fingerprints are user collected based on paths with known location and collecting a fixed number of measurements. A limitation is that a GPS device (and a car) is needed to practically collect fingerprints. 

The MoteTrack system proposed by Lorincz et al. \cite{Lorincz2005,Lorincz2006} targeted for sensor networks aims at building-scale deployment and provides spatial locations as output. The system has been tested in two setups, here named A and B. Setup A measures BSI, Power level, and signal strength for 916 MHz communication and setup B measures BSI, LQI, and signal strength for IEEE 802.15.4 communication. The roles are assigned matching infrastructure-less: collaborate with beacon nodes taking the role as base stations. The estimation method is ratio-nearest neighbor with Manhattan distance to lower computational needs. The radio map is constructed using deterministic aggregation using the mean from empirically collected fingerprints. An administrator collects fingerprints at known locations standing at each point collecting a fixed number of measurements. A limitation is the needed deployment and maintenance of beacon nodes.

\section{Main Contribution}
\label{radio:contrib}

The main contribution of [Paper 1] is the taxonomy itself. It contains eleven main taxons and 88 subtaxons that in more detail classifies LF systems as described in Section \ref{radio:intro}. The taxonomy has been constructed based on a literature study of 51 papers and articles. The 51 papers and articles propose 30 different systems which have been analyzed and methods and techniques grouped to form taxons for the taxonomy. The analysis results for all of the 30 systems are available online at \cite{lfwebsite}. The taxonomy allows researchers to make detailed comparison of systems and methods and help scope out new research paths within this area. However, the quality of the taxonomy can only be jugged by how valuable it will be for other's work.

To use the taxonomy for detailed comparison, one approach would be first to find classifications for existing systems. As mentioned earlier a starting point for finding such classifications is to look at the classifications online at \cite{lfwebsite}. Second, one would make a classification of the new system for each of the eleven taxons for the new system's methods and assumptions according to the subtaxons. Third, one would make a comparison of the new and the existing systems. For evaluation of LF systems, the taxonomy can also be used to highlight the evaluated system's assumptions and methods. This can be done by providing a classification for the evaluated system which explicitly states what methods and assumptions are used. 

The taxonomy can also help scope future research by illustrating what research topics have not yet been covered. One way to analyse this is to group systems in terms of some of the taxons. A grouping of the taxons scale and radio map is shown in Table \ref{radio:tab:scalevsradiomap}. The table shows that only one system aims at a campus-size scale. The table also shows that generally systems either use empirical or model-based radio maps and not a combination. So an open research topic is exploring the boundary between building and city-wide systems by for example combining empirical and model-based radio maps\footnote{However, a lack of papers can also be an indication of that the specific combination is a bad idea.}.

\begin{table}[h]
\centering
\begin{tabular}{l|p{4cm}|p{4cm}}
 & \textit{Empirical} & \textit{Model-based} \\
\hline
\textit{Building} & \cite{Youssef2005b,Otsason2005,Lorincz2005,Bahl2000,Prasithsangaree2002,Battiti2002,Krumm2004,Castro2001,Ladd2002,Haeberlen2004,Bahl2000b,Krishnan2004,Roos2002b,Smailagic2002,Brunato2005,Berna2003,Saha2003,Chai2005,Varshavsky2007,Kushki2007,Seshadri2005,Agiwal2004,Elnahrawy2004,Yin2005} & \cite{Bahl2000,Brunato2005,Wallbaum04,Ji2006,Madigan2005,Elnahrawy2004} \\ 
\hline
\textit{Campus} & \cite{Bhasker2004} & \\
\hline
\textit{City} & \cite{Laitinen2001,LaMarca2005} & \cite{Roos2002a} \\
\end{tabular}
\caption{Grouping in terms of scale and radio map}
\label{radio:tab:scalevsradiomap}
\end{table}

\section{Related Work}
\label{radio:relatedwork}
Related taxonomies cover location systems in general and are therefore of limited use when answering the many questions specific to LF. An example is the taxonomy proposed by Hightower et al. \cite{Hightower2001}, only covering four of the proposed taxonomy's eleven taxons. Their concepts for these four taxons differ slightly in output being split over the four concepts of physical, symbolic, absolute, and relative, in measurements being indirectly described by their technique concept, and in roles being partly described by their concept of localized location computation.

The focus of the proposed taxonomy is on methods for LF and therefore the taxonomy does not cover evaluation properties for LF systems. Evaluation properties for all kinds of location systems have for instance been suggested by Muthukrishnan et al. \cite{Muthukrishnan2005}, who list: precision, accuracy, calibration, responsiveness, scalability, cost, and privacy. The taxonomy proposed by Hightower et al. \cite{Hightower2001} also lists several evaluation properties: precision, accuracy, scale, cost, and limitations. The analysis in [Paper 1] includes the following evaluation properties: precision, accuracy, evaluation setup, and limitations. These four were chosen because these informations are available from most papers. Responsiveness and cost were not included because the first is only available from very few papers and the second from none. Calibration, privacy, scalability, and scale are partly covered by the taxons scale, roles, and collection method.

A limitation of the proposed taxonomy is that it does not cover non-functional properties. One reason for this is that work has not yet matured in these directions for LF systems. Non-functional properties of LF systems have been addressed by several recent papers, such as system robustness by Lorincz et al. \cite{Lorincz2005}, server scalability by Youssef et al. \cite{Youssef2005b}, and minimal communication in [Paper 4] and [Paper 5]. Also, the taxonomy does not cover the application of LF techniques with other types of sensor measurements such as sound and light.

\clearemptydoublepage
\chapter{Handling Heterogeneous Clients}
\label{chap:hdlf}
\mycitation
{\textbf{heterogeneous} \textit{(adj)} consisting of many different kinds of people or things.}
{\textit{Oxford Advanced Learner's Dictionary}}
This chapter discusses [Paper 2] (\emph{Automatic Mitigation of Sensor Variations for Signal Strength Based Location Systems}) and [Paper 3] (\emph{Hyperbolic Location Fingerprinting: A Calibration-Free Solution for Handling Differences in Signal Strength}). Section \ref{heterogeneous:intro} introduces and motivates the contributions. Section \ref{heterogeneous:contrib} summarises the main contributions of the papers and in section \ref{heterogeneous:relatedwork} related work is discussed.

\section{Introduction}
\label{heterogeneous:intro}
A fundamental problem for LF systems is the heterogeneity of clients referred to as a cause of sensor variations in Chapter \ref{chap:rlf}. The heterogeneity is due to different radios, antennas, and firmwares of clients, causing measurements for LF not to be directly comparable among clients. For instance, signal strength measurements or radio sensitivity can be different. For IEEE 802.11 signal strength differences above 25 dB have been measured for same-place measurements with different clients by Kaemarungsi \cite{Kaemarungsi2006}. Such differences have a severe impact on LF systems' accuracy. The results published in [Paper 3] show that signal-strength and sensitivity differences can make room-size accuracy for the Nearest Neighbor algorithm \cite{Bahl2000} drop to unusable 10\%.

For IEEE 802.11-based clients, signal-strength differences can mainly be attributed to the standard's lack of specification of how clients should measure signal strength \cite{Kaemarungsi2006}. The standard specifies signal strength as the received signal-strength index with an integer value between $0$ and $255$ with no associated measurement unit. The standard also states that this quantity is only meant for internal use by clients and only in a relative manner. The internal use of the value is for detecting if a channel is clear or for detecting when to roam to another base station. Therefore IEEE 802.11 client manufacturers are free to decide their own interpretation of signal-strength values. Most manufacturers have chosen to base signal-strength values on dBm values. However, different mappings from dBm values to the integer scale from $0$ to $255$ have been used. The result of this is that most signal-strength values represent dBm values with different limits and granularity. However, differences in hardware also contribute to the problem. The sensitivity differences are mainly due to hardware constraints.

Current solutions for handling signal-strength differences are based on manually collecting measurements to find mappings between signal strength reported by different clients. Such manual solutions are: (i) time consuming because measurements have to be taken at several places for each client; (ii) error prone because the precise location of each place has to be known; (iii) unpractical considering the huge number of different IEEE 802.11 and GSM clients on the market. For instance, due to such issues the company Ekahau maintains lists of supported clients \cite{ekahauwebsite}. To the author's knowledge there has, so far, not been any solutions published for addressing sensitivity differences.

An additional problem is that some clients are only able to provide measurements with very low quality for LF. Measurement quality can be defined by a set of client characteristics. Clients with high measurement quality have some of the following characteristics: 
\begin{itemize}
	\item High sensitivity so that the client can measure many base stations.
	\item No artificial limits in the signal strength values.
	\item Does not cache the signal strength measurements.
	\item Support a high update frequency of measurements.
\end{itemize}

On the other hand, clients with low measurement quality have: 

\begin{itemize}
	\item Low sensitivity.
	\item Limit the signal strength values.
	\item Signal strength values do not represent signal strength but some other measure.
	\item Caches measurements.
	\item Support only a low update frequency of measurements.
\end{itemize}

\begin{figure}[h]
	\centering
		\includegraphics[viewport=0 245 810 565,width=1.0\textwidth,clip]{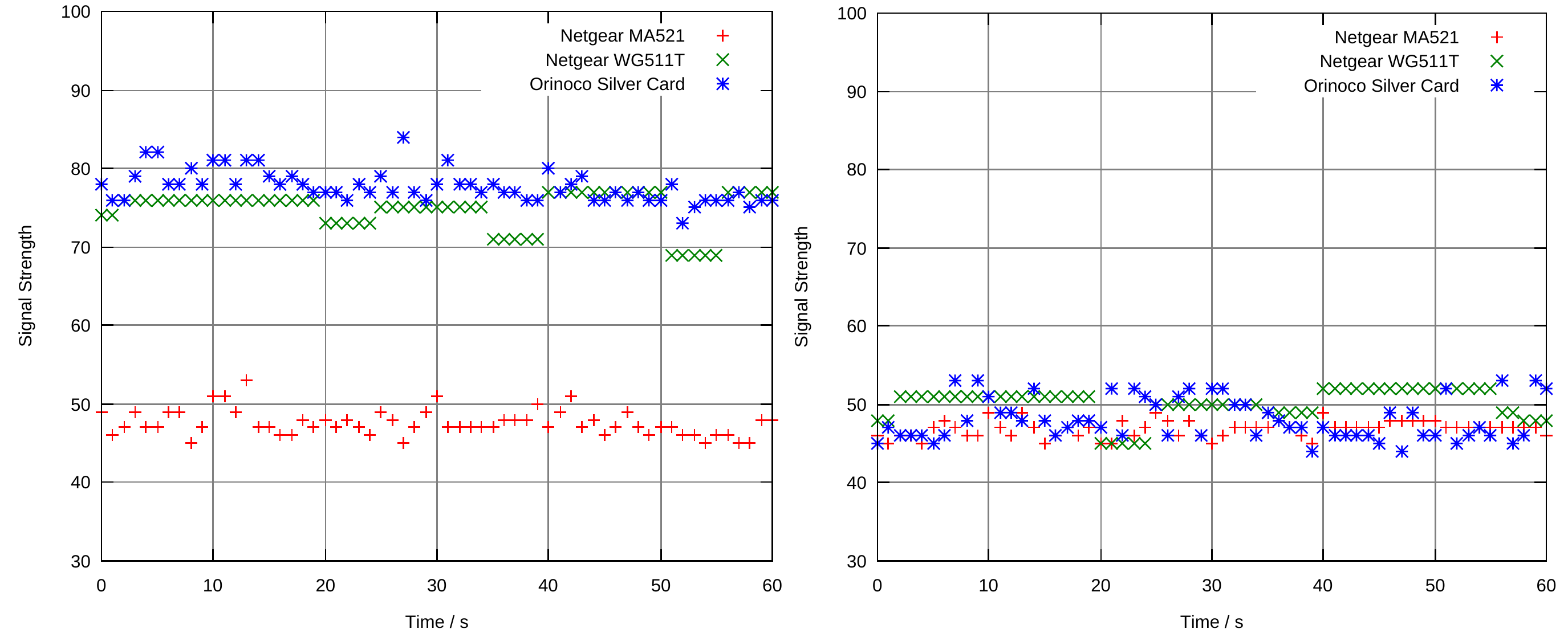}
	\caption{Plots of signal strength measurements from different clients and base stations at the same location.}
	\label{hete:fig:plotsforsensoreffects}
\end{figure}

To illustrate the effects of low and high measurement quality, Figure \ref{hete:fig:plotsforsensoreffects} shows signal strength measurements for different clients taken at the same position and at the same time, but for two different 802.11 base stations. On the first graph the effect of caching or low update rate for the Netgear WG511T card can be seen, since the signal strength only changes every five seconds. By comparing the two graphs, the effect of signal strength values not corresponding to the actual signal strength can be seen for the Netgear MA521 card. This is evident from the fact that the signal strength values for the Netgear MA521 card do not change when the values reported by the other cards change for specific base stations (cf. the second graph). 

\section{Main Contribution}
\label{heterogeneous:contrib}
[Paper 2] and [Paper 3] make the following four contributions.

The first contribution is two classifiers that can classify a client's measurement quality which are published in [Paper 2]. Quality is classified in terms of if a client is caching, has a low measurement frequency, or if it provides measurements that do not correspond to signal strength measurements. Each of the classifiers uses a naive Bayesian estimator for the classification. The classifiers have been evaluated by emulation using 14-fold cross validation on triple data sets for 14 heterogeneous IEEE 802.11 clients. The result of the evaluation was that the classifiers could classify client quality correctly in 96.2\% of the tested cases.

The second contribution is a method that uses a linear mapping to transform one client's measurements to match another client's measurements which is published in [Paper 2]. The method is automatic, but requires a learning period to find the parameters for the linear mapping. The solution is based on movement detection which is used to group same-place measurement into calibration fingerprints. The parameters are then estimated from the calibration fingerprints using weighted least squares. The method has been evaluated by emulation using three-fold cross validation on triple data sets for 14 heterogeneous clients and using a fingerprint set collected with one client. The method improved overall LF accuracy with 13.1 percentage points from 32.6\% to 45.7\%. In comparison a method using linear mapping with parameters found with manually collected calibration fingerprints was able to improve the accuracy with 19.2 percentage points to 52.1\%.

The third contribution is a method named \emph{Hyperbolic Location Fingerprinting (HLF)} published in [Paper 3]. The key idea behind HLF is that fingerprints are recorded as signal-strength ratios between pairs of base stations instead of as absolute signal strength. A client's location can be estimated from the fingerprinted ratios by comparing these with ratios computed from currently measured signal-strength values. The advantage of HLF is that it can resolve the signal-strength differences \textit{without} requiring any extra calibration by the use of ratios. The method has been evaluated by extending two well-known LF techniques to use signal-strength ratios: \emph{Nearest Neighbor} \cite{Bahl2000} and \emph{Bayesian Inference} \cite{Haeberlen2004}. The HLF-extended techniques have been evaluated by emulation on ten-hour-long signal-strength traces collected with five heterogeneous IEEE 802.11 clients and using a fingerprint set collected with one client. The HLF-extended Bayesian inference technique improves the overall accuracy with 15 percentage points from 31\% to 46\% and in comparison the manual improved it with 17 percentage points to 48\%.

The fourth contribution is a filter for handling sensitivity differences which is published in [Paper 3]. The problem is that if clients do not see the same base stations at similar locations then the accuracy of a LF system is decreased. To address this problem a K-strongest filter is proposed in [Paper 3]. The rationale behind this filter is that if a client makes more observations because of higher sensitivity these can be filtered out by only keeping the K-strongest measurements in each sample. K should here be set to match the sensitivity of the fingerprint client. The filter has been evaluated by emulation on the traces collected for five heterogeneous IEEE 802.11 clients and using a fingerprint set collected with one client. With the sensitivity filter the HLF-extended Bayesian inference technique further improves it's accuracy from 46\% to 52\% and the manual improves it's accuracy from 48\% to 51\%.

\begin{figure}[h]
	\centering
		\includegraphics[viewport=40 450 420 800,width=1.0\textwidth,clip]{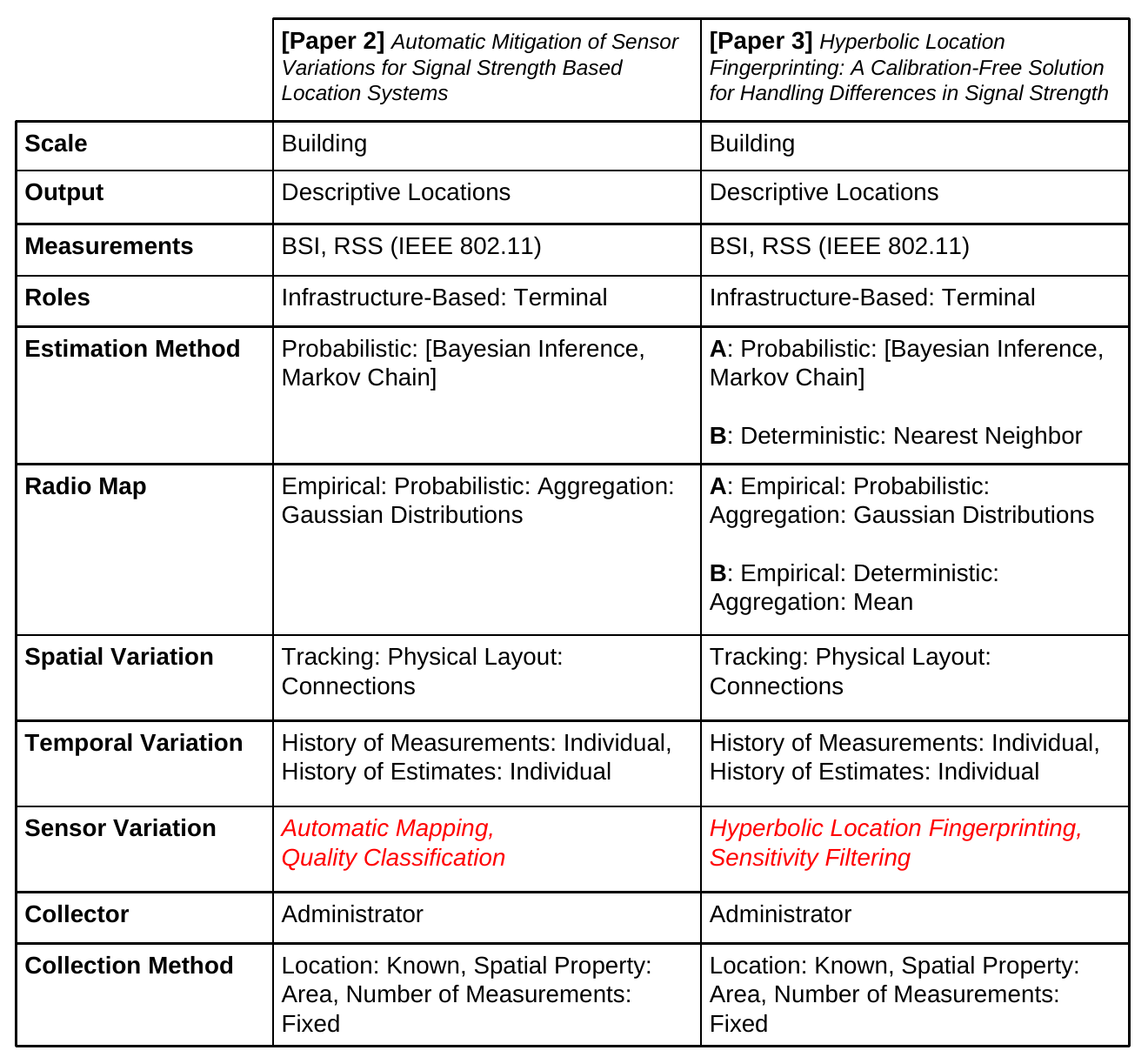}
	\caption{Taxonomy entries for [Paper 2] and [Paper 3]}
	\label{hete:fig:papers}
\end{figure}

To discuss the types of LF techniques that can be extended with the four contributions, Figure \ref{hete:fig:papers} classifies the used LF techniques according to the proposed taxonomy of [Paper 1]. The purpose of this classification is to highlight what assumptions from the underlying LF system the contributions depend on. Therefore most of the taxonomy entries in Figure \ref{hete:fig:papers} are specific for the LF system that was choosen to be extended with the contributions. The classification reveals that one LF technique was extended with the contributions in [Paper 2] and two techniques (A and B) with the contributions in [Paper 3]. However, the contributions are not limited to the extended types of LF techniques. The four contributions were designed for terminal-based and terminal-assisted techniques and can therefore not be applied to network-based systems. For network-based systems sensor variations are also not a major issue because all client measurements from a specific base station will be affected by the same systematic error that therefore does not need to be removed. With respect to the other dimensions of the taxonomy there are no major limitations for applying the contributions. 

\section{Related Work}
\label{heterogeneous:relatedwork}
In Kaemarungsi \cite{Kaemarungsi2006}, a study is presented of the properties of the signal strength measurements from different IEEE 802.11 clients. However, the paper does not propose any methods for handling the differences or study the impact on LF accuracy.

Haeberlen et al. \cite{Haeberlen2004} propose the use of a linear mapping for transforming a client's samples to match another client's samples. They propose three different methods for finding the two parameters in the linear mapping. The first method is a manual one, where a client has to be taken to a couple of known locations to collect fingerprints and parameters are found using least squares estimation. The second method is a quasi-automatic one, for which a client has to be taken to a couple of arbitrary locations to collect fingerprints. For finding the parameters, the authors propose the use of confidence values from Markov localization and find parameters that maximize this value. The third method is an automatic one requiring no user intervention. Here they propose the use of an expectation-maximation algorithm combined with a window of recent measurements. For the manual method, the authors have published results which show a gain in accuracy for three clients; for the quasi-automatic method it is stated that the performance is comparable to that of the manual method and for the automatic one it is stated that it does not work as well as the two other methods. In comparison, the contributed automatic method in [Paper 2] has a performance that is 7.4 percentage points worse than the manual method but requires a short learning period to work. The HLF-extended LF method in [Paper 3] has a performance that is one percentage point better than the manual method and does not involve any extra steps of collecting additional fingerprints.

In addition to systems which estimate the location of clients, a number of systems, such as NearMe \cite{Krumm2004b}, have been studied for which the calibration step is only carried out by users for tagging relevant places. The system uses simple metrics based on signal strength to quantify when clients are in proximity of calibrated places. One of the strengths of these simple metrics is that they overcome the problem of signal-strength differences.


\clearemptydoublepage
\chapter{Scalability to Many Clients}
\label{chap:scallf}
\mycitation
{\textbf{scale} \textit{(verb)} to change the size of something.}
{\textit{Oxford Advanced Learner's Dictionary}}
This chapter discusses [Paper 4] (\emph{Zone-based RSS Reporting for Location Fingerprinting}) and [Paper 5] (\emph{Efficient Indoor Proximity and Separation Detection for Location Fingerprinting}). Section \ref{scalability:intro} introduces and motivates the contributions. Section \ref{scalability:contrib} summarises the main contributions of the papers and section \ref{scalability:relatedwork} discusses related work.

\section{Introduction}
\label{scalability:intro}

When resource-constrained clients are used for LF they are unable to store the fingerprinting radio map and therefore have to be supported by a \emph{location server} for terminal-assisted positioning. The server accesses the radio map and estimates their location based on signal strength measurements conducted by the client. Measured signal strength values are by exisiting systems either transmitted over a wireless link on request, or the client updates them periodically with the location server, according to a pre-defined update interval. The associated problem is that periodic updating generates an excessive number of messages if the client changes its location only sporadically. The periodic protocol performs especially bad if it only has to be observed when the client enters or leaves certain pre-defined update zones.

The excessive number of messages is both a problem for the wireless link, the server, and the client. For the wireless link, an excessive number of messages use valuable bandwidth and might increase the monetary costs clients have to spend for mobile data services. The latter aspect is of special importance for cross-organizational scenarios, when the update messages can not be directed over the network that is used for the signal strength measurements, but, e.g., only by using public bearer services like GPRS or UMTS (packet switched). For the server the excessive number of messages reduces the number of clients that the server is able to support. For the client the excessive number of messages consumes battery power and increases the need of IEEE 802.11 clients to continuously switch back and forth between communication mode for sending messages and scanning mode for observing signal strength values. The latter aspect is discussed in more detail in Chapter \ref{chap:lfcom}.

In the above case one client uses a location server to estimate its position for use by applications either on the client or in connection with an application server. In other cases the end goal might not be to calculate the clients' positions but the detection of some relationship between the clients. One example of such a relationship is proximity detection which is defined as the capability to detect when two mobile clients approach each other closer than a pre-defined proximity distance. Analogously, separation detection discovers when two clients depart from each other by more than a pre-defined separation distance. The detection of such events can be used in manifold ways, for example, in the context of community services for alerting the members of a community when other members approach or depart. To detect such events a location server needs to continuously monitoring the position of clients and then compare their positions. Implementing such monitoring using a periodic protocol again creates the same problems as described above. Existing methods such as that proposed by K{\"u}pper et al. \cite{KuTr06} for proximity and seperation detection address the inefficiency of periodic protocols for terminal-based positioning for outdoor scenarios. However, these methods are not directly applicable indoors because they are based on line-of-sight distances which are in many cases meaningless in indoor environments. Furthermore they do not address the protocol issues for terminal-assisted positioning.

\section{Main Contribution}
\label{scalability:contrib} 
[Paper 4] and [Paper 5] make the following three contributions.

The first contribution is an efficient zone-based signal strength protocol for terminal-assisted LF published in [Paper 4]. The protocol works as follows: a location server dynamically configures a client with update zones defined in terms of signal strength patterns. Only when the client detects a match between its current measurements and these patterns, that is, when it enters or leaves the zone, it notifies the server about the fact. The associated challenge is the adequate definition of signal strength patterns for which [Paper 4] proposes several methods. The proposed methods have been evaluated by emulation for correct detection of zones with different shapes and sizes and message efficiency. The emulation uses traces and fingerprints collected with one IEEE 802.11 client. Furthermore the methods' computational overheads have been analyzed. As it turns out, an adaptation of classical Bayes estimation is the best suited method. This method has the best detection accuracy, a low computational overhead, and is able in the evaluated scenarios to reduce the number of messages with a factor of 15 compared to a periodic protocol.

The second contribution is a novel semantic for indoor distances for proximity and separation detection published in [Paper 5]. Checking for proximity and separation under consideration of Euclidean distances do not make much sense indoors, because several clients could be located on top of each other on different floors of a building, to give only one example. Applying both detection functions for walking distances is therefore a more reasonable, but also a more sophisticated approach. A location model that allows the modelling and calculation of such walking distances in buildings is presented in the paper. 

The third contribution is an efficient method for walking-distance-based proximity and separation detection for LF published in [Paper 5]. The method uses a modified version of the dynamic centred circles strategy proposed by K{\"u}pper et al. \cite{KuTr06}. The proposed method modifies the dynamic centred circles strategy for working with walking distances and combines it with the zone-based signal strength protocol. The \textit{dynamic centred circles strategy} dynamically assigns each client update zones in order to correlate the positions of multiple clients. In indoor environments such update zones can be effectively realized with the zone-based signal strength protocol and walking distances between mobile clients are used instead of Euclidean ones. The method has been evaluated in terms of efficiency and application-level accuracy based on numerous emulations on experimental data. The data set used consists of six sets of traces, each comprising three 40-minutes-walks simultaneously performed with three clients, totalling about 12 hours of data and a fingerprint set. The result of the evaluation was that the method decreased the number of transmitted messages with a factor of 9 compared to a periodic protocol while achieving an application level-accuracy above 94.5\%. Furthermore an implementation of the method was validated in a real-world deployment.

\begin{figure}[h]
	\centering
		\includegraphics[viewport=40 450 420 800,width=1.0\textwidth,clip]{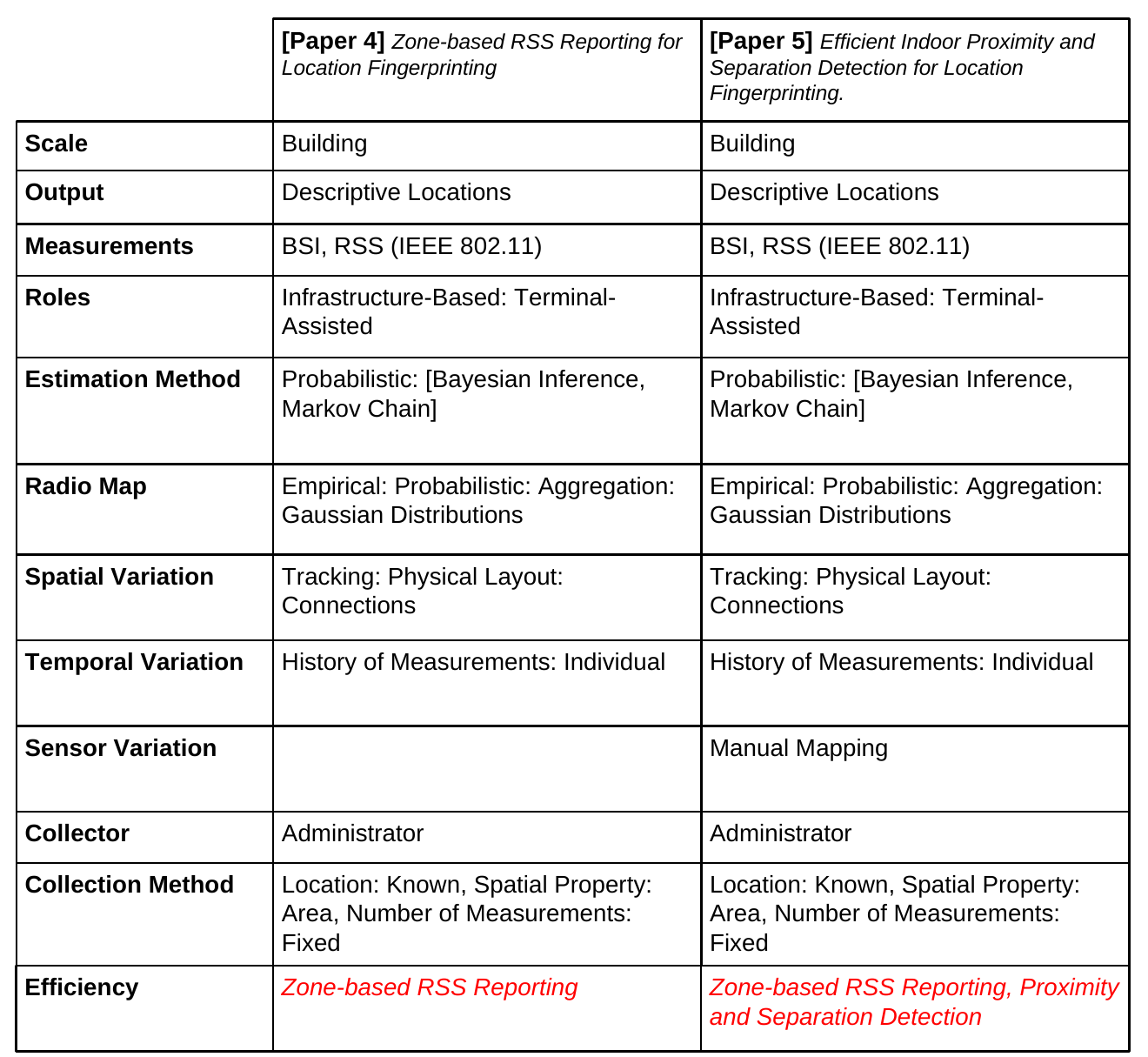}
	\caption{Taxonomy entries for Paper 4 and Paper 5}
	\label{scalability:fig:papers}
\end{figure}

To discuss the types of LF techniques that can be extended with the three contributions Figure \ref{scalability:fig:papers} classifies the used LF techniques according to the proposed taxonomy in [Paper 1]. The classification reveals that for both the contributions in [Paper 4] and [Paper 5] a single LF technique was extended. However, the contributions are not limited to this LF technique but can be applied with a range of LF techniques. For the contribution of zone-based signal strength reporting the main limitation is that the protocol is designed for only terminal-assisted systems. The method for proximity and separation on the other hand can be applied for both terminal-based and terminal-assisted. However, both contributions can not be applied with network-based systems because in this case the clients' only output are beacons for base stations to measure and therefore the clients are not able to handle zone updates.

\section{Related Work}
\label{scalability:relatedwork}
In this section related work is discussed, first, for zone-based signal strength reporting and, second, for proximity and separation detection.

\subsubsection{Zone-based Signal Strength Reporting}
From a perspective of resource-constrained clients, existing LF systems such as \cite{Roos2002b,Youssef2005b,Krumm2004,Haeberlen2004,Castro2001} are not optimal with respect to the overhead induced by only using poll or periodic update protocols. In addition to the these systems, which estimate the location of clients, a number of systems, such as NearMe \cite{Krumm2004b}, have been studied where fingerprint collection is only carried out by users for tagging relevant places. The systems propose simple metrics based on signal strength measurements to quantify when clients are in proximity of calibrated places. Such systems are relevant to this work with respect to the methods they propose for proximity detection. However, such systems can only detect presence at a single point and not within zones with specific shapes and sizes as addressed by zone-based signal strength reporting.

A system which has addressed the needs of resource-constrained clients for LF, by using additional sensors, is published by You et al. \cite{You2006}. The authors propose a communication protocol between a location server and a client, which dynamically adapts the signal strength update rate of the client based on the distance to the last reported update using measurements from an accelerometer. In comparison, the methods proposed in this paper do not require any extra sensors and are therefore usable for a broader range of clients where such extra sensors are not present or too expensive to include. In addition to this, the proposed methods in [Paper 4] can also be used with arbitrary shaped zones and not just zones defined by a distance to a specific point.

A later LF system for resource-constrained clients has been proposed by King et al. \cite{KingWowMom2008}. This system is terminal-based and works by caching a part of the fingerprint radio map on clients. Two algorithms are proposed for how to fill the cache where both are based on observed base stations. Compared to the approach proposed in [Paper 4] this system requires that a client carry out computations for LF positioning and stores a fingerprint cache whereby clients' resource demands are increased.

Infrastructure-less systems are based on protocols which are more energy-efficient than for instance IEEE 802.11, such as IEEE 802.15.4 or communication over the 433/916 MHz telemetry bands. Bulusu et al. \cite{Bulusu00a} propose a system which senses the proximity of a mobile client to static beacon clients which output their id and position. The position of the mobile client is then estimated by finding the centroid of the positions of the proximate clients. A system that proposes methods for infrastructure-less localization inspired by infrastructure-based techniques is MoteTrack \cite{Lorincz2005}. The system consists of a number of wireless clients where some have the role as static beacon clients and others are mobile clients which the system should locate. The system is based on LF using signal strength to the static beacon clients. The fingerprints are distributively stored on the static beacon clients and provided to the mobile clients when in proximity. The system's method for location estimation is based on weighted nearest neighbors based on the Manhattan distance instead of the Euclidian distance to lower computation needs. The computing of the location estimates can be carried out either by the mobile clients or by the beacon clients, depending on which of the proposed sharing techniques is used. These systems are related to the methods proposed in [Paper 4] in terms of how they achieve energy-efficiency and do decentralized estimation. However, since all such systems assume that there is no infrastructure, they do not address how to combine decentralized estimation with the capabilities of infrastructure-based solutions.

\subsubsection{Proximity and Seperation Detection}
In recent years, LF has been evaluated and used mainly for positioning of single clients, therefore not addressing proximity and separation detection \cite{Bhasker2004,Haeberlen2004,Roos2002b,Youssef2005b}, with NearMe \cite{Krumm2004} as an exception. NearMe supports a short-distance proximity detection, which only takes signal strength measurements and Euclidean distances into consideration, as well as a long distance mode, which applies a base station coverage-graph analysis. NearMe is a client-server approach with periodic signal strength updating between mobile clients and a location server, which causes significant overhead when a client does not move for a long period of time.

Applications have been built and evaluated for usability that apply LF on IEEE 802.11 networks and that use proximity information. The location-based messaging system InfoRadar \cite{Rantanen2004}, for example, uses the LF technique proposed by Roos et al. \cite{Roos2002b}. In the system, a location server polls signal strength measurements from clients to estimate their positions and checking them for proximity subsequently. The ActiveCampus \cite{Griswold2004} system provides a set of applications to foster social interactions in a campus setting. One of these services can list nearby buddies and show maps overlaid with information about buddies, sites, and current activities. Clients are located using a terminal-assisted LF technique proposed by Bhasker et al. \cite{Bhasker2004} and a combination of poll-based and periodic signal strength updating, which, however, turned out to be a bottleneck in this system when trying to scale beyond 300 concurrent users. The strategies proposed in [Paper 5] scale much better and are novel in the sense that they consider walking instead of Euclidean distances which better reflects the needs of indoor location-based applications.

Several systems support the realization of location-based applications based on LF in general. Many of the systems have been proposed for integrating position estimates produced by different positioning technologies, among them LF, thus easing implementation and improving server-side efficiency. Examples of such systems are the Rover system \cite{Banerjee2002}, the Location Stack \cite{Hightower2002}, and its implementation in the Universal Location Framework (ULF) \cite{Graumann2003}. They provide means to integrate and fuse information from several positioning methods, query location information, improve scalability, and define location-based triggers. The systems have been integrated with LF techniques such as Horus \cite{Youssef2005b} and RADAR \cite{Bahl2000}. Position estimates are obtained from the location sources by push, pull, and periodic location updating methods. The Rover system has been evaluated for server-side efficiency in terms of CPU load based on simulated inputs. In comparison to these systems, [Paper 5] proposes strategies for an efficient message transfer over the wireless link, which also improves server-side efficiency and saves client resources.


\clearemptydoublepage
\chapter{Interference between Communication and Positioning}
\label{chap:lfcom}
\mycitation
{\textbf{Interference} \textit{(noun)} interruption of a radio signal by another signal on a similar wave-length, causing extra noise that is not wanted.}
{\textit{Oxford Advanced Learner's Dictionary}}
This chapter discusses [Paper 6] (\emph{ComPoScan: Adaptive Scanning for Efficient Concurrent Communications and Positioning with 802.11}). Section \ref{communication:intro} introduces and motivates the contributions. Section \ref{communication:contrib} summarises the main contributions of the paper and related work is discussed in Section \ref{communication:relatedwork}.

\section{Introduction}
\label{communication:intro}
Back in 1999, when IEEE~802.11 was being standardized, the researchers and engineers working on the standard probably never thought about the new ways we use this technology today. Real-time applications such as voice over IP and video conferencing were a rarity years ago but are a common phenomenon nowadays. 

Even the newer sub-standard 802.11b and 802.11g do not satisfy these requirements. Furthermore, several workarounds and novel approaches (e.g., \cite{Forte2006,Mhatre2006, Shin2004}) have been proposed to make 802.11 ready for many of these new demands. However, still unsolved remains the problem that occurs when 802.11 wireless clients are utilized for positioning and communicating at the same time. On the one hand, the positioning system requires a steady stream of measurements from active scans to be able to deliver accurate position estimates to location-based applications. Especially if the positioning system is used to track users as, e.g., required for indoor navigation systems in huge buildings. Performing an active scan means that the wireless client switches through all the different channels in search of base stations. Dependent on the wireless client this takes about 600~milliseconds. During this time no communication is feasible.  On the other hand, there are the demanding real-time applications that use communication. For instance, a video conference requires around 512~KBit/s of bandwidth and a round trip delay of less than 200~milliseconds, depending on the video and voice quality~\cite{Varshney2002}.

\begin{figure}[h]
  \centering{
    {\includegraphics[width=0.95\textwidth]{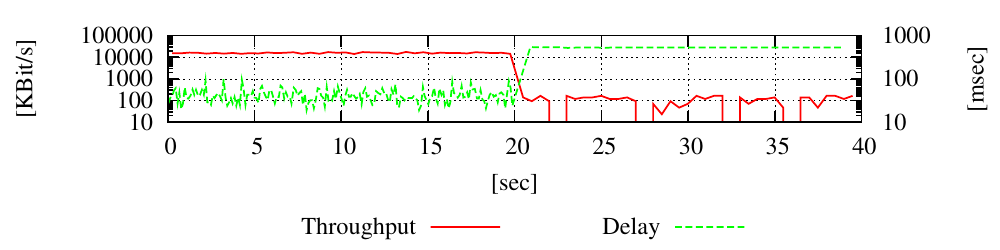}}
  }
  \caption{Throughput and delay.}
  \label{communication:fig:throughputdelayscanning}
\end{figure}

Figure~\ref{communication:fig:throughputdelayscanning} depicts what happens to a wireless client's throughput and delay if requested to perform an active scan every 600~milliseconds. During the first 20~seconds communication is untroubled, which means a throughput of about 20~MBit/s on average and that a round trip delay of less than 45~milliseconds is achievable. In the 20$^{th}$ second active scanning starts. The remaining seconds only provide 0.1~MBit/s of throughput and 532~milliseconds of delay, because active scans are performed so often. Due to variations in the execution time of scans, on some rare occasions no data transmission is possible at all.

\section{Main Contribution}
\label{communication:contrib}
[Paper 6] makes the following two contributions.

The first contribution is a novel solution for the scanning problem named ComPoScan. The ComPoScan system is based on movement detection to switch between light-weight monitor sniffing and invasive active scanning based on adaptability. Only in cases where the system detects movement of the user active scans are performed to provide the positioning system with the signal strength measurements it needs. If the system detects that the user is standing still, it switches to monitor sniffing to allow communications to be uninterrupted. Monitor sniffing is a novel scanning technique proposed in~\cite{KingPercom2007}. It works with most 802.11 wireless clients available today. Monitor sniffing allows a wireless client to recognize base stations operating on channels close to the one it is using for communication. It has been shown that up to seven channels can be overheard without any disturbance of the actual communication. For evaluating the system by validation, ComPoScan was implemented and this prototype was used in several real-world deployments. The validation provided results for ComPoScan's impact on communication showed that it increases throughput by a factor of 122, decreases the delay by a factor of ten, and the percentage of dropped packages by 73\%. Additionally, the results show that ComPoScan does not harm the positioning accuracy of LF.

The second contribution is a novel movement detection system that utilizes monitor sniffing and active scanning. The movement detection approach is also based on signal strength measurements. However, the measurements provided by monitor sniffing are sufficient to detect reliably whether the user is moving or standing still. We designed the movement detection system to be configurable so that depending on the user's preferences, communication capabilities or positioning accuracy can be favoured. A \emph{Hidden Markov Model(HMM)}-based detector turned out to be the best suited method given these requirements. The movement detection system has been evaluated by means of emulation to show that it works independently of the environment, the wireless client, the signal strength measurement method, and the number and placement of base stations. Furthermore ComPoScan was implemented and used in a real-world deployment to gather validation results showing that the real system works as predicted by the emulation. 

\begin{figure}[h]
	\centering
		\includegraphics[viewport=90 250 370 600,width=0.7\textwidth,clip]{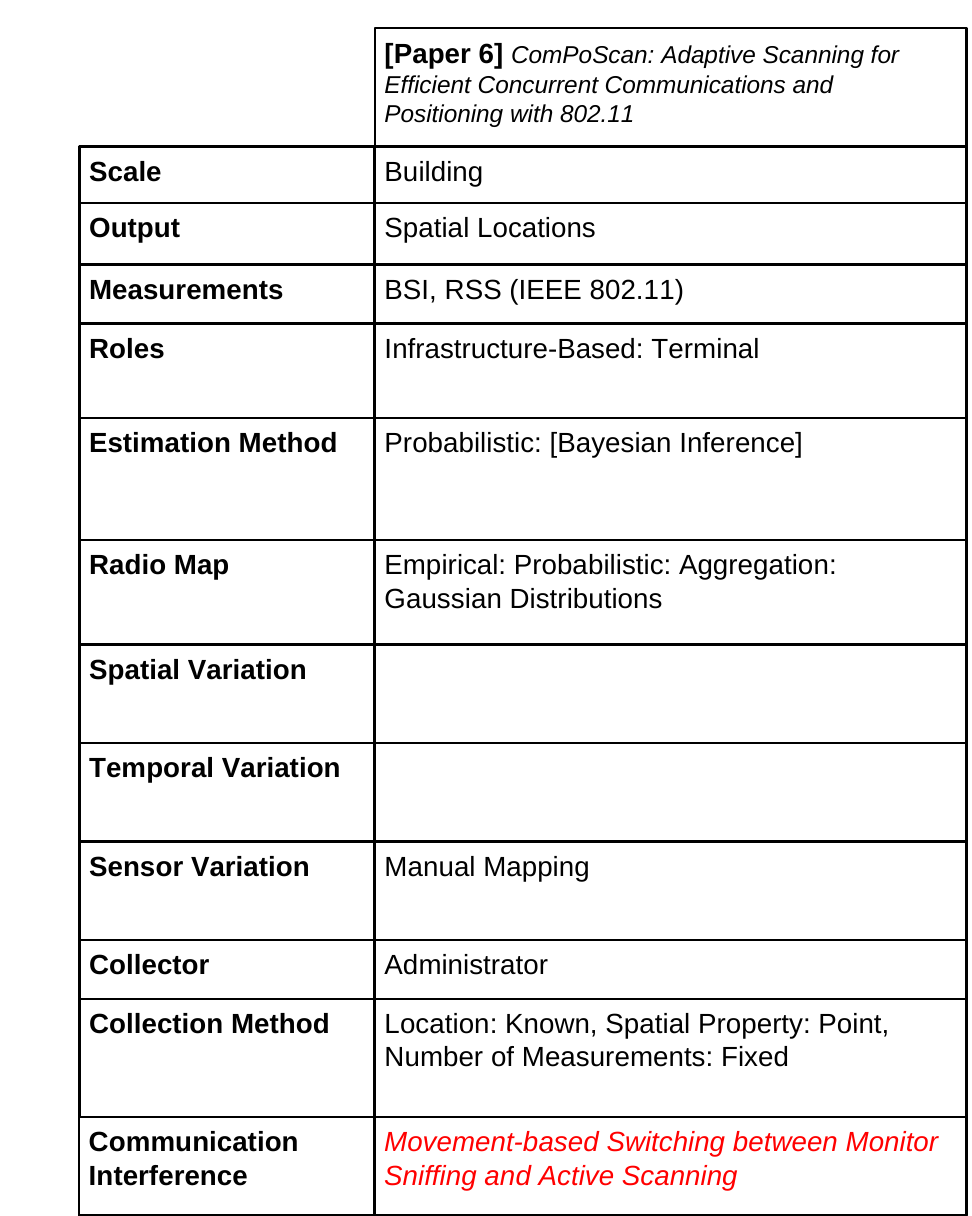}
	\caption{Taxonomy entries for Paper 6}
	\label{communication:fig:papers}
\end{figure}

To discuss the types of LF techniques that can be extended with the two contributions Figure \ref{communication:fig:papers} classifies the used LF technique according to the proposed taxonomy in [Paper 1]. The main restriction of the contributions is that they can not be applied with network-based systems. This is because network-based systems do not measure signal strength using active scanning but measuring the strength of incoming packets. The contributions also impact methods for addressing spatial and temporal variations because when ComPoScan switches to active scanning, no history of either estimates or measurements are available for the methods to use when trying to improve LF accuracy.

\section{Related Work}
\label{communication:relatedwork}

Existing 802.11 LF systems (e.g., \cite{Bahl2000, Haeberlen2004}) have not considered the problem of concurrent communication and positioning. As a central part of the ComPoScan system movement detection was applied to deal with this problem.

The first, and as far as the literature goes, the only 802.11-based system that focuses on movement detection is the LOCADIO
system~\cite{Krumm2004}. In their paper, the authors propose an
algorithm that exploits the fact that the variance of signal strength
measurements increases if the mobile device is moved compared to if it
is kept still. To smooth the high frequency of state transitions, an HMM is
applied. The results in the paper show that the system detects whether the mobile device is in motion or
not in 87~percent of all cases. Compared to the approach proposed in [Paper 6], the authors do not compare their system
to other movement detection algorithms. Furthermore, the results are
only based on emulation which means that the signal strength data is
collected in a first step and then, later on, analyzed and processed to
detect movement. This is a valid approach, but some real-world effects
might be missed. Another fact that the authors of the aforementioned
paper do not look at is the impact of periodic scanning to the
communication capabilities of mobile devices. The authors just assume that a
802.11 wireless client is solely used for movement detection. Finally,
all results are based on one single client, which means that
variations in signal strength measurements caused by different
wireless clients are not taken into consideration.

Two GSM-based systems have also been proposed by Sohn et
al. \cite{Sohn2006} and Anderson et al. \cite{Anderson2006}. The
system by Sohn et al. is based on several features including variation
in Euclidean distance, signal strength variance, and correlation of
strength ranking of cell towers. The system classifies data into the
three states of still, walking, and driving. By emulation on collected
data, the authors achieve an overall accuracy of 85~percent. The system by
Anderson et al. detects the same states, but uses the features of
signal strength fluctuation and number of neighbouring cells. Using
these features the authors achieve a comparable overall accuracy compared to the
former system. As for LOCADIO the results for both systems are only
based on emulation, they do not consider communication, and the
results are based on one client.

  
\clearemptydoublepage
\chapter{Conclusions and Future Work}
\label{chap:conclusions}
\mycitation
{\textbf{future} \textit{(noun)} the time that will come after the present or the events that will happen then.}
{\textit{Oxford Advanced Learner's Dictionary}}
This chapter concludes Part I of this thesis. Section \ref{conandfuture:sumcont} summarises the main contributions of this thesis and Section \ref{conandfuture:futurework} presents a number of directions for future work.

\section{Summarizing the Contributions}
\label{conandfuture:sumcont}
As stated in Section \ref{introduction:resobj} the research goal of this thesis has been to address the limitations of current indoor LF systems. In particular the aim is to advance LF for the challenges of handling heterogeneous clients, scalability to many clients, and interference between communication and positioning. The research presented here contributes to the conceptual foundation, methods, protocols, and techniques for LF. The main contributions of the thesis are summarised below.

\begin{itemize}
	\item A taxonomy to improve the conceptual foundation of LF. The taxonomy consists of eleven main taxons and 88 subtaxons that in more detail classifies LF systems. The taxonomy has been constructed based on a literature study of 51 papers and articles. The 51 papers and articles propose 30 different systems which have been analyzed and methods and techniques grouped to form taxons for the taxonomy. The taxonomy allows researchers to make detailed comparison of systems and methods and can help scope out new research paths in the area.
	
	\item Several methods for handling the heterogeneity of clients. First, methods for classifying a client's measurement quality that when evaluated by emulation were able to classify clients' quality correctly in 96.2\% of the tested cases. Second, an automatic linear-mapping method for handling signal-strength differences that was able, with automatically collected calibration data, to improve LF accuracy with 13.1 percentage points for the evaluated data set. Third, the method of hyperbolic location fingerprinting which addresses signal-strength differences by recording fingerprints as signal-strength ratios between pairs of base stations. The method was able, without any calibration data, to improve LF accuracy with 15 percentage points for the evaluated data set. Fourth, a method in the form of a filter to handle sensitivity differences among clients that improved LF accuracy with 6 percentage points for the evaluated data set.
	
	\item Several methods and protocols for increasing the scalability of LF systems. First, an efficient zone-based signal-strength protocol for terminal-assisted LF that reduces the number of messages needed to track the positions of wireless clients. The protocol has been evaluated by emulation and was able to reduce the number of messages with a factor of 15 compared to a periodic protocol. Second, an efficient method for walking-distance-based proximity and separation detection that reduces the number of messages needed to monitor proximity and separation relationships among clients. The method is based on a novel semantic for indoor distances that considers the walking distances in buildings. The method has been evaluated by emulation where it decreased the number of transmitted messages with a factor of 9 compared to a periodic protocol while achieving an application level-accuracy above 94.5\%.

	\item A solution to address interference between communication and positioning. The solution, named ComPoScan, is based on movement detection to switch between light-weight monitor sniffing and invasive active scanning. Only in the case that the system detects movement of the user, active scans are performed to provide the positioning system with the signal strength measurements it needs. If the system detects that the user is standing still it switches to monitor sniffing to allow communications to be uninterrupted. The movement detection system has been evaluated by means of emulation and validation to show that it works independently of the environment, the wireless client, the signal strength measurement method, and number and placement of base stations. The validation results for ComPoScan's impact on communication showed that it increases throughput by a factor of 122, decreases the delay by a factor of ten, and the percentage of dropped packages by 73 \%. Additionally, the results show that ComPoScan does not harm the positioning accuracy of LF.

\end{itemize}

\section{Future Work}
\label{conandfuture:futurework}

The contributions open up several paths for future work.

The proposed taxonomy lays the groundwork for several interesting extensions. First, the taxonomy could be extended to cover non-functional properties. Non-functional properties such as computational efficiency and robustness are important properties for a production-ready LF system and therefore also important to cover in a taxonomy for LF. Second, the taxonomy can be used for several kinds of synthesis of new research paths by comparing and grouping the all ready taxonomized systems. Third, the foundation for the taxonomy could be broadened by taxonomizing more systems to increase the confidence that no aspects of existing systems have been missed.

The proposed techniques for handling heterogeneous clients provide a good foundation for addressing the heterogenity problem. However, it would be relevant to have classifiers that could detect if signal strength measurements have artificial limits or are measured by a client that has poor sensitivity. Furthermore it would be relevant to further analyse how sensitivity affects accuracy. For instance, evaluating if a recommendation such as \textit{always use a client which maximizes the number of measured base stations} could limit the sensivity problem. In addition it would be interesting to apply the proposed techniques to technologies such as GSM where signal-strength differences are also present.

A technique was proposed for proximity and separation detection. However, in addition to this problem there are other equally important relationships that would be interesting to detect efficiently. For instance, a possible extension to the described community service, which recognizes targets closer than a static threshold would be a buddy tracker that constantly shows the user a sorted list of the n-nearest-neighbors among his buddies. One piece of future work could therefore be how such a service can be realized efficiently by dynamically applying proximity and separation detection to pairs of clients. There are also other problems such as detection of when clients' cluster. A related issue is that LF systems are generally evaluated for single target accuracy but what matters when detecting relationships is the multi-client accuracy which is the accuracy of the distance between the clients computed from the estimated positions of the clients. Very little knowledge exists about multi-client accuracy and what impacts it.

For some technologies, such as IEEE 802.11, scanning for signal strength measurements is rather resource consuming, which makes it desirable to minimize the needed scans. The ComPoScan system goes some of the way by trading high consuming active scans to less consuming monitor sniffs. However, a further improvement could become possible by integrating ComPoScan with the zone-based idea. One possible method, which, however, only applies to large zones, would be to subdivide a zone in a way where central parts could use long scanning intervals, while short intervals could be applied at the borders of the zones. Between the scans the wireless client could be powered-off and thereby save resources.

Another path of future work is error estimation for LF. For an user or an administrator it is important to know how large position errors to expect. The question is therefore how to estimate errors for indoor LF systems. A solution for this problem should be able to both estimate the error in each estimate and to generate information for map-based visualizations that can highlight the expected errors in different building parts.

A further challenge is to decrease LF's dependency on an installed infrastructure. For instance, is it possible to base LF on sensor inputs such as natural light, the chemical-components in the air or ionizing radiation such as gamma radiation. If realized such system could work without depending on an installed infrastructure.



\clearemptydoublepage
\part{Papers}

\clearemptydoublepage
\chapter{Paper 1}
\label{chap:loca2007}

The paper \emph{A Taxonomy for Radio Location Fingerprinting} presented in this
chapter has been published as a conference paper~\cite{KjaergaardLoca2007}.

\begin{publist}{\cite{KjaergaardLoca2007}}
  \item[\cite{KjaergaardLoca2007}] M.\ B.\ Kjærgaard. A Taxonomy for Radio Location Fingerprinting. In \emph{Proceedings of the Third International Symposium on Location and Context Awareness}, pages~139--156, Springer, 2007.
\end{publist}

\noindent
The analysis results for all of the surveyed systems are available online at \texttt{wiki.daimi.au.dk/mikkelbk}.

\clearemptydoublepage


\mytitle{A Taxonomy for Radio Location Fingerprinting}{ 
  Mikkel Baun Kj\ae rgaard\footnotemark[1]}{  
  \footnotetext[1]{Department of Computer Science, University of
    Aarhus, IT-parken, Aabogade 34, DK-8200 Aarhus N, Denmark. E-mail:
    \texttt{mikkelbk@daimi.au.dk}.}
    } 
    

\begin{myabstract}
  \emph{Location Fingerprinting (LF)} is a promising location technique for many awareness applications in pervasive computing. However, as research on LF systems goes beyond \textit{basic methods} there is an increasing need for better comparison of proposed LF systems. Developers of LF systems are also lacking good frameworks for understanding different options when building LF systems. This paper proposes a taxonomy to address both of these problems. The proposed taxonomy has been constructed from a literature study of 51 papers and articles about LF. For researchers the taxonomy can also be used as an aid when scoping out future research in the area of LF.
\end{myabstract}


\section{Introduction}
\label{paper1:sec:introduction}
A popular location technique is \emph{Location Fingerprinting (LF)}, having the major advantage of exploiting already existing network infrastructures, like IEEE 802.11 or GSM, which avoids extra deployment costs and effort. Based on a database of pre-recorded measurements of network characteristics from different locations, denoted as \emph{fingerprints}, a wireless client's location is estimated by inspecting currently measured network characteristics. Network characteristics are typically base station identifiers and the received signal strength.

LF is different by the use of fingerprints to other location techniques such as lateration, angulation, proximity detection and dead reckoning \cite{Kuep05}. Lateration and angulation techniques estimate location from measurements to fixed points with known locations. A technology example is the \emph{Global Positioning System (GPS)} which estimate a GPS client's location from measurements to GPS satellites with known locations. Proximity detection identifies the location of clients when in proximity of fixed points. A technology example is \emph{Radio-Frequency IDentification (RFID)} where a passive RFID tag's location is known when in proximity of a RFID scanner. Dead reckoning estimates location by advancing previous estimates by known speed, elapsed time and direction. A technology example is dead reckoning based on accelerometer measurements.

Many different LF systems have been proposed. When surveying LF systems one has to answer many different questions. For instance, how do systems differ in scale; can they be deployed to cover a single building or an entire city? What network characteristics are measured? What are the roles of the wireless clients, base stations, and servers in the estimation process? Which estimation method is used? How are fingerprints collected and used? These questions are not only important for researchers surveying LF but also developers of LF systems who have to understand the different possibilities. We believe that a taxonomy will aid LF system developers and researchers better survey, compare, and design LF systems. Being able to better survey and compare existing work also makes it possible to use the taxonomy as an aid when scoping out future research. This is especially important as research more and more moves from understanding the basic mechanisms to optimizing existing methods for non-functional properties such as robustness and scalability. Existing taxonomies such as that proposed by Hightower et al. \cite{Hightower2001} cover location systems in general and are therefore not too much help when answering the many questions specific to LF.

The taxonomy we have chosen to propose has been constructed based on a literature study of 51 papers and articles. The 51 papers and articles propose 30 different systems which have been analyzed and methods and techniques grouped to form taxons for the taxonomy. The analyses of four of the 30 systems are covered as case studies in Section \ref{paper1:sec:caserw}. The analysis results for all of the 30 systems are available online at \cite{lfwebsite}. 

The structure of the paper is as follows. The taxons of the proposed taxonomy are discussed in Section \ref{paper1:sec:tax}. The individual taxons are then presented in Sections \ref{paper1:sec:lf} to \ref{paper1:sec:collection}. Four case studies are afterwards presented in Section \ref{paper1:sec:caserw} and a discussion is given in Section \ref{paper1:sec:discussion}. Finally, conclusions are given in Section \ref{paper1:sec:conclusion}. Due to the limited size of this paper, the presentation level is advanced; for introductions to LF refer to books such as K{\"u}pper \cite{Kuep05} and papers such as Krishnakumar et al. \cite{Krishnakumar2005}.

\section{Taxonomy}
\label{paper1:sec:tax}
The proposed taxonomy is built around eleven taxons listed with definitions in Table \ref{tab:deftaxon}. These were partly inspired by earlier work on taxonomies for location systems in general and from our literature study. The four taxons: \emph{scale}, \emph{output}, \emph{measurements}, and \emph{roles} describe general properties of LF systems. We mean by scale the size of the deployment area and by output the type of provided location information. Measurements means the types of measured network characteristics and roles means the division of responsibilities between wireless clients, base stations, and servers. Only these four of our eleven taxons are covered by existing taxonomies such as Hightower et al. \cite{Hightower2001}. Their concepts for these four taxons differ by output being split over the four concepts of physical, symbolic, absolute, and relative, measurements being indirectly described by their technique concept and roles being partly described by their concept of localized location computation.

\emph{Estimation method} and \emph{radio map} describe the location estimation process. Estimation method denote a method for predicting locations from a radio map and currently measured network characteristics and radio map a model of network characteristics in a deployment area. The division into estimation method and radio map is used by many papers about LF, for instance Youssef et al. \cite{Youssef2005b}. However, some papers use a slightly different naming for instance Otsason et al. \cite{Otsason2005} use \emph{localization algorithm} and \emph{radio map}. 

How changing network characteristics over time, space and sensors can be handled is described by \emph{spatial, temporal and sensor variations}. The spatial and temporal dimensions were introduced by Youssef et al. \cite{Youssef2005b}. The sensor dimension was introduced in our earlier work, Kj\ae rgaard \cite{KjaergaardLoca2006}. The taxons \emph{collector} and \emph{collection method} describe how fingerprints are collected. These two taxons have been introduced to characterize the assumptions systems put on fingerprint collection.

\begin{table}[h]

\centering
\begin{tabular}{p{3.7cm}p{8.3cm}}
\textit{Taxon} & \textit{Definition} \\
\hline
\textbf{Scale} & Size of deployment area.\\ 
\textbf{Output} & Type of provided location information.\\
\textbf{Measurements} & Types of measured network characteristics.\\
\textbf{Roles} & Division of responsibilities between wireless clients, base stations, and servers.\\
\textbf{Estimation Method} & Method for predicting locations from a radio map and currently measured network characteristics.\\
\textbf{Radio Map} & Model of network characteristics in a deployment area.\\
\textbf{Spatial Variations} & Observed differences in network characteristics at different locations because of signal propagation characteristics.\\
\textbf{Temporal Variations} & Observed differences in network characteristics over time at a single location because of continuing changing signal propagation.\\
\textbf{Sensor Variations} & Observed differences in network characteristics between different types of wireless clients.\\
\textbf{Collector} & Who or what collects fingerprints.\\
\textbf{Collection Method} & Procedure used when collecting fingerprints.\\
\hline
 & \\
\end{tabular}
\caption{Taxon definitions}
\label{tab:deftaxon}
\end{table}

The focus of the proposed taxonomy is on methods for LF and therefore the taxonomy does not cover evaluation properties for LF systems. Evaluation properties for all kinds of location systems have for instance been suggested by Muthukrishnan et al. \cite{Muthukrishnan2005}, who list: precision, accuracy, calibration, responsiveness, scalability, cost, and privacy. The taxonomy proposed by Hightower et al. \cite{Hightower2001} also lists several evaluation properties: precision, accuracy, scale, cost, and limitations. In our analysis we have included the following evaluation properties: precision, accuracy, evaluation setup, and limitations. These four were chosen because this information is available from most papers. Responsiveness and cost were not included because the first is only available from very few papers and the second from none. Calibration, privacy, scalability, and scale are partly covered by our taxons scale, roles and collection method. These four properties are also listed in our case studies in Section \ref{paper1:sec:caserw}.

The taxonomy does not cover non-functional system properties, because work has not yet matured in these directions for LF systems. Non-functional properties of LF systems have been addressed by several recent papers, such as system robustness by Lorincz et al. \cite{Lorincz2005}, server scalability by Youssef et al. \cite{Youssef2005b}, and minimal communication by Kj\ae rgaard et al. \cite{KjaergaardPervasive2007}. Also, the taxonomy does not cover the application of LF techniques to other types of sensor measurements such as sound and light.

\section{General Taxons}
\label{paper1:sec:lf}
\begin{figure}[h]
	\centering
		\includegraphics[viewport=60 310 770 570,width=0.85\textwidth,clip]{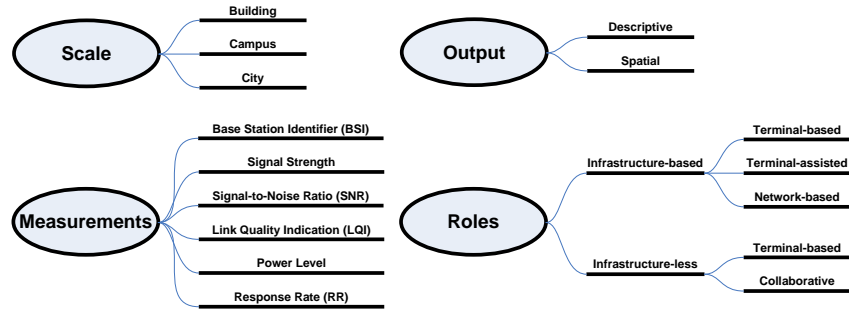}
	\caption{Scale, output, measurements and roles.}
	\label{fig:LF}
\end{figure}

The proposed general taxons for LF systems are: \emph{scale}, \emph{output}, \emph{measurements} and \emph{roles}. These taxons are shown including subtaxons in Figure \ref{fig:LF}. In this and the following sections when taxons are presented up to four references are given to papers or articles that propose systems that are grouped below the particular taxon. Therefore not all papers groupped under a taxon are listed, this type of information can be found online at \cite{lfwebsite}.

\textit{Scale} describes a system's size of deployment. Scale is important because size of deployment impacts how fingerprints can be collected and some systems are limited in scale because of specific assumptions. Scale is proposed to be classified as \emph{building}, \emph{campus}, or \emph{city}. Many LF systems have been proposed for a \emph{building} scale of deployment \cite{Bahl2000,Prasithsangaree2002,Battiti2002,Roos2002a}. Some systems are limited to this scale because they assume knowledge about the physical layout of buildings \cite{Krumm2004,Castro2001,Ladd2002,Haeberlen2004}; others because they assume the installation of a special infrastructure \cite{Bahl2000b,Krishnan2004}. \emph{Campus}-wide systems \cite{Bhasker2004} scale by proposing more practical schemes for fingerprint collection. \emph{City}-wide systems \cite{Laitinen2001,Roos2002b,LaMarca2005} scale even further by not assuming that a system is deployed by or for a single organization. City wide systems could scale to any area that is covered by base stations.

\textit{Output} denotes the type of provided location information. The subtaxons for output are proposed to follow the notion introduced in K{\"u}pper \cite{Kuep05} of dividing location information into \emph{descriptive} and \emph{spatial} information. Descriptive locations are described by names, identifiers or numbers assigned to natural geographic or man-made objects\footnote{Some authors refer to this as symbolic locations}. Spatial locations are described by a set of coordinates stated with respect to a spatial reference system. Many LF systems output \emph{spatial} locations \cite{Bahl2000,Roos2002a,LaMarca2005,Smailagic2002} but systems have also been proposed that output \emph{descriptive} locations \cite{Castro2001,Haeberlen2004,Bhasker2004}. However, a location outputted as either of the two types can be mapped to the other type given a suitable location model.

\textit{Measurements} are the types of measured network characteristics. The following network characteristics have been used in existing systems: \emph{Base Station Identifiers (BSI)}, \emph{Received Signal Strength (RSS)}, \emph{Signal-to-Noise Ratio (SNR)}, \emph{Link Quality Indicator (LQI)}, \emph{power level}, and \emph{Response Rate (RR)}. BSI is a unique name assigned to a base station. RSS, SNR, and LQI are signal propagation metrics collected by radios for handling and optimizing communication. The power level is information from the signal sender about current sending power. The response rate is the frequency of received measurements over time from a specific base station. Many LF systems are based on BSI and RSS \cite{Bahl2000,Roos2002a,Haeberlen2004,Smailagic2002}; other systems have used RR in addition to RSS \cite{Krumm2004,Ladd2002,LaMarca2005}. BSI and SNR have also been used \cite{Castro2001} and the combination BSI, LQI, RSS, and Power level \cite{Lorincz2005,Lorincz2006}. 

\textit{Roles} denote the division of responsibilities between wireless clients, base stations, and servers. How roles are assigned impact both how systems are realized, but also important non-functional properties like privacy and scalability. The two main categories for roles are \textit{infrastructure-based} and \textit{infrastructure-less}. Infrastructure-based systems depend on a pre-installed powered infrastructure of base stations. Infrastructure-less systems consist of ad-hoc-installed battery-powered wireless clients where some of them act as "base stations". Infrastructure-based systems are following K{\"u}pper \cite{Kuep05}, being further divided into \emph{terminal-based}, \emph{terminal-assisted} and \emph{network-based} systems. The infrastructure-less systems are divided into \emph{terminal-based} and \emph{collaborative} systems. The different types of systems differ in who sends out beacons, who makes measurements from the beacons and who stores the radio map and runs LF estimation, as shown in Figure \ref{fig:LFRoles}. Most LF systems have been built as infrastructure-based and terminal-based \cite{Youssef2005b,Prasithsangaree2002,LaMarca2005}, which is attractive because this setup supports privacy. Terminal-assisted \cite{Castro2001,Bhasker2004} and network-based systems \cite{Bahl2000,Krishnan2004} have also been built offering better support for resource-weak wireless clients\footnote{However, when only considering the basic method of each system, most can be realized in all of the three setups.}. Infrastructure-less LF-systems have to be optimized for the resource-weak wireless clients, which is addressed by the collaborative setup \cite{Lorincz2005,Lorincz2006}.

\begin{figure}[h]
	\centering
		\includegraphics[viewport=20 200 630 570,width=0.85\textwidth,clip]{paper1/Images/LFRoles.pdf}
	\caption{Different assignments of responsabilities to wireless clients, base stations, and servers.}
	\label{fig:LFRoles}
\end{figure}

\section{Estimation Taxons}
\label{paper1:sec:estimation}
\begin{figure}[h]
	\centering
		\includegraphics[viewport=20 460 480 790,width=0.85\textwidth,clip]{paper1/Images/LFEstimation.pdf}
	\caption{Estimation method}
	\label{fig:LFEstimation}
\end{figure}

The following two taxons describe the location estimation process: \emph{estimation method} and \emph{radio map}. The two taxons are shown including subtaxons in Figure \ref{fig:LFEstimation}.

\begin{figure}[h]
	\centering
		\includegraphics[viewport=20 110 530 450,width=0.85\textwidth,clip]{paper1/Images/LFEstimation.pdf}
	\caption{Radio map}
	\label{fig:LFRadioMap}
\end{figure}

A central part of a LF system is the \textit{estimation method} used for predicting locations from a radio map and currently measured network characteristics. It would, however, be very challenging to taxonomize all possible methods because nearly all methods developed for machine learning (see Witten et al. \cite{Witten2005} for a list of methods) or in the field of estimation (see Crassidis et al. \cite{Crassidis2004} for a list of methods) are applicable to the problem of LF estimation. Here we follow Krishnakumar et al. \cite{Krishnakumar2005} and divide methods only into deterministic and probabilistic methods. \emph{Deterministic methods} estimate location by considering measurements only by their value \cite{Bahl2000,Prasithsangaree2002,Laitinen2001,Smailagic2002}. \emph{Probabilistic methods} estimate location considering measurements as part of a random process \cite{Youssef2005b,Krumm2004,Castro2001,Haeberlen2004}. In Figure \ref{fig:LFEstimation} examples of applied methods for LF are shown for each of the two categories, including number of identified varieties in our literature study\footnote{However, even this simple classification is fuzzy for instance when considering the machine learning technique of support vector machines (SVMs) as applied for LF \cite{Brunato2005}. Because SVMs are defined on a probabilistic foundation but when applied for LF SVMs only consider the actual values of measurements.}. For example, the classical deterministic technique of Nearest Neighbor was identified during the literature study in twelve different variations. A comment is that many of the studied LF systems use more than one of the listed methods.

A \textit{radio map} provides a model of network characteristics in a deployment area. Radio maps can be constructed by methods which can be classified as either \emph{empirical} or \emph{model-based}. Empirical methods work with collected fingerprints to construct radio maps \cite{Youssef2005b,Bahl2000,Krumm2004,Haeberlen2004}. Model-based methods use a model parameterised for the LF-system covered area to construct radio maps \cite{Bahl2000,Roos2002b,Wallbaum04,Ji2006}.

\begin{figure}[h]
	\centering
		\includegraphics[viewport=55 320 265 506,width=0.6\textwidth,clip]{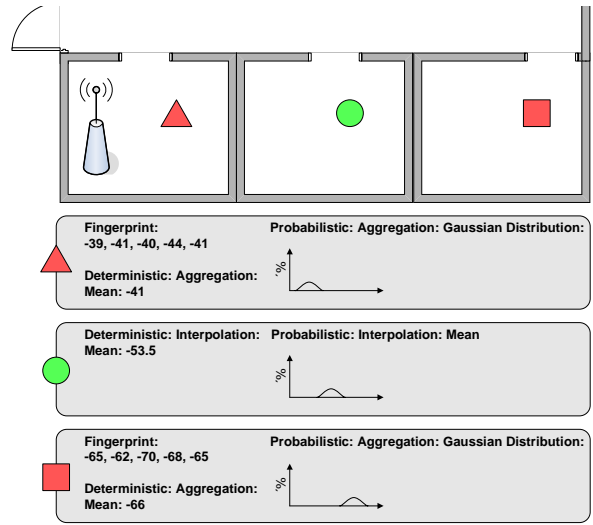}
	\caption{Deterministic and probabilistic aggregation and interpolation}
	\label{fig:LFAggInt}
\end{figure}

Empirical methods can be subdivided into \emph{deterministic} and \emph{probabilistic} methods in the same manner as estimation methods, depending on how they deal with fingerprint-collected measurements. Deterministic methods represent entries in a radio map as single values and probabilistic methods represent entries by probability distributions. Both of these can be further subcategorised into \emph{aggregation} and \emph{interpolation} methods. An aggregation method creates entries in a radio map by summarising fingerprint measurements from a single location \cite{Bahl2000,Roos2002a,Haeberlen2004,Berna2003}. Figure \ref{fig:LFAggInt} illustrates two aggregation methods for five RSS measurements at two locations marked with a triangle and a square on the figure. The first aggregation method is a deterministic mean method which takes the five measurements and finds the mean and put this value as this location's entry in the radio map. The second aggregation method is a probabilistic Gaussian distribution method which takes the five measurements and fits them to a Gaussian distribution and puts the distribution as the location's entry in the radio map. An interpolation method generate entries in a radio map at unfingerprinted locations by interpolating from fingerprint measurements or radio map entries from nearby locations \cite{Krumm2004,Krishnan2004,LaMarca2005}. Figure \ref{fig:LFAggInt} illustrates two interpolation methods at the location marked with a circle using the square-marked and triangle-marked locations as nearby locations. The first interpolation method is a deterministic mean interpolation which finds the mean of nearby radio-map entries and put this value as the entry in the radio map. The second interpolation method is a probabilistic mean method that finds the mean of nearby radio-map entries' gaussian distributions and put the mean distribution as the entry in the radio map. Two other deterministic methods are \emph{outlier removal} filtering away outliers \cite{Saha2003} and \emph{direct} creating a radio map using a direct one-to-one mapping to measurements \cite{Otsason2005}.

Model-based methods can be categorized based on how \emph{parameters} for the model are specified, how signal \emph{propagation} is modeled, and what type of \emph{representation} is used by the generated radio map. Parameters can either be given \emph{a priori} \cite{Bahl2000} or they can be \emph{estimated} from a small set of parameter-estimation fingerprints \cite{Ji2006}. Propagation can either be modeled by only considering the \emph{direct path} between a location and a base station \cite{Bahl2000} or by considering multiple paths categorized as \emph{ray tracing} \cite{Ji2006}. The representation of the generated radio map can either be \emph{deterministic} (using single values) \cite{Bahl2000} or \emph{probabilistic} (using probability distributions) \cite{Madigan2005}.

\section{Variation Taxons}
\label{paper1:sec:variation}
\begin{figure}[h]
	\centering
		\includegraphics[viewport=20 365 550 810,width=0.85\textwidth,clip]{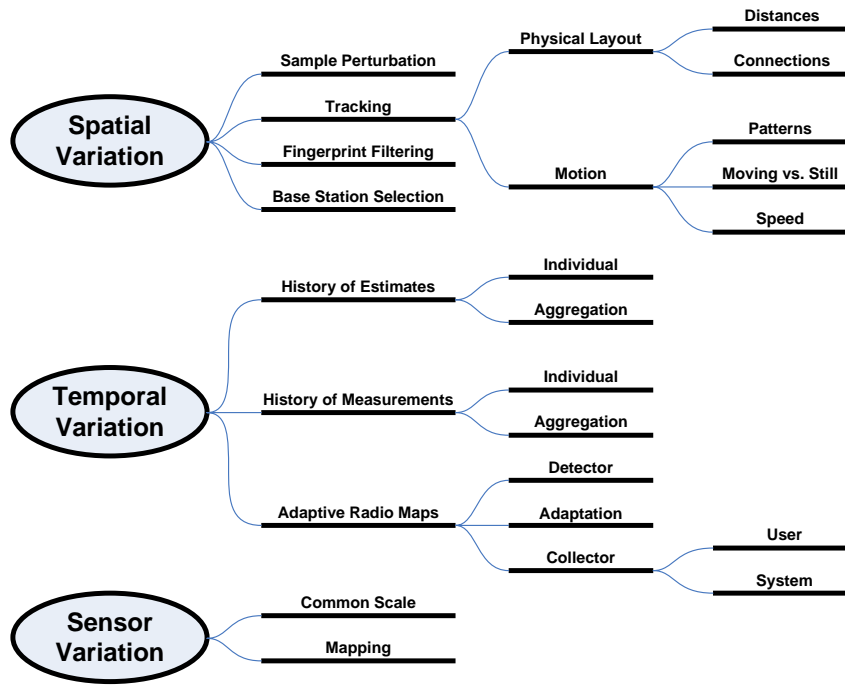}
	\caption{Spatial variations, temporal variations, and sensor variations.}
	\label{fig:LFVariation}
\end{figure}

The three taxons for variations are: \emph{spatial variations}, \emph{temporal variations}, and \emph{sensor variations}. The three taxons are shown including subtaxons in Figure \ref{fig:LFVariation}.

\textit{Spatial variations} are the observed differences in network characteristics at different locations because of signal propagation characteristics. Because of how signals propagate even small movements can create large variations in the measured network characteristics. The main method for addressing spatial variations  is \emph{tracking}: the use of constraints to optimize sequential location estimates. Tracking can be based on motion in terms of target \emph{speed} \cite{LaMarca2005,Chai2005}, target being \emph{still versus moving} \cite{Krumm2004}, and knowledge about motion \emph{patterns} \cite{Chai2005}. Tracking can also be based on physical constraints such as how \emph{connections} exist between locations \cite{Castro2001} and the \emph{distance} between them \cite{Krumm2004,Bahl2000b}. Tracking using one or several of the listed constraints is implemented using an estimation method (such as the ones listed in Section \ref{paper1:sec:estimation}) that is able to encode the constraints. Spatial variations can also be addressed by \textit{base station selection}, \textit{fingerprint filtering}, and \emph{sample perturbation}. Base station selection filters out measurements to base stations that are likely to decrease precision and accuracy \cite{Varshavsky2007,Kushki2007}. Fingerprint filtering limits the set of used fingerprints to only those that are likely to optimize precision and accuracy \cite{Kushki2007}. \emph{Sample perturbation} apply perturbation of measurements to mitigate spatial variations \cite{Youssef2005b}.

\textit{Temporal variations} are the observed differences in network characteristics over time at a single location because of continuing changing signal propagation. On a large-scale, temporal variations are the prolonged effects observed over larger periods of time such as day versus night. On a small-scale, temporal variations are the variations implied by quick transient effects, such as a person walking close to a client. Methods for handling temporal variations can be divided into methods that are based on a \emph{history of estimates}, a \emph{history of measurements}, or \emph{adaptive radio maps}. A history of either measurements or estimates here denotes a set of estimates or measurements inside a defined time window. The alternative to a history is only to use the most recent estimate or measurements. The history of either measurements or estimates can either be used as \emph{individual} \cite{Krumm2004,Haeberlen2004} measurements or estimates or, using some \emph{aggregation} \cite{Youssef2005b,Roos2002a}, can be combined to one measurement or estimate. The adaptive radio map method introduces the idea of handling temporal variations by making the radio map adapt to the current temporal variations \cite{Bahl2000b,Krishnan2004,Berna2003}. For this idea to work, some \emph{collector} has to make measurements that can be used by a \emph{detector} to control if some adaptation should be applied to the current radio map. The measurements can either be collected from the measurements a \emph{user} collects \cite{Berna2003} to run LF estimation on or it can be collected by some specially-installed \emph{system} infrastructure \cite{Bahl2000b,Krishnan2004}.  

\textit{Sensor variations} are the observed differences in network characteristics between different types of wireless clients. On a large-scale, variations can be observed between clients from different manufactures. On a small-scale, variations can be observed between different examples of similar clients. One method for addressing sensor varations is to define a \emph{common scale} and then, for each type of sensor, find out how this sensor's measurements can be converted to the common scale. A second approach is to use a single sensor to fingerprint with and then find a mapping from new sensors to the sensor that was used for fingerprinting \cite{KjaergaardLoca2006,Haeberlen2004}.

\section{Collection Taxons}
\label{paper1:sec:collection}
\begin{figure}[h]
	\centering
		\includegraphics[viewport=50 500 470 810,width=0.85\textwidth,clip]{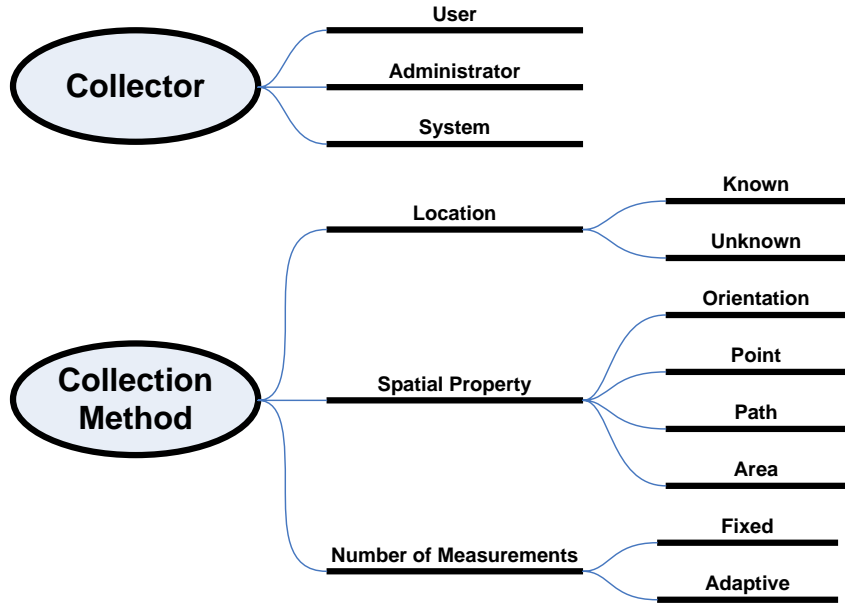}
	\caption{Collector and collection method.}
	\label{fig:LFCollection}
\end{figure}

The two taxons for fingerprint collection are \emph{collector} and \emph{collection method} as shown in Figure \ref{fig:LFCollection}.

\textit{Collector} describes who or what collect fingerprints. There are three categories: \emph{user}, \emph{administrator}, and \emph{system}. A user is a person who is either tracked by or uses information from a LF system \cite{Bhasker2004,LaMarca2005}. An administrator is a person who manages a LF system \cite{Bahl2000,Haeberlen2004,Seshadri2005} and a system is a specially-installed infrastructure for collecting fingerprints \cite{Krishnan2004}.

The fingerprints are collected following some \textit{collection method}. A collection method places assumptions on if fingerprints are collected on a \emph{location} that is either \emph{known} \cite{Otsason2005} or \emph{unknown} \cite{Madigan2005,Chai2005}. If fingerprints are collected to match a \emph{spatial property} such as: \emph{orientation} \cite{Bahl2000}, at a \emph{point} \cite{Krumm2004}, covering a \emph{path} \cite{LaMarca2005}, or covering an \emph{area} \cite{Haeberlen2004,Varshavsky2007}. If the collected \emph{number of measurements} for each fingerprint is \emph{fixed} \cite{Youssef2005b,Roos2002a} or determined based on some \emph{adaptive} strategy.

\section{Case Studies}
\label{paper1:sec:caserw}
To show the use of the proposed taxonomy, this section presents our analysis using the taxonomy on four of the 30 different systems identified in the literature study. Figure \ref{fig:LFCases} shows the analysis results in a compact form. The four systems have been selected to highlight different parts of the taxonomy. As mentioned earlier, the analysis of the rest of the analyzed systems are available online at \cite{lfwebsite} in a similar format. In addition to the eleven taxons, four extra categories describe the systems from an evaluation perspective; these are: \emph{accuracy, precision, evaluation setup and limitations}. The listed evaluation results have been taken from the original papers. Evaluation setup is grouped into \emph{stationary} (meaning that the authors' test data was collected while keeping a wireless client at a static position) or \emph{moving} (for which the wireless client was moved around mimicking normal use). 

\begin{figure}[!]
	\centering
		\includegraphics[viewport=20 170 460 830,width=1.0\textwidth,clip]{paper1/Images/LFCase.pdf}
	\caption{Analysis results for the four case studies.}
	\label{fig:LFCases}
\end{figure}

The RADAR system proposed by Bahl et al. \cite{Bahl2000} is aimed at a building scale of deployment and provides spatial locations as output. The system measures BSI, and RSS for the WaveLAN technology and roles are assigned as infrastructure-based: network. The estimation method is the deterministic k-nearest neighbor algorithm. They propose two setups, here named A and B. For A the radio map is constructed using deterministic aggregation using the mean from empirical-collected fingerprints. For B the radio map is deterministically constructed model-based considering the direct path of transmission using a priori parameters. For A an administrator will collect fingerprints at known locations standing at one point with different orientations collecting a fixed number of measurements and for B no fingerprints are collected. A limitation for setup B is that knowledge is needed of spatial locations of base stations and walls. 

The Horus system proposed by Youssef et al. \cite{Youssef2005b,Youssef2003a,Youssef2003b,Youssef2004a,Youssef2005a} also aims at a building scale of deployment and provide spatial locations as output. The system measures BSI, and RSS for the IEEE 802.11 technology and the assigned roles match infrastructure-based: terminal. The estimation method is a combination of two probabilistic techniques: discrete space estimator and center of mass. The radio-map is built using probabilistic aggregation, either based on a histogram method or on a kernel distribution method; in addition, a method for correlation modeling is also applied. To handle spatial variations sample perturbation is applied and temporal variations are handled by both mean aggregating measurements and estimates. An administrator collects fingerprints at known locations standing at one point collecting a fixed number of measurements.

The Place Lab system proposed by LaMarca et al. \cite{LaMarca2005,Cheng2005,Hightower2004} aims at a city-wide deployment and provides spatial locations as output. The system measures BSI, RSS, and RR for both IEEE 802.11 and GSM and the assigned roles match infrastructure-based: terminal. The most advanced of the system's estimation methods uses a particle filter. The radio map is built in two steps, first applying deterministic interpolation based on means and then probabilistic interpolation based on the histogram method. Spatial variations are addressed by tracking based on motion by speed constraints. The fingerprints are user collected based on paths with known location with a fixed number of measurements. A limitation is that a GPS device (and a car) is needed to practically collect fingerprints. 

The MoteTrack system proposed by Lorincz et al. \cite{Lorincz2005,Lorincz2006} targeted for sensor networks aims at building-scale deployment and provides spatial locations as output. The system has been tested in two setups, here named A and B. Setup A measures BSI, Power level, and RSS for 916 MHz FSK communication and setup B measures BSI, LQI, and RSS for IEEE 802.15.4. The roles are assigned matching infrastructure-less: collaborate with beacon nodes taking the role as base stations. The estimation method is ratio-nearest neighbor with Manhattan distance to lower computational needs. The radio map is constructed using deterministic aggregation using the mean from empirical-collected fingerprints. An administrator collects fingerprints at known locations standing at one point collecting a fixed number of measurements. A limitation is the needed deployment and maintenance of beacon nodes.

\section{Discussion}
\label{paper1:sec:discussion}
During the literature study both many similarities and differences were identified between studied systems. This can be seen from just the four included case studies in Section \ref{paper1:sec:caserw}. For instance, the well-known nearest-neighbor estimation method were identified in many variations of the basic method. The differences were not only in terms of improvements to the basic estimation method but also how systems address spatial and temporal variations. One system use a history of measurements and mean-aggregate them before applying nearest neighbor \cite{Bahl2000}. Another system use the measurements directly and use a history of estimates and aggregate these instead \cite{Varshavsky2007}. By using the proposed taxonomy these differences become clear when classifying systems. Another example also for systems based on nearest neighbor is how the radio map is built. For instance Krishnan et al. \cite{Krishnan2004} builds the radio map by applying advanced aggregation and interpolation methods where as the original system proposed by Bahl et al. \cite{Bahl2000} only use a simple aggregation based on mean values. The taxonomy also here creates a better starting point when comparing and evaluating systems.

To use the proposed taxonomy for comparison too a new system, one approach would be to, first, find classifications for compared-to existing systems. As mentioned earlier a starting point for finding such classifications is to look at our classifications online at \cite{lfwebsite}. Second, one would make a classification for the new system by classifying for each of the eleven taxons the new system's methods and assumptions according to the subtaxons. Third, one would make the comparison of the new and the existing systems. For evaluation of LF systems the taxonomy can also be used to highlight the evaluated system's assumptions and methods. This can be done by providing a classification for the evaluated system which makes it explicit what methods and assumptions are evaluated. For instance, as mentioned in the discussion above many systems have been evaluation in comparison to the nearest neighbor estimation method. But this estimation method has been implemented with many different choices when considering the used radio map and methods for addressing spatial and temporal variations. This means that it is not the same baseline method that is compared-to making results incomparable.

The taxonomy can also help scoping out future research by illustrating what research topics have not yet been covered. One way to analyse this is to group systems in terms of some of the taxons. A grouping for the taxons scale and radio map is shown in Table \ref{tab:scalevsradiomap}. The table shows that only one system aims at a campus-size scale was identified. The table also shows that generally systems either use empirical or model-based radio maps not a combination. So an open research topic is exploring the boundary between building and city-wide systems maybe by combining empirical and model-based radio maps. A grouping for the taxons spatial and temporal variations is also shown in Table \ref{tab:temporalvsspatial}. The table shows that for these taxons most systems only address one of the variations. Few systems combine them and several combinations of the different methods remain unexplored. 

\begin{table}[h]
\centering
\begin{tabular}{l|p{4cm}|p{4cm}}
 & \textit{Empirical} & \textit{Model-based} \\
\hline
\textit{Building} & \cite{Youssef2005b,Otsason2005,Lorincz2005,Bahl2000,Prasithsangaree2002,Battiti2002,Krumm2004,Castro2001,Ladd2002,Haeberlen2004,Bahl2000b,Krishnan2004,Roos2002b,Smailagic2002,Brunato2005,Berna2003,Saha2003,Chai2005,Varshavsky2007,Kushki2007,Seshadri2005,Agiwal2004,Elnahrawy2004,Yin2005} & \cite{Bahl2000,Brunato2005,Wallbaum04,Ji2006,Madigan2005,Elnahrawy2004} \\ 
\hline
\textit{Campus} & \cite{Bhasker2004} & \\
\hline
\textit{City} & \cite{Laitinen2001,LaMarca2005} & \cite{Roos2002a} \\
\end{tabular}
\caption{Grouping in terms of scale and radio map}
\label{tab:scalevsradiomap}
\end{table}

\begin{table}[h]
\centering
\begin{tabular}{p{3.2cm}|p{2cm}|p{2cm}|p{2cm}|p{2cm}|}
 & \textit{None} & \textit{History of} & \textit{History of} & \textit{Adaptive}\\
 & & \textit{Measurements} & \textit{Estimates} & \textit{Radio Maps}\\
 \hline
\textit{None} & \cite{Otsason2005,Lorincz2005,Prasithsangaree2002,Battiti2002,Roos2002a,Bhasker2004,Brunato2005,Wallbaum04,Madigan2005} & \cite{Bahl2000,Laitinen2001,Smailagic2002,Saha2003,Elnahrawy2004} & \cite{Roos2002b} & \cite{Krishnan2004,Yin2005}\\
\hline
\textit{Sample Perturbation} & & \cite{Youssef2005b} & \cite{Youssef2005b} & \\
\hline
\textit{Tracking} & \cite{Castro2001,LaMarca2005,Ji2006,Berna2003,Seshadri2005,Agiwal2004} & \cite{Haeberlen2004,Bahl2000b,Chai2005} & \cite{Krumm2004,Ladd2002,Haeberlen2004} & \cite{Haeberlen2004,Bahl2000b}\\
\hline
\textit{Fingerprint Filtering} & \cite{Kushki2007} & & & \\
\hline
\textit{Base Station Selection} & \cite{Kushki2007} & & & \\
\hline
\end{tabular}
\caption{Grouping in terms of spatial and temporal variations}
\label{tab:temporalvsspatial}
\end{table}

We do not expect that the proposed taxonomy is complete in its current form. Instead, it is intended to enable better and more complete understanding of LF and to evolve as that understanding improves. At the same time, we feel that our eleven main taxons and many of the subtaxons are fairly stable. During the process of creating the taxonomy, analyzing papers and classifying systems, we found that all 30 systems and their methods could be classified. On the other hand, some of the subtaxons are likely to evolve as our understanding of LF evolves. An area for which it would be interesting to extend the taxonomy is for non-functional properties as mentioned in Section \ref{paper1:sec:tax}. However, only a limited number of papers have so far been published in this direction \cite{Youssef2005b,Lorincz2005,KjaergaardPervasive2007}.

\section{Conclusion}
\label{paper1:sec:conclusion}
This paper presented a taxonomy for location fingerprinting. The proposed taxonomy was constructed from a literature study of 51 papers and articles about LF. The taxonomy consists of the following eleven taxons: \textit{scale}, \textit{output}, \textit{measurements}, \textit{roles}, \textit{estimation method}, \textit{radio map}, \textit{spatial variations}, \textit{temporal variations}, \textit{sensor variations}, \textit{collector}, and \textit{collection method}. The 51 analyzed papers described 30 LF systems of which four were presented as case studies.

Valuable taxonomies can account for everything that is known so far and can predict things to come, as variations of parameters accounted for and enumerated in the taxonomy. A taxonomy first and foremost shows the depth and the breadth of our understanding. We would like others to join and based on inputs from the community further improve the proposed taxonomy.

\section*{Acknowledgements}
The author would like to thank Doina Bucur, Azadeh Kushki and the reviewers for their insightful comments on earlier drafts of this paper. The research reported in this paper was partially funded by the software part of the ISIS Katrinebjerg competency centre http://www.isis.alexandra.dk/software/.


\clearemptydoublepage
\chapter{Paper 2}
\label{chap:loca2006}

The paper \emph{Automatic Mitigation of Sensor Variations for Signal Strength Based Location Systems} presented in this
chapter has been published as a workshop paper~\cite{KjaergaardLoca2006}.

\begin{publist}{\cite{KjaergaardLoca2006}}
  \item[\cite{KjaergaardLoca2006}] M.\ B.\ Kjærgaard. Automatic Mitigation of Sensor Variations for Signal Strength Based Location Systems. In \emph{Proceedings of the Second International Workshop on Location and Context Awareness}, pages~30--47, Springer, 2006.
\end{publist}

\noindent

\clearemptydoublepage



\mytitle{Automatic Mitigation of Sensor Variations for Signal Strength Based Location Systems}{ 
  Mikkel Baun Kj\ae rgaard\footnotemark[1]}{  
  \footnotetext[1]{Department of Computer Science, University of
    Aarhus, IT-parken, Aabogade 34, DK-8200 Aarhus N, Denmark. E-mail:
    \texttt{mikkelbk@daimi.au.dk}.}
      } 
        

\begin{myabstract}
In the area of pervasive computing a key concept is context-awareness. One type of context information is location information of wireless network clients. Research in indoor localization of wireless network clients based on signal strength is receiving a lot of attention. However, not much of this research is directed towards handling the issue of adapting a signal strength based indoor localization system to the hardware and software of a specific wireless network client, be it a tag, PDA or laptop. Therefore current indoor localization systems need to be manually adapted to work optimally with specific hardware and software. A second problem is that for a specific hardware there will be more than one driver available and they will have different properties when used for localization. Therefore the contribution of this paper is twofold. First, an automatic system for evaluating the fitness of a specific combination of hardware and software is proposed. Second, an automatic system for adapting an indoor localization system based on signal strength to the specific hardware and software of a wireless network client is proposed. The two contributions can then be used together to either classify a specific hardware and software as unusable for localization or to classify them as usable and then adapt them to the signal strength based indoor localization system.
\end{myabstract}

\section{Introduction}
\label{paper2:sec:introduction}
In the area of pervasive computing a key concept is context-awareness. One type of context information is location information of wireless network clients. Such information can be used to implement a long range of location based services. Examples of applications are speedier assistance for security personnel, health-care professionals or others in emergency situations and adaptive applications that align themselves to the context of the user. The implementation of speedier assistance could, for example, come in the form of a tag with an alarm button that, when pressed, alerts nearby persons to come to assistance. The alarm delivered to the people nearby would contain information on where in the physical environment the alarm was raised and by whom. Applications that adapt themselves to the context they are in are receiving a lot of attention in the area of pervasive computing, where they can solve a number of problems. One type of context information is location which can be used in its simplest form to implement new services optimized based on the location information.

One type of indoor location system, which can be used to support the above scenarios, is systems based on signal strength measurements from an off-the-shelf 802.11 wideband radio client (WRC). The WRC can be in the form of either a tag, phone, PDA or laptop. Such systems need to address several ways in which the signal strength can vary. The variations can be grouped into \textit{large} and \textit{small-scale} \textit{spatial}, \textit{temporal}, and \textit{sensor} variations as shown in Table \ref{paper2:tab:ssvar}. The spatial variations can be observed when a WRC is moved. Large-scale spatial variations are what makes localization possible, because the signal strength depends on how the signals propagate. The small-scale spatial variations are the variations that can be observed when moving a WRC as little as one wave length. The temporal variations are the variations that can be observed over time when a WRC is kept at a static position. The large-scale temporal variations are the prolonged effects observed over larger periods of time; an example is the difference between day and night where during daytime the signal strength is more affected by people moving around and the use of different WRCs. The small-scale temporal variations are the variations implied by quick transient effects such as a person walking close to a WRC. The sensor variations are the variations between different WRCs. Large-scale variations are the variations between radios, antennas, firmware, and software drivers from different manufactures. Small-scale variations are the variations between examples of the same radio, antenna, firmware, and software drivers from the same manufacture. The chosen groupings are based on the results in \cite{Haeberlen2004,Youssef2005b}.

\begin{table}[h]
\begin{tabular}{|p{2.8cm}|p{3cm}|p{3cm}|p{3cm}|}
\hline
 & Spatial & Temporal & Sensor\\
\hline
Small-scale & Movement around one wavelength & Transient effects & Different examples of the same WRC combination \\
\hline
Large-scale & Normal movement & Prolonging effects & Different WRC combinations \\
\hline
\end{tabular}
\caption{Signal strength variations}
\label{paper2:tab:ssvar}
\end{table}

Most systems based on signal strength measurements from off-the-shelf 802.11 wideband radio clients do not address the above variations explicitly, with \cite{Haeberlen2004} and \cite{Youssef2005b} as exceptions. Especially the handling of sensor variations has not been given much attention. Therefore current location systems have to be manually adapted by the provider of the location system for each new type of WRC to work at its best. This is not optimal considering the great number of combinations of antennas, firmware, and software drivers for each radio. To the users the large-scale sensor variation poses another problem, because the different implementations of firmware and software drivers have different properties with respect to localization. To the users it would therefore be of help if the system could automatically evaluate if the firmware and software drivers installed could be used for localization.

The contribution of this paper is twofold. To solve the problem of large-scale sensor variations, an automatic system is proposed for adapting an indoor localization system based on signal strength to the specific antenna, radio, firmware, and software driver of a WRC. To solve the problem of evaluating different sensors, an automatic system for evaluating the fitness of a specific combination of antenna, radio, firmware, and software driver is proposed. The two contributions can then be used together to either classify a combination of antenna, radio, firmware, and software drivers as unusable for localization or to classify them as usable and then adapt them to the signal strength based indoor localization system.

The methods proposed for providing automatic classification and adaptation are presented in Section 2. The results of applying these methods to 14 combinations of antennas, radios, firmware, and software are given in Section 3. Afterwards the results are discussed in Section 4 and finally conclusions are given in Section 5.
		
\subsection{Related Work}
Research in the area of indoor location systems, as surveyed in \cite{Sun2005,Muthukrishnan2005}, spans a wide range of technologies (wideband radio, ultra-wideband radio, infrared,...), protocols (IEEE 802.11,802.15.1,...), and algorithm types (least squares, bayesian, hidden markov models, ...). Using these elements the systems estimate the location of wireless entities based on different types of measurements such as time, signal strength, and angles. Systems based on off-the-shelf 802.11 wideband radio clients using signal strength measurements have received a lot of attention. One of the first systems was RADAR \cite{Bahl2000}, that applied different deterministic mathematical models to calculate the position in coordinates of a WRC. The mathematical models used had to be calibrated for each site where the systems had to be used. In comparison to RADAR, later systems have used probabilistic models instead of mathematical models. This is because a good mathematical model which can model the volatile radio environment has not been found. As in the case of the mathematical models in RADAR, the probabilistic models should also be calibrated for each site. Examples of such systems determining the coordinates of a WRC are published in \cite{Youssef2005b,Krumm2004,Roos2002b,Ladd2002} and systems determining the logical position or cell of a WRC are published in \cite{Haeberlen2004,Castro2001,Locher2005}\footnote{The system in \cite{Castro2001} uses the signal to noise ratio instead of the signal strength}. Commercial positioning systems also exist such as Ekahau \cite{Ekahau} and PanGo \cite{PanGo}. In the following, related work is presented with respect to how the systems address the signal strength variations introduced above.

Small-scale spatial variations are addressed by most systems using a method to constrain how the location estimate can evolve from estimate to estimate. The method used for the system in \cite{Roos2002b} is to average the newest estimate with previous estimates. In \cite{Haeberlen2004,Krumm2004,Ladd2002,Kontkanen2004} more advanced methods based on constraining the estimates using physical properties are proposed. The constraints include both the layout of the physical environment and the likely speed by which a WRC can move. One way these constraints can be incorporated in a probabilistic model is to use a Hidden Markov Model to encode the constraints with. In \cite{Youssef2005b} another method is proposed which in the case of movement triggers a perturbation technique that addresses the small-scale variations. In \cite{Bahl2000b} a graph-inspired solution is presented which weights measurements based on the physical distance between location estimates. Large-scale spatial variations are, as stated in the introduction, the variation which makes indoor location system using signal strength possible. The different methods for inferring the location are a too extensive area to cover here in detail. Some examples of different types of systems were given above.

Small-scale temporal variations can be addressed using several techniques. The first concerns how the probabilistic model is build from the calibration measurements. Here several options exist: the histogram method \cite{Krumm2004,Roos2002b,Ladd2002}, the Gaussian kernel method \cite{Roos2002b}, and the single Gaussian distribution \cite{Haeberlen2004}. The second technique is to include several continuous measurements in the set of measurements used for estimating the location. By including more measurements quick transient effects can be overcome. This can be done as in \cite{Haeberlen2004,Roos2002b}, where the measurements are used as independent measurements or as in \cite{Youssef2005b}, where a time-averaging technique is used together with a technique which addresses the correlation of the measurements. Large-scale temporal variations have been addressed in \cite{Bahl2000b} based on extra measurements between base stations, which were used to determine the most appropriate radio map. In \cite{Haeberlen2004} a method is proposed were a linear mapping between the WRC measurements and the radio map is used. The parameters of this mapping can then be fitted to the characteristics of the current environment which addresses the large-scale temporal variations.

Small-scale sensor variations have not been explicitly addressed in earlier research. One reason for this is that the small variations between examples often are difficult to measure, because of the other variations overshadowing it. Therefore there exist no general techniques, but possibly the techniques for the large-scale sensor variations could be applied. For large-scale sensor variations \cite{Haeberlen2004} proposed applying the same linear approximation as in the case of large-scale temporal variations. They propose three different methods for finding the two parameters in the linear approximation. The first method is a manual one, where a WRC has to be taken to a couple of known locations to collect measurements. For finding the parameters they propose to use the method of least squares. The second method is a quasi-automatic one where a WRC has to be taken to a couple of locations to collect measurements. For finding the parameters they propose using the confidence value produced when doing Markov localization on the data and then find the parameters that maximize this value. The third is an automatic one requiring no user intervention. Here they propose using an expectation-maximation algorithm combined with a window of recent measurements. For the manual method they have published results which show a gain in accuracy for three cards; for the quasi-automatic method it is stated that the performance is comparable to that of the manual method, and for the automatic one it is stated that it does not work as well as the two other techniques. 

The methods proposed in this paper to solve the problem of large-scale sensor variations are a more elegant and complete solution than the method proposed in \cite{Haeberlen2004}. It is more elegant, because it uses the same type of estimation technique for both the manual, quasi-automatic, and automatic case. It is more complete, because it can recognize WRCs that cannot be used for localization. Also it has been shown to work on a larger set of WRC combinations with different radios, antennas, firmware, and software drivers.

\section{Methods for classification and normalization}
\label{paper2:sec:methods}
A cell based indoor localization system, such as the ones proposed in \cite{Haeberlen2004,Castro2001}, should estimate the probability of a WRC being in each of the cells which the system covers. A cell is here normally a room or part of a room in larger rooms or a section of a hallway. Formally a set S = \{s$_1$,...,s$_n$\} is a finite set of states where each state corresponds to a cell. The state $s^*$ is the state of the WRC that should be located. The location estimate of the WRC can then be denoted by a probability vector $\vec{\pi}$ with each entry of the vector denoting the probability that the WRC is in this particular state $\vec{\pi}_i = P(s^* = s_i)$.

To solve the localization problem the vector $\vec{\pi}$ has to be estimated, which is addressed by infrastructure-based localization using two types of measurements. First, there are the measurements M = \{m$_1$,...,m$_s$\} reported by the WRC, which is to be located. Second, there is a set C = \{c$_1$,...,c$_t$\} of calibration measurements collected prior to the launch of the location service. Each measurement is defined as $M = V \times B$ where B = \{b$_1$,...,b$_k$\} is the set of base stations and V = \{0,...,255\} is the set of signal strength values for 802.11 WRCs. The calibration measurements are collected to overcome the difficulties in localizing clients in the volatile indoor radio environment. 

The estimation of the vector $\vec{\pi}$ based on the two types of measurements can be divided into three sub-problems. The first problem is the normalization problem, which adresses how WRC-dependent measurements are transformed into normalized measurements. The reason the measurements need to be normalized is that otherwise they cannot be combined with the calibration measurements which have most often not been collected by the same WRC. The next problem, state estimation, is how the normalized measurements are transformed into a location estimate. The last problem, tracking, is how the physical layout of the site and prior estimates can be used to enrich the location estimate. In respect to these problems, it is the problem of normalization made in an automatic fashion that this paper addresses. For evaluating the proposed methods in the context of a localization system an implementation based on the ideas in \cite{Haeberlen2004} without tracking is used.

In the following sections methods are proposed for solving the problem of automatic normalization (Section 2.3-2.6) and the problem of classifying the fitness of a WRC for localization automatically (Section 2.2). The solutions are stated in the context of indoor localization system using signal strength measurements from off-the-shelf 802.11 wideband radio clients. However, the solutions could be applied to other types of radio clients which can measure signal strength values.

\subsection{Automatic Still Period Analyzer}
In the proposed methods an analyzer, called an automatic still period analyzer, is used to divide measurements into groups of measurements from single locations. The idea behind the analyzer is that, if we can estimate if a WRC is still or moving, we can place a group of still measurements in one location. One thing to note here is that localization cannot be used to infer this information, because the parameters for adapting the WRC to the localization system have not yet been found. The still versus moving estimator applied is based on the idea in \cite{Krumm2004} of using the variations in the signal strength to infer moving versus still situations. To do this, the sample variation is calculated for the signal strength measurements in a window of 20 seconds. The estimation is then based on having training data from which distributions of the likelihood of the WRC being still or moving at different levels of variations is constructed. To make a stable estimate from the calculated variations and likelihood distributions a Hidden Markov Model (HMM) is applied as estimator with the parameters proposed in \cite{Krumm2004}. To evaluate the implemented estimator two walks were collected with the lengths of 44 minutes and 27 minutes, respectively, where the person collecting the walks marked in the data when he was still or moving. These two walks were then used in a simulation, where one was used as training data to construct the likelihood distributions and the other as test data. The results were that 91\% of the time the estimator made the correct inference and with a small number of wrong transitions between still and moving because of the HMM as experienced in \cite{Krumm2004}. However, the estimator performs even better when only looking at still periods, because the errors experienced are often that the estimator infers moving when the person is actually still.

The estimator used here differs in two ways with respect to the method proposed in \cite{Krumm2004}. First, weighted sample variations for all base stations in range are used instead of the sample variation for the strongest base station. This was chosen because our experiments showed this to be more stable. Second, the Gaussian kernel method is used instead of the histogram method to construct the likelihood distributions. One thing to note is that the estimator does not work as well with WRC combinations, which cache measurements or have a low update frequency.

\subsection{Fitness classifier}
Methods for classifying the fitness of a single combination of antenna, radio, firmware, and software drivers for localization are presented. To make such a classifier, it first has to be defined what makes a combination fit or unfit. A good combination has some of the following characteristics: the radio has high sensitivity so that it can see many bases, has no artificial limits in the signal strength values, does not cache the signal strength values, and has a high update frequency.\footnote{Pure technical constraints, such as cards that can not return signal strength values, are not addressed in this paper.} On the other hand, a bad combination has low sensitivity, limits the signal strength values, the signal strength values reported do not represent the signal strength but some other measurements, such as the link quality, caches the measurements, and has a low update frequency.

\begin{figure}[h]
	\centering
		\includegraphics[viewport=0 245 810 565,width=0.84\textwidth,clip]{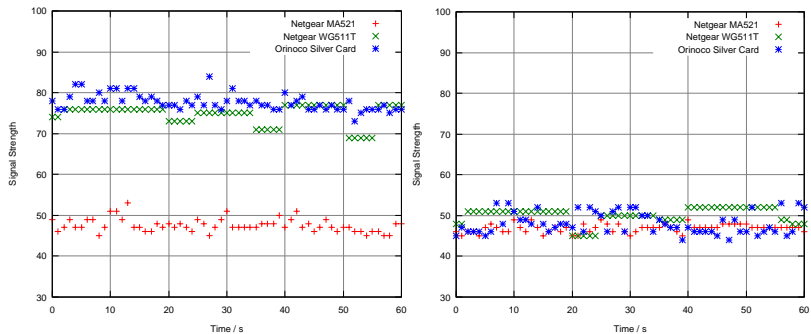}
	\caption{Plots of signal strength measurements from different cards and base stations at the same location.}
	\label{paper2:fig:plotsforsensoreffects}
\end{figure}

To illustrate the effects of good and bad combinations on data collected from several WRCs, Figure \ref{paper2:fig:plotsforsensoreffects} shows signal strength measurements for different WRCs taken at the same location and at the same time, but for two different 802.11 base stations. On the first graph the effect of caching or low update rate for the Netgear WG511T card can be seen, because the signal strength only changes every five seconds. By comparing the two graphs, the effect of signal strength values not corresponding to the actual signal strength can be seen for the Netgear MA521 card. This is evident form the fact that the signal strength values for the Netgear MA521 card does not change when the values reported by the other cards change for specific base stations. 

In the following it is assumed that, for evaluating the fitness of a WRC combination, five minutes of measurements are available. The measurements should be taken in an area where at least three base stations are in range at all times. The measurements should be taken over five minutes and the WRC combination should be placed at four different locations for around 30-60 seconds. Of course, the techniques could be applied without these requirements. The system could, for instance, collect measurements until it had inferred that the WRC combination had been placed at four locations. Then it would of course depend on the use of the WRC combination when enough measurements have been collected.

To automatically evaluate the fitness of a specific combination, methods for finding the individual faults are proposed. For caching or low update frequency a method using a naive Bayesian estimator \cite{Witten2005} based on the autocorrelation coefficient is proposed. For measurements that do not correspond to the signal strength a method using a naive Bayesian estimator based on the variations between measurements to different base stations at the same place is proposed. For artificial limits a min/max test can be applied, but it is difficult to apply in the five minutes scenario, because data for a longer period of time is needed. For sensitivity a test based on the maximum number of bases can be used, but requires data for a longer period of time. The evaluation of the two last methods has not been carried out and is therefore left as future work.

\subsubsection{Caching or low update frequency}
To evaluate if a combination is caching or has a low update frequency the signal strength measurements for each base station are treated as time series. Formally, let $m_{t,j}$ be the signal strength measurement of time $t$ and for base station $b_j$. The autocorrelation coefficient\cite{Chatfield2003} $r_{k,j}$ is then for base station $b_j$ with lag $k$ where $\overline{m_j}$ is the mean of the signal strength measurements for base station $b_j$:

\begin{equation}
	r_{k,j} = \frac{\sum_{t=1}^{N-k} (m_{t,j} - \overline{m_j})(m_{t+k,j} - \overline{m_j})}{\sum_{t=1}^{N} (m_{t,j} - \overline{m_j})^{2}}
\end{equation}

\begin{figure}[h]
	\centering
		\includegraphics[viewport=0 230 470 590,width=0.46\textwidth,clip]{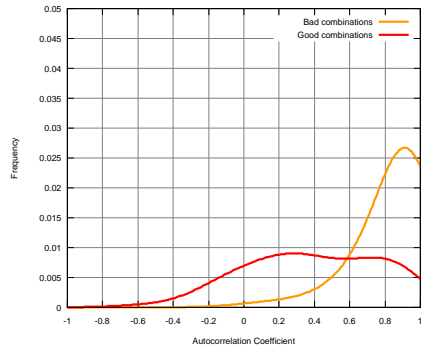}
	\caption{Plot of the likelihood for different autocorrelation coefficients that the WRC combination is good or bad}
	\label{paper2:fig:plotfrequenceAutoCoor}
\end{figure}

$r_{k,j}$ is close to 1.0 when the measurements are in perfect correlation and close to -1.0 when in perfect anticorrelation. This can be used to detect WRC combinations that are caching or has a low update frequency because the autocorrelation coefficient will in these cases be close to 1.0. The autocorrelation coefficient is then calculated from signal strength measurements for different base stations and different lags. Based on initial experiments lag 1 and 2 were used in the evaluations. These coefficients are then used with a naive Bayesian estimator to calculate the probability of the WRC combination is caching or having a low update frequency. To construct the likelihood function for the naive Bayesian estimator, a training set of known good and bad combinations with respect to caching or low update frequency are used. The examples in the training set were classified by the author. A likelihood function constructed from the training data used in one of the evaluations is plotted in Figure \ref{paper2:fig:plotfrequenceAutoCoor}. The Figure shows the likelihood for different autocorrelation coefficients that the WRC combination is good or bad.

\subsubsection{Measurements do not correspond to signal strength values}
The best test to determine if measurements do not correspond to signal strength measurements is to calculate if the measurements at a known location correlate with measurements from a known good combination. However, this can not be used in an automatic solution. Another way to automatically test this is to calculate the average sample variation for measurements to different base stations. It is here assumed that if the measurements do not correspond to signal strength values they will be more equal for different base stations. One example of this is the Netgear MA521 as shown in the plot in Figure \ref{paper2:fig:plotsforsensoreffects}.

\begin{figure}[h]
	\centering
		\includegraphics[viewport=0 230 470 590,width=0.46\textwidth,clip]{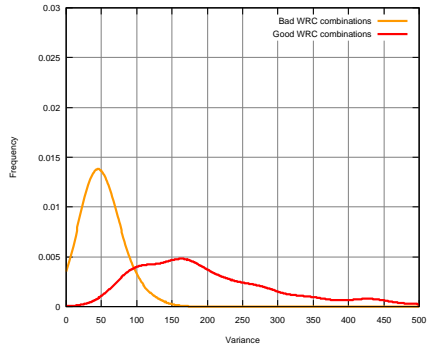}
	\caption{Plot of the likelihood for different average sample variations that the WRC combination is good or bad}
	\label{paper2:fig:plotfrequenceSelfCoor}
\end{figure}

The calculated average sample variation is used as input to a naive Bayesian estimator. The estimator calculates the probability that a combination's measurements do not correspond to the signal strength. It is assumed in the evaluation that measurements are collected for at least three base stations at each location. To construct the likelihood function for the naive Bayesian estimator, a training set of known good and bad combinations with respect to correspondence to signal strength is used. A likelihood function constructed from the training data used in one of the evaluations is plotted in Figure \ref{paper2:fig:plotfrequenceSelfCoor}. The Figure shows the likelihood for different average sample variations that the WRC combination is good or bad.

\subsection{Normalization}
In the following sections the methods proposed for normalizing the measurements reported by WRC combinations are presented. The measurements are normalized with respect to the measurements reported by the WRC combination that was used for calibrating the deployment site of the localization system. The first method is a manual method in which a user has to take a WRC to a number of known locations and collect measurements. The second is a quasi-automatic method where the user has to take the WRC to some unknown locations and collect measurements. The third is an automatic solution where there is no need for initial data collection, the user can just go to locations and use the WRC. The formulation of these three types of methods is the same as in \cite{Haeberlen2004}, however, this work applies other techniques to solve the problems. As done in \cite{Haeberlen2004}, it is assumed that a linear model can be used to relate measurements from one combination to another. The reason this is a reasonable assumption is that most WRC combinations use a linearized scale for the reported signal strength values. Formally, $c(i) = c1*i+c2$, where c1 and c2 are two constants, $i$ is the normalized signal strength that can be compared with the calibration observations, and c(i) is the signal strength of the combination. 

\subsection{Manual Normalization}
To solve the problem of manual normalization, the method of linear least squares \cite{Crassidis2004} is used. In stead of applying this method to the individual signal strength measurements, the mean $\mu_{o_{i,j}}$ and the standard deviation $\sigma_{o_{i,j}}$ of the measurements for some state $s_i$ and base station $b_j$ are used. For the calibration measurements also the the mean $\mu_{c_{i,j}}$ and the standard deviation $\sigma_{c_{i,j}}$ of the measurements for some state $s_i$ and base station $b_j$ are used. Formally, a linear observation model is assumed, where $x$ is the true state, $\widetilde{y}$ is the measurement vector and v the measurement error:
\begin{equation}
	\widetilde{y} = Hx + v
\end{equation}

To make an estimate of $c1$ and $c2$ denoted by $\widehat{x}$, the following definitions are used for $\widehat{x}$, $\widetilde{y}$ and $H$. It is assumed that a set of observations for some subset of $S$ denoted by $1$ to $r$ and some subset of base stations for each location denoted by $1$ to $s$ are given.

\begin{equation}
\widehat{x} = [c1,c2] \ \ \
\widetilde{y} = \left[\begin{array}{c}
\mu_{o_{1,1}}\\
\sigma_{o_{1,1}}\\
\vdots\\
\mu_{o_{1,s}}\\
\sigma_{o_{1,s}} \\
\vdots\\
\mu_{o_{r,1}}\\
\sigma_{o_{r,1}}\\
\vdots\\
\mu_{o_{r,s}}\\
\sigma_{o_{r,s}}
\\\end{array} \right] \ \
H = \left[\begin{array}{cc}
\mu_{c_{1,1}} & 1.0\\
\sigma_{c_{1,1}} & 0.0\\
\vdots & \vdots \\
\mu_{c_{1,s}} & 1.0\\
\sigma_{c_{1,s}} & 0.0\\
\vdots & \vdots \\
\mu_{c_{r,1}} & 1.0\\
\sigma_{c_{r,1}} & 0.0\\
\vdots & \vdots \\
\mu_{c_{r,s}} & 1.0\\
\sigma_{c_{r,s}} & 0.0\\
\end{array} \right]
\end{equation}

The relations between $c_1$ and $c_2$ and the mean and deviations comes from the following two equations \cite{Berry1996}.
\begin{equation}
\mu_{o_{i,j}} = c_1*\mu_{c_{i,j}} + c_2 \ \ \
\end{equation}
\begin{equation}
\sigma_{o_{i,j}} = c_1*\sigma_{c_{i,j}}
\end{equation}

By using linear least squares an estimate of $\widehat{x}$ is found using:
\begin{equation}
	\widehat{x} = (H^{T}H)^{-1}H^{T}\widetilde{y}
\end{equation}

\subsection{Quasi-automatic Normalization}
To solve the problem of quasi-automatic normalization, the method of weighted least squares \cite{Crassidis2004} is used. Since the locations of the measurements are unknown they have to be compared to all possible locations. But some locations are more likely than others and therefore weights are use to incorporate this knowledge. It is assumed that a set of observations for some unknown subset of $S$ denoted by $1$ to $r$ and some subset of base stations for each unknown location denoted by $1$ to $s$ are given.

First $\widetilde{y_i}$ and $H_i$ are defined as:
\begin{equation}
\widetilde{y_i} = \left[\begin{array}{c}
\mu_{o_{i,1}}\\
\sigma_{o_{i,1}}\\
\vdots\\
\mu_{o_{i,1}}\\
\sigma_{o_{i,1}}\\
\vdots\\
\mu_{o_{i,s}}\\
\sigma_{o_{i,s}}\\
\vdots\\
\mu_{o_{i,s}}\\
\sigma_{o_{i,s}}\\
\end{array} \right] \ \
H_i = \left[\begin{array}{cc}
\mu_{c_{1,1}} & 1.0\\
\sigma_{c_{1,1}} & 0.0\\
\vdots & \vdots \\
\mu_{c_{n,1}} & 1.0\\
\sigma_{c_{n,1}} & 0.0\\
\vdots & \vdots \\
\mu_{c_{1,s}} & 1.0\\
\sigma_{c_{1,s}} & 0.0\\
\vdots & \vdots \\
\mu_{c_{n,s}} & 1.0\\
\sigma_{c_{n,s}} & 0.0\\
\end{array} \right]
\end{equation}
With these definitions $\widehat{x}$, $\widetilde{y}$ and $H$ can be defined as:
\begin{equation}
\widehat{x} = [c1,c2] \ \ \
\widetilde{y} = \left[\begin{array}{c}
\widetilde{y_1}\\
\vdots\\
\widetilde{y_r}\\
\end{array} \right] \ \
H = \left[\begin{array}{c}
H_1 \\
\vdots \\
H_r \\
\end{array} \right]
\end{equation}
The weight matrix W is then defined as:
\begin{equation}
W = diag(w_{1,1},...,w_{1,n},...,w_{r,1},...,w_{r,n})
\end{equation}
Two methods are proposed for the definition of $w_{i,j}$, where $i$ is an observation set from an unknown location and $j$ denotes a known location. The first method is to attempt to apply bayesian localization with the ith observation set from an unknown location and to define $w_{i,j} = \vec{\pi_j}$. The second method is a comparison method which tries to match the means and standard deviations of the observations and calibration observations using the following definition, where $O_{i,k}\sim  \mathcal{N}(\mu_{o_{i,k}},\sigma_{o_{i,k}})$ and $C_{j,k}\sim  \mathcal{N}(\mu_{c_{j,k}},\sigma_{c_{j,k}})$, where $w_{i,j}$ can be defined as:

\begin{equation}
w_{i,j} = \frac{1}{s}\sum_{k=1}^s\sum_{v=0}^{255} \min (P(v-0.5<O_{i,k}<v+0.5), P(v-0.5<C_{j,k}<v+0.5))
\end{equation}

By using weighted least squares an estimate of $\widehat{x}$ is then found using:
\begin{equation}
	\widehat{x} = (H^{T}WH)^{-1}H^{T}W\widetilde{y}
\end{equation}

\subsection{Automatic Normalization}
To solve the problem of automatic normalization, the automatic still period analyzer is used. Given signal strength measurements from five minutes, the analyzer is used to divide the data into parts which come from the same location. These data are then used with the solution for quasi-automatic normalization. If, however, the automatic still period analyzer is unable to make such a division the complete set of measurements from the five minutes is used.

\section{Results}
\label{paper2:sec:evaluation}
\begin{figure}
	\centering
		\includegraphics[width=0.80\textwidth]{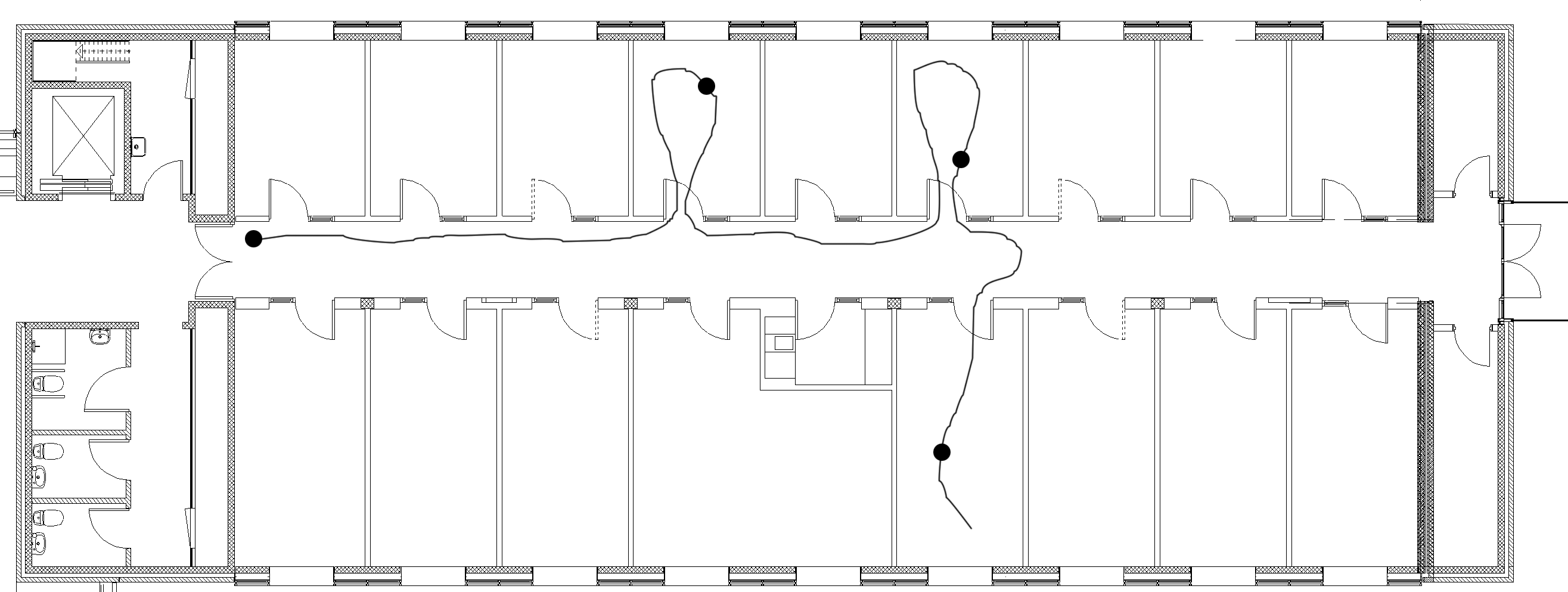}
	\caption{Floor layout with walking path}
	\label{paper2:fig:ada-floor2}
\end{figure}

In this section evaluation results are presented for the proposed methods based on collected measurements. The measurements used in the evaluation were collected in an 802.11 infrastructure installed at the Department of Computer Science, University of Aarhus. Two types of measurements were collected, and for both types the signal strength to all base stations in range was measured every second. The first type was a set of calibration measurements collected using WRC combination number 11 from Table \ref{paper2:tab:wrccombination}. The calibration set covers 18 cells spread out over a single floor in a office building as shown on Figure \ref{paper2:fig:ada-floor2}. The second type of measurements were walks collected by walking a known route on the same floor where the calibration set was collected. Each walk lasted for around 5 minutes and went through 8 of the cells; in four cells the WRC combination was placed at a single spot, each shown as a dot in Figure \ref{paper2:fig:ada-floor2}, for around a minute. Two walks were collected for each of the WRC combinations listed in Table \ref{paper2:tab:wrccombination} on different days. For collecting the measurements on devices running Windows XP, Mac OS X or Windows Mobile 2003 SE, the Framework developed as part of the Placelab\cite{placelab} project was used. For the single WRC combination installed on a device running Linux a shell script was used to collect the measurements.

\begin{table}[h]
\begin{flushleft}
\tiny
\begin{tabular}{|p{2.5cm}|c|c|c|c|}
\hline
\small Product name & \small Antenna & \small Firmware/Driver & \small OS & \small Classification\\
\hline
1. AirPort Extreme (54 Mbps) & In laptop & OS provided & Mac OS X (10.4) & Good \\
\hline
2. D-Link Air DWL-660 & In card & D-Link 7.44.46.450 & Windows XP & Good \\
\hline
3. Fujitsu Siemens Pocket Loox 720 & In PDA & OS provided & Windows Mobile 2003 & Caching/Low Freq \\
\hline
4. Intel Centrino 2100 3B & In laptop & Intel 1.2.4.35 & Windows XP & Caching/Low Freq \\
\hline
5. Intel Centrino 2200BG & In laptop & Intel 9.0.2.31 & Windows XP & Caching/Low Freq \\
\hline
6. Intel Centrino 2200BG & In laptop & Kernel provided(ipw2200) & Debian (2.6.14) & Caching/Low Freq \\
\hline
7. Netgear MA521 & In card & Netgear 5.148.724.2003 & Windows XP & Not SS \\
\hline
8. Netgear WG511T & In card & Netgear 3.3.0.156 & Windows XP & Caching/Low Freq \\
\hline
9. Netgear WG511T (g disabled)& In card & Netgear 3.3.0.156 & Windows XP & Caching/Low Freq \\
\hline
10. NorthQ-9000& In dongle & ZyDAS ZD1201 & Windows XP & Good \\
\hline
11. Orinoco Silver & In card & OS provided (7.43.0.9) & Windows XP & Good \\
\hline
12. Ralink RT2500 & In dongle & Ralink 2.1.10.0 & Windows XP & Good \\
\hline
13. TRENDnet TEW-226PC & In card & OEM 5.140.521.2003 & Windows XP & Not SS \\
\hline
14. Zcom XI-326HP+ & In card & Zcom 4.0.7 & Windows XP & Good \\
\hline
\end{tabular}
\end{flushleft}
\caption{WRC combinations with classification, where \textit{Not SS} means that the reported values do not correspond to signal strength values.}
\label{paper2:tab:wrccombination}
\end{table}

\subsection{Classifier}
To evaluate the proposed classifiers for evaluating the fitness of a WRC combination for localization, the walks collected as explained above were used. In Table \ref{paper2:tab:wrccombination} the different classifications for the WRC combinations are shown. These classifications were made by the author by inspecting the measured data from the WRC combinations.

Two evaluations were made to test if the proposed method can predict if a WRC combination caches measurements or has a long scanning time. For the first evaluation for each of the WRC combinations, one of the walks was used as training data and the other as test data. This tests if the methods can make correct predictions regardless of the influence of small and large-scale temporal variations. The results from this evaluation are given in Table \ref{paper2:tab:classificationresults} and show that the method was able to classify all WRC combinations correctly.

In the second evaluation it was tested if the method worked without being trained with a specific WRC combination. This was done by holding out a single WRC combination from the training set and then using this to test the method. The results are given in Table \ref{paper2:tab:classificationresults} and the method were in this case also able to classify all the WRC combinations correctly.

To test the method for predicting if a WRC combination is not returning values corresponding to signal strength values, the same two types of evaluations were made. The results are given in Table \ref{paper2:tab:classificationresults} and in this case the method was able to classify all the WRC combinations correctly in the time case. For the holdout evaluations there were, however, two WRC which were wrongly classified as not returning signal strength measurements.
\begin{table}[h]
\centering
\begin{tabular}{|l|c|c|}
\hline
 & Correct & Wrong \\
\hline
Caching/Low Freq (Time) & 24 & 0 \\
\hline
Caching/Low Freq (Holdout) & 24 & 0 \\
\hline
Correspond to Signal Strength (Time) & 28 & 0 \\
\hline
Correspond to Signal Strength (Holdout) & 26 & 2 \\
\hline
\end{tabular}
\caption{Classification results}
\label{paper2:tab:classificationresults}
\end{table}
	
\subsection{Normalization}
\begin{figure}[h]
	\centering
		\includegraphics[viewport=0 230 480 640,width=0.8\textwidth,clip]{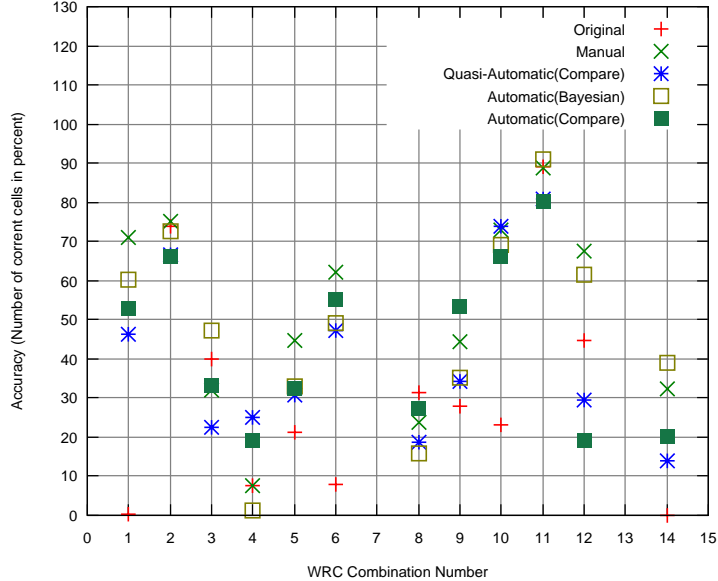}
	\caption{Results of the localization accuracy with correct localization in percent for the different WRC combinations.}
	\label{paper2:fig:plotResultsAutomaticComb}
\end{figure}

To evaluate the performance of the proposed methods for normalization, the walks and calibration set collected as explained above were used. In the evaluation of a specific WRC combination one of the walks was used to find the normalization parameters and the other was used to test how well the WRC combination could predict the route of the walk with normalized measurements. In the test the location accuracy in terms of correctly estimated cells and the average likelihood of the measurements with respect to the probabilistic model of the localization system were collected. The probabilistic model used was constructed from the calibration set. The average likelihood was collected to show how close the actual measured values come to the calibration measurements after they have been normalized. The average likelihoood is calculated by averaging the likelihood for each measurement looked up in the probabilistic model. The higher these values are the more equal the normalized measurements are to the measurements that was used to construct the probabilistic model. The localization results and the average likelihood results are given in Table \ref{paper2:tab:norlocresults}. For single WRC combinations localization results are given in Figure \ref{paper2:fig:plotResultsAutomaticComb}.

\begin{table}[h]
\centering
\begin{tabular}{|p{3.5cm}|p{2.3cm}|p{2.3cm}|p{3cm}|}
\hline
 & All & Good & Caching/Low frequency \\
\hline
Original & 32.6\% (1.83\%)& 41.7\% (2.08\%) & 24.5\% (1.87\%)\\
\hline
Manual & 52.1\% (2.80\%)& 73.6\% (3.40\%)& 38.8\% (2.66\%)\\
\hline
Quasi-Automatic(Compare) & 41.0\% (2.13\%)& 56.1\% (2.67\%) & 32.2\% (1.93\%)\\
\hline 
Automatic(Bayesian) & 45.7\% (2.52\%)& 64.3\% (2.81\%)& 33.6\% (2.61\%)\\
\hline
Automatic(Compare) & 43.4\% (2.20\%)& 55.1\% (2.47\%)& 39.8\% (2.29\%)\\
\hline
\end{tabular}
\caption{Results for evaluating the normalization methods with respect to localization accuracy and average likelihood. The location accuracy given are the correct localizations in percent and the likelihoods are given in the parentheses.}
\label{paper2:tab:norlocresults}
\end{table}

The results show that the manual normalization method gives the highest gain in localization accuracy. Among the automatic methods, the Bayesian method gives the highest gain for all and the good WRC combinations. However, for the caching/low frequency WRC combinations the method based on comparison gives the best results. One reason for this is that the Bayesian method does not work well with highly correlated measurements. The likelihood results show that there is some correspondence between the gain in localization accuracy and the average likelihood. However there are also exceptions as for the Caching/Low Frequency WRC combinations, where the automatic Bayesian method gives the highest average likelihood but has a lower accuracy than the automatic comparison method which has a lower average likelihood. The results in Figure \ref{paper2:fig:plotResultsAutomaticComb} also highlight that the accuracy a indoor location system can achieve is highly dependent on the WRC combination used.

\section{Discussion}
\label{paper2:sec:discussion}
\subsection{Application of classifiers}
The method for classifying if a WRC combination is caching or has a low update frequency were, as presented in the result section, able to classify all combinations correctly. The method for classifying if a WRC combination is not returning values corresponding to signal strength value were, however, not able to classify all correctly. One method for improving the last method is maybe to use another estimator as for example a linear classifier\cite{Witten2005}.

\subsection{Application of normalizer}
The results showed that the manual method made the highest improvement in accuracy. However, the automatic method was also able to considerably improve the accuracy. A method for addressing that the automatic method for some cases did not give as good a result as the manual is to integrate the two. This could for instance be done so a user of a localization system with automatic normalization could choose to do manual normalization if the automatic method failed to improve the accuracy. The results also showed that the two automatic methods were best for different types of WRC combinations. A solution to this was to use the proposed classifiers to find out what kind of automatic method to apply. The results for normalization reported in this paper are, however, not directly comparable to \cite{Haeberlen2004} because their results concerns temporal variations. Therefore they make different assumptions about the data they use in their evaluation.

An interesting question is, how the proposed methods perform over a longer period of time. For instance if a location system could run normalization several times and then try to learn the parameters over a longer period of time, some improvement in accuracy might be observed. To do this some sequential technique has to be designed that makes it possible to include prior estimates. Such a technique could also be used to address large-scale temporal variations.

\subsection{The still period analyzer}
The use of the still period analyzer solved the problem of dividing measurements into groups from different locations. This actually made the automatic normalizer perform better than the quasi-automatic normalizer because noisy measurements were filtered off. However, the still period analyzer also had problems with some of the WRC combinations such as WRC combination 1 for which signal strength values did not vary as much as for WRC combination 11, which the still period analyzer was trained with. Also generally the caching/low frequency WRC combinations made the period analyzer return too many measurements. This was because the variations were too low due to the low update rate at all times making the still period analyzer unable to divide the measurements into different parts. A solution to these problems might be to include some iterative step in the method so that the automatic normalization is run several times on the measurements. This would also normalize the variations so they would be comparable to the variations for which the still period analyzer was trained for.

\subsection{The linear approximation}
\begin{figure}
	\centering
		\includegraphics[viewport=0 235 470 585,width=0.44\textwidth,clip]{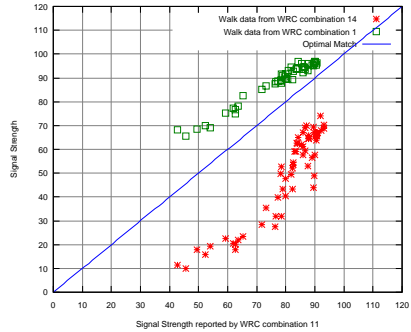}
	\caption{Plots of signal strength values reported by different WRC combinations relative to the values reported by WRC combination 11 which was used for calibration.}
	\label{paper2:fig:plotszcom}
\end{figure}

The use of a linear approximation for normalization gave good results in most cases. However, for WRC combinations that do not report signal strength values which are linearized, the linear approximation does not give as good results. One example of this is WRC combination 14 which was classified as good but only reached a location accuracy of 32\% with manual normalization. The reason is that the signal strength values reported by WRC combination 14 are not linear as can be seen on Figure \ref{paper2:fig:plotszcom} (Because the manufacture did not implement a linearization step of the signal strength values in either the firmware or software driver). To illustrate the linearity of the measurements reported by other WRC combinations, results from WRC combination 1 have also been included in the Figure. The optimal match line in the Figure shows what the measurements should be normalized to. To address this issue an option is to include a linearization step in the methods for WRC combinations that do not return linearized signal strength values, such as WRC combination number 14.

\section{Conclusion}
\label{paper2:sec:conclusion}
In this paper methods for classifying a WRC combination in terms of fitness for localization and methods for automatic normalization were presented. It was shown that the proposed classifiers were able to classify WRC combinations correctly in 102 out of 104 cases. The proposed methods for normalization were evaluated on 14 different WRC combinations and it was shown that manual normalization performed best with a gain of 19.2\% over all WRC combinations. The method of automatically normalization was shown also able to improve the accuracy with 13.1\% over all WRC combinations. The applicability of the methods for different WRC combinations and scenarios of use was also discussed. Possible future extensions to the methods include: extending the fitness classification to the last two cases of artificial limits and sensitivity, adding a linearization step to the normalization methods, and make normalization iterative to address some of the issues of applying the automatic still period analyzer.



\section*{Acknowledgements}
The research reported in this paper was partially funded by the software part of the ISIS Katrinebjerg competency centre http://www.isis.alexandra.dk/software/. Carsten Valdemar Munk helped collecting signal strength measurements and implementing the facilities for collecting these.


\clearemptydoublepage
\chapter{Paper 3}
\label{chap:percom2007}

The paper \emph{Hyperbolic Location Fingerprinting: A Calibration-Free Solution for Handling Differences in Signal Strength} presented in this
chapter has been published as a conference paper~\cite{KjaergaardPercom2008}.

\begin{publist}{\cite{KjaergaardPercom2008}}
  \item[\cite{KjaergaardPercom2008}] M.\ B.\ Kjærgaard and C.\ V.\ Munk. Hyperbolic Location Fingerprinting: A Calibration-Free Solution for Handling Differences in Signal Strength. In \emph{Proceedings of the Sixth Annual IEEE International Conference on Pervasive Computing and Communications}, pages~110--116, IEEE, 2008.
\end{publist}

\noindent

\clearemptydoublepage



\mytitle{Hyperbolic Location Fingerprinting: A Calibration-Free Solution for Handling Differences in Signal Strength}{ 
  Mikkel Baun Kj\ae rgaard\footnotemark[1] \and Carsten Valdemar Munk\footnotemark[1]}{ 
  \footnotetext[1]{Department of Computer Science, University of
    Aarhus, IT-parken, Aabogade 34, DK-8200 Aarhus N, Denmark. E-mail:
    \texttt{mikkelbk@daimi.au.dk}.}
      } 
        

\begin{myabstract}
Differences in signal strength among wireless network cards, phones and tags are a fundamental problem for location fingerprinting. Current solutions require manual and error-prone calibration for each new client to address this problem. This paper proposes hyperbolic location fingerprinting, which records fingerprints as signal-strength ratios between pairs of base stations instead of absolute signal-strength values. The proposed solution has been evaluated by extending two well-known location fingerprinting techniques to hyperbolic location fingerprinting. The extended techniques have been tested on ten-hour-long signal-strength traces collected with five different IEEE 802.11 network cards. The evaluation shows that the proposed solution solves the signal-strength difference problem without requiring extra manual calibration and provides a performance equal to that of existing manual solutions.
\end{myabstract}

\section{Introduction}
\label{paper3:sec:introduction}
\emph{Location Fingerprinting (LF)} based on signal strength is a promising location technique for many awareness applications in pervasive computing. LF has the advantage of exploiting already existing network infrastructures, like IEEE 802.11 or GSM, and therefore avoiding extra deployment costs and effort. LF is based on a database of pre-recorded measurements of signal strength, denoted as location fingerprints. A client's location can be estimated from the fingerprints by comparing these with the current measured signal strength. Clients can be in the form of, e.g., a tag, a phone, a PDA, or a laptop.

A fundamental problem for LF systems is the differences in signal strength between clients. Such signal-strength differences can be attributed to inequalities in hardware and software and lack of standardization. For IEEE 802.11 differences above 25 dB have been measured for same-place measurements with different clients by Kaemarungsi \cite{Kaemarungsi2006}. Such differences have a severe impact on LF systems' accuracy. Our results show that signal-strength differences can make room-size accuracy for the Nearest Neighbor algorithm \cite{Bahl2000} drop to unusable 10\%.

Current solutions for handling signal-strength differences are based on manually collecting measurements to find mappings between signal strength reported by different clients. Such manual solutions are: (i) time consuming because measurements have to be taken at several places for each client; (ii) error prone because the precise location of each place has to be known; (iii) unpractical considering the huge number of different IEEE 802.11 and GSM clients on the market. For instance, due to such issues the company Ekahau maintains lists of supported clients \cite{ekahauwebsite}. Solutions have been proposed by Haeberlen et al. \cite{Haeberlen2004} and Kj\ae rgaard \cite{KjaergaardLoca2006} that avoid manual measurement collection by learning from online-collected measurements. However, both of these solutions require a learning period and they perform considerably worse in terms of accuracy than the manual solutions.

This paper proposes \emph{Hyperbolic Location Fingerprinting (HLF)} to solve the signal-strength difference problem. The key idea behind HLF is that fingerprints are recorded as signal-strength ratios between pairs of base stations instead of as absolute signal strength. A client's location can be estimated from the fingerprinted ratios by comparing these with ratios computed from currently measured signal-strength values. The advantage of HLF is that it can solve the signal-strength difference problem \textit{without} requiring any extra calibration. The idea of HLF is inspired from hyperbolic positioning, used to find position estimates from time-difference measurements \cite{Chan1994}. The method is named \emph{hyperbolic} because the position estimates are found as the intersection of a number of hyperbolas each describing the ratio difference between unique pairs of base stations. We have evaluated HLF by extending two well-known LF techniques to use signal-strength ratios: \emph{Nearest Neighbor} \cite{Bahl2000} and \emph{Bayesian Inference} \cite{Haeberlen2004}. The HLF-extended techniques have been evaluated on ten-hour-long signal-strength traces collected with five different IEEE 802.11 clients. The traces have been collected over a period of two months in a multi-floored building. In our evaluation the HLF-extended techniques are compared to LF versions and LF versions extended with a manual solution for signal-strength differences.

We make the following contributions: (i) we show that signal-strength ratios between pairs of base stations are more stable among IEEE 802.11 clients than absolute signal strength; (ii) we propose the novel idea of HLF and show that the HLF-extended LF techniques perform clearly better than their LF versions and equal to their manual-solution-extended LF versions; and (iii) we show that the HLF-extended techniques place the same requirements as LF techniques on common parameters.

The paper is structured as follows: signal-strength ratios are quantified to be more stable than absolute signal strength among IEEE 802.11 clients in Section \ref{paper3:sec:absvsratio}. The definition of HLF and the extension of two well-known LF-techniques are presented in Section \ref{paper3:sec:hyplf}. The results of evaluating the HLF-extended techniques for five different IEEE 802.11 clients are then given in Section \ref{paper3:sec:evaluation}. Afterwards, a discussion of the results are given in Section \ref{paper3:sec:discussion} and Section \ref{paper3:sec:relatedwork} discuss related work. A conclusion and a discussion of further work are given in Section \ref{paper3:sec:conclusion}.

\section{Signal-Strength Differences}
\label{paper3:sec:absvsratio}
For IEEE 802.11 signal-strength differences can mainly be attributed to the standard's lack of specification of how clients should measure signal strength \cite{Kaemarungsi2006}. In the standard, signal strength is specified as the received signal-strength index with an integer value between $0,\ldots,255$ with no associated measurement unit. The standard also states that this quantity is only meant for internal use by clients and only in a relative manner. The internal use of the value is for detecting if a channel is clear or for detecting when to roam to another base station. Therefore, IEEE 802.11 client manufacturers are free to decide what their interpretation of signal-strength values is. Most manufacturers have chosen to base signal-strength values on dBm values. However, different mappings from dBm values to the integer scale from $0,\ldots,255$ have been used. The result of this is that most signal-strength values represent dBm values with different limits and granularity. However, inequalities in hardware also attribute to the problem.

This paper explores the use of signal-strength ratios between pairs of base stations. The following definitions are needed: $B$ = $\{b_1,...,b_n\}$ is an ordered set of visible base stations and $O = \{o_1, ..., o_m\}$ a finite observation space. Each observation $o_i$ being a pair of a base station $b \in B$ and a measured signal-strength value $v \in V = \{v_{min}, ..., v_{max}\}$ according to a discrete value range. For the range of $V$ the following restriction is necessary: $v_{min},v_{max} > 0$. The signal-strength ratio $r$ is defined for a unique base station pair $b_i \times b_j$ $\in$ $B \times B$ with the constraint $i < j$ for uniqueness. The signal-strength ratio $r$ can be computed from two observations $o_i=(b_i,v) \in O$ and $o_j = (b_j,y) \in O$ as follows:

\begin{equation}
	r(o_i,o_j) = \frac{v}{y}
\end{equation}

However, because the signal-strength ratios are non-linear with respect to changes in either of the signal-strength measurements, normalized log signal-strength ratios are used. These are calculated from the signal-strength ratios as follows:

\begin{equation}
	nlr(o_i,o_j) = \log(r(o_i,o_j)) - \log(\frac{1}{v_{max}})
\end{equation}
where the last term normalizes the ratios in order to keep them on a positive scale. When we refer to signal-strength ratios in the rest of the paper it will be in their log-normalized form.

\subsection{Data Collection}
\label{paper3:sec:datacollection}
\begin{figure}
\begin{center}
   \includegraphics[width=0.8\linewidth]{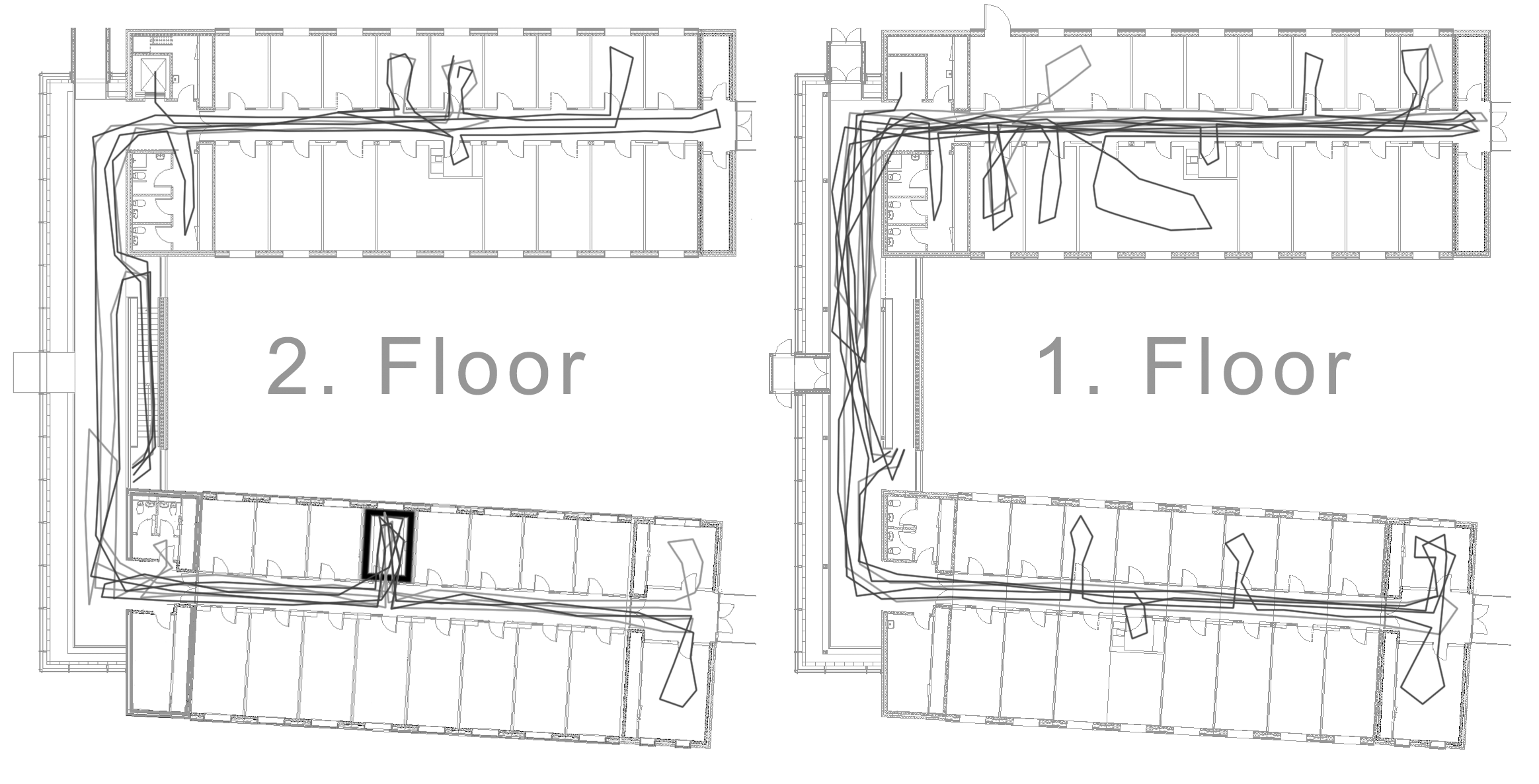}
    \caption{Path for one 40-minute client trace.}
    \label{paper3:fig:floors}
\end{center}
\end{figure}

For our analysis and evaluation data have been collected at a two-floored test site covering 2256 $m^2$ and offering an 802.11 infrastructure with 26 reachable base stations. Signal-strength data have been collected as continuous traces with five different IEEE 802.11 clients, which are listed in Table \ref{paper3:tab:clients}. The five clients have been picked to cover different manufactures, options of antennas and operating systems. For each client three separate 40-minute traces have been collected, totaling about 10 hours of data. The traces were collected over two months and for each client the three separate traces were collected at different days and time of day to make sure the data was affected by temporal variations. Each entry in the traces consist of a time stamp, measured signal strength to surrounding base stations, and current ground truth. The ground truth was manually specified by the person collecting the trace by clicking on a map. The area of the test site were divided up into 126 clickable cells, with an average size of 16 m$^{\textrm{2}}$, corresponding to rooms or parts of hallways, and spanning two floors. The cells approximately represent a coarse grained four meter fingerprinting grid. The people collecting the traces walked at moderate speeds, with several pauses through the test site on both floor levels, as illustrated for one trace in Figure \ref{paper3:fig:floors}. Signal strength were measured with a sampling rate of 0.5 Hz for the Fujitsu Siemens Pocket Loox 720 and 1 Hz for the four other clients.

\begin{table}[h]
\caption{Evaluated IEEE 802.11 clients}
\begin{flushleft}
\footnotesize
\begin{tabular}{lll}
\small \textit{Client name} & \small \textit{Antenna} & \small \textit{OS / Driver}\\
\hline
Apple AirPort Extreme & In laptop & Mac OS X (10.4) / OS provided\\
D-Link Air DWL-660 & In card & Windows XP / D-Link 7.44.46.450 \\
Fujitsu Siemens Pocket Loox 720 & In PDA & Windows Mobile 2003 SE / OS provided\\
Intel Centrino 2200BG & In laptop & Windows XP / Intel 10.5.0.174\\
Orinoco Silver & In card & Windows XP / OS provided (7.43.0.9) \\
\hline
\end{tabular}
\end{flushleft}
\label{paper3:tab:clients}
\end{table}

\subsection{Stability of Signal-Strength Ratios}
If normalized log signal-strength ratios should be able to solve the signal-strength difference problem they have to be more stable than absolute signal-strength values among IEEE 802.11 clients. To quantify if this is the case the variations in absolute signal strength and signal-strength ratios have been analysed among different IEEE 802.11 clients. The analysis is based on statistics calculated from the collected traces. To make the statistics directly comparable the presented values have been converted to percentages of mean values.

\begin{figure}[h]
 \centering
	\includegraphics[width=0.7\textwidth]{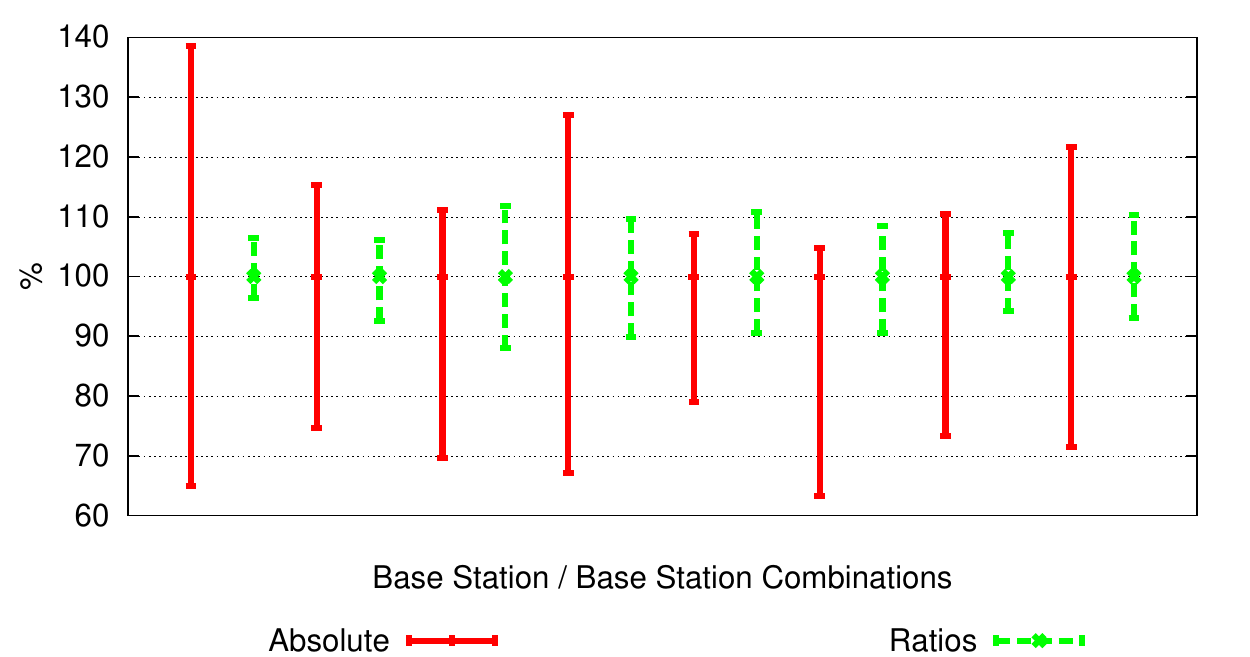}
	\caption{Absolute versus Ratios}
 \label{paper3:fig:plotAbs} 
 \end{figure}

The analysis uses trace data for all five clients from the black-rectangle-highlighted cell on Figure \ref{paper3:fig:floors}. The calculated statistics from this trace data are shown in Figure \ref{paper3:fig:plotAbs}. The figure shows the minimum and maximum values of absolute signal strength and signal-strength ratios for base stations and combinations, respectively. For the first base station the clients' absolute signal-strength values are at anytime at most 35.1\% below and 38.6\% above the mean absolute signal strength for this base station. For the first base station combination the signal-strength ratios are at any time only 4.5\% below and 6.5\% above the mean signal-strength ratio for this combination. Looking at all base stations and combinations the results show that the variations are only +/- 10\% for signal-strength ratios but +/- 20\% for absolute signal strength. Similar results were obtained in an analysis using data from all cells contained in the traces. The results confirm that signal-strength ratios vary less between IEEE 802.11 clients than absolute signal strength. Furthermore, because the used signal-strength traces were collected spread out over two months the signal-strength ratios are also shown to be stable over time.

\section{Hyperbolic Location Fingerprinting}
\label{paper3:sec:hyplf}
This section presents the extension of two well-known LF-techniques to HLF. The main change is the replacement of absolute signal-strength with signal-strength ratios. This change affects both the representation of location fingerprints and the calculation of location estimates. The extended techniques are the techniques of Nearest Neighbor \cite{Bahl2000} and Bayesian inference \cite{Haeberlen2004}. Both techniques are in this paper applied for cell-based localization, i.e. locations are represented as cells. A cell may correspond to a room or a part of it, or a section of a hallway. The following definitions are needed: $C = \{c_1,...,c_n\}$ is a finite set of \emph{cells} covered by the location system, a \emph{sample} $s$ is a set of same-time same-place observations, one for each visible base station and a \emph{fingerprint} $f$ is a set of samples collected within the same cell.

\subsection{Nearest Neighbor}
A common \emph{deterministic} LF technique calculates the nearest neighbor in Euclidian space between a client's measured samples and the fingerprints in the database \cite{Bahl2000}. The cell with the lowest Euclidian distance is picked to be the current one of the client. In the nearest-neighbor calculations each fingerprint is represented as a vector with entries for each visible base station. Each entry contains the average signal-strength for a base station computed from the samples of the fingerprint.

To extend this technique to HLF, both the fingerprint representation and the nearest-neighbor calculation have to be changed. The HLF fingerprint representation has entries for each unique pair of visible base stations in the fingerprint. The entries of the vector are computed as the average signal-strength ratio from the fingerprint's sample set. Let $f_{c_x,b_i}$ denote the set of observations from the fingerprint taken in cell $c_x$ that refers to base station $b_i$. Each entry of a fingerprint representation vector $v$ for a cell $c_x$ and unique base station pair $b_i \times b_j$ can be computed as follows:

\begin{equation}
	v_{c_x,b_i \times b_j} = \frac{1}{n}\sum_{o_i \in f_{c_x,b_i}}\sum_{o_j \in f_{c_x,b_j}}(nlr(o_i,o_j))
\end{equation}
where n is the number of observation combinations. An example with three base stations is given in Table \ref{paper3:tab:frnn}. The table includes both the LF average absolute signal-strength and the HLF average signal-strength ratios.

\begin{table}[h]
\caption{Example of representation}
\centering
\begin{tabular}{p{1.5cm}|p{1.5cm}|p{1.5cm}}
& \textit{Entry} & \textit{Average}\\
\hline
\multirow{3}{*}{\textit{LF}} & $b_1$ & 81.8\\
& $b_2$ & 62.1\\
& $b_3$ & 85.1\\
\hline
\multirow{3}{*}{\textit{HLF}} & $b_1 \times b_2$ & 2.12\\
& $b_1 \times b_3$ & 1.98\\
& $b_2 \times b_3$ & 1.86\\
\end{tabular}
\label{paper3:tab:frnn}
\end{table}

The HLF location estimation step computes the nearest-neighbor with Euclidian distances in signal-strength ratio space. Euclidian distances are computed using the set of signal-strength ratios $R$ calculated from the currently measured sample. The following formula is used with $B_o$ as the set of base stations currently observed by the client:

\begin{equation}
    E(c_x) = \sqrt{\sum_{b_i \times b_j \in B_o \times B_o, i<j}(R_{b_i \times b_j} - v_{c_x, b_i \times b_j})^2}
\end{equation}

\subsection{Bayesian Inference}
Several LF systems use Bayesian inference \cite{Haeberlen2004,Roos2002b}, which represents a \emph{probabilistic} method. In simple terms, for each cell in the system a probability is calculated based on the currently measured sample. The probabilities are computed using Bayesian inference. The cell associated with the highest probability is picked to be the current location of the client. In Bayesian inference each fingerprint for each base station $b \in B$ is represented as a probability distribution over the range of absolute signal-strength values $V$.

\begin{figure}[h]
\centering
\includegraphics[width=0.7\textwidth]{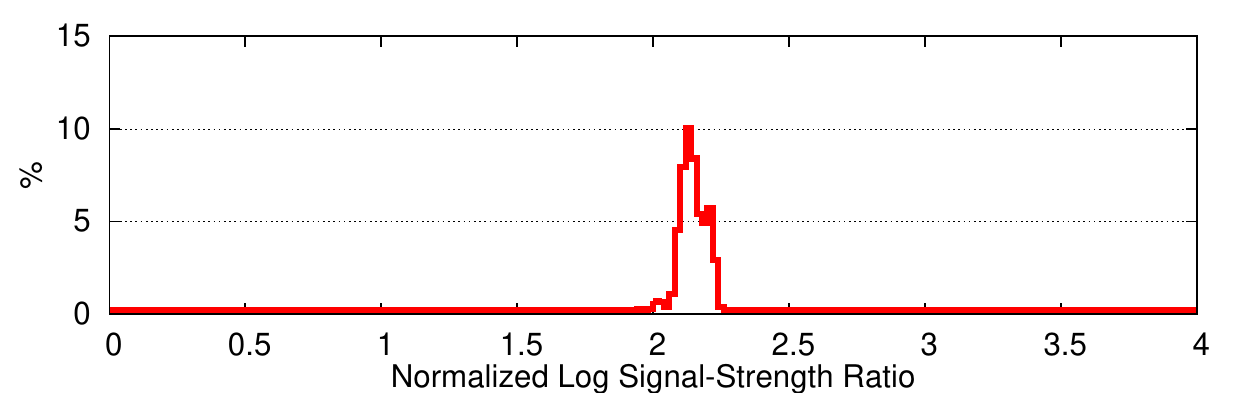}
\caption{HLF Histogram}
\label{paper3:fig:plotHistogramRatio}
\end{figure}

To extend this technique to HLF both the fingerprint representation and the Bayesian inference calculation have to be changed. The HLF fingerprint representation is for each unique pair $b_i \times b_j \in B \times B$ a probability distribution over the range of signal-strength ratios $V^{'} = [0 : nlr(v_{max})]$. The probability distributions over $V^{'}$ are computed using the histogram method \cite{Roos2002b} from the fingerprints' samples. An example of a distribution is shown in Figure \ref{paper3:fig:plotHistogramRatio} for a specific fingerprint and a unique base station pair. A parameter that can be used to tune the histogram method is the size of the discrete steps; a size of 0.02 was used for the histogram on Figure \ref{paper3:fig:plotHistogramRatio} and for the evaluation in Section \ref{paper3:sec:evaluation}. This value was chosen by the authors based on evaluations that showed that larger values would deteriorate accuracy and smaller values would not improve it. 

The HLF location estimation step performs Bayesian inference from signal-strength ratios computed from currently measured samples. The HLF fingerprint representation is used to describe the conditional probability of measuring a specific signal-strength ratio in a specific cell. The conditional probabilities over all cells are defined for a finite observation space $O^{'} = \{o^{'}_1, ..., o^{'}_m\}$ with each \emph{observation} $o^{'}_i$ being a tuple with a unique pair of \emph{base stations} $b_i \times b_j$ and a normalized log signal-strength ratio $v^{'} \in V^{'}$. The probabilities are calculated for a observation $o^{'}_j \in O^{'}$ within a cell $c_x \in C$ with fingerprint $f_{c_x}$ as: 
\begin{equation}
 P(o^{'}_j|c_x) = Histogram(o^{'}_j,f_{c_x})	
\end{equation}
where the function \emph{Histogram} is the probability of the observation computed from the HLF-histogram fingerprint representation. The HLF location estimation step follows the LF procedure and returns the cell with the highest probability as the current cell of the client.

\section{Evaluation}
\label{paper3:sec:evaluation}
Our evaluation uses the traces collected as described in Section \ref{paper3:sec:datacollection}. In addition to traces a set of fingerprints have been collected for the test site's 126 cells one month before the traces. Each cell was fingerprinted by a person walking around in the cell for 60 seconds using a laptop with an Orinoco client. The evaluation uses this set of fingerprints for each technique's database of fingerprints. The evaluation is performed as emulated localization. This means that trace samples are given as input to a technique and the returned cell estimates are compared with trace ground truth. The evaluation results are given in terms of accuracy: the percentage of samples where the ground truth and the estimated cell matched. Both the algorithms and the emulation environment were implemented by the authors in Java.

Our evaluation covers the techniques of \emph{Nearest Neighbor (NN)} \cite{Bahl2000} and \emph{Bayesian Inference (BI)} \cite{Haeberlen2004} implemented in three setups: a HLF version (implemented as presented in Section \ref{paper3:sec:hyplf}), a LF version, and a LF version extended with a manual solution for signal-strength differences. The manual solution handles signal-strength differences using linear mapping, as described in Kj\ae rgaard \cite{KjaergaardLoca2006}. The linear mapping transforms one client's samples to match another client's samples. The parameters for the linear mapping are found by comparing fingerprints collected with both clients using least squares estimation. The linear mapping is then applied to all samples before they are forwarded to a LF technique. The linear mapping parameters used in the evaluation were calculated from separate data collected with each of the clients.

\begin{figure*}[!]
 \begin{minipage}[t]{0.5\linewidth}
 		\centering
		\includegraphics[width=1.0\textwidth]{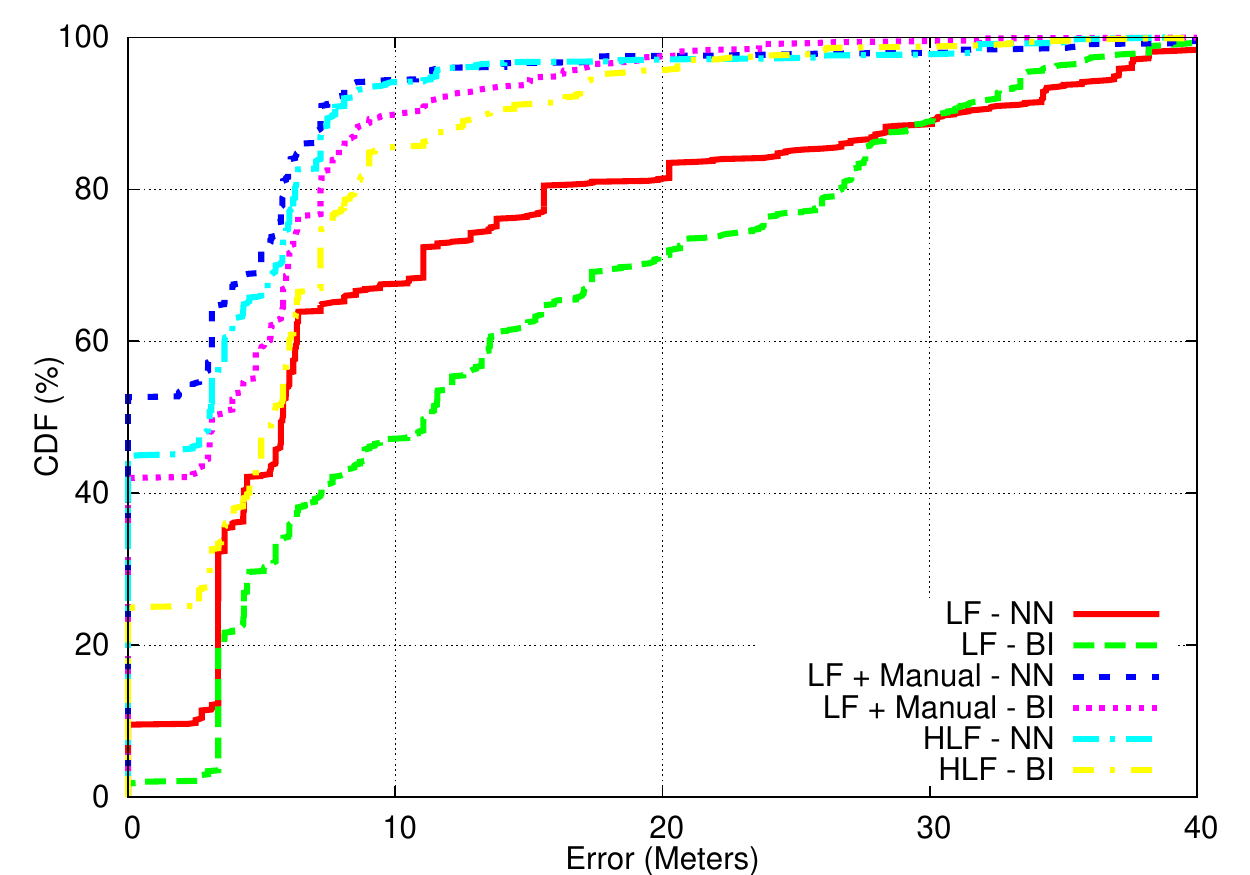}
		\caption{Error for Intel}
		\label{paper3:fig:plotIntel}
 \end{minipage}
 \vspace{0.1cm}
 \begin{minipage}[t]{0.5\linewidth}
 		\centering
 		\includegraphics[width=1.0\textwidth]{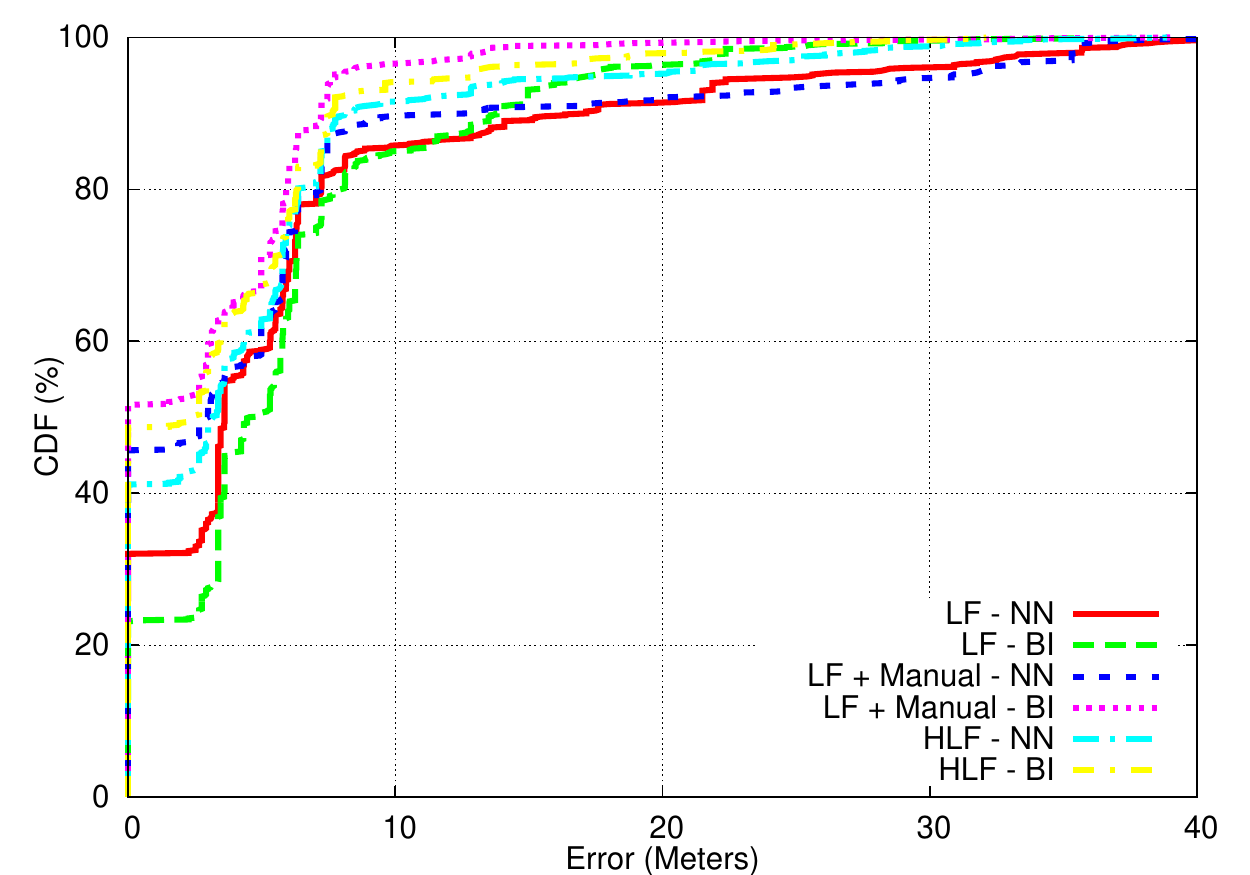}
		\caption{Error for Fujitsu}
		\label{paper3:fig:plotFujitsu}
 \end{minipage}
 \vspace{0.1cm}
 \begin{minipage}[t]{0.5\linewidth}
 		\centering
 		\includegraphics[width=1.0\textwidth]{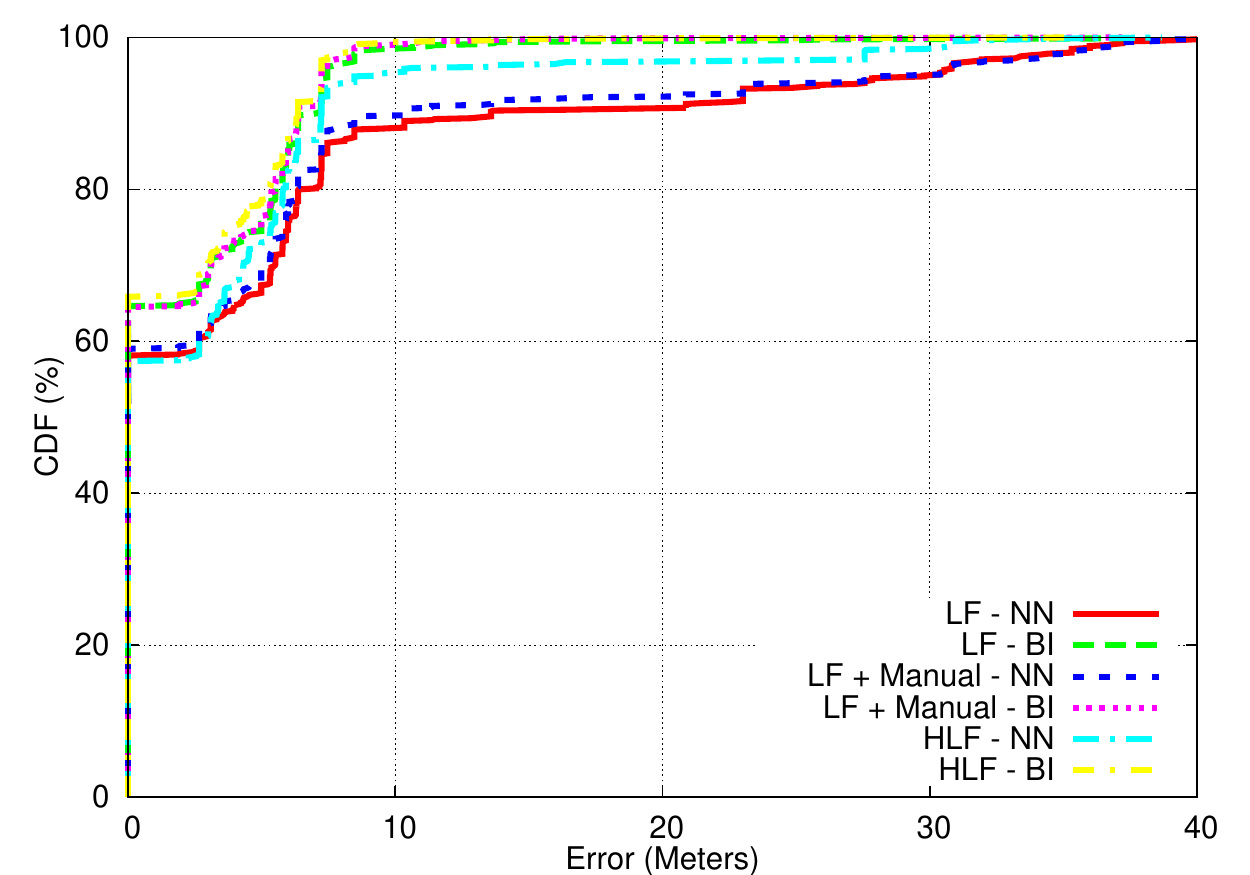}
		\caption{Error for Orinoco}
		\label{paper3:fig:plotOrinoco}
 \end{minipage}
\end{figure*}

Results of emulated localization with traces are given in Table \ref{paper3:tab:movingresults} for each client and as an average over all clients. Accuracy for LF (first column) was highest for Orinoco (65\% for BI) which can be attributed to the absence of signal-strength differences. However, for Intel and Apple BI accuracy is only 2\% and 12\%, respectively. The Fujitsu and D-link clients have higher accuracy and the NN accuracies are generally also a bit higher across all clients but for Intel only 10\%. The results demonstrate that signal-strength differences have a large impact on LF accuracy for both NN and BI. Accuracy for LF extended with a manual solution (second column) is again highest for Orinoco. However, accuracy improves on average compared to LF for Apple, Fujitsu and Intel with 27\% for BI and 22\% for NN. For D-Link and Orinoco no improvement can be observed. One thing that can be noticed is that the BI accuracy for Apple and Intel do not improve as much as one could expect. This issue will be further analysed below. Accuracy with HLF (third column) improves on average compared to LF for Apple, Fujitsu and Intel with 22\% for BI and 14\% for NN. For D-Link there is a small improvement and no improvement for Orinoco. However, again it can be noticed that the BI accuracy for Apple and Intel do not improve as much as one could expect.

To give a more detailed analysis error distributions are shown in Figure \ref{paper3:fig:plotIntel} to \ref{paper3:fig:plotOrinoco}. The error distributions for Apple and D-Link have been omitted because they are nearly similar to Intel and Orinoco, respectively. For Intel the distributions reveal a high percentage of large errors for LF, in comparison, both LF + Manual and HLF have much less large errors. The distributions also show that HLF for Intel recovers from the low accuracy in terms of percentage of large errors. For Fujitsu the better performance of LF is also apparent in lower errors which converge towards the distributions for LF + Manual and HLF. The lower accuracy of NN compared to BI is also visible as larger errors for NN than for BI. For Orinoco the distributions form a narrow band again with BI having the smallest percentage of large errors.


\begin{table}[h]
\caption{\% of correct estimations}
\hspace{0.01cm}
\centering
\footnotesize
\begin{tabular}{p{1.2cm}|p{0.7cm}p{0.7cm}|p{0.7cm}p{0.7cm}|p{0.7cm}p{0.7cm}}
 & \multicolumn{2}{c|}{\textit{LF}}  & \multicolumn{2}{c|}{\textit{LF + Manual}} & \multicolumn{2}{c}{\textit{HLF}}\\
 & \textit{BI} & \textit{NN} & \textit{BI} & \textit{NN} & \textit{BI} & \textit{NN} \\
\hline
\textit{Apple} & 12  & 31 & 28 & 42 & 32 & 30 \\
\textit{D-Link} & 55  & 55 & 56 & 55 & 59 & 56 \\
\textit{Fujitsu} & 23  & 32 & 51 & 45 & 48 & 41 \\
\textit{Intel} & 2  & 10 & 39 & 53 & 25 & 45 \\
\textit{Orinoco} & 65  & 58 & 65 & 59 & 65 & 57 \\
\hline
\textit{All} & 31  & 37 & 48 & 51 & 46 & 46 \\
\end{tabular}
\label{paper3:tab:movingresults}
\end{table}

Further analysis has shown that the smaller improvement for Apple and Intel can be attributed to a difference in the number of measured base stations at similar locations. Statistics calculated from the traces and fingerprints reveal that each D-Link and Fujitsu sample contains on average one extra observation than the Orinoco's samples. Apple and Intel samples contain on average approximately three extra base station observations. To address this problem we propose to use a K-strongest filter. The rationale behind this filter is that if a client makes more observations because of higher sensitivity we can filter out these by only keeping the K strongest measurements in each sample. K should here be set to match the sensitivity of the fingerprint client, from statistics calculated from the Orinoco fingerprints K was set to seven in our case. To evaluate this idea two emulations have been run for which results are given in Table \ref{paper3:tab:filtermovingresults} for BI. The first emulation applies a K-strongest filter to each sample before it is passed on to one of the techniques. The second emulation applies a ground-truth filter. This filter removes from each sample any extra observations that the Orinoco client did not observe at this location. For Apple and Intel the K-strongest filter has a large impact by improving BI accuracy with 15\% and 20\%, respectively, and reducing the percentage of large errors. The BI accuracy of the other clients is not improved by the K-strongest filter, which is consistent with the above calculations. The ground-truth filter improved BI accuracy for all clients except the Orinoco client. However, the ground-truth filter cannot be implemented in practice and are included to indicate an upper limit of performance for any filter. An interesting line of future work would be to develop a filter that using a prediction step could predict the base stations to sort out instead of only selecting the K strongest observations. Emulations were also run for LF where BI accuracy did not improve and LF + Manual where the filter made a small improvement in BI accuracy. For NN neither of the filters had a noticeable impact on accuracy.

\begin{table}[h]
\caption{\% of correct estimations for BI}
\hspace{0.01cm}
\centering
\footnotesize
\begin{tabular}{p{1.2cm}|c|c|c}
 & \textit{HLF} & \textit{HLF + K-Strongest} & \textit{HLF + GT} \\
\hline
\textit{\textbf{Apple}} & \textbf{32}  & \textbf{47} & \textbf{72} \\
\textit{D-Link} & 59  & 59 & 65 \\
\textit{Fujitsu} & 48  & 48 & 53 \\ 
\textit{\textbf{Intel}} & \textbf{25}  & \textbf{45} & \textbf{73} \\
\textit{Orinoco} & 65  & 64 & 65 \\
\hline
\textit{All} & 46  & 52 & 66 \\
\end{tabular}
\label{paper3:tab:filtermovingresults}
\end{table}

For the preceding results a history of five samples were used. This means that, in addition to the current sample, the four preceding samples are supplied with each trace sample to the techniques. The preceding samples are treated by the Bayesian inference techniques in the same manner as the current sample. For the nearest neighbor method, samples are aggregated to the mean value for each base station. Additional emulations have shown that consistently for both LF, LF + Manual and HLF a history of samples smaller than five make accuracy slowly drop and larger histories does not improve accuracy. For the preceding results the size of fingerprints have been 60 samples. Additional emulations have shown that consistently for both LF, LF + Manual and HLF a size of fingerprints below 20 samples make accuracy drop. The number of deployed base stations needed for techniques to work is an important number in practice. The preceding results were based on using data for all 26 base stations reachable in some parts of the two-floored 2256 $m^2$ test site. Additional emulations have shown that consistently for both LF, LF + Manual and HLF if we randomly remove base stations accuracy drops.

\section{Discussion}
\label{paper3:sec:discussion}
The results of the evaluation were that the average accuracy for BI (with K-strongest filter) was 51\% for LF + Manual and 52\% for HLF and for NN it was 51\% for LF + Manual and 47\% for HLF. These results show that the accuracy of HLF and LF + Manual are nearly similar and improvements compared to LF. Distributions of errors also revealed that HLF and LF + Manual lower the percentage of large errors compared to LF. In this paper two HLF techniques were proposed and evaluated but the use of signal-strength ratios are possible with other LF techniques. The results in this paper are based on data from five IEEE 802.11 clients, which are representative in terms of hardware and antenna options for many other clients. However, clients also exist that cannot be used for LF and also for HLF because of faulty or poor signal-strength measuring capabilities, for lists of such clients see Ekahau \cite{ekahauwebsite} and Kj\ae rgaard \cite{KjaergaardLoca2006}.

The evaluation also revealed that accuracy depends on clients making same-place measurements to the same set of base stations. Because the client used for fingerprinting collection in our data measured least base stations we cannot evaluate if this also is a problem if fingerprints are collected with a client that measure the most base stations. But it is an interesting line of future work to collect such data to see if a recommendation could be to always use a client that collect measurements to a maximum number of base stations for fingerprinting. From our analysis we can conclude that if the client is not maximal you have to filter the samples of other clients to maximize accuracy.

The evaluation of the common parameters showed that the HLF-extended techniques have the same sensitivity as LF techniques to the history of samples, the size of the fingerprints and the number of deployed base stations.

\section{Related Work}
\label{paper3:sec:relatedwork}
One of the first IEEE 802.11 LF systems was RADAR \cite{Bahl2000}, which applied different deterministic mathematical models to calculate a client's position (in coordinates). Similar methods have also been applied to GSM by Otsason et al. \cite{Otsason2005}. In comparison to RADAR, later systems have used probabilistic models instead of deterministic models, following the definitions in Kj\ae rgaard \cite{KjaergaardLoca2007}.  An example of a probabilistic system, which determine the coordinates of a client, is published by Youssef et al. \cite{Youssef2005b}. A probabilistic system determining the logical position or cell of a client is published by Haeberlen et al. \cite{Haeberlen2004}. The basic LF systems do not address the issue of signal-strength differences.

Haeberlen et al. \cite{Haeberlen2004} propose using a linear mapping for transforming a client's samples to match another client's samples. They propose three different methods for finding the two parameters in the linear mapping. The first method is a manual one, where a client has to be taken to a couple of known locations to collect fingerprints and parameters are found using least squares estimation. The second method is a quasi-automatic one, for which a client has to be taken to a couple of unknown locations to collect fingerprints. For finding the parameters, they propose using confidence values from Markov localization and find parameters that maximize this value. The third is an automatic one requiring no user intervention. Here they propose using an expectation-maximation algorithm combined with a window of recent measurements. For the manual method, they have published results which show a gain in accuracy for three clients; for the quasi-automatic method it is stated that the performance is comparable to that of the manual method, and for the automatic one it is stated that it does not work as well as the two other methods. In comparison, HLF has a performance comparable or better than the manual method and does not involve any extra steps of collecting additional fingerprints.

The method proposed by Kj\ae rgaard \cite{KjaergaardLoca2006} is also based on a linear mapping. This method is automatic, but it requires a learning period to find the parameters for the linear mapping. The solution is based on movement detection which is used to group same-place measurement into fingerprints. The parameters are then estimated from the grouped fingerprints using least squares estimation. The method, however, does only achieve lower or comparable performance to the manual approach, and it requires a learning period.

In addition to the above systems, which estimate the location of clients, a number of systems, such as NearMe \cite{Krumm2004b}, have been studied, for which the calibration step is only carried out by users for tagging relevant places. The systems propose simple metrics based on signal strength to quantify when clients are in proximity of calibrated places. One of the strengths of these simple metrics is that they overcome the problem of signal-strength differences. To summarize, HLF address signal-strength differences without requiring any extra steps.

\section{Conclusion and Further Work}
\label{paper3:sec:conclusion}
We showed that the proposed solution of HLF was able to address signal-strength differences. HLF records fingerprints as signal-strength ratios between pairs of base stations instead of as absolute signal-strength values. Signal-strength ratios factor out scaling differences in signal strength between clients. HLF is an improvement over existing solutions that require either error-prone manual steps or a learning period to work. Two LF techniques were extended to HLF and evaluated for five different IEEE 802.11 clients. The evaluation showed that the accuracy of HLF techniques is similar to that of existing manual solutions.

Two further issues subject to future work are proposed in the following. First, it would be interesting to evaluate other LF techniques with HLF and other technologies such as GSM where signal-strength differences are also present. Second, a further analysis is also interesting of how sensitivity affects the same-place measured base stations across clients. Here more data has to be collected to evaluate if a recommendation such as \textit{always use a client which maximizes the number of measured base stations} can address the problem.

%

\section*{Acknowledgements}
The research reported in this paper was partially funded by the ISIS Katrinebjerg competency centre.

\clearemptydoublepage
\chapter{Paper 4}
\label{chap:pervasive2007}

The paper \emph{Zone-based RSS Reporting for Location Fingerprinting} presented in this
chapter has been published as a conference paper~\cite{KjaergaardPervasive2007}.

\begin{publist}{\cite{KjaergaardPervasive2007}}
  \item[\cite{KjaergaardPervasive2007}] M.\ B.\ Kjærgaard, G. Treu, and C. Linnhoff-Popien. Zone-based RSS Reporting
for Location Fingerprinting. In \emph{Proceedings of the 5th International Conference on Pervasive Computing}, pages~316--333, Springer, 2007.
\end{publist}

\noindent

\clearemptydoublepage



\mytitle{Zone-based RSS Reporting for Location Fingerprinting}{ 
  Mikkel Baun Kj\ae rgaard\footnotemark[1] \and 
  Georg Treu\footnotemark[2] \and
  Claudia Linnhoff-Popien\footnotemark[2] }{  
  \footnotetext[1]{Department of Computer Science, University of
    Aarhus, IT-parken, Aabogade 34, DK-8200 Aarhus N, Denmark. E-mail:
    \texttt{mikkelbk@daimi.au.dk}.}
  \footnotetext[2]{Mobile and Distributed Systems Group, Institute for Informatics, Ludwig-Maximilian University Munich, Germany. E-mail:
    \texttt{[georg.treu$|$linnhoff]@ifi.lmu.de}.}
    } 
        

\begin{myabstract}
In typical location fingerprinting systems a tracked terminal reports sampled \emph{Received Signal Strength (RSS)} values to a location server, which estimates its position based on a database of pre-recorded RSS fingerprints. So far, poll-based and periodic RSS reporting has been proposed. However, for supporting proactive \emph{Location-based Services (LBSs)}, triggered by pre-defined spatial events, the periodic protocol is inefficient. Hence, this paper introduces zone-based RSS reporting: the location server translates geographical zones defined by the LBS into RSS-based representations, which are dynamically configured with the terminal. The terminal, in turn, reports its measurements only when they match with the configured RSS patterns. As a result, the number of messages exchanged between terminal and server is strongly reduced, saving battery power, bandwidth and also monetary costs spent for mobile bearer services. The paper explores several methods for realizing zone-based RSS reporting and evaluates them simulatively and analytically. An adaption of classical Bayes estimation turns out to be the best suited method.
\end{myabstract}

\section{Introduction}
\label{paper4:sec:introduction}
\emph{Location-based Services (LBSs)} compile information for their
users based on the position of one or several target persons. LBSs
can be initiated on request by the user, e.g., for being informed
about nearby \emph{Points of Interest (PoIs)}, or they can be
initiated on the arrival of certain spatial events, such as the
target person entering or leaving a pre-defined geographic zone.
Services of the first type are called \emph{reactive}, while the
latter ones are \emph{proactive}.

Another distinction of fundamental technical concern is whether an
LBS is used \emph{indoors} or \emph{outdoors}. So far, there is no
single positioning system that supports both environments in an
acceptable quality. While high-quality receivers for the
\emph{Global Positioning System (GPS)} are meanwhile integrated in
mass market cellular phones, GPS only works outdoors and not inside
buildings.

The most popular indoor localization technique to-date is
\emph{Location Fingerprinting (LF)}, having the major advantage to
exploit already existing network infrastructures, like IEEE 802.11
or GSM, which avoids extra deployment costs and effort. Based on a
database of pre-recorded measurements of \emph{Received Signal
Strength (RSS)} values sampled from different locations within a
building, denoted as fingerprints, a mobile terminal's location is
estimated by inspecting the RSS values it currently measures.

Resource-constrained terminals which are unable to store the
fingerprinting database, such as mobile phones or active badges, are
supported by a central \emph{location server}. The server accesses
the database and estimates their location based on RSS measurements
conducted at the terminal. So far, measured RSS values are either
transmitted on request, or the terminal updates them periodically
with the location server, according to a pre-defined update
interval. The associated problem is that periodic updating generates
an excessive number of messages, if the target person changes her
location only sporadically.

The periodic protocol performs especially badly if it
only needs to be observed when the target enters or leaves certain
pre-defined update zones, which is the case for proactive LBSs: As
it turns out, by automatically detecting update zones, not only
proactive single-target LBSs can be realized, e.g., for notifying
the LBS user as soon as she is near a PoI. Also proactive community
services, which consider the positions of multiple targets, are
possible. An example is proximity detection \cite{KuTr06}, which
automatically detects when two mobile targets have entered below a
pre-defined proximity distance. In this case the update zones for
each target are dynamically configured based on the current distance
to the other.

This paper explores a novel, more efficient approach for realizing
zone detection based on LF: The location server dynamically
configures the terminal with update zones defined in terms of RSS
patterns. Only when the terminal detects a match between its current
measurements and these patterns, that is, when it enters or leaves
the zone, it notifies the server about the fact. The associated
challenge is the adequate definition of RSS patterns, for which the
paper proposes several methods and compares them with respect to
message efficiency, computational overhead, and detection accuracy.
Also, the methods' support for different shapes and sizes of the
zones are evaluated. As it turns out, the approach strongly reduces
the message exchange at the air-interface, which has
the following advantages:

First, by avoiding excessive messages exchanged with the location
server, the power consumption of the tracked terminals is
significantly lowered. Second, valuable bandwidth is saved and
monetary costs the targets have to spend for mobile data services
are reduced. The latter aspect is of special importance for
cross-organizational scenarios, when the update messages can not be
directed over the network that yields the RSS measurements, but,
e.g., only by using public bearer services like GPRS or UMTS
packetswitched. Third, the approach avoids that the terminals need
to continuously switch back and forth between communication mode for
sending messages and scanning mode for observing RSS values, which
is an actual problem for many 802.11 adapters.
Finally, by reducing the general amount of location
information collected about the terminal, privacy of the target
person is enhanced.

The paper is structured as follows. The next section
discusses alternatives ways of organizing LF systems and motivates
and explains the chosen architecture and protocol for zone-based RSS
reporting. Several methods for representing geographical zones in
terms of RSS patterns are devised in Section \ref{paper4:sec:methods} and
compared analytically and by simulation in Section
\ref{paper4:sec:evaluation}. Section \ref{paper4:sec:relatedwork} overviews
related work. A conclusion and a discussion of further
work is given in Section \ref{paper4:sec:conclusion}.

\section{Architecture and Protocol}
\label{paper4:sec:protocols}
This work assumes LF systems to be organized in a
\emph{terminal-assisted} fashion, i.e., the terminal conducts the
RSS measurements and the location server estimates its location
based on the fingerprinting database. Alternatively, LF could also
be done in a \emph{network-based} as well as a \emph{terminal-based}
way, see \cite{Kuep05} for a classification of positioning methods.
This section first discusses the pros and cons of these two
alternatives. Then, an overview about efficient position update
methods devised for terminal-based positioning like GPS, which motivated this work, is given.
Finally, the novel protocol proposed for terminal-assisted LF is
presented.
\subsection{Alternative LF architectures}
In network-based LF systems the base stations
measure the RSS values of their clients and forward them to the
server, which, in turn, estimates the terminal's location. Thus, the
whole procedure, including measuring as well as location estimation,
takes place in the network. Network-based LF, however,
comes with several pitfalls. First, the base
stations need to be especially configured and attached to the
location server, which hinders cross-organizational operation.
Second, the target person's privacy control is very limited, because
all of her movements are observed at the location server. Third,
there is no obvious way for saving the energy of the terminal, which
continuously has to emit radio beacons for being tracked.

In terminal-based LF the RSS measurements and the
location estimation takes place at the mobile terminal, which caches
the fingerprinting database. The approach enhances the privacy of
the target person, because less data is collected about her than in
the network-based scenario. Also, terminal-based LF enables
cross-organizational operation "in the wild" \cite{LaMarca2005}, i.e.,
base stations not controlled by the location server can be included.
Finally, terminal-based LF can be combined with the existing
position update methods described below, where the position is
determined at the device and reported to the LBS only when needed.
From an architectural viewpoint this is similar to using GPS. A
drawback of terminal-based LF not present with GPS, however, is that
the fingerprinting database has to be stored at the device, which is
not an option for resource-constrained terminals like mobile phones
and active badges. Also, sophisticated location estimation
algorithms conducted at the device may overstrain its computational
capacities. Finally, every time the fingerprinting database is
changed the terminals have to be re-synchronized, which creates
severe scalability problems, independent of the terminal type.
\subsection{Existing position update methods}
For supporting proactive LBSs as well as services
which continuously track the position of a target, different position
update methods have been proposed and compared. The goal is to
provide for an efficient transmission of position data between a
location server in the Internet and a mobile device using
terminal-based positioning like GPS \cite{KTL06,WSCY99,LNR02}. The
methods are motivated by periodic reporting, according to a
pre-defined \emph{update interval}, being inefficient. As it turns
out, long update intervals increase the server's uncertainty about
the mobile's position, which negatively affects the quality of the
LBS. On the other hand, short intervals generate an excessive number
of messages in case the target person changes her location only
sporadically. Messages are also wasted when the target never
approaches the locations that are relevant for interaction with the
LBS.

A more efficient technique is distance-based
position reporting: The terminal is dynamically configured with a
certain \emph{update distance}, which prescribes the line-of-sight
distance between two consecutive position reports. A way to further
reduce messages is dead reckoning: Based on observed movement
parameters like speed and direction, the location server estimates
the mobile's current position. The most flexible method is
zone-based reporting: Position updates are only reported when the
terminal enters or leaves a pre-defined geographical \emph{update
zone}.
\subsection{Zone-based updating for terminal-assisted LF}
This paper explores zone-based updating for
terminal-assisted LF, enabling the efficient realization of
proactive LF-based LBSs.

\begin{figure}[h]
    \centering
        \includegraphics[width=0.85\textwidth]{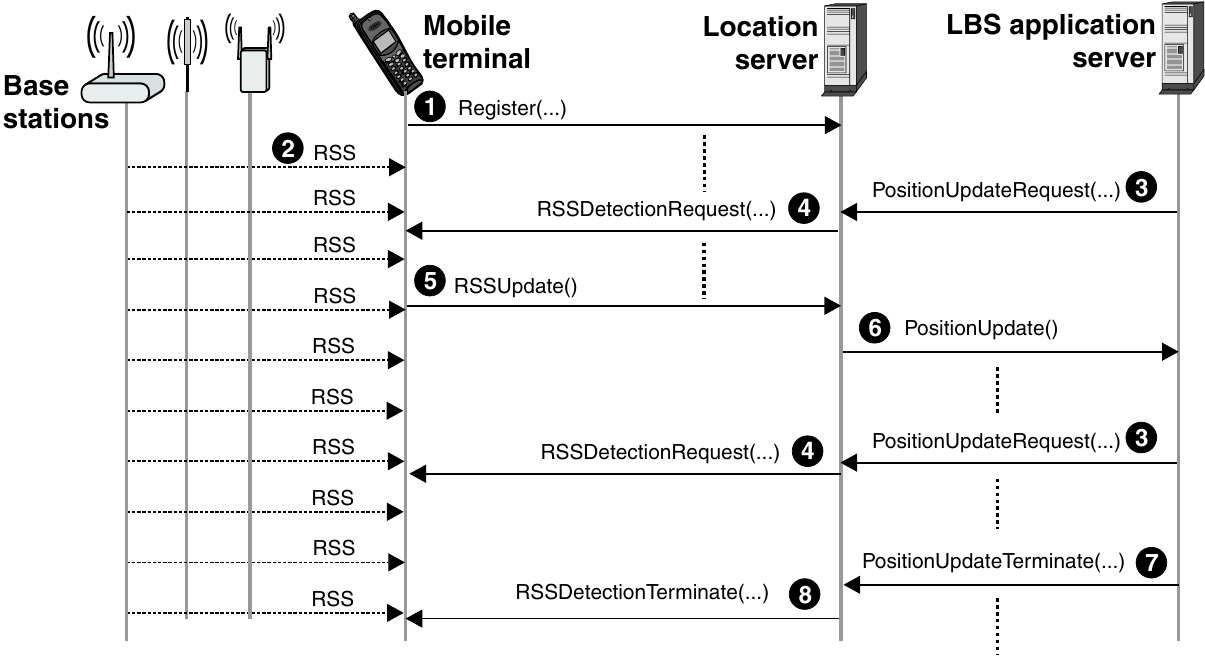}
    \caption{Proposed Tracking Protocol.}
    \label{paper4:fig:Example-floor}
\end{figure}

Figure \ref{paper4:fig:Example-floor} illustrates the proposed procedure:
First, the mobile terminal registers with the
location server (1) and then starts observing the RSS values of the
surrounding base stations (2). An LBS application server can
subscribe to zone-based updates by sending a respective request
message to the location server (3). The request carries the zone
definition, either in terms of geographical coordinates, e.g. as a
circle or a polygon, or symbolically, e.g. as a floor section. The
location server then translates the geographical update zone into an
RSS-based representation, which parameterizes one of the
\emph{detection methods} presented in Section \ref{paper4:sec:methods}. The
configuration is passed on to the mobile device (4), where it is
continuously compared to measured RSS values. Only when the current
measurements match the zone representation, they are reported (5).
At the location server, it is checked whether the updated RSS values
correctly correspond to entering or leaving the update zone. If so,
a position update is sent to the LBS application server (6). If a
position update request is canceled by the LBS (7), the location
server notifies the terminal about the fact (8).

It can be seen that terminal-assisted LF in the described
configuration has all the advantages of terminal-based LF, including
update efficiency and enhanced privacy due to the
reduced amount of collected data. However, the problem of carrying
and synchronizing the database is avoided. The main challenge
associated with the new approach is to translate geographical zones
into RSS-based representations. The next section explores several
methods for that.

\section{Detection Methods}
\label{paper4:sec:methods}
This section presents several methods for
implementing the proposed procedure. In order to be executable on
resource-constrained terminals, space and computational requirements
are kept as low as possible. Therefore, the methods mainly
constitute simplifications of classical LF techniques. They are
defined in terms of cell-based localization, i.e. locations are
represented as cells. A cell may correspond to a room or a part of
it, or a section of a hallway.

The following definitions are needed:
\begin{itemize}
\item $C = \{c_1,...,c_n\}$ is a finite set of \emph{cells} covered by the location system.
\item $Z = \{c_a,...,c_b\}$ is a subset of $C$ that corresponds to an \emph{update zone}.
\item A finite observation space $O = \{o_1, ..., o_m\}$ is assumed, with each \emph{observation} $o_i$ being a pair of a \emph{base station} $b$ and a \emph{measured RSS value} $v \in V = \{v_{min}, ..., v_{max}\}$ according to a discrete value range.
\item A \emph{sample} $s$ is a set of same-time
same-place observations, one for each visible base station.
\item A \emph{fingerprint} $f$ is a set of samples collected within the same cell.
\end{itemize}

\subsection{Common Base Stations}
A simple detection method, which does not even consider RSS values,
is to inspect the base stations occurring in the samples taken by
the terminal and compare them with those found in the fingerprints
for the cells of the update zone $Z$. If the number of common base
stations $n_{\cap}$ exceeds a certain threshold, the terminal is
assumed to be within $Z$.

\subsection{Ranking}
A possible improvement can be achieved by ranking common base
stations according to their RSS values. Instead of considering the
whole update zone at once, for each fingerprint within $Z$, the
common base stations' ranking is compared to their ranking in the
terminal's samples. The comparison is done using the spearman rank-order
correlation coefficient as proposed by
\cite{Krumm2004b}. If for any of the fingerprints a
certain threshold is exceeded, the mobile terminal
is assumed to be within the zone.
\subsection{Manhattan Distance}
A common \emph{deterministic} method in LF systems calculates the Euclidian distance in RSS space between a
terminal's measured samples and the fingerprints in the database
\cite{Bahl2000}. A simplified version can be applied
for the envisioned zone detection: First, instead of the Euclidian
distance, using the Manhattan distance as proposed by
\cite{Lorincz2005} comes with less computational overhead. Second,
current LF systems compare the distances of a measured sample to all
collected fingerprints and yield as a result the location associated
with the minimum distance. However, in our approach this would
require the whole fingerprinting database to be available at the
terminal. As an alternative fixed distance thresholds are proposed,
one associated with each fingerprint of $Z$. The thresholds are
independent of the remaining fingerprints in the database and are
based merely on the experienced deviations in a cell. The standard
deviations $\sigma_{c_i,b_j}$ of the RSS values experienced in cell
$c_i$ regarding all visible base stations $b_j \in B_{c_i}$ can be
easily derived from a cell's fingerprint. Upon the deviations, for
each cell contained in the update zone a distance threshold
$T_{c_i}$ is calculated as follows:
\begin{equation}
    T_{c_i} = \sum_{b_j \in B_{c_i}}{\sigma_{c_i,b_j}}
\end{equation}
$T_{c_i}$ is computed for each cell $c_i$ of $Z$.
Also for each cell, the means $\mu_{c_i,b_j}$ of the
base station's RSS values are provided. Thus, at the terminal for
each cell $c_i \in Z$ the Manhattan distance $manDist(c_i)$ is
calculated based on the means of the measured RSS values $m_{b_j}$,
with $b_j$ being in the set of base stations $B_o$ observed by the
terminal, as follows:
\begin{equation}
    manDist(c_i) = \sum_{b_j \in B_o \cap B_{c_i}}{|m_{b_j} - \mu_{c_i,b_j}|}
\end{equation}
A mobile terminal is estimated to be within $Z$, if and only if at least one of 
the cells $c_i \in Z$ satisfies the Manhattan distance:
$manDist(c_i) < T_{c_i}$.

A problem of the ranking method and the one based on Manhattan
distance is that often the terminal's samples and the fingerprints
only have a few base stations in common. As a possible solution,
both methods detect a terminal to be out of a cell, if there are
less than three base stations in common.

\subsection{Bayes Estimator}
Several LF systems use Bayesian estimation
\cite{Roos2002b,Youssef2005b,Haeberlen2004}, which represents a
\emph{probabilistic} method. In simple terms, for
each cell in the system a probability is calculated based on the
current samples taken by the terminal. The cell
associated with the highest probability is picked to be the current
one of the terminal. In the following the method is adapted for zone
detection by collapsing the underlying probabilistic model to a
simpler one:

Instead of testing one hypothesis for each cell in the system, only
two hypothesis are tested: $H_0$ states that the terminal is located
within the zone, while hypothesis $H_1$ states that it is located
out of it\footnote{Two hypotheses are used to ease notion instead of
one hypothesis and the negation.}. The probability
vector $\vec{\pi}$ describes the probabilities of these two
hypotheses being true, defined as follows:
\begin{equation}
\vec{\pi} = \left[\begin{array}{c}
P(H_0) \\
P(H_1) \\
\end{array} \right]
\end{equation}
To estimate the probabilities of the two
hypotheses, a Bayes estimator is used. The estimator calculates a
probability vector $\vec{\pi}$ based on a previous probability
vector $\vec{\pi}^{'}$ and a measurement which corresponds to an
element $o_j$ in the finite observation space. Initially, both entries of
$\vec{\pi}^{'}$ have the same probability. Then, $\vec{\pi}$ is continuously updated by the
following equation, where $P(o_j|H_i)$ is looked up in the simple
model provided by the location server:

\begin{equation}
    \vec{\pi}_{i} = \frac{P(o_j|H_i)\vec{\pi}^{'}_{i}}{P(o_j|H_0)\vec{\pi}^{'}_{0}+P(o_j|H_1)\vec{\pi}^{'}_{1}}
\end{equation}

The simple model is created as follows: The
probabilities $P(o_j|H_0)$ are calculated based on a set of
fingerprints taken from cells in the zone. In turn, the
probabilities $P(o_j|H_1)$ are calculated based on a set of
fingerprints of cells not in the zone. For that the histogram
method \cite{Roos2002b} is used.

In addition to the Bayes estimator, a simple Markov model is used to
guard the transitions of the detector over different time steps.
Thus, in a new time step $t+1$, $\vec{\pi}^{t+1}$ is calculated based on the previous estimate $\vec{\pi}^{t}$ at time $t$ as follows:

\begin{equation}
    \vec{\pi}^{t+1} = A \vec{\pi}^{t}
\end{equation}

where the Markov model A is defined as follows:
\begin{equation}
A = \left[\begin{array}{cc}
P_{s} & P_{ch} \\
P_{ch} & P_{s} \\
\end{array} \right]
\end{equation}
$P_{s}$ is the probability of sustaining the same hypothesis and
$P_{ch}$ is the probability of changing to another hypothesis. The
probabilities could be defined based on the sizes of the zones or
the expected movement behavior of the mobile
terminals.

\section{Evaluation}
\label{paper4:sec:evaluation}
In this section evaluation results are presented for the proposed
detection methods concerning their accuracy and efficiency. The
results have been achieved based on collected IEEE 802.11 RSS
measurements. Two scenarios are considered. One concerns the
accuracy of the methods and is based on correctly recognizing the
entering and exiting of single update zones randomly placed in an
indoor environment. The methods' efficiency is evaluated in the
second scenario, where a terminal is continuously tracked while
moving around in the same indoor environment, i.e., whenever the
terminal notifies the server about leaving an update zone, it is
configured with a neighboring one. In addition to these simulative
evaluations, an analysis of the computational and space requirements
for each of the proposed methods is given. As a
benchmark for comparison, a reference strategy based on
terminal-assisted LF with periodic RSS reporting according to
\cite{Haeberlen2004} was used.

\begin{figure}[h]
    \centering
        \includegraphics[width=0.45\textwidth]{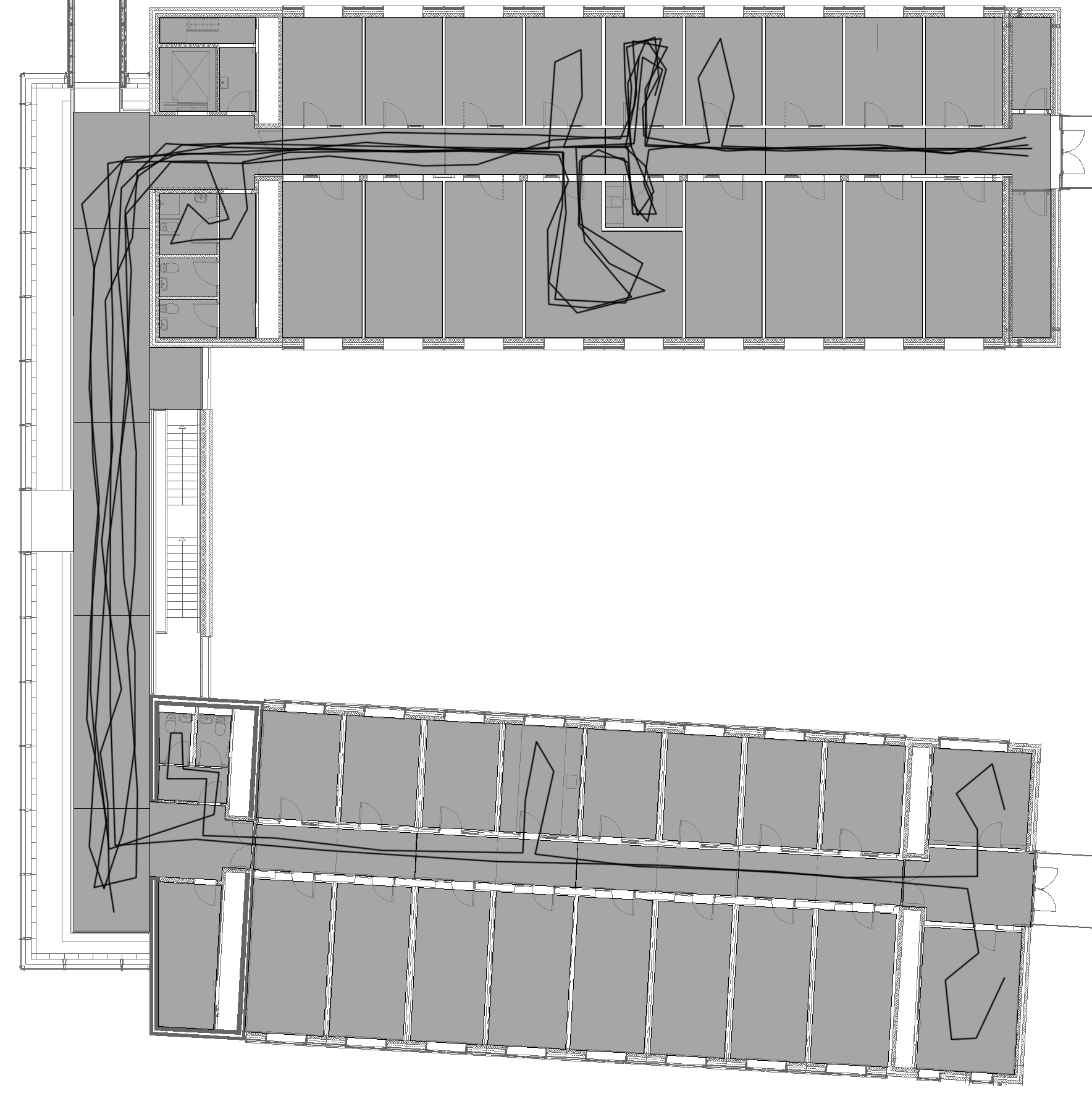}
    \caption{Layout of sampled area, covered by 63 cells}
    \label{paper4:fig:BabAda2Layout_2}
\end{figure}

All observations used in the evaluation were
collected in an 802.11 infrastructure with 22 reachable base
stations by a laptop with an Orinoco Silver 802.11 card. The
evaluation does not address the issue that different 802.11 cards
may measure RSS values differently. However, a possible solution
that could be applied for the Manhattan and the Bayes detector is
proposed in \cite{KjaergaardLoca2006}. The Common Base Station and the
Ranking detector are already designed to overcome the problem,
compare \cite{Krumm2004b}. Samples underlying the fingerprints as
well as those for the terminal's localization were taken
at 1 Hz. The set of fingerprints covers 63 cells in an office
building, compare Figure \ref{paper4:fig:BabAda2Layout_2}.
The building was broken up into cells with an
average size of 16 $m^2$ matching rooms or parts of hallways. Each
fingerprint consists of 60 seconds of samples collected by a person
walking around in the fingerprinted cell. The observations taken for
the localization were collected during 5 walks, totaling
34 minutes. They were taken on different days along different routes
as shown in Figure \ref{paper4:fig:BabAda2Layout_2}. The framework for
taking the samples is partly based on software by the Placelab
project \cite{LaMarca2005}.

\subsection{Accuracy}
To assess the detectors' accuracy, each of them was
tested by 50 different circular zones placed randomly in each of the 5 walks,
yielding a total of $5 \times 50 = 250$ tested zones per detector.
The circle radii were randomly selected between
4-10 meters.

The parameters used by the detectors in the
evaluation were chosen based on the results of a number of initial
experiments. For the common base station detector the threshold for
being in a zone was set to 70\% overlap. For the ranking detector a
threshold of 0.9 for the spearman rank-order correlation coefficient
was used. For the Bayes estimator detector the probabilities for the
Markov model were set to $P_s = 99\%$ and $P_{ch} = 1\%$.

The detectors' accuracies are compared at a time frame level, with
each frame being one second long. Therefore, the three measures:
\textit{sensitivity}, \textit{specificity} and \textit{global
accuracy} are calculated as described below. The calculations are
based on the following metrics: $TP$ (true positives) equals the
number of time frames the terminal stays in a zone and correctly
detects to do so. $FP$ (false positives) is the number of time
frames the terminal does not stay in a zone, yet wrongly a
zone-containment is detected, $TN$ (true negatives) is the number of
frames out of the zone correctly documented by a detector. Finally,
$FN$ (false negatives) equals the number of frames spent
within the zone, but falsely assumed to be out of
the zone. The \textit{sensitivity} is then defined as $Sn = TP/(TP +
FN)$. The \textit{specificity} is defined as $Sp =
TN/(TN + FP)$. Neither $Sn$ nor $Sp$ alone constitute a good
measure of global accuracy. For calculating \textit{global accuracy}
the \emph{correlation coefficient (CC)} is used, a well-known
mathematical concept which is normally used for
mapping two random variables onto one and which has been applied in
gene prediction \cite{Burset1996} for combining specificity and
sensitivity. This application of the CC is adopted in this work and
thus the global accuracy quantifies how much the sensitivity and the
specificity agree about a detector's performance:
\begin{equation}
    CC = \frac{TP \cdot TN - FP \cdot FN}{\sqrt{(TP + FP) \cdot (TN + FN) \cdot (TP + FN) \cdot (TN + FP)}}
\end{equation}
All three measures take their values between 0 and 100 percent, where values close to 100 indicate good detection accuracy.

\begin{figure}[h]
\centering
\includegraphics[viewport=20 410 470 770,width=0.6\textwidth,clip]{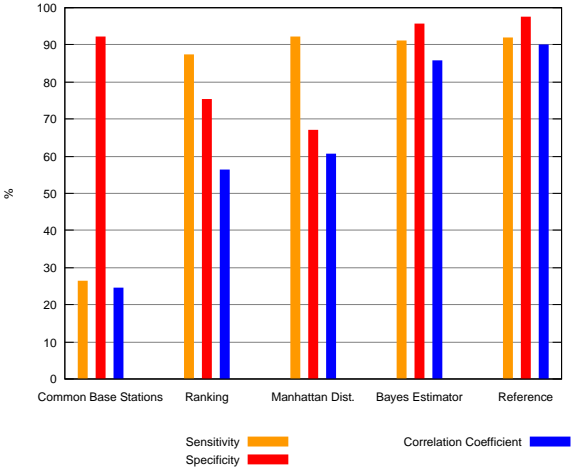}
\caption{Results for a single sample}
\label{paper4:fig:plotResultsZonePlot}
\end{figure}

\begin{figure}[h]
\centering
\includegraphics[viewport=20 410 450 770,width=0.6\textwidth,clip]{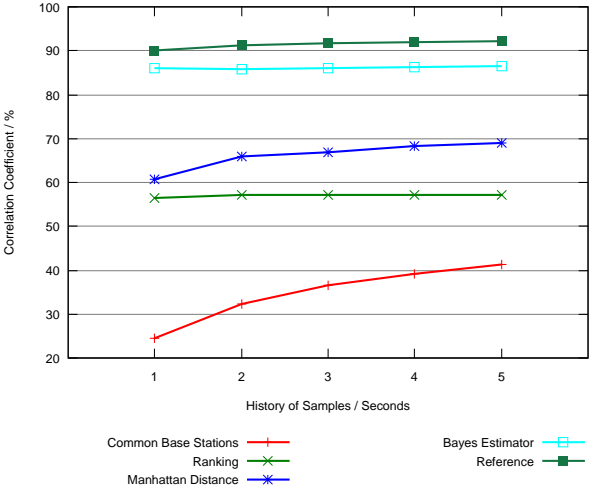}
\caption{Results for increasing sampling times}
\label{paper4:fig:plotResultsSamples}
\end{figure}

The first evaluation assumes that the terminal
provides the detector with single samples as an input value,
corresponding to a sampling time of one second, compare Figure
\ref{paper4:fig:plotResultsZonePlot}. The results show that the common base
station detector and the ranking detector are the least accurate
detectors with a global accuracy of 24.55\% and 56.54\%
respectively. The ranking detector performs better than the common
base station detector, which indicates that taking the ranking of
the RSS measurements into account gives a gain in accuracy. The low
sensitivity of the common base station detector shows that the low
global accuracy is caused by a tendency to not detect zone presence.
The Manhattan distance detector yields a global accuracy of 60.73\%.
The most accurate of the detectors is the Bayes estimator detector
with a global accuracy of 85.96\%. The reason may be its detailed
model for representing RSS values. In comparison, the reference
strategy yields a global accuracy of 90.12\%, which is
only slightly better than the Bayes detector. 
Evaluations were also run based on longer sampling
times at the terminal-side, compare Figure
\ref{paper4:fig:plotResultsSamples}. For the ranking and the Manhattan
distance detector multiple samples taken for each base station were
aggregated to their mean value. The evaluation shows that the
accuracy of the common base stations and Manhattan distance
detectors increases to respectively 41.35\% and 68.96\% with five
samples. The accuracy of the ranking detector, the Bayes estimator
detector and reference system only increase with a small gain to
respectively 57.06\%, 86.55\%, and 92.28\% with five samples. Again,
the Bayes estimator is the best of the detectors, even when using
single samples. Such short sampling times are desirable in order to
increase the responsiveness of the system.

\begin{figure}[h]
\begin{minipage}[b]{0.5\linewidth} 
\centering
\includegraphics[viewport=20 410 450 770,width=1.0\textwidth,clip]{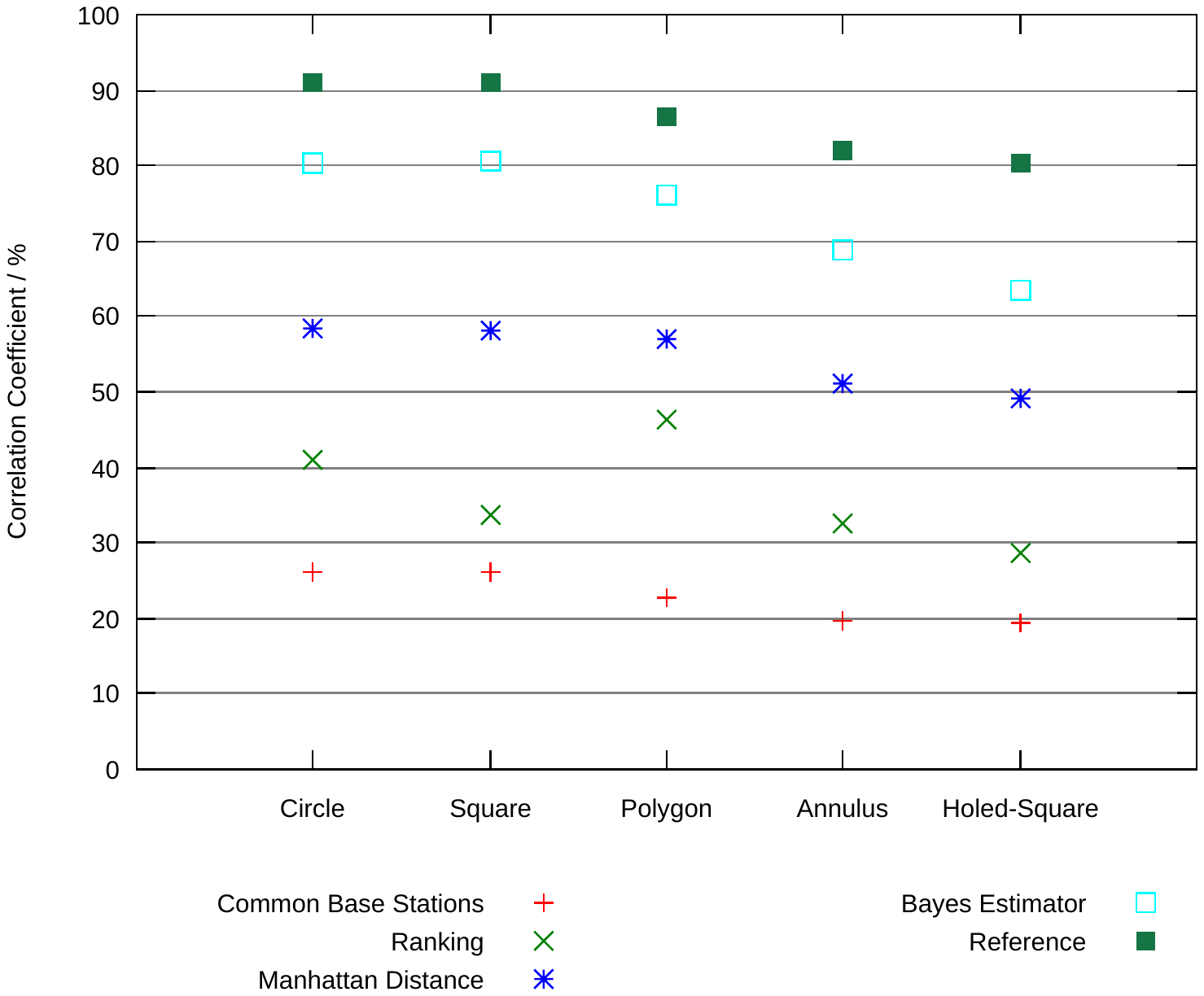}
\caption{Results for different zone shapes}
\label{paper4:fig:plotResultsShapes}
\end{minipage}
\hspace{0.1cm} 
\begin{minipage}[b]{0.5\linewidth}
\centering
\includegraphics[viewport=20 410 450 770,width=1.0\textwidth,clip]{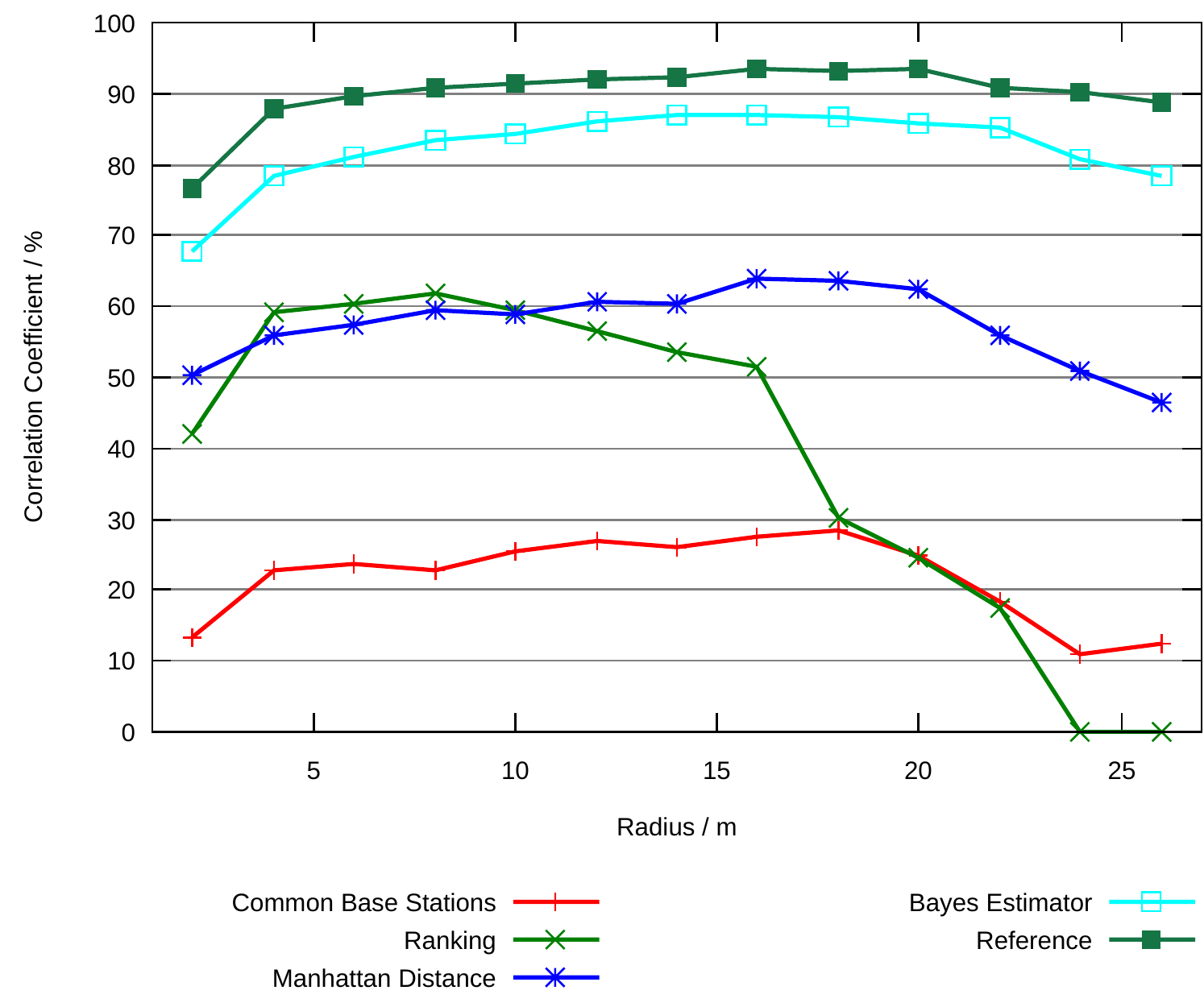}
\caption{Results for different zone sizes}
\label{paper4:fig:plotResultsSize}
\end{minipage}
\end{figure}

It was also important to evaluate whether the proposed detectors
could handle zones of different shapes and sizes. Therefore,
simulations based on five different shapes of approximately equal
sizes were conducted. The evaluated shapes were circles, squares,
annuli, holed-squares and polygons with between 4 to 8 edges. Figure
\ref{paper4:fig:plotResultsShapes} shows the obtained results, which
indicate that all detectors perform best with closed shapes,
however, with little accuracy losses for the more irregular-shaped
polygons. For both of the holed shapes there is about a 10\%
decrease, showing that the detectors are still able to handle such
complex zones. The results of the ranking detector differ from these
trends as they indicate a better support for polygon-shaped zones.

To evaluate the impact of the size of the shapes, evaluations were
run with circle-shaped zones of different radii. The results are
shown in Figure \ref{paper4:fig:plotResultsSize}. It can be seen that all
the detectors' accuracy drops for very small zones, primarily
because the detectors have very little fingerprinting data to base
their estimates on. One can also see that the threshold selected for
the ranking detector is not optimal for larger zones. All detectors,
however, experience a decrease in accuracy for radii above 20
meters. This fact can be attributed to the detectors being
pessimistic, that is, they prefer estimating a terminal to be out of
a zone over being contained in it. The pessimism shows up as an
increase in errors when more and more space of the evaluated walks
is covered by a zone. The collected data did not enable us to
correctly evaluate circle-shaped zones with radii above 24 meters,
because in this case more than 70\% of the time frames of the walks
would be contained by the zone. Based on the accuracy evaluations it
can be concluded that the Bayes estimator detector is the most
accurate and robust of the proposed detectors.

\subsection{Efficiency}
To evaluate the efficiency of the proposed protocols and detectors,
another evaluation simulating the continuous tracking of a terminal
has been carried out. The evaluation is based on the same collected
walks as before and a simple tracking protocol: First, a
circle-shaped zone detector of 10 meter radius is set up with its
center located at the starting cell of the walk. When the detector
reports that the terminal has moved out of the zone, a second
detector is set up with a new zone, now with the just-estimated
location being its center. This process is repeated until the end of
the walk. To be able to use the same collected walk data several
times each evaluation is run several times with the first five
different locations in the collected walks as starting points.
During the evaluation the following statistics are collected: the
\emph{correctly saved updates}, which count the time frames when the
detector correctly estimates that it is in a zone and therefore an
RSS update is avoided; the \emph{wrongly saved updates}, which count
the frames where the detector wrongly estimates that it is in a zone
and therefore does not send an RSS update; and the \emph{RSS
updates}, which are actually sent when the detector has estimated
that the terminal may have moved out of the current zone. The used
walks in the evaluation actually represent a worse-than-average
scenario, because the terminal is moving most of the time. In a
scenario with a more static movement pattern a larger number of RSS
updates would be saved.

\begin{figure}[h]
\centering
\includegraphics[viewport=20 410 490 770,width=0.6\textwidth,clip]{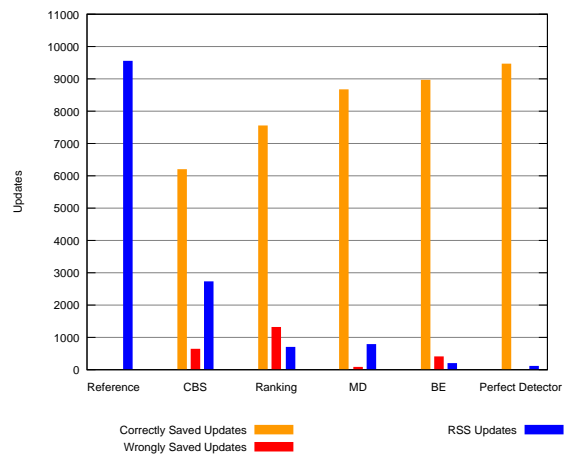}
\caption{Efficiency evaluation results}
\label{paper4:fig:plotTrackResults}
\end{figure}

The results show that for all of the detectors the
number of RSS updates is considerably lowered in comparison to the
9572 RSS updates produced by secondwise RSS reporting, which was
assumed for the reference system, compare Figure
\ref{paper4:fig:plotTrackResults}. The common base stations (CBS) detector,
the ranking detector, and the Manhattan distance (MD) produce the
most updates with respectively 2721, 693, and 803 RSS updates. The
RSS updates produced by the Bayes estimator (BE) detector is 192
which is close to the efficiency of a perfect detector, which would
produce 114 RSS updates. The Bayes estimator shows the fewest RSS
updates but generates more wrongly saved updates than the Manhattan
distance detector respectively 423 and 89. However, the detectors' performance can be fine tuned by changing some
of the parameters. For instance, wrongly saved updates can be traded
for generating a few excessive RSS updates, which in turn can be
filtered out at the location server, thus ensuring better overall
accuracy. In summary, considering all three metrics the Bayes estimator detector is the best choice.

\begin{figure}[h]
\centering
\includegraphics[viewport=20 410 490 770,width=0.6\textwidth,clip]{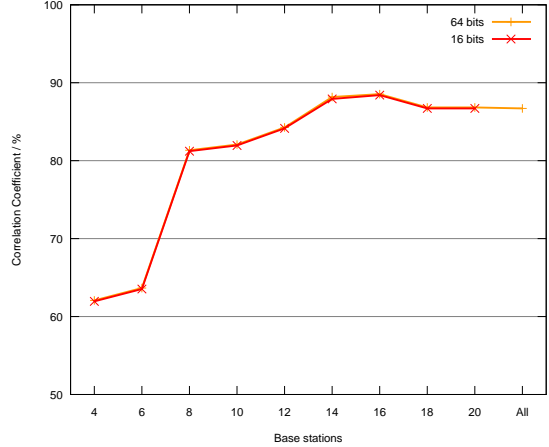}
\caption{Effect of the number of base stations on the accuracy of the Bayes estimator}
\label{paper4:fig:zoneresults}
\end{figure}

\subsection{Space and computation analysis}
In this section the space and computation requirements of the different detectors are analyzed. The analysis is based on the following parameters: $M$ is the number of observations provided by the terminal to the detector; $B_{zone}$ is the number of base stations visible from cells in the zone; $B_{all}$ is the number of all base station covered by the system; $Z$ is the number of cells in the zone; $V$ is the number of possible RSS values. For each of the detectors the results of the analysis are given in Table \ref{paper4:tab:compspace}.

\begin{table}[h]
\center
\begin{tabular}{|l|c|c|}
\hline
Detector & Computations & Space \\
\hline
Common Base Stations & $O(M)$ & $O(B_{zone})$ \\ \hline
Ranking & $O(M + Z \times B_{zone} \times \log(B_{zone}))$ & $O(B_{zone} \times Z)$ \\
\hline
Manhattan Distance & $O(M + B_{zone} \times Z)$ & $O(B_{zone} \times Z)$ \\
\hline
Bayes Estimator & $O(M)$ & $O(B_{all} \times V)$ \\
\hline
Reference System & $O(1)$ & $O(1)$ \\
\hline
\end{tabular}
\caption{Space and computational requirements on mobile terminals}
\label{paper4:tab:compspace}
\end{table}
The computation and space requirements are low for both the common base stations detector and the reference system, the latter because it does not perform any extra calculations or use any additional space on the mobile terminal. The ranking detector has higher space requirements and computation requirements, because it needs to sort the measurements and also store the calculated rankings for each cell in the zone. The Manhattan distance detector has lower computation but the same space requirements. Computations are needed for calculating the Manhattan distances to all cells in the zone and each distance computation considers all base stations visible in the zone. Its space use is attributed to storing mean values for all cells in the zone. The Bayes estimator detector has low computation requirements, but the highest space requirements because it needs to store the simple probabilistic model.

To further reduce the space consumption of the Bayes estimator three techniques are proposed. First, a lossless compression technique for representing repeated entries is applied, which just counts repetitions of the same values. Because 802.11 RSS measurements in practice only span a small range of V and because the entries are generated using the histogram method, the entries contain a lot of repetitions. Second, the representation of the entries is constrained to only 16 bits. Third, the number of base stations used for the entries can be reduced.

For example, without these techniques the space consumption of the detector on the collected data, with $V=255$, $B_{all}=47$, two hypotheses, and a 64 bit representation of probabilities, the memory needed for representing one zone would be $2 \times 47 \times 255 \times 8b = 95,9Kb$. However, when the first two techniques are applied and all base stations are kept, the data can be compressed to $1Kb$. If the number of base stations is also reduced to a maximum of 12, even $0.5Kb$ are possible. Both values seem fairly acceptable.

To learn whether the reduction of base stations and bit representation negatively affects the accuracy of the Bayes detector, an extra accuracy evaluation was run and the results are shown in Figure \ref{paper4:fig:zoneresults}. They indicate that the reductions do not have a major impact on the accuracy, as long as the maximum number of base stations is not limited to fewer than 8. However, this number is only valid for the zone sizes used in the evaluation because for larger zones more base stations might be needed for a whole zone to be covered.

To subsume, the Bayes estimator turns out as the best of the presented methods for all considered aspects: accuracy, responsiveness, support for different sizes and shapes, as well as efficiency. With respect to the reference system, it yields a comparable accuracy, while the number of exchanged messages is strongly reduced. As discussed, the little lack of accuracy can be counterbalanced by slightly reducing the number of saved update messages.

\section{Related Work}
\label{paper4:sec:relatedwork}

\subsection{Infrastructure-based}
One of the first infrastructure-based systems was RADAR \cite{Bahl2000}, that applied different deterministic mathematical models to calculate the position (in coordinates) of a terminal based on IEEE 802.11 measurements. Similar methods have also been applied to GSM \cite{Otsason2005}. The mathematical models used had to be calibrated for each site where the systems had to be used. In comparison to RADAR, later systems have used probabilistic models instead of deterministic models. This is because a good deterministic model for the volatile radio environment has not been found. As in the case of the deterministic models in RADAR, the probabilistic models are calibrated for each site. Examples of systems, which determine the coordinates of a terminal, are published in \cite{Roos2002b,Youssef2005b,Krumm2004}. Systems determining the logical position or cell of a terminal are published in \cite{Haeberlen2004,Castro2001}. From a perspective of resource-constrained terminals, existing systems are not optimal with respect to the overhead induced by using poll or periodic update protocols only, as discussed in Section \ref{paper4:sec:protocols}. However, from an accuracy perspective the proposed zone updating protocol has the drawback that history tracking algorithms cannot be applied to improve LF accuracy. A possible solution is to report RSS values sampled over the last $n$ seconds whenever a zone update is due. This way, a possible historical analysis and the decision whether the update is really in the zone or not could still be done at the server-side.

In addition to the above systems, which estimate the location of terminals, a number of systems, such as \cite{Krumm2004b}, have been studied where the calibration step is only carried out by users for tagging relevant places. The systems propose simple metrics based on signal strength measurements to quantify when terminals are in proximity of calibrated places. One of the strengths of these simple metrics is that they overcome the problem of 802.11 cards returning different RSS values. Such systems are relevant to this work with respect to the methods they propose for proximity detection. However, such systems can only detect presence at a single point and not within zones with specific shapes and sizes, as addressed in this paper.

A system which has addressed, by using additional sensors, the needs of resource-constrained terminals when used with fingerprinting-based indoor location systems is \cite{You2006}. They propose a communication protocol between the location server and the terminal, which dynamically adapts the RSS update rate of the terminal based on the distance to the last reported update using measurements from an accelerometer. In comparison, the methods proposed in this paper do not require any extra sensors and are therefore usable for a broader range of terminals where such extra sensors are not present or too expensive to include. In addition to this, the proposed methods in this paper can also be used with arbitrary shaped zones and not just zones defined by a distance to a specific point.

Thus, in comparison to existing infrastructure-based solutions the proposed approach represents an improvement, because it enables efficient tracking and accurate zone detection based on  RSS measurements only.

\subsection{Infrastructure-less}
Most infrastructure-less systems are based on protocols which are more energy-efficient than for instance IEEE 802.11, such as IEEE 802.15.4 or communication over the 433/916 MHz bands reserved for telemetry. In \cite{Bulusu00a} a system is presented which senses the proximity of a mobile node to static beacon nodes which output their id and position. The position of the mobile node is then estimated by finding the centroid of the positions of the proximate beacon nodes. A system that proposes methods for infrastructure-less localization inspired by infrastructure-based techniques is MoteTrack \cite{Lorincz2005}. The system consists of a number of wireless sensor network nodes where some have the role as static beacon nodes and other are mobile nodes which the system should locate. The system is based on location fingerprinting using RSS to the static beacon nodes. The fingerprints are stored distributely over the static beacon nodes and provided to the mobile nodes when in proximity. The system's method for location estimation is based on weighted nearest fingerprints based on the Manhattan distance instead of the Euclidian distance to lower computation needs. The computing of the location estimates can be carried out by either the mobile nodes or the beacon nodes, depending on which of the proposed sharing techniques is used. These systems are related to the proposed methods in terms of how they achieve energy-efficiency and do decentralized estimation. However, because all such systems assume that there is no infrastructure, they do not address how to combine decentralized estimation with the capabilities of infrastructure-based solutions.

\section{Conclusion and Further Work}
\label{paper4:sec:conclusion}
The paper proposed the novel approach of zone-based RSS reporting for location fingerprinting, where the terminal is dynamically configured with RSS-based representations of geographical update zones. Only when the terminal detects a match to the RSS patterns, it reports its measurements to the server.
Several methods for realizing zone-based RSS reporting were proposed and profoundly compared. As it turned out, an adaption of classical Bayes estimation is a promising approach, which, in comparison to the assumed reference system, strongly reduces message overhead while yielding a high accuracy and responsiveness. Given the mechanisms described in this paper, existing approaches for efficiently realizing proactive LBSs -- which, so far, assume terminal-based positioning like GPS -- can be easily applied to LF systems. This concerns not only single-target LBSs, but also proactive multi-target LBSs, compare \cite{KTL06}.
Two further issues subject to future work are discussed in the following.

First, with some technologies, such as IEEE 802.11, already the RSS scanning is rather resource consuming, which makes it desirable to minimize the needed scans. One possible method, which, however, only applies to big zones, is to subdivide a zone in a way that in the central part of it a long scanning interval is used, while short intervals are applied at the borders of the zone. Another method is using an moving-versus-still estimator based on RSS measurements, such as the one proposed in \cite{Krumm2004}, to estimate whether the terminal is moving or not, and then adapt the scanning intervals to this information. However, the proposed estimator is rather expensive in terms of needed samples and computations, so a scaled-down version would have to be developed.

A second issue this work has not addressed is how the building
layout in terms of floors affects the
detection methods. LF techniques evaluated for both GSM and 802.11
in \cite{Otsason2005} have shown good performance, at least in
office-like buildings, for estimating the floor level. So, at least
for the Manhattan distance detector and the Bayes estimator, floor
errors should not be a major issue. The presented detectors also
allow zones to be defined over several floors. 

\subsubsection{Acknowledgments.} We appreciate the comments, advice, and insights of our reviewers and especially our shepherd John Krumm. We thank Carsten Valdemar Munk for helping collecting signal strength measurements. M. B. Kj\ae rgaard is partially funded by the software part of the ISIS Katrinebjerg competency center http://www.isis.alexandra.dk/software/.


\clearemptydoublepage
\chapter{Paper 5}
\label{chap:mobilware2008}

The paper \emph{Efficient Indoor Proximity and Separation Detection for Location Fingerprinting} presented in this
chapter has been published as a conference paper~\cite{KjaergaardMobilware2008}.

\begin{publist}{\cite{KjaergaardMobilware2008}}
  \item[\cite{KjaergaardMobilware2008}] M.\ B.\ Kjærgaard, G. Treu, P. Ruppel and A. Küpper. Efficient Indoor Proximity and Separation Detection for Location Fingerprinting. In \emph{Proceedings of the First International Conference on MOBILe Wireless MiddleWARE, Operating Systems, and Applications}, pages~1--8, ACM, 2008.
\end{publist}

\noindent

\clearemptydoublepage



\mytitle{Efficient Indoor Proximity and Separation Detection\\ for Location Fingerprinting}{ 
  Mikkel Baun Kj\ae rgaard\footnotemark[1] \and 
  Georg Treu\footnotemark[2] \and
  Peter Ruppel\footnotemark[2] \and
  Axel Küpper\footnotemark[2] }{  
  \footnotetext[1]{Department of Computer Science, University of
    Aarhus, IT-parken, Aabogade 34, DK-8200 Aarhus N, Denmark. E-mail:
    \texttt{mikkelbk@daimi.au.dk}.}
  \footnotetext[2]{Mobile and Distributed Systems Group, Institute for Informatics, Ludwig-Maximilian University Munich, Germany. E-mail:
    \texttt{[georg.treu$|$peter.ruppel|axel.kuepper]@ifi.lmu.de}.}
    } 
        

\begin{myabstract}
Detecting proximity and separation among mobile targets is a basic mechanism for many location-based services (LBSs) and requires continuous positioning and tracking. However, realizing both mechanisms for indoor usage is still a major challenge. Positioning methods like GPS cannot be applied there, and for distance calculations the particular building topology has to be taken into account. To address these challenges, this paper presents a novel approach for indoor proximity and separation detection, which uses location fingerprinting for indoor positioning of targets and walking distances for modeling the respective building topology. The approach applies efficient strategies to reduce the number of messages transmitted between the mobile targets and a central location server, thus saving the targets' battery power, bandwidth, and other resources. The strategies are evaluated in terms of efficiency and application-level accuracy based on numerous emulations on experimental data.
\end{myabstract}


\section{Introduction}
Location-based Services (LBSs) take into consideration the current positions of users or other targets in order to support navigation, to deliver a list of nearby points of interest like restaurants or to show buddies being in close proximity. LBSs can be realized in a \emph{reactive} or \emph{proactive} fashion. In the former category, location-based data is delivered to the user only on request, while proactive services are automatically triggered as soon as a pre-defined \emph{location event} occurs, for example, when a target enters or leaves a city, district, building or another geographic zone. The user can then be informed about that event and receive additional information. Unlike reactive LBSs, proactive ones are much more difficult to realize, because targets need to be permanently tracked for checking the occurrence of location events. This paper focuses on two special problems that belong to the class of multi-target location events, where the positions of several targets need to be determined and compared on a permanent basis. \emph{Proximity detection} is defined as the capability of an LBS to detect when two of a group of mobile targets approach each other closer than a pre-defined \emph{proximity distance}. Analogously, \emph{separation detection} discovers when two targets depart from each other by more than a pre-defined \emph{separation distance}. The detection of such events can be used in manifold ways, for example, in the context of community or dating services for alerting the members of these communities when other members approach or depart. The solutions presented in this paper have been especially tailored for indoor environments like offices, factory floors, university campuses, hospitals, or railway stations. 

In earlier work, mechanisms for proactive proximity and separation detection have been included into the LBS middleware TraX, see also \cite{KuTr06} and \cite{KTL06}. These mechanisms control the positioning process within GPS-capable mobile devices carried by the targets and coordinate the transfer of the derived position fixes to a central location server for checking for proximity and separation with other targets. This transfer is referred to as \emph{position updating}, and it may happen periodically, when the target has covered a certain distance with respect to the last reported position or if she has entered or left a certain zone. Proximity and separation checks are based on the \emph{line-of-sight} or \emph{Euclidean distance}, which can be simply calculated from the geographic positions of the involved targets. TraX applies a combination of different position updating and polling strategies with the goal to reduce the number of messages that pass the GPRS or UMTS air interface, to lower the battery consumption of the mobile phones, and to disburden the location server. Unfortunately, the use of GPS makes TraX applicable only in outdoor environments, because GPS signals typically do no penetrate buildings. Alternative outdoor positioning technologies, for example cellular methods like Cell-Id, may work indoors, but lack in providing a sufficient degree of accuracy of position fixes as required for both detection schemes. Therefore, the only solution to offer proximity and separation detection within buildings is to use an indoor positioning scheme.

In the recent years, many indoor positioning schemes have been developed differing from each other in the kinds of signals used (infrared, radio, ultrasound), the type of signal measurements (signal traveling time, received signal strength, coverage) and the mathematical methods (fingerprinting, lateration, angle of arrival) for deriving a position fix from the measurements. One of the most prominent schemes is called \emph{location fingerprinting} (LF). It estimates the position of a target from measuring the strength of radio beacons (\emph{received signal strength}, RSS) emitted by several WLAN 802.11 access points in the close surrounding. The location of the target is then determined by mapping the measured values onto RSS patterns, which are called \emph{fingerprints} and which have been pre-recorded at well-defined positions for storage in a map database. LF has been selected for extending the TraX framework, because it provides a comparatively high accuracy of location data when compared to other technologies. Another advantage is that it does not require dedicated hardware, that is, it works with existing WLAN 802.11 installations available in many buildings as well as with conventional WLAN-capable mobile devices.    

Unfortunately, replacing GPS by LF in the TraX middleware is not enough. Unlike GPS, where mobile devices can determine their geographic position, LF only delivers a vector of RSS measurements as observed by the device on the spot. As a consequence, position updating cannot be triggered when the target has covered a certain distance or left a zone, but it requires a new position updating scheme, which carries RSS values and which is triggered by a certain change of RSS values. Another novelty concerns the semantic of distance. Checking for proximity and separation under consideration of Euclidean distances does not make much sense indoors, because several targets could be located on top of each other on different floors of a building, to give only one example. Applying both detection functions for walking distances is therefore a more reasonable, but also a more sophisticated approach. 

This paper proposes different strategies for efficiently performing proactive proximity and separation detection in indoor environments based on walking distances and by using LF. Similar to its outdoor counterparts, the goal of these strategies is to lower the battery consumption of mobile WLAN devices carried by the targets, to reduce the workload of the server performing the checks and to keep the amount of messages passing the air interface as low as possible. The latter especially makes sense in cross-organizational scenarios, where position update and polling messages are not sent over the WLAN network used for performing LF, but by using public bearer services like GPRS or UMTS.

LF and advanced functions for LBSs have been a hot topic in research during the recent years. The following section gives an overview about related work and explains differences to and similarities with the approaches presented in this paper. Section \ref{paper5:sec:Trax} introduces the TraX middleware from a conceptual point of view and explains how to extend it for the purposes of indoor proximity and separation detection. Section \ref{paper5:sec:Approach} then describes position updating and polling strategies for both detection functions that work in combination with LF and walking distances. Finally, Section \ref{paper5:sec:Evaluation} presents the results achieved by prototype evaluation and emulation for the proposed strategies, followed by the conclusions and discussion of further work in Section \ref{paper5:sec:Conclusion}.

\section{Related Work}
\label{paper5:sec:Related_Work}
In the recent years, LF has been evaluated and used mainly for single target location determination, therefore not addressing proximity and separation detection \cite{Bhasker2004,Haeberlen2004,Roos2002b,Youssef2005b}, with NearMe \cite{Krumm2004} as an exception. NearMe supports a short-distance proximity detection, which takes into consideration RSS measurements and Euclidean distances only, as well as a long distance mode, which applies a base station coverage-graph analysis. NearMe is a client-server approach with periodic RSS updating between mobile device and location server, which causes significant overhead when a target does not move for a longer period of time. 

LBSs applying LF in IEEE 802.11 networks and using proximity information have been built and evaluated for usability. The location-based messaging system InfoRadar \cite{Rantanen2004}, for example, uses an LF technique proposed by Roos et al. \cite{Roos2002b}. A location server polls RSS measurements from the targets' devices for estimating their positions and checking them for proximity subsequently. The ActiveCampus \cite{Griswold2004} system provides a set of LBSs to foster social-interactions in a campus setting. One of these services can list nearby buddies and show maps overlaid with information about buddies, sites and current activities. Targets are located using a terminal-assisted LF method proposed by Bhasker et al. \cite{Bhasker2004} and a combination of poll-based and periodic RSS updating, which, however, turned out to be a bottleneck in this system when trying to scale beyond 300 concurrent users. The strategies proposed in this paper scale much better and are novel in that they consider walking instead of Euclidean distances, which, as mentioned before, better reflects the needs of indoor LBSs.

Several systems support the realization of LBSs based on LF in general. Many have been proposed for integrating position fixes produced by different positioning technologies, among them LF, thus easing implementation and improving server-side efficiency. Examples of such systems are the Rover system \cite{Banerjee2002}, the Location Stack \cite{Hightower2002} and its implementation in the Universal Location Framework (ULF) \cite{Graumann2003}. They provide means to integrate and fuse information from several positioning methods, query location information, improve scalability and define location-based triggers. The systems have been integrated with LF techniques applied in Horus \cite{Youssef2005b} and RADAR \cite{Bahl2000}. Position fixes are obtained from the location sources by push, pull and periodic location updating methods. The Rover system has been evaluated for server-side efficiency in terms of CPU-load based on simulated inputs. In comparison, this paper proposes strategies for an efficient message transfer over the air interface, which also improves server-side efficiency and saves battery resources at the client-side.
\section{TraX}
\label{paper5:sec:Trax}
\begin{figure}
\begin{center}
	\includegraphics{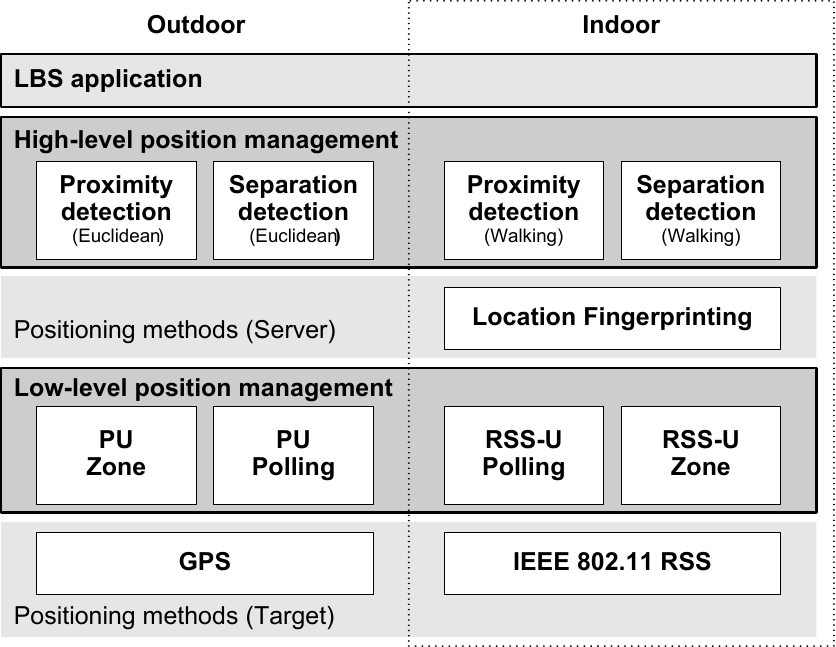}
	\caption{TraX}
	\label{paper5:fig:TraX}
\end{center}
\end{figure}

The strategies proposed in this paper for proximity and separation detection are part of the LBS middleware TraX \cite{KuTr06}, which has been developed for efficiently exchanging position fixes and for collecting, processing, and interrelating position fixes of several targets. The framework provides a set of basic building blocks, which can be applied for a broad range of LBS applications and which can be dynamically configured, for example in order to meet accuracy and up-to-dateness demands on position fixes. The position management framework is arranged between a layer representing the on-target parts of one or several positioning methods and the LBS application, as illustrated in Figure \ref{paper5:fig:TraX}. It is subdivided into so-called \emph{low-level} and \emph{high-level functions} and the on-server parts of positioning methods. The layer of the low-level functions sits on top of the on-target positioning methods and provides different methods for exchanging position fixes or position measurements between a mobile device and a location server. The high-level position management offers advanced functions for LBSs, for example proximity and separation detection as treated in this paper or k-nearest neighbor search and clustering. They apply the low-level functions according to a certain strategy. The on-server positioning methods sit in between the low-level and high-level layers and provide estimation of position fixes from position measurements.

TraX was originally tailored for outdoor use and for Eu\-cli\-dean-distance proximity and separation detection in conjunction with GPS, see the left of Figure \ref{paper5:fig:TraX}. The low-level methods for exchanging position fixes include: position updating based on dynamically configuration of terminals for updating their positions when leaving a geographical update zone (\emph{PU Zone}), and explicit \emph{polling} of terminals for immediate reports of their positions (\emph{PU Polling}). The high-level layer implements the functions of Euclidean-distance proximity and separation detection based on the so-called \emph{Dynamic Centered Circles (DCC)} strategy \cite{KuTr06}.

In this paper, the middleware is extended for indoor use of walking-distance proximity and separation detection in conjunction with LF, see the right of Figure \ref{paper5:fig:TraX}. The low-level methods for exchanging IEEE 802.11 RSS measurements include: RSS updating for sending RSS measurements when leaving a pre-configured update zone (\emph{RSS-U Zone}), and explicit \emph{polling} of terminals for immediate reports of RSS position measurements (\emph{RSS-U Polling}). The high-level layer implements the functions of walking-distance proximity and separation detection based on the strategy proposed in Section \ref{paper5:sec:Approach}.

LF positioning is supported in a \emph{terminal-assisted} mode: the terminal conducts the RSS measurements and reports it to the location server, the latter usually on request or by sending periodic updates. The estimation of the target's location then happens at the server, which relieves the terminal from carrying the fingerprinting database and from applying complex estimation algorithms, thus enabling LF on resource-constrained terminals. In comparison, other LF architectures such as \emph{network-based} or \emph{terminal-based} setups can either not support resource-constrained devices or cannot be efficiently optimized in terms of message overhead as discussed in Kj{\ae}rgaard et al. \cite{KjaergaardPervasive2007}.

The RSS-U Zone method as presented in Kj{\ae}rgaard et al. \cite{KjaergaardPervasive2007} is an RSS updating protocol that replaces the periodic updating of RSS measurements as usually practiced for terminal-assisted LF. Update zones are translated into compact RSS patterns, which can be passed to the terminal as a so-called \emph{RSS detection request}. Based on its current RSS measurements and these patterns, the mobile device can decide whether it stays within or without the zone. Hence, RSS values are transmitted to the server only when needed and the overhead associated with periodic updating or polling is avoided. For deciding whether the terminal is within or without the zone with reasonable computational costs, a Bayes estimator is used that collapses the big probabilistic model over all locations available at the location server into a simpler one (maximum of 500 bytes), which distinguishes only between being within or without a configurable set of locations (the update zone). It turned out that this approach only induces little computational burden on the device and significantly saves the amount of messages passing the air interface when compared to periodic RSS updating. Despite of these advantages, it showed that the accuracy of the Bayes estimator is comparable to the classical approach. 

\section{Approach}
\label{paper5:sec:Approach}

The presented approach for indoor proximity and separation detection modifies the DCC strategy for working with walking distances and combines it with zone-based RSS reporting. The DCC strategy dynamically assigns each target update zones in order to correlate the positions of multiple targets. In indoor environments, such update zones can be effectively realized with zone-based RSS reporting, and walking distances between mobile users are much more relevant than Euclidean ones.

%
\begin{figure}
	\begin{center}
\includegraphics[width=0.85\linewidth]{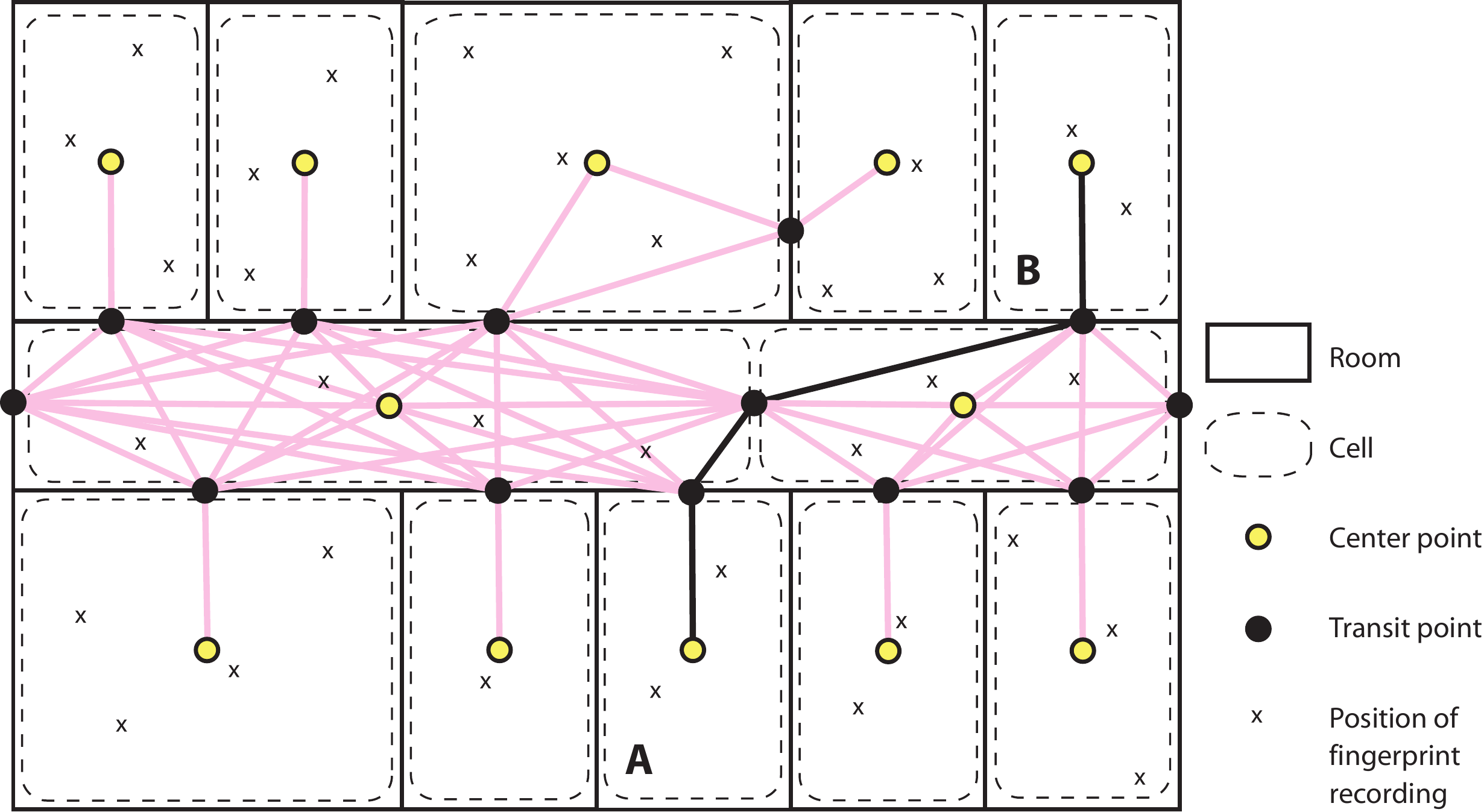}
	\caption{Walking distance between two cells.}
	\label{paper5:fig:WalkingDistance2D}
\end{center}
\end{figure}

\subsection{Walking Distances}
\label{paper5:sec:topological_distances}

For calculating walking distances, a topological building model must be constructed. A building can be described by a set of elements (rooms, corridors, stairways, etc.), all of which have a certain spatial expansion and one or more connection points to neighboring elements. A \emph{cell} is defined as the basic unit of location the LF system can distinguish, that is, it is assumed that localization happens in terms of cells instead of coordinates. A cell usually covers small rooms or parts of a corridor. A more fine-grained discrimination is unrealistic, because of the moderate accuracy of current LF systems. Hence, building elements are always fully covered by one or more cells, and no cell can be part of more than one element. For simply calculating walking distances, the location of a target within a cell is always assumed to be the center point of the cell's enclosing rectangle. This model also solves the determination of walking distances between rooms on different floors. 

However, a problem of this approach is that a target does not necessarily cross the center points of interjacent cells when walking from a source to a destination cell. To give an example, in Figure \ref{paper5:fig:WalkingDistance2D} cells on different sides of the corridor should be reachable directly and not by passing through the corridor cell's center point. As a solution, in addition to the center point, each cell is associated with a set of transit points, which connect a cell to neighboring cells. The topological model of a building is then defined as an undirected connected graph $B = \{P, E\}$, where $P$ is the set of all center and transit points of all cells. The set of weighted edges $E$ represents the distances between connected points. The center and transit points of one cell are always fully connected. Thus, the walking distance $d_{walk}:C \times C \rightarrow \mathbb{R}$ between two cells is defined as the length of the shortest path between their center points, which, however, may include passing interjacent cells through their transit points only. 

\subsection{DCC with Euclidian Distances}
\label{paper5:sec:ProximityAndSeparationDetection}
The classical DCC strategy includes a location server for monitoring the positions of several targets in order to detect when a pair of them gets closer to each other than a \emph{proximity distance} $d_p$ or when it separates by more than a \emph{separation distance} $d_s$. The basic message flow between location server and device is as follows: when proximity or separation detection is requested for a pair of targets, their positions are first \emph{polled} and compared. If the detection condition is already met, the requesting application is notified and the procedure stops. Otherwise, \emph{position update requests}, which carry the definition of the update zones, are sent to both of the devices. The zones are chosen in a way that without any of the two devices triggering an update proximity and separation respectively cannot occur. The devices then continuously check generated position fixes against the update zone. In case of a match, a \emph{position update} is sent to the location server. There, the reported position is compared to the update zones placed on the other target's device, which may or may not result in a need to poll it for its exact position as well. If, based on the exact positions, proximity or separation is detected, the application is notified and the procedure stops. Otherwise, new position update requests are sent to the devices.

\begin{figure}
\begin{center}
   \includegraphics[width=0.65\linewidth]{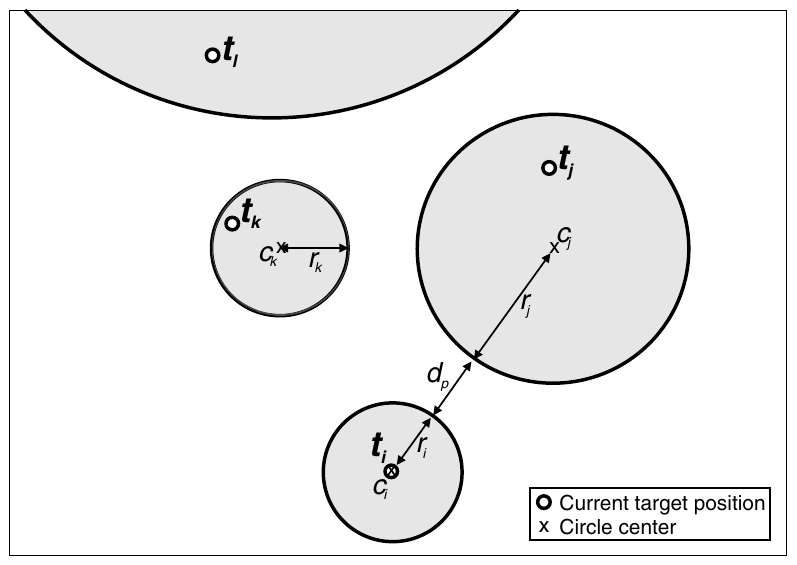}
    \caption{DCC with Euclidean distances.}
    \label{paper5:fig:DCC}
\end{center}
\end{figure}

The update zones in the DCC strategy are circle-shaped and centered around the terminal's last reported position. Positions are reported only when leaving the circle. For proximity detection, the circle computation works as follows, compare Figure \ref{paper5:fig:DCC}: suppose $t_i$ reports its current position and the neighbor of $t_i$ with the closest circle turns out to be $t_j$. Assuming the circle of $t_j$ has the radius $r_j$ and the center point $c_j$, then $t_i$ is assigned a new circle with center point $c_i$ set to its current position and with radius $r_i := dist(c_j,c_i)  - r_j - d_p$. In this way it is impossible that the distance between $t_i$ and $t_j$ can get below $d_p$ without either of the two leaving its circle and reporting a position update. 

For separation detection, suppose that from all targets $t_j$ is farthest away from $t_i$, assuming that $t_j$ is located at the border of its circle in opposite direction to $t_i$, which leads to the so-called maximum distance between both targets. The circle computed for $t_i$ again has the center point $c_i$ set to its current position, but the radius is set to $r_i := d_s - dist(c_j,c_i)  - r_j$. Analogous to before, the distance between $t_i$ and $t_j$ can thus not exceed $d_s$ without sending a position update. By choosing the neighbor $t_j$ as described, the proximity and separation conditions are also guaranteed with respect to other possible neighbors $t_i$ is tracked with.

\subsection{DCC with Walking Distances}
\label{paper5:sec:MappingDCC}
\begin{figure}
\begin{center}
	\includegraphics[width=0.8\linewidth]{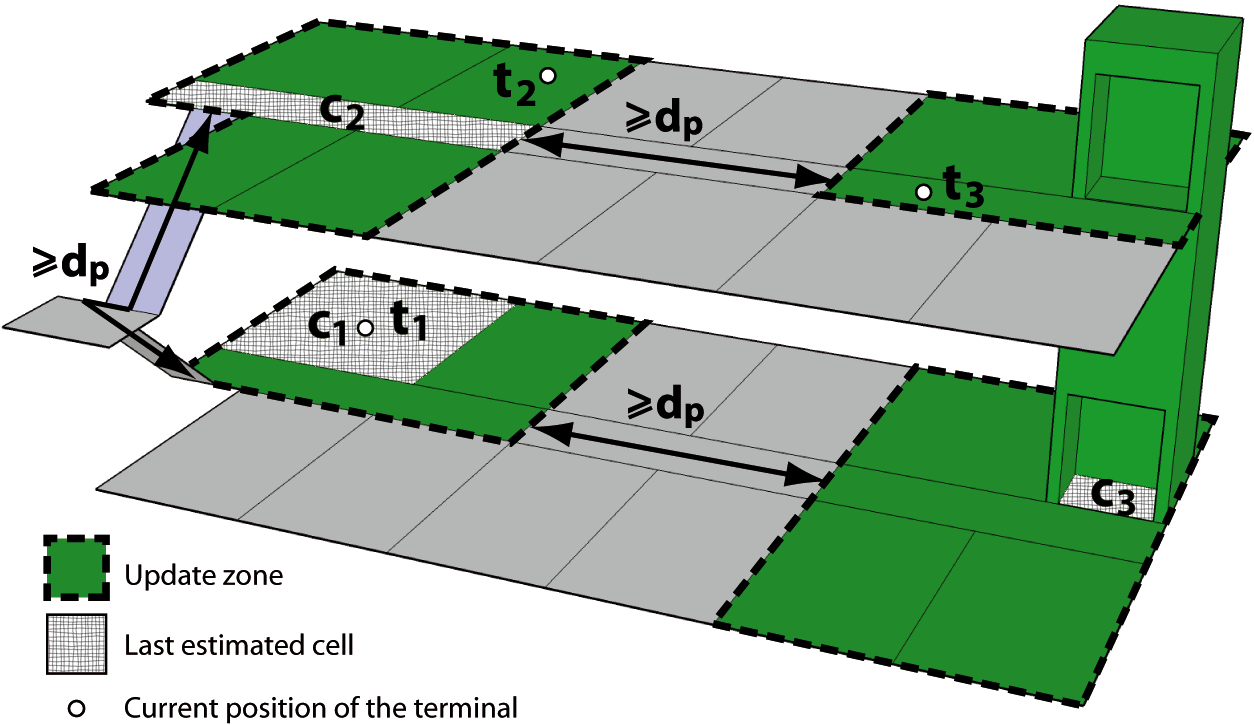}
	\caption{DCC for cells and walking distances}
	\label{paper5:fig:WalkingDistance3D}
\end{center}
\end{figure}%

Indoor proximity detection based on walking distances uses the proximity distance $d_p >0$ and an associated borderline tolerance $b>=0$. Let $c_i$ be the current cell of target $t_i$ and $c_j$ the cell of $t_j$. Furthermore, let $d_{walk}(c_i,c_j)$ be the walking distance between the targets' current cells as defined before. Then, proximity is checked by the following conditions:
\begin{enumerate}
	\item If $d_{walk}(c_i,c_j) < d_p$, then proximity \textit{must be} detected.
	\item If $d_p \leq d_{walk}(c_i,c_j) \leq d_p + b$, then proximity \textit{may be} detected.
	\item If $d_{walk}(c_i,c_j) > d_p + b$, then proximity \textit{must not be} detected.
\end{enumerate}
For separation detection based on the separation distance $d_s>0$ the conditions are defined analogous. The purpose of the fuzziness interval given by the borderline tolerance $b$ is to avoid excessive location reporting when the distance between $t_i$ and $t_j$ is approaching $d_p$. Without $b$, it would be necessary to track the devices on a very fine-grained level just to determine the exact moment when $d_{walk}(t_i,t_j)$ meets $d_p$. Put differently, the parameter $b$ enables a trade off between desired detection accuracy and costs in terms of transmitted messages. In any way, it would not make sense to specify a higher detection accuracy than the accuracy of position fixes delivered by the used LF system. The reason for the gain in efficiency when using a bigger value for $b$ is that, as described more extensively in \cite{KuTr06}, the minimum radius of the update circles used by the DCC strategy can be limited to $\frac{b}{2}$. Obviously, bigger circles lead to less position updates on average. 

In order to apply the DCC strategy to the topological indoor model, the \emph{walking distance space} (WDS) of a cell is introduced. Given a radius $r$, $WDS(c_i,r)$ of a cell $c_i$ equals the set of all cells $c_j$ whose walking distance $d_{walk}(c_i,c_j)$ to $c_i$ is smaller than or equal to $r$. 
Hence, instead of geographical circle-shaped update zones centered around the last reported position, our adaption of DCC for indoors calculates the WDS with respect to a target's last estimated cell based on the calculated radius. This update zone, which is defined in terms of cells, is then configured at the targets' terminals by a respective \emph{RSS detection request} using the RSS pattern technique described in \cite{KjaergaardPervasive2007}. 
The rest of the DCC algorithm basically remains the same:
when a target $t_i$ leaves its update zone, an \emph{RSS update} is reported to the server. Based on the update, the current cell $c_i$ of $t_i$ is estimated. In case of proximity detection, the minimum walking distance $m$ between $c_i$ and the closest cell of the current update zones of all other targets $t_j$ is calculated. If $m$ is small enough so that proximity could occur, an \emph{RSS polling} is issued to the respective target(s) $t_j$ and its (their) current cell(s) $c_j$ is (are) estimated as well. If, based on the cell estimates, the trigger condition is fulfilled, the application is notified. Otherwise, the minimum distance the targets $t_i$ and $t_j$ may walk without conflicting with one another, or with a zone of the other targets, is calculated. From these distances, two update zones (WDSs based on the estimated cells) are computed and assigned to the targets' terminals by means of new RSS detection requests. In case $m$ was not too small before, only $t_i$ is assigned a new update zone, reflecting a WDS with radius $r_i:=m-d_p$. For separation detection the procedure is analogous.

As an example for proximity detection, Figure \ref{paper5:fig:WalkingDistance3D} shows a scenario inside a building, where the devices of three targets are configured with update zones (dark areas). Device $t_1$ has just reported an RSS update and its new update zone has been calculated as follows: the closest neighboring update zone to $t_1$'s estimated cell was the one of $t_3$, so that the distance between the update zone assigned to $t_1$ and $t_3$ is as close to $d_p$ as possible. As a consequence, the walking distance between the zone of $t_1$ and the zone of $t_2$ is larger than $d_p$ (in the model distances along stairs are weighted heavier than horizontal ones).

\section{Experimental Results}
\label{paper5:sec:Evaluation}
For evaluating the approach, a simple location-based community service was implemented, which keeps the users of an office environment up-to-date about which persons of their buddy list are currently staying within a walking distance of $p$ or smaller. Each possible pair of buddies is either observed for proximity or separation events. When a proximity event is detected, the buddy's name appears on the user's proximity list and separation detection is started for both of them. If, in turn, separation is detected, the person is removed from the list and proximity detection is restarted.

The fuzziness intervals for separation and proximity detection are made non-overlapping in order to avoid possible ping-pong effects. For a borderline tolerance of $b$, proximity detection is initialized with $d_p=p-b$ and separation detection with $d_s=p$. Thus, if the walking distance $d_{walk}(t_i,t_j)$ between two target persons $t_i$ and $t_j$ is below $p-b$, then they \emph{must} appear on each other's proximity list. If $p-b \leq d_{walk}(t_i,t_j) \leq p+b$, then they \emph{may} appear on the list. Finally, if $d_{walk}(t_i,t_j) > p+b$, then they \emph{must not} be on the list.

\subsection{Prototype}
In order to show the practical feasibility of our approach with state-of-the-art equipment, a prototype was im\-ple\-men\-ted and tested with Fujitsu Siemens Pocket LOOX 720 PDAs with built-in WiFi (IEEE 802.11) functionality. At the PDA, the functions for measuring RSS and evaluating RSS detection requests are implemented as a .NET application for Windows Mobile 2003 SE. The TraX server is implemented as a Java application, passing RSS detection requests to the PDAs and receiving RSS updates from the PDAs. Connectivity to the terminals was provided by a WiFi infrastructure using a proprietary protocol on top of TCP. For estimating locations from RSS updates and for computing RSS detection requests from sets of cells the TraX server utilizes an existing LF server.  

A field test with two targets and an area spanning two floors with about 30 cells and 14 reachable base stations was conducted. After experimenting with different configurations, the proximity distance of the community service $p$ was set to 12 m and the borderline tolerance $b$ to 5 m. First, the targets walked in different patterns on the two floors. During one walk, a target went to the second floor while the other stayed on the first one. Then both targets walked to the second floor and back together. Finally, both walked up and back again, however, with the second target following at a certain distance. 

From our experiences, it can be stated that the system worked properly and most of the time correct proximity and separation states were reported. However, also wrong or missing detections were experienced, which, apart from general LF inaccuracy, had two reasons: first, some communication delays happened as a result of roaming between the base stations used in the experiment. With the used combination of WiFi driver on the PDAs and type of WiFi access points, these delays amounted to several seconds, which made the system miss some detections and also report several detections in a bulk after the event had already passed.
Second, the sampling rate of the used PDA is only 0.5 Hz, and hence the position derived at a device is delayed by up to 2 seconds. Considering both devices, the true distance between two targets then deviates from the measured one by up to 4 seconds of walking.
\subsection{Emulation}
\begin{figure}
\begin{center}
   \includegraphics[width=0.95\linewidth]{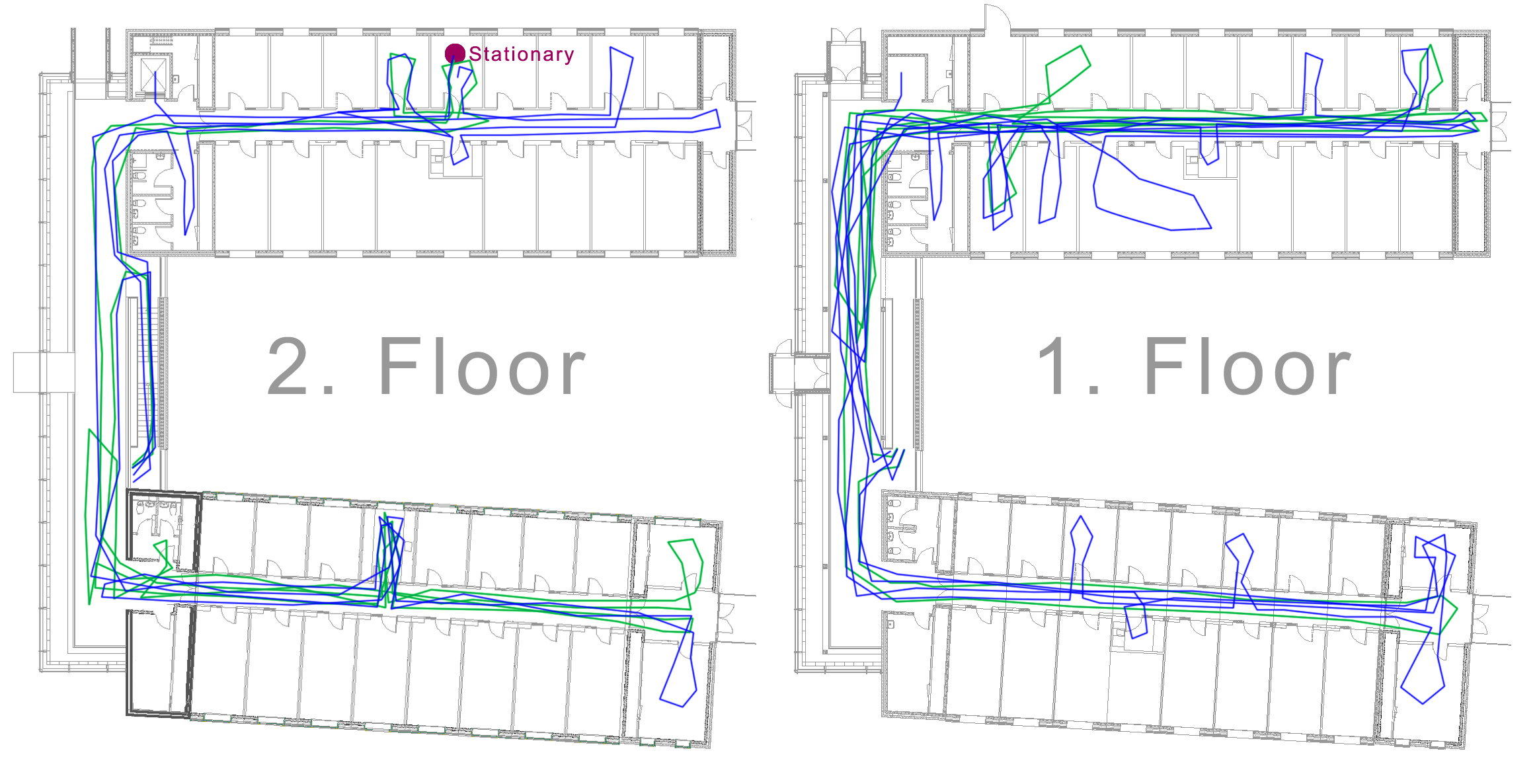}
    \caption{Walks recorded at two floors.}
    \label{paper5:fig:floors}
\end{center}
\end{figure}

In addition to the prototype and in order to obtain quantitative results, emulations were run based on data collected from a second test site.
This test site offers 31 reachable WiFi base stations. It was divided up into 126 cells with an average size of 16 m$^{\textrm{2}}$ matching rooms or parts of hallways, spanning two floors. Each cell was fingerprinted by walking around in the cell for 60 seconds with a laptop that was equipped with an Orinoco Silver 802.11 card. After that, six sets of walks were collected, each comprising three 40-minutes-walks simultaneously performed with three devices, totaling about 12 hours. The fingerprinting and walk collection were separated by several weeks. Three of the six walk sets were recorded by the PDAs also used for the prototype. The other three used the laptops with the Orinoco cards. The RSS values were collected at a sampling rate of 0.5 Hz and 1 Hz respectively. Each sample of a walk contains a time-stamp, the measured RSS values of the
surrounding base stations, as well as the current ground truth, which
was manually specified on a laptop-shown map. During the recording of a set
of walks always one of the three devices was kept stationary, while
the other two were carried along different routes through the
building. The targets walked at moderate speeds, with several pauses
and over two alternating floor levels, compare Figure \ref{paper5:fig:floors}. 

Based on the recorded data the approach was examined in terms of efficiency and accuracy. For that, from the zone detection methods presented in \cite{KjaergaardPervasive2007} the Bayes estimator was selected. As a benchmark for comparison, a \emph{reference strategy} based on terminal-assisted LF with periodic RSS reporting at 1 Hz was assumed. In this way, for all possible pairs of targets and at every moment in time the location server can decide whether the proximity
criterion is met or not. For location estimation from reported RSS values at the server-side the same LF system, which is based on the techniques described in \cite{Haeberlen2004}, was used by the proposed DCC strategy as well as by the reference strategy. The PDA's RSS measurements were normalized to match the fingerprints collected with the Orinoco cards using the method proposed in \cite{KjaergaardLoca2006}.

As explained before, three operations are needed for target tracking: RSS detection requests, RSS updates, and RSS pollings. While DCC combines all three operations, the reference strategy only uses RSS updates. Each of these operations causes one message in the uplink and another one in the downlink. The only exception are RSS updates in the DCC strategy. They need no explicit acknowledgement in the downlink, because they are always confirmed by a new position RSS update request message. Technically, up- and downlink have different re\-sour\-ce-consuming properties and should be treated separately. For brevity, however, they are not distinguished in the following and the total number of messages transferred per target is summed up.

\begin{figure}[h]
\begin{center}
\includegraphics{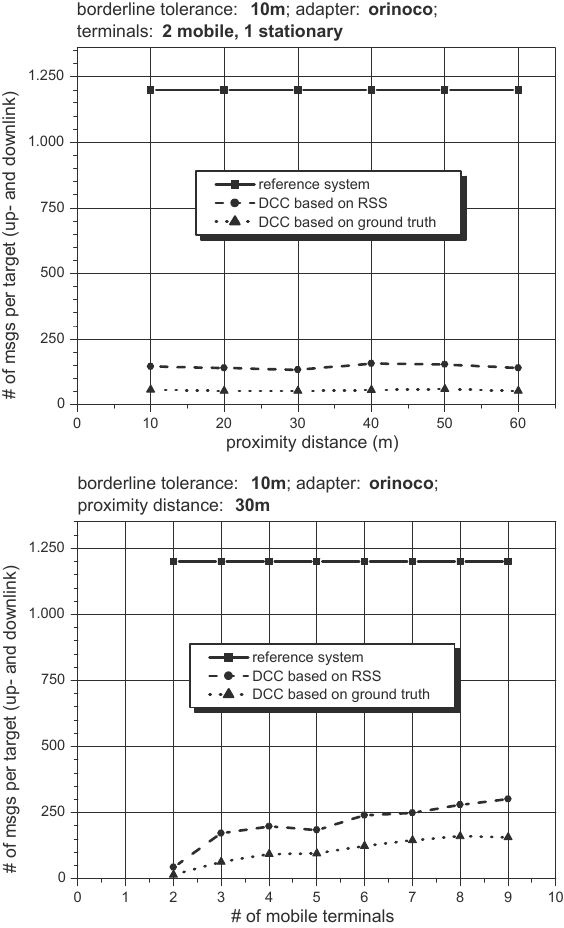}
    \caption{(a) \# of messages dependent on proximity distance $p$, (b) \# of messages dependent on number of terminals}
    \label{paper5:fig:perf_orin}
\end{center}
\end{figure}

Another issue is the amount of transferred data. While message acknowledgments as well as polling requests (the downlink message of an RSS polling) are very lightweight, RSS updates as well as polling responses carry measured RSS values, which amounts to more data. For example, the Orinoco and the PDA walks contain on average around 5-7 base stations per sample. Furthermore, experiments with an Apple Airport Express card yielded about 14 visible stations at a time. However, in practice only the 5-7 strongest stations need to be reported, because including more stations will not significantly increase the accuracy. Thus, the size of an RSS update has an upper limit, which, however, is dependent on the underlying technology. The RSS detection request messages (downlink)
have the biggest size, which, according to \cite{KjaergaardPervasive2007}, can be limited to 500 bytes for the Bayes estimator. For the other (more inaccurate) RSS detection methods, the size is typically smaller.

Whether the goal is to save transferred bytes or messages depends on the constraints considered. Monetary costs for transmission over public bearer services like GPRS or UMTS are typically billed according to data volume in bytes. On the other hand, server scalability is rather constricted by the number of messages that have to be handled in the uplink. Considering physically limited resources like the air-interface or the battery power at the device used for message sending and receiving, the number of transmitted frames seems most critical. For IEEE 802.11 this figure equals the number of transferred messages, because all described message types are small enough to fit within one 802.11 frame. Therefore and also because the number of bytes per message can be specified rather arbitrarily, the following evaluation only discusses the number of transferred messages.

For evaluating message efficiency, three parameters were varied: the proximity distance $p$, the number of terminals observed in a pairwise fashion (i.e., the size of the buddy list), and the borderline tolerance $b$. Additionally to the DCC and the reference strategy based on collected RSS values, DCC was also performed on ground truth, which behaves as if the RSS detection requests worked with perfect accuracy.

Figure \ref{paper5:fig:perf_orin}a shows the number of messages transferred per target dependent on $p$ averaged for the three walk sets collected with the Orinoco cards. The time was normalized to 10 minutes. Three things become apparent: first, in comparison to the reference strategy, DCC based on RSS reduces the amount of messages strongly (about factor 9). Second, the performance of all three approaches is rather independent from the chosen proximity distance. While this was expected for the reference strategy, which steadily sends 120 messages per minute, for DCC this can be explained by the fact that independent of the current distance of a pair of targets and $p$, both of them are permanently observed either for proximity or for separation events. The third observation is the difference between the performance of DCC based on RSS and DCC based on ground truth. The former triggers about 2.5 times as much messages as the latter. 
Obviously, the employed RSS detector (Bayes estimator) triggers a number of wrongly sent RSS updates, which do still belong to the cells contributing to the update zone and which are therefore correctly not sent by DCC based on ground truth. However, it can be stated that the difference between the real and
the ideal DCC detector is still acceptable when taking into account the savings compared to the reference strategy. Also, it must be stated that the collected walks represent a mobility pattern presumably more mobile than in a typical office scenario. 

Figure \ref{paper5:fig:perf_orin}b shows the number of messages per target dependent on the number of pairwise observed targets. For this, all of the $3*3=9$ walks collected with the Orinoco cards were aligned in time and played simultaneously. Expectedly, the number of messages per target used by the reference strategy stays the same, while for DCC it increases. The proportion between messages sent by DCC based on RSS and DCC based on ground truth starts with a value of 2.8:1 for two targets, then slowly decreases with an increasing number of targets and settles at a value of about 1.8:1 for five to nine targets. The slope of the DCC curves is not too steep, so that the approach seems practicable even for bigger buddy lists. Note that the number of targets tracked pairwise (equals the size of the buddy list) is not equal to the number of users of the community services. While our aim is to make the service scalable to thousands of users, this examination was related to the size of a single user's buddy list, that is, the number of users she constantly wants to keep track of, a figure which is assumed to be rather small. Thus, by limiting the number of messages per user as described before, server scalability in terms of the number of users is improved.

\begin{figure}[h]
\begin{center}
    \includegraphics{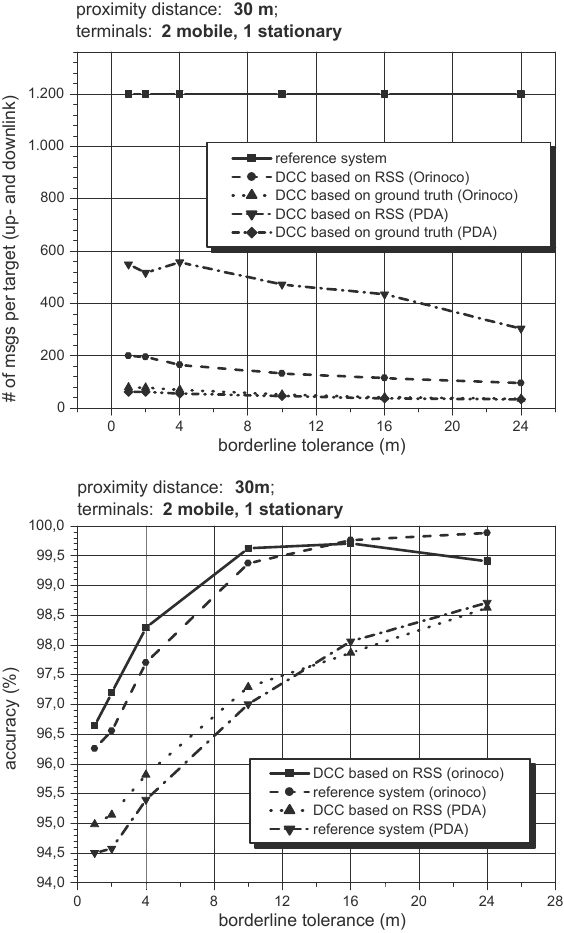}
    \caption{(a) \# of messages dependent on borderline tolerance $b$, (b) Accuracy dependent on borderline tolerance $b$}
    \label{paper5:fig:borderline}
\end{center}
\end{figure}

Figure \ref{paper5:fig:borderline}a depicts the message overhead dependent on the borderline tolerance $b$. For the Orinoco cards as well as for the PDAs, all three-person-walk sets were averaged. Two observations are noteworthy here: first, the number of messages in all configurations decreases by roughly the same factor of about 50 \% from $b=1$ to $b=24$. This can be explained by taking into account that the minimum radius measured in walking distance of a DCC zone is limited to $\frac{b}{2}$. Thus, with an increasing $b$ the minimum zone size increases, which leads to a decreasing number of RSS updates. The second observation is that DCC with RSS performs considerably worse for the PDAs than for the Orinoco cards (the factor ranges between 2.6 and 3.8). One reason for this may be that the PDA's RSS measurements need to be normalized as described before to match the fingerprints in the database, which were collected with the Orinoco card. The normalization function does, however, not perfectly account for the difference in RSS measuring between the Orinoco card and the PDA, which degrades accuracy in general. Hence, the RSS detectors at the PDAs produce more wrongly sent RSS updates. 

The application-level accuracy of the presented strategies is analyzed according to a simple metric: based on the ground truth at each moment in time and for
each pair of tracked targets $t_i$ and $t_j$, the current walking distance $dist(t_i,t_j)$ is computed. It is mapped onto a state $X \in \{P,F,S\}$ with $X=P$ if $dist(t_i,t_j) < p-b$ ($t_i$ and $t_j$ are in proximity), $X=F$ if $p-b \leq dist(t_i,t_j) \leq p+b$ (they are within the fuzziness interval), or $X=S$ if $dist(t_i,t_j) > p+b$ (they are separated). Based on this mapping, the number of situations (time frames of one second) are counted where the DCC and the reference strategy indicate a wrong state information, that is, when the state $X_{DCC}$ or $X_{ref}$ deviates from the ground truth $X_{gt}$. However, a wrong
state information is only logged when $X_{gt}=P$ or $X_{gt}=S$, because within the fuzziness interval both states are allowed. The metric is very simple, because in the tested service there is an interplay between proximity and separation detection. For testing the events separately, it would be necessary to consider false and true positives and negatives respectively and derive from that metrics like sensitivity and precision. In this case, however, a positive with respect to proximity detection is a negative for separation detection. Since both situations ($X=P$ and $X=S$) have a comparable probability (dependent on the building layout and the proximity distance), the two event types actually cancel each other out and hence one accuracy metric suffices.

Figure \ref{paper5:fig:borderline}b plots the achieved accuracy (that is, the percentage of situations where no wrong state information is given) for the DCC as well as for the reference strategy. First, for all curves the accuracy increases with an increasing borderline tolerance, which is due to the decreasing impact of LF inaccuracy on distinguishing the states $S$ and $P$. Second and confirmatory for the good applicability of the DCC strategy, its accuracy is generally not worse than that of the reference strategy. It performs even slightly better for a low borderline tolerance and slightly worse for higher borderline values. Third, the Orinoco measurements yield a higher accuracy than those of the PDAs. However, it can be stated that in general a high accuracy is achieved (all four strategies are always above 94.5 \%), even for a low borderline tolerance.

\section{Conclusion and Further Work}
\label{paper5:sec:Conclusion}
The paper has demonstrated that proactive proximity and separation detection can be effectively realized for indoor environments, while being re\-sour\-ce-aware at the same time. The evaluation showed that the presented approach can decrease the number of transmitted messages with a factor of 9. The approach is feasible for very resource-limited devices like mobile phones or active tags and makes use of state-of-the-art LF technology and device hardware. Also, despite the general inaccuracy of LF, it turned out that at an application level a rather high detection accuracy above 94.5\% can be achieved. A possible extension to the described community service, which recognizes targets closer than a static threshold, would be a buddy tracker that constantly shows the user a sorted list of the n-nearest-neighbors among his buddies. One piece of future work is to show how such a service can be realized by dynamically applying proximity and separation detection to pairs of targets. 

\section*{Acknowledgements}
M. B. Kj\ae rgaard is partially funded by the software part of the ISIS Katrinebjerg competency centre.

\clearemptydoublepage
\chapter{Paper 6}
\label{chap:mobisys2008}

The paper \emph{ComPoScan: Adaptive Scanning for Efficient Concurrent Communications and Positioning with 802.11} presented in this
chapter has been published as a conference paper~\cite{KingMobisys2008}.

\begin{publist}{\cite{KingMobisys2008}}
  \item[\cite{KingMobisys2008}] T.\ King and M.\ B.\ Kjærgaard. ComPoScan: Adaptive Scanning for Efficient Concurrent Communications and Positioning with 802.11. In \emph{Proceedings of the 6th ACM International Conference on Mobile Systems, Applications, and Services}, ACM, 2008.
\end{publist}

\noindent

\clearemptydoublepage



\mytitle{ComPoScan: Adaptive Scanning for Efficient Concurrent Communications and Positioning with 802.11}{ 
  Thomas King\footnotemark[1] \and
  Mikkel Baun Kj\ae rgaard\footnotemark[2]}{  
  \footnotetext[1]{Department of Computer Science, University of Mannheim, Germany. E-mail:
    \texttt{king@informatik.uni-mannheim.de}.}
  \footnotetext[2]{Department of Computer Science, University of
    Aarhus, Denmark. E-mail:
    \texttt{mikkelbk@daimi.au.dk}.}
    }
        

\begin{myabstract}
Using 802.11 concurrently for communications and positioning is
problematic, especially if location-based services (e.g., indoor
navigation) are concurrently executed with real-time applications
(e.g., VoIP, video conferencing).  Periodical scanning for measuring
the signal strength interrupts the data flow.  Reducing the scan
frequency is no option because it hurts the position
accuracy. For this reason, we need an adaptive technique to mitigate
this problem.

This work proposes ComPoScan which, based on movement detection, adaptively
switches between light-weight monitor sniffing and invasive active scanning to
allow positioning and to minimize the impact on the data flow. The system is
configurable to realize different trade-offs between position accuracy and the
level of communication interruption.

We provide extensive experimental results by emulation on data collected
at several sites and by validation in several real-world deployments. Results
from the emulation show that the system can realize different trade-offs by
changing parameters. Furthermore, the emulation shows that the system works
independently of the environment, the network card, the signal strength
measurement technology, and number and placement of access points. We also
show that ComPoScan does not harm the positioning accuracy of a positioning
system. By validation in several real-world deployments, we provided evidence
for that the real system works as predicted by the emulation. In addition, we
provide results for ComPoScan's impact on communication where it increases
throughput by a factor of 122, decreases the delay by a factor of ten, and
the percentage of dropped packages by 73~percent.
\end{myabstract}

\section{Introduction}
\label{paper6:sec:introduction}
Back in 1999, when IEEE~802.11 was being standardized, the
researchers and engineers working on the standard probably never
thought about the new ways we use this technology today. Real-time
applications such as voice over IP and video conferencing were a
rarity years ago but are a common phenomenon
nowadays. 
These real-time applications have hard requirements in terms of
bandwidth, delay, and packet loss to be functional. An even more
extreme new way of usage is to utilize the signal strength measurement
capabilities of 802.11 network cards as a basis for indoor positioning
systems to enable location-based services.  Initially, signal
strength measurements are performed during a so-called active scan to
let a network card decide which access point might be the best to
connect to. Many indoor positioning systems (e.g., \cite{Bahl2000,
  Haeberlen2004}) make use of 802.11, because almost all modern
cell phones and laptops are equipped with this wireless
technology. Therefore, the devices can be used for positioning as they
come out of the box, which means that no additional hardware is required.

Even the newer sub-standard 802.11b and 802.11g do not satisfy all
these requirements. Furthermore, many workarounds and novel
approaches (e.g., \cite{Forte2006,Mhatre2006, Shin2004}) have been
proposed to make 802.11 ready for many of these new demands.
However, still unsolved remains the problem that occurs when a
802.11 network card is utilized for positioning and communicating
at the same time. On the one hand, the positioning system requires a
steady stream of active scans to be able to deliver accurate
position estimates to location-based services. Especially, if the
positioning system is used to track users as e.g., required for
indoor navigation systems in huge buildings. Performing an active
scan means that the network card switches through all the different
channels in search of access points. Dependent on the network card,
this takes about 600~milliseconds. During this time no communication
is feasible.  On the other hand, there are the demanding real-time
applications. For instance, a video conference requires around
512~KBit/s of bandwidth and a round trip delay of less than
200~milliseconds, depending on the video and voice
quality~\cite{Varshney2002}.

\begin{figure}[h]
  \centering{
    {\includegraphics[width=0.85\textwidth]{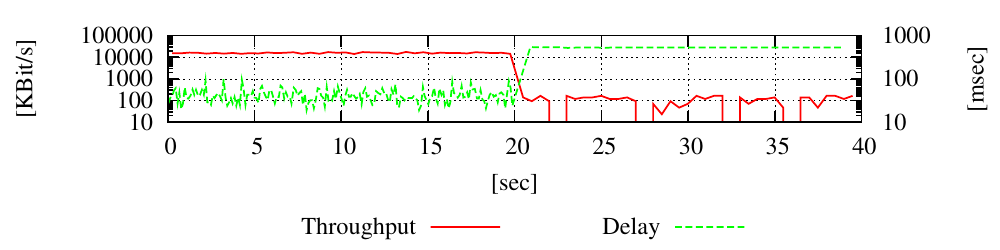}}
  }
  \caption{Throughput and delay.}
  \label{paper6:fig:throughputdelayscanning}
\end{figure}

Figure~\ref{paper6:fig:throughputdelayscanning} depicts what happens to
throughput and delay of a 802.11g-enabled mobile device if the
network card is requested to perform an active scan every
600~milliseconds. During the first 20~seconds communication is
untroubled, which means a throughput of about 20~MBit/s on average
and that a round trip delay of less than 45~milliseconds is
achievable. In the 20$^{th}$ second active scanning kicks in. The
remaining seconds only provide 0.1~MBit/s of throughput and
532~milliseconds of delay, because active scans are performed so
often. Due to variations in the execution time of scans, on some
rare occasions no data transmission is possible at all.

In this paper, we propose a novel solution to this problem which is
called ComPoScan. It is based on movement detection to switch, on the
basis of adaptability, between light-weight monitor sniffing and
invasive active scanning. Only in case that the system detects
movement of the user, active scans are performed to provide the
positioning system with the signal strength measurements it needs. If
the system detects that the user is standing still, it switches to
monitor sniffing to allow communications to be uninterrupted.

Monitor sniffing is a novel scanning technique proposed
in~\cite{KingPercom2007}. It works with most 802.11 network cards around
today. Monitor sniffing allows a mobile device to recognize access points
operating on channels close to the one it is using for communications with the
access point it is associated with. It has been shown that up to seven
channels can be overheard without any disturbance of the actual communication.

Our movement detection approach is also based on signal strength
measurements. However, the measurements provided by monitor sniffing
are sufficient to detect reliably whether the user is moving or
standing still. We designed the movement detection system to be
configurable so that, depending on the user's preferences,
communication capabilities or positioning accuracy can be favored.

We make the following contributions in this work: First of all, we are
the first who present a system to mitigate the effect of scanning on
concurrent communications. Secondly, we are the first utilizing
monitor sniffing and active scanning to build a reliable indoor
movement detection system. Thirdly, we provide a deep investigation by
means of emulation to show that our movement detection system works
independently of the environment, the network card, the signal
strength measurement technology, and number and placement of access
points. Additionally, we show that it does not harm the positioning
accuracy of the positioning system. Fourthly, we implement ComPoScan
and use this prototype in a real-world deployment to gather results
showing that the real system works as predicted by the emulation. The
results show that our goal of mitigating the effect of scanning on
communications is full-filled.

The remainder of this paper is structured as follows: In
Section~\ref{paper6:sec:relatedwork}, we present the relevant related work.
Subsequently, we introduce our novel ComPoScan system.  The details of
our movement detection approach are discussed and evaluated by means
of emulation in Section~\ref{paper6:sec:mobilitydetection}.
Section~\ref{paper6:sec:prototype} discusses our prototype implementation of
ComPoScan in detail.  The results of our real-world deployment are
presented in Section~\ref{paper6:sec:validation}. Finally,
Section~\ref{paper6:sec:discussion} provides a discussion and
Section~\ref{paper6:sec:conclusion} concludes the paper and provides
directions for future work.

\section{Related Work}
\label{paper6:sec:relatedwork}
As mentioned earlier, existing 802.11 positioning systems
(e.g., \cite{Bahl2000, Haeberlen2004}) have not considered the problem of
concurrent communication and positioning. As a central part of the ComPoScan
system we apply movement detection to deal with this problem.

The first, and as far as we know the only, 802.11-based system that
emphatically focuses on movement detection is the LOCADIO
system~\cite{Krumm2004}. In their paper, the authors propose an
algorithm that exploits the fact that the variance of signal strength
measurements increases if the mobile device is moved compared to if it
is still. To smooth the high frequency of state transitions an HMM is
applied. The results in the paper show that the system detects in
87~percent of all cases whether the mobile device is in motion or
not. Compared to our approach, the authors do not compare their system
to other movement detection algorithms. Furthermore, the results are
only based on emulation which means that the signal strength data is
collected in a first step and then, later on, analyzed and processed to
detect movement. This is a valid approach, but some real-world effects
might be missed. Another fact that the authors of the aforementioned
paper do not look into is the impact of periodic scanning to the
communication capabilities of mobile devices. They just assume that a
802.11 network card is solely used for movement detection. Finally,
all results are based on one single client, which means that
variations in signal strength measurements caused by different
wireless network cards are not taken into account.

Two GSM-based systems have also been proposed by Sohn et
al. \cite{Sohn2006} and Anderson et al. \cite{Anderson2006}. The
system by Sohn et al. is based on several features including variation
in Euclidean distance, signal strength variance and correlation of
strength ranking of cell towers. The system classifies data into the
three states of still, walking and driving. By emulation on collected
data they achieve an overall accuracy of 85~percent. The system by
Anderson et al. detects the same states, but uses the features of
signal strength fluctuation and number of neighbouring cells. Using
these features, they achieve a comparable overall accuracy to the
former system. As for LOCADIO the results for both systems are only
based on emulation, they also do not consider communication and the
results are based on one client.


\section{ComPoScan System}
\label{paper6:sec:composcan}
For our system we assume that the mobile device that should be
ComPoScan-enabled contains a 802.11 network card. This card should be
able to perform active scans and monitor sniffs on a high rate (e.g.,
every 600~milliseconds). Further, the card should not include buffered
results from a previous scan into the current scan result. For the
area where ComPoScan should be deployed we assume that at least one
access point is recognizable at all times by monitor sniffing and
active scanning.

Our main goal for ComPoScan is to minimize the impact of scanning on
concurrent communications. For this, we want to build a movement
detection system that, based on signal strength measurements provided
by monitor sniffing or active scanning, detects correctly whether the
user is standing still or moving. If this is possible only active
scans are required in case that the user is roaming around. However,
we expect that it might be impossible to build a completely perfect
movement detection system with 802.11. So this brings up a sub-goal:
The movement detection system should be configurable in such a way
that the user can define the kind of the error the movement detection
system is producing. In case that the user is more interested in
precise position estimates than in uninterrupted communications this
scenario should be configurable. The other way around should also be
supported.
    
\begin{figure}[h]
	\centering
		\includegraphics[viewport=30 275 460 525,width=0.65\textwidth,clip]{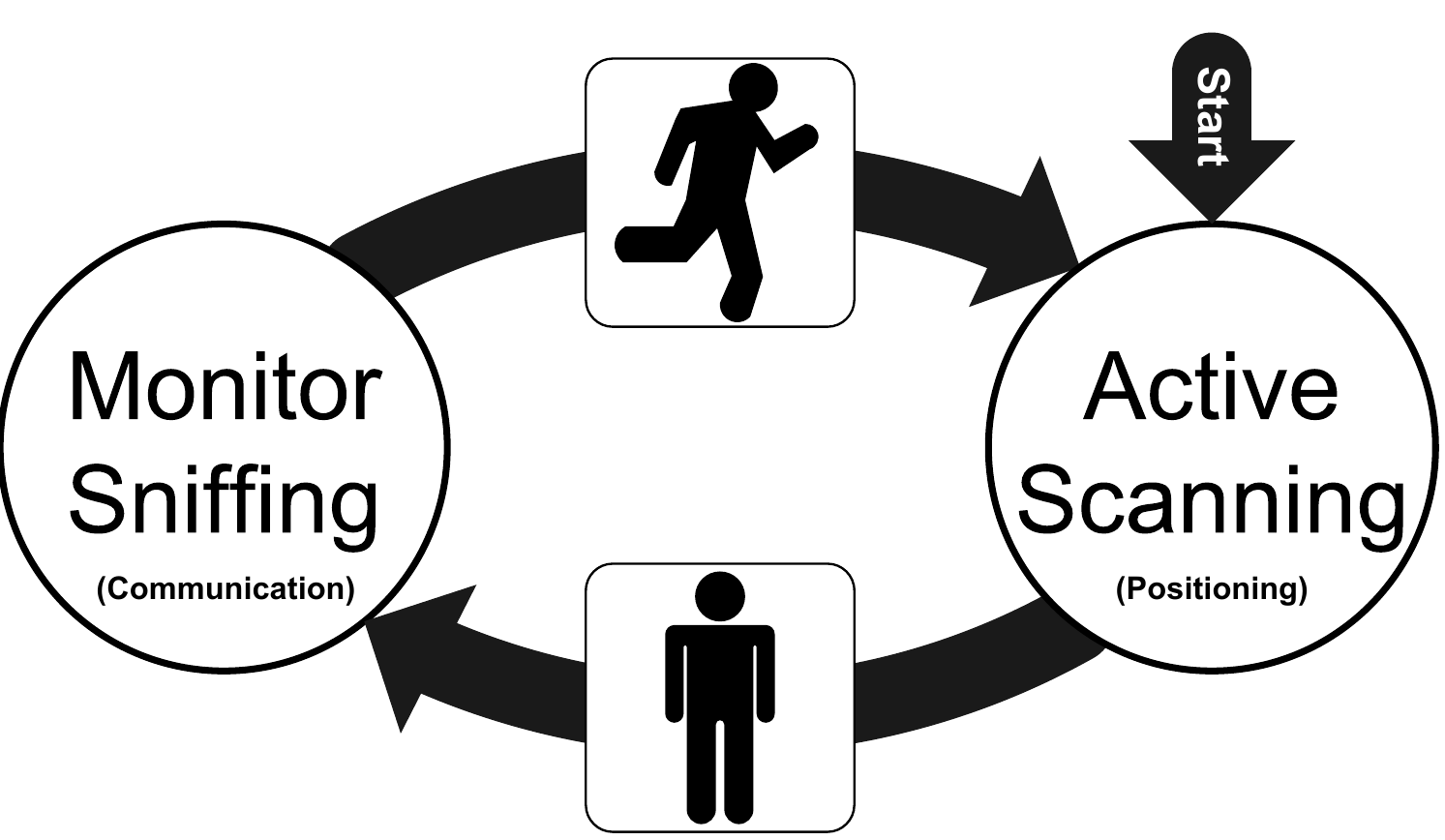}
	\caption{The ComPoScan system.}
	\label{paper6:fig:composcansystem}
\end{figure}

ComPoScan works as illustrated in Figure~\ref{paper6:fig:composcansystem}. At startup,
active scans are performed to collect signal strength values from as many
access points as possible. Based on this data, the current state is
calculated. If the system detects movement, it performs another active scan. In
case that the system draws the conclusion that the user is standing still it
switches to monitor sniffing for signal strength measurement. Based on this
data, the current state is reevaluated and the system starts over again.

\section{Mobility Detection}
\label{paper6:sec:mobilitydetection}
A central part of the ComPoScan system is movement detection based on signal
strength. This section describes our experimental setup, gives an analysis of
features used for movement detection, presents the used method and discusses
our emulation results.

\subsection{Experimental Setup}
\label{paper6:sec:experimentalsetup}
For our experimental setup, we describe the used hardware and software setup,
the test environments and the details of the data collection process.

\subsubsection{Hardware and Software Setup}
\label{paper6:sec:hardwaresoftwaresetup}
To collect the signal strength measurements, we used an IBM Thinkpad R51 laptop
running Linux kernel 2.6.22.12 and Wireless Tools 29pre22. To show that our
approach works independent of a particular card, we use different network
cards. For this, three network cards were chosen that are all quite
frequently used today. We selected a Lucent Orinoco Silver PCMCIA card, a
TRENDnet TEW-501PC PCMCIA card, and an Intel Centrino 2200 mini-pci card. The
Lucent Orinoco card is a 802.11b only card. The TRENDnet card is based on the
widely used Atheros AR5006XS chip-set and supports 802.11b, 802.11g, and
802.11a. Only 802.11b and 802.11g are supported by the Intel Centrino
chip. However, all three network cards can be used for our purposes, because
they all support monitor sniffing and active scanning.

For the Intel Centrino 2200 card, we used the ipw2200 driver in
version~1.2.0\footnote{\url{http://ipw2200.sf.net}}. In the default
settings, the driver caches a scan result for 3.45~seconds which means
that an access point, that has been seen during the last 3.45~seconds,
will appear in a subsequent scan result and even that it might be out
of communication range. We modified the driver to discard old scan
results before a new scan is performed because this property harms our
movement detection system.

The driver of the TRENDnet card needed modifications, too. For this
card, we used the madwifi driver
version~0.9.3.3\footnote{\url{http://www.madwifi.org}}. In the default
settings, the driver caches scan results in the same way as the
ipw2200 driver. The difference here is that the cache timeout is even
longer and set to 60~seconds. With our modifications the driver purges
the cache before initiating a new scan. Since the TEW-501PC card
supports three 802.11 sub-standards, it scans all the channels
provided by 802.11b/g and 802.11a if a scan is initiated. As 802.11a
access points are quite rare and not deployed at all at the
environments where we collected signal strength measurements, we
wanted to stop the card from scanning 802.11a channels. For this, we
restricted the driver to scan only 802.11b/g channels. During our
analysis, we realized that the driver scans only these channels
actively which have been recently used by access points. The recently
unused channels are only scanned passively. This behavior disturbs our
approach, because it might happen that access points which moved into
communication range will not instantly be recognized. We solved this
problem by forcing the driver to scan all channels actively. In order
to improve the scanning speed, we reduced the dwelling time during
which the card is waiting for responses from access points at each
channel up to 10~milliseconds. The default settings chose randomly
between 5~and 50~milliseconds. Furthermore, the driver cancels an
ongoing scan as soon as application data emerges to be
transmitted. During our bandwidth measurements, the driver stopped
scanning completely, because data was always available to be
delivered. To stop this habit of the driver, we completely disabled
this feature and modified the driver so that it performs a scan
whenever it is asked to do so.

The orinoco\_cs driver
version~0.15\footnote{\url{http://www.nongnu.org/orinoco/}} for the Lucent
Orinoco card is unchanged, because it behaves as required for our purposes.

The signal strength measurements are collected by using Loclib and
Locana~\cite{King2007e}. Loclib is a library that provides methods to
invoke a scan and returns signal strength measurements collected from
the driver of the selected network card. This data then is forwarded
to the so-called Tracer application of the Locana software
suite. Tracer visualizes signal strength measurements while they are
taken. Furthermore, Tracer stores the measurements together with user
generated data, such as position information, into a file for further
processing. We enhanced Tracer to update position information while
scans are performed. This was required to be able to take measurements
while roaming around.

\subsubsection{Local Test Environments}
\label{paper6:sec:localtestenvironment}

\begin{figure*}[!th]
  \centerline{%
    \subfigure[The second floor of the Hopper building. The fingerprint database
      is marked in blue and the movement track is depicted in red.\label{paper6:fig:hopper}]%
    {\includegraphics[width=0.65\textwidth]{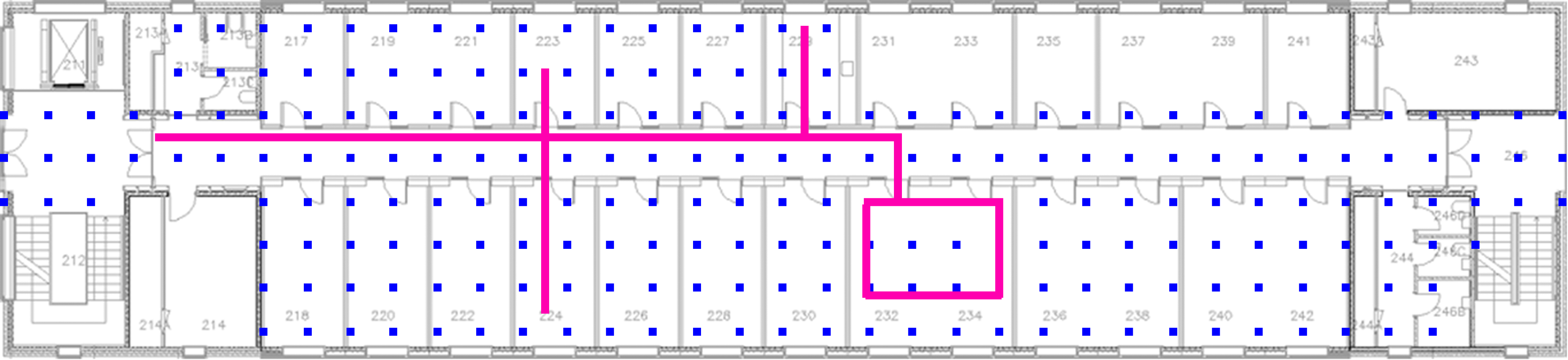}}
    \hspace{0.5cm}
    \subfigure[A wide open lecture hall in the Benjamin building. The red line depicts the movement track.\label{paper6:fig:hall}]
        {\includegraphics[viewport=0 70 840 530,width=0.32\textwidth,clip]{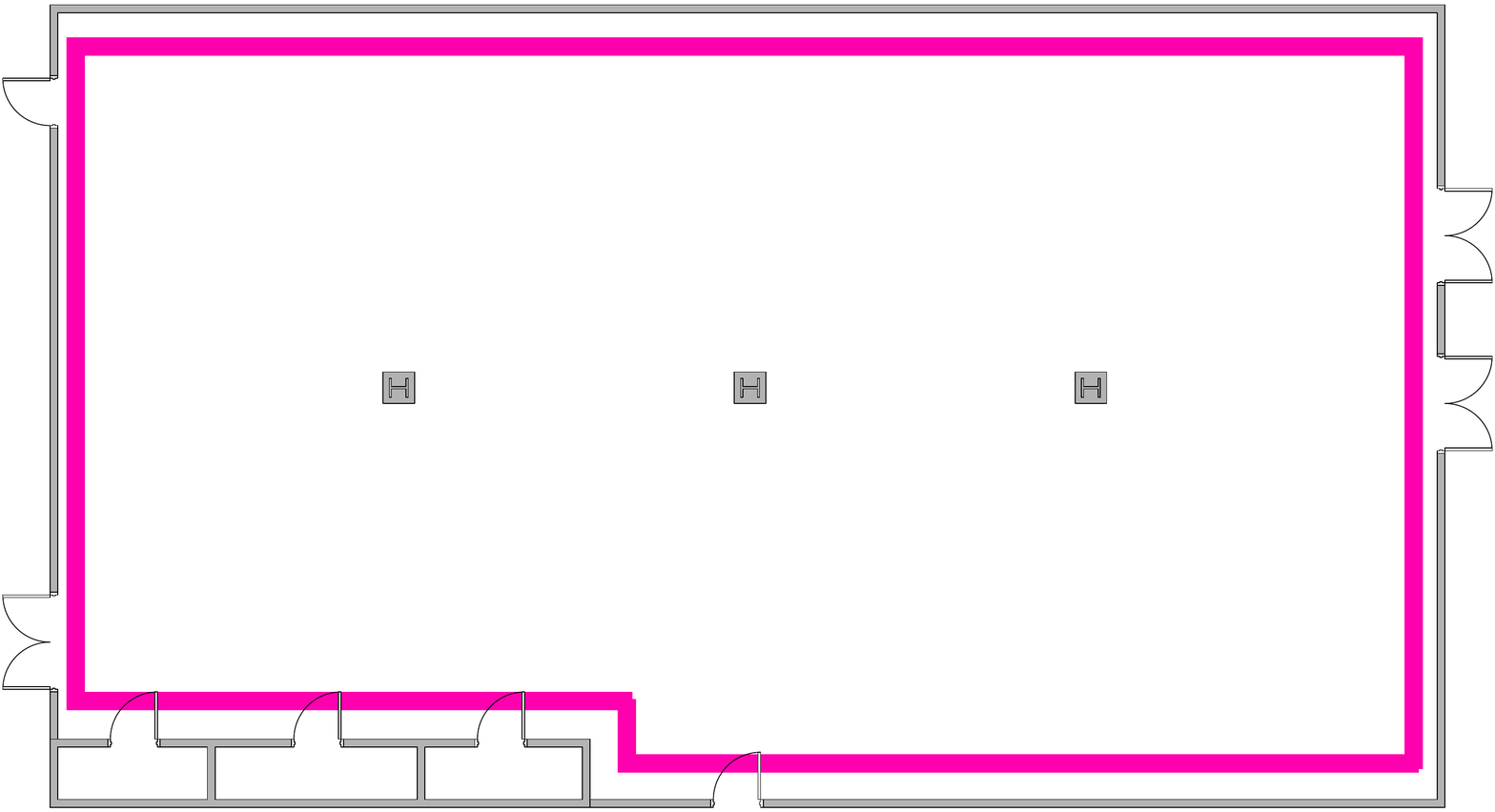}}
  }
  \caption{Ground plans for the two local test environments.}
\end{figure*}

We collected signal strength measurements in two different
environments: On the second floor of the Hopper building and in a
large hall at the ground floor of the Benjamin building at the
University of Aarhus. The former environment is a newly built office
building consisting of many offices (see
Figure~\ref{paper6:fig:hopper}). During a typical day, many people move
around. The area is covered by 23~access points of different vendors
whereas only five of these access points can be detected in half of
the measurements. Nine far-off access points are detectable in less
than ten percent of all measurements. We also deployed a
802.11-based positioning system on this environment covering an area
of 55.7~times 12.7~meters. The blue dots in Figure~\ref{paper6:fig:hopper}
depict the positions where data for the fingerprint database has
been collected.

The latter environment is an old warehouse building refitted to a lecture hall,
which means that the place is scattered with tables and chairs (see
Figure~\ref{paper6:fig:hall}). The hall is 26.3~meters in length and 15~meters in
width. During our measurements, only the people who collected the data were
inside the room. The place is covered with 33~access points but only six are
available in more than half of the measurements. In fact, 19~access points
weakly cover small parts of the hall and hence are only available in less
than ten percent of all measurements.

\subsubsection{Data Collection}
\label{paper6:sec:datacollection}

For the two test environments, we collected signal strength data with two
network cards at the same time. One network card uses monitor sniffing, the
other one active scanning. This allows us to directly compare signal strength
measurements taken by monitor sniffing and active scanning, because they are
collected at the same time in exactly the same scenario.  The network cards
perform an active scan or a monitor sniff every 600~milliseconds.

To be able to compare different network cards, we collected data for each
environment with two different hardware configurations. The first
configuration uses the Intel Centrino and the Lucent Orinoco network
cards. The Centrino network card is performing monitor sniffing and the Orinoco
card carries out active scans. For the second configuration, the Centrino
card is configured to perform active scans and the TRENDnet card collects data
by using monitor sniffing.

For each test environment and each hardware configuration, we collected
four different movement scenarios. The four scenarios consist of two
slow walking scenarios and two fast walking scenarios. \emph{Slow}
walking is defined as an average movement speed below
0.7~m/s. \emph{Fast} walking is defined if the movement speed is above
this threshold.  One of the slow walking and one of the fast walking
scenarios comprise two movement transitions and the other scenarios
include nine. A \emph{movement transition} is defined as start walking
or stop walking. The percentage of time where the person remains still
is varied between different scenarios. We selected the parameters for
these scenarios in such a way that the parameter space containing
movement speed, number of transitions, and percentage of time where
the person stands still is masked. All the scenarios are listed and
described in Table~\ref{paper6:tab:movementscenarios}. This table also names
typical representatives for these kind of movements. The
representation was what we had in mind when defining the movement
scenarios to collect data.

\begin{table}[!h]
\center
\caption{Description of the different scenarios used for data collection.}
{\small
\begin{tabular}{cccc}
Speed & Transitions & \% being still & Example\\
\hline
Slow & 2 & 90 & Meeting attendant \\
Fast & 2 & 40 & Student working \\
 & & & in a lab \\
Slow & 9 & 40 & Student during \\
 & & & lunch break \\
Fast & 9 & 90 & Office worker \\
\end{tabular}
}
\label{paper6:tab:movementscenarios}
\end{table}

To be able to investigate the impact of different times of the day, we
collected all possible options once during typical office hours and
once during evenings. In total, we collected more than eleven hours of
signal strength measurements.

We used the aforementioned Tracer application to trace the walks of
the persons collecting data. For this, we stuck labels on the floor so
that each corner and each dead end of the walking track was marked.
Each time the person carrying out signal strength measurements reached
one of these labels the Tracer application was notified of the arrival
at this particular landmark by a push of a button. Based on the trace
of button clicks, we calculated the average movement speed between two
landmarks. To be able to recognize still periods, another button was
pressed each time the person started moving again.

The data for the fingerprint database was also collected using the
Tracer application. We applied a grid of reference points to the
operation area which includes 225~points with a spacing of 1.5~meter
(see the blue markers in Figure~\ref{paper6:fig:hopper}). During
fingerprinting, we collected 120 signal strength samples at each
reference point, resulting in a total of 27,000 samples. For
the data collection of the fingerprint database, we used the Orinoco
network card.

\subsection{Feature Analysis}
Movement detection using signal strength can be based on several
features calculated from a sliding window of signal strength
measurements. Previous research has explored the features of Euclidean
distance, signal strength variance and rank correlation. The Euclidean
distance feature can be calculated in several forms. First, the gap
form where the Euclidean distance is calculated between the first and
the last signal strength measurement in a sliding window. Second, the
average form where the Euclidean distance is calculated between each
consecutive signal strength measurement and then average
together. Compared to previous work, in this paper we assume that
signal strength measurements are only available for one access point
to support movement detection with monitor sniffing. This assumption
means that the rank correlation feature cannot be used, because this
feature requires measurements from several access points to be able to
rank them in terms of signal strength. The Euclidean distance feature
collapses to the absolute difference in signal strength for one access
point. The goal of this section is to both analyze how such features
behave under movement and whether this behavior can satisfy a number of
reliability requirements which are listed in
Table~\ref{paper6:tab:designrequirements}.

\begin{figure}[h]
	\centering
		\includegraphics[width=0.65\textwidth]{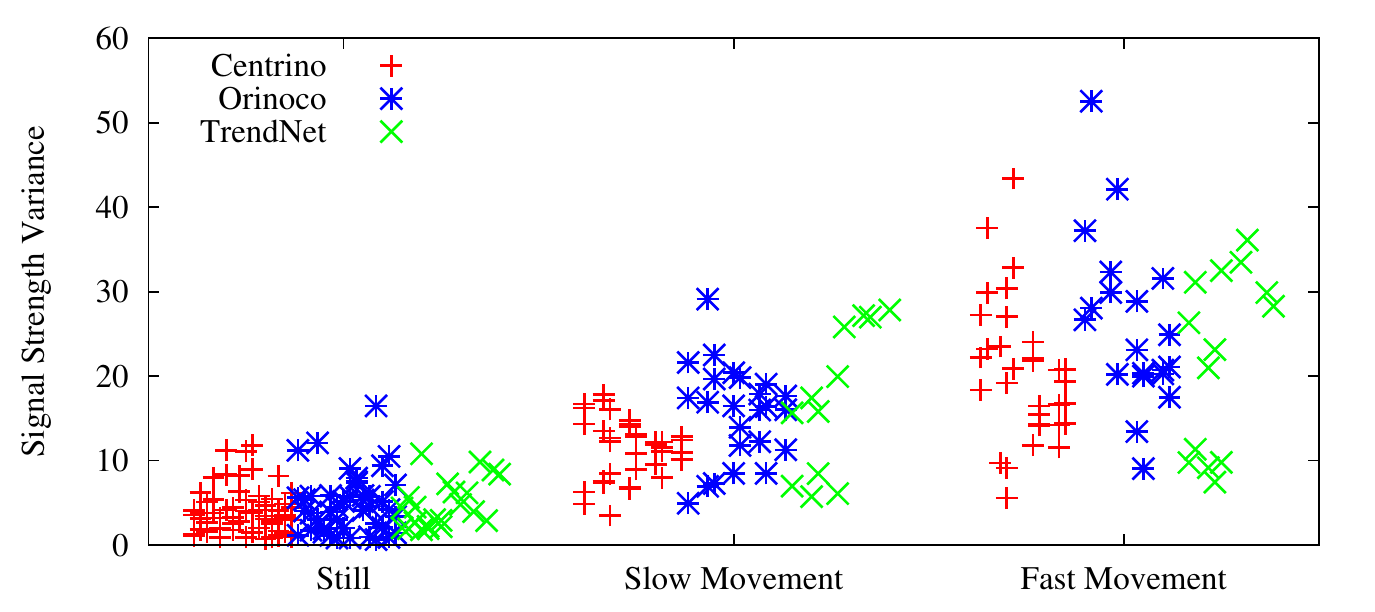}
	\caption{Network cards.}
	\label{paper6:fig:plotAnalysisClients}	
\end{figure}

\begin{figure}[h]
	\centering
		\includegraphics[width=0.65\textwidth]{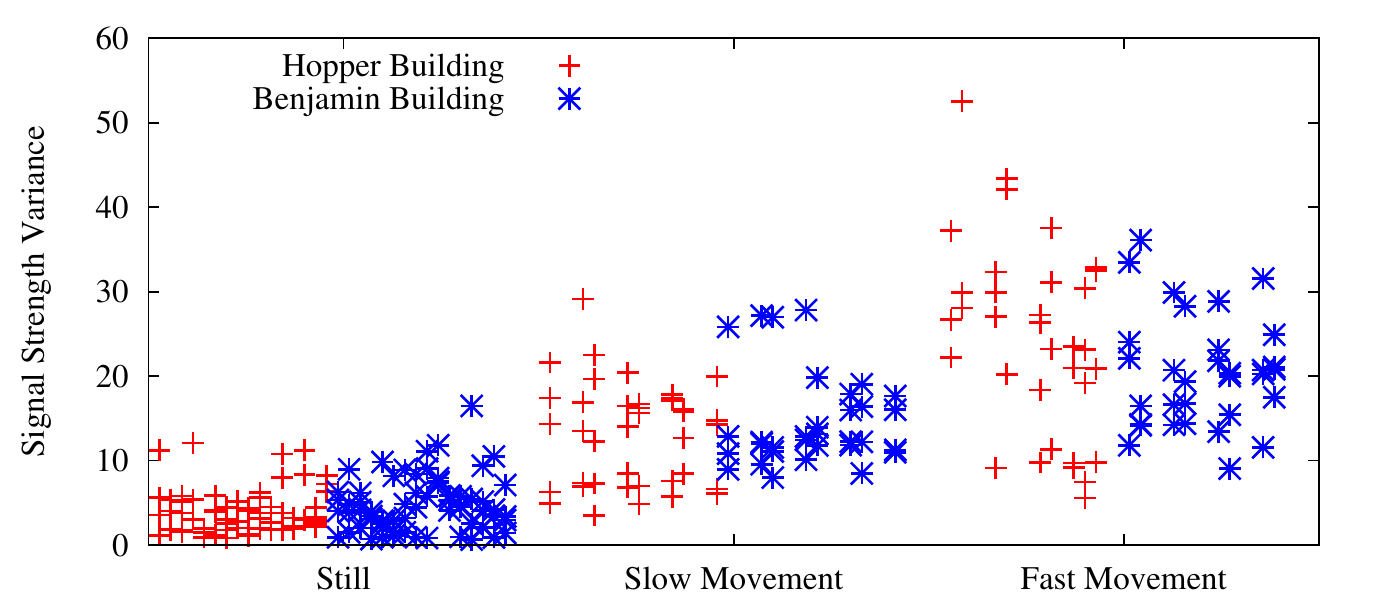}
	\caption{Environments.}
	\label{paper6:fig:plotAnalysisEnvironments}	
\end{figure}

\begin{figure}[h]
	\centering
		\includegraphics[width=0.65\textwidth]{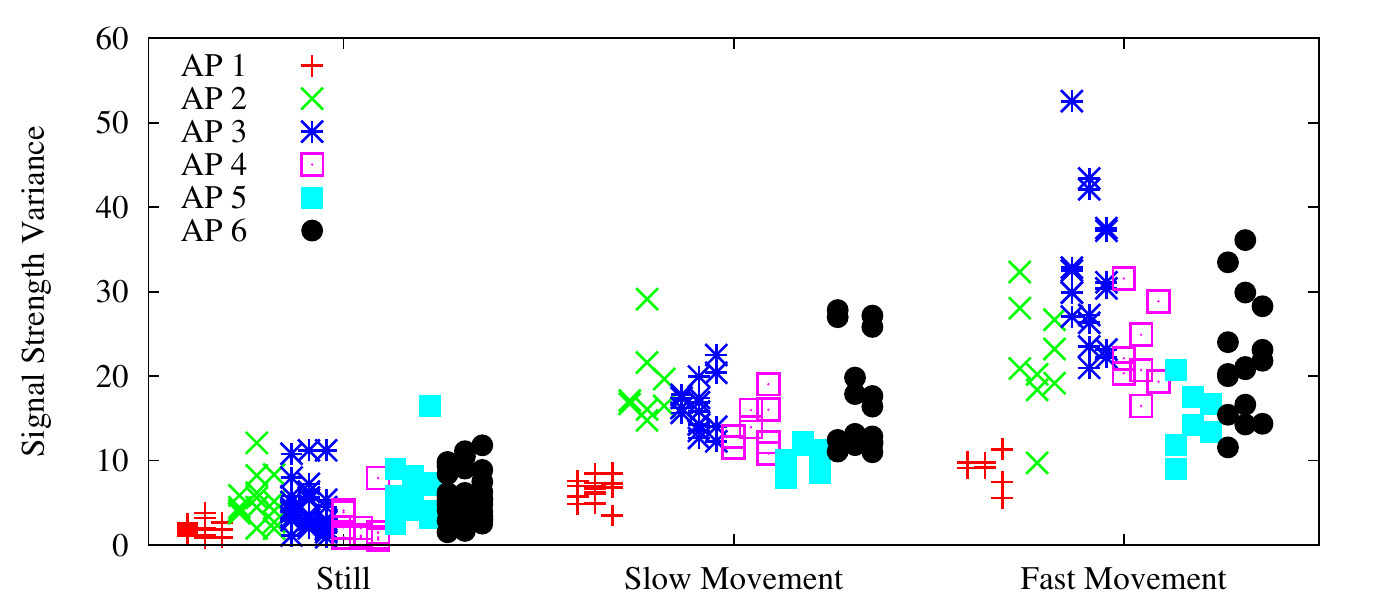}
	\caption{Access points.}
	\label{paper6:fig:plotAnalysisAPs}	
\end{figure}

The data collected as described in Section~\ref{paper6:sec:experimentalsetup}
allows us to evaluate, by means of emulation, the mentioned features
with respect to these requirements. In the following, we focus on the
feature of signal strength variance which is the feature that we
later, based on our emulation results, will choose for our validation
of the system. For the analysis, Figure \ref{paper6:fig:plotAnalysisClients}
to \ref{paper6:fig:plotAnalysisAPs} plot the median signal strength variance
for each of the 64~different options in our data set within the
categories still, slow and fast movement. The feature values are
calculated using a ten measurements long sliding window. They are
calculated for three access points from the Hopper building and for
three access points from the Benjamin building. These six access
points were chosen out of all measured access points to be
representative in order to increase the readability of the
graphs. Each single data point of the graph therefore represents the
median feature value for a specific access point in one of the
64~options. Within each category, the data points are distributed based
on when they were collected for readability of the graphs.

\begin{table}[!h]
\center
\caption{Requirements.}
{\small
\begin{tabular}{ll}
1. & Detect using both monitor sniffing and \\
   & active scanning measurements. \\ 
2. & Work with different network cards. \\ 
3. & Work in different physical environments. \\ 
4. & Work with access points with different \\
   & physical replacement. \\ 
5. & Work both at day and at night. \\ 
6. & For people with different mobility patterns. \\ 
\end{tabular}
}
\label{paper6:tab:designrequirements}
\end{table}

A central feature of the ComPoScan system is the use of light-weight
monitor sniffing in addition to invasive active scanning. Therefore it
is important that the chosen feature works with signal strength
measured using either monitor sniffing or active scanning. Our
data analysis confirms that both with active scanning and monitor sniffing
measurements the signal strength variance increases with
movement. Another important consideration for the system is that it
should work with different network cards. In
Figure~\ref{paper6:fig:plotAnalysisClients}, data is plotted marked and
grouped with respect to the collecting network card. From the figure
it can be noticed that the median signal strength variance changes
similarly across the categories for all three network cards.

\begin{figure}[h]
	\centering
		\includegraphics[width=0.65\textwidth]{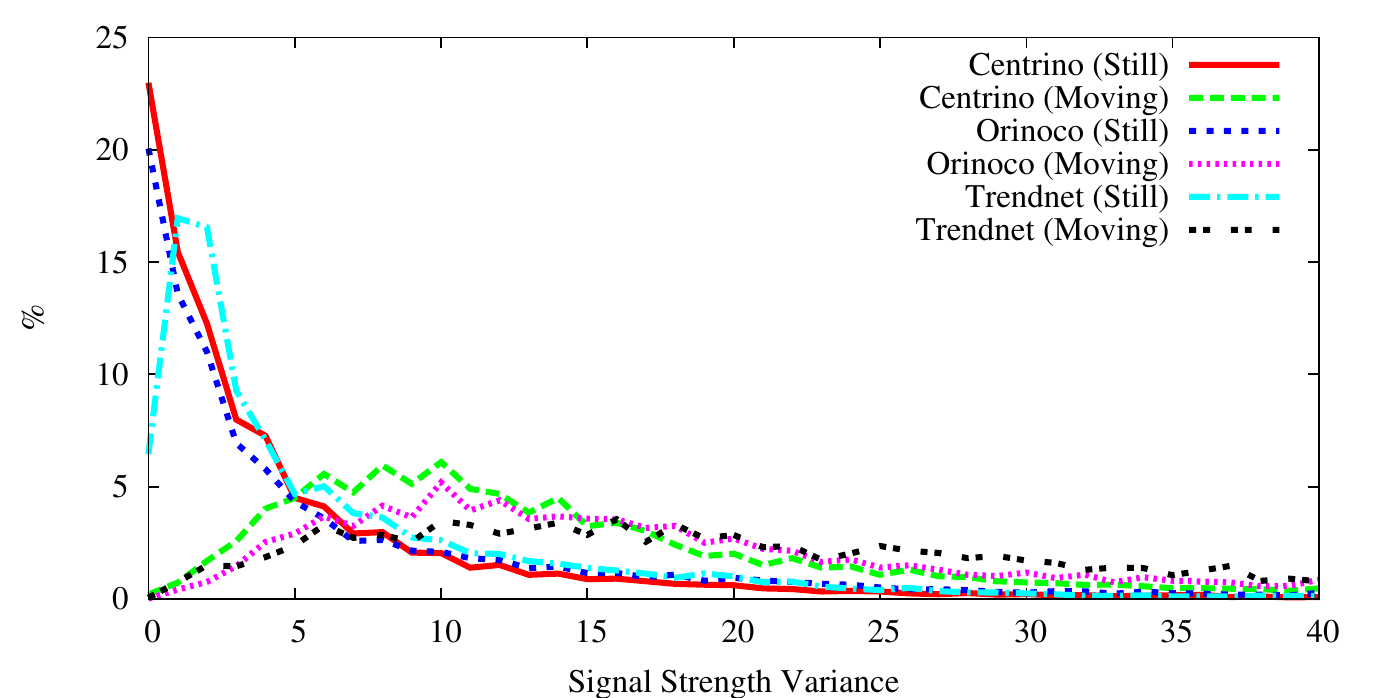}
	\caption{Network card distributions.}
	\label{paper6:fig:plotAnalysisDistClients}	
\end{figure}

\begin{figure}[h]
	\centering
		\includegraphics[width=0.65\textwidth]{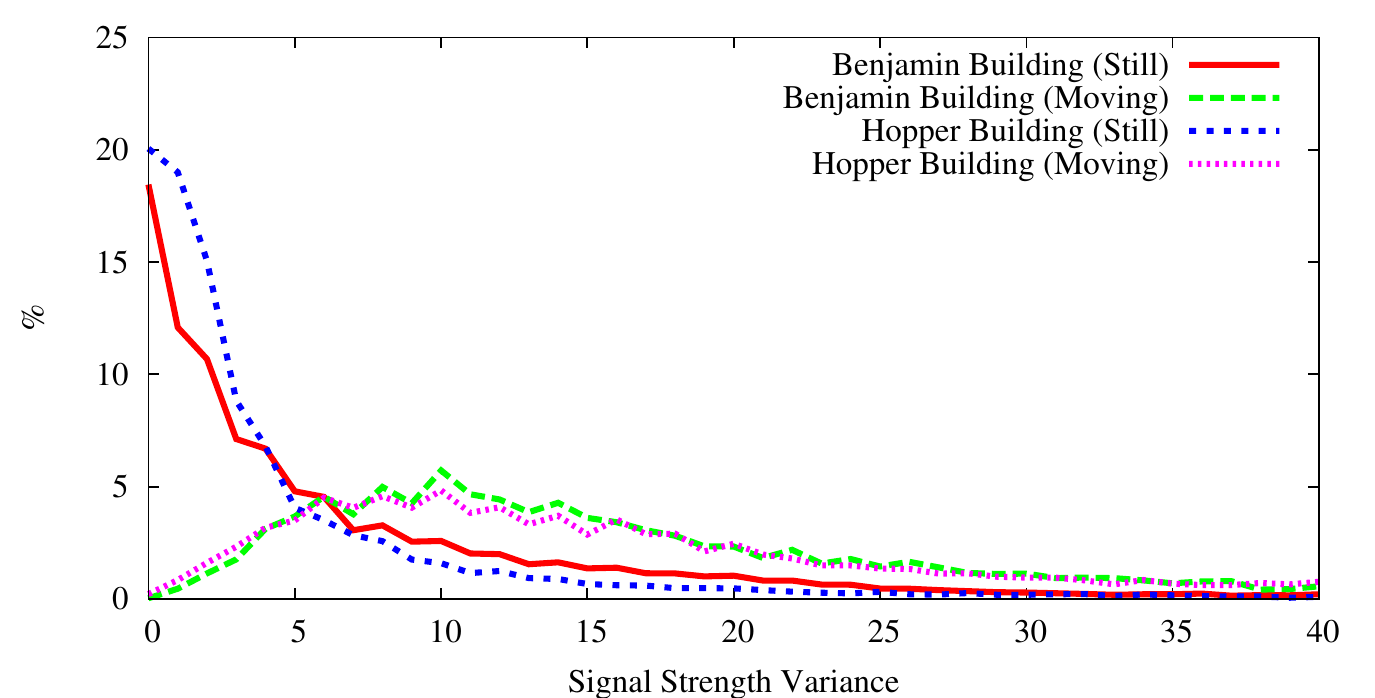}
	\caption{Environment distributions.}
	\label{paper6:fig:plotAnalysisDistEnvironments}	
\end{figure}

To consider the impact of different environments,
Figure~\ref{paper6:fig:plotAnalysisEnvironments} plots the data marked and
regrouped depending on where they were collected. From the plot it can
be noticed that in both environments the median signal strength
variance increases with movement. However, the spread in values is
higher in the Hopper building than in the Benjamin building. In order
to further analyze why this is the case,
Figure~\ref{paper6:fig:plotAnalysisAPs} plots the data marked and regrouped
for the six access points. The access points one to three are from the
Hopper building and four to six are from the Benjamin building. For
each environment, the access points are ordered by how often they were
measured and their median signal strength. From the graph it can be
observed that the signal strength variance is higher for frequently
measured access points with a high median signal strength. Therefore
the difference between the two environments can be explained by
different distributions of weak and strong access points. This means
that signal strength variance does not only depend on speed, but also on the
strength of measured access points.

The median signal strength variance has also been analyzed with respect to
the different mobility scenarios and the time of collection, but the graphs for
these have been omitted. For mobility scenarios, the median signal strength
variance showed the same behavior across the categories and had the same
spread. In respect to time the same change in signal strength variance could
be noticed across categories between day and night data.

The above analysis was based on the median signal strength variance split into
three categories. Below, we will only consider the two categories of still and
moving which are our target categories. Above, the median signal strength
variance was considered. In the following the complete distribution of the
signal strength variance is considered for the two categories. Figure
\ref{paper6:fig:plotAnalysisDistClients} plots such distributions for still and
moving for the three network cards. This plot shows that the still
distributions for Orinoco and Centrino tops at zero and TRENDnet at one. The
moving distributions tops for Orinoco at ten, Centrino at eighth, and TRENDnet
at~14. However, the overall shapes of the distributions are the same but
with the TRENDnet distributions tending to include larger values.

In respect to the environment,
Figure~\ref{paper6:fig:plotAnalysisDistEnvironments} plots the still and
moving distributions for the two environments. The still distributions
both top at zero, but the still distribution of the Benjamin building
has a higher percentage of larger values than the Hopper
distribution. The moving distributions both top at ten. The small
differences in the distributions can again be attributed to the
presence of different access points.
 
To summarize, from this analysis we can make several conclusions with
regard to the listed requirements. First, signal strength variance is
consistent when calculated from signal strength values measured using
either monitor sniffing or active scanning. Second, signal strength
variance calculated from measurements collected with different network
cards share the same difference with respect to being still or
moving. Third, with respect to different physical environments a minor
variation was observed and further analysis identified this difference
to be attributed to the physical replacement of access points in the
areas. It was identified that strong access points show the largest
signal strength variance in the different categories. Finally, no
significant differences were identified at different times of a day
and for different mobility scenarios.

\subsection{Methods}

Several detectors have been applied by earlier work to detect movement using
the aforementioned features. Using our data, we have evaluated several of these
and finally selected a Hidden-Markov Model (HMM) following Krumm
et~al.~\cite{Krumm2004} as the best option. In addition to the HMM detector
we evaluated a simple Naive Bayes detector and two AdaBoost detectors
instantiated with ZeroR and Naive Bayes as basic detectors. In
the next section, we will provide some emulation results that will support the
choice of the HMM. The primary difference between the HMM detector and the
other three detectors is that HMM is able to take previous feature values into
account and can thereby minimize that the detector is immediately flipping back
and forth between detecting movement and detecting still. A drawback with the
HMM method is that it is only able to work with one feature type whereas the
others can use several.

The used HMM has two states: still and moving. Probabilities are
assigned to each state for staying or transition to the other
state. The probability of being in either states are initially set to
be equal. Each state also has a distribution associated with it that
gives the probability of observing a feature value in this state. In
each prediction step, a set of consecutive feature values within a
sliding window is used by a Viterbi algorithm to calculate the most
likely sequence of state changes in the model. The estimated state
then is the ending state of the calculated sequence. The distributions
are calculated from a set of training data. In this work, the Gaussian
kernel method is applied, because it creates more generalizable
distributions than the histogram method. Based on initial experiments,
a standard deviation of 0.1 was selected for generating the still
distributions and a standard deviation of 1.5 for generating the
movement distributions by the kernel method. An important point is
that the feature analysis showed that such distributions can be used
for different network cards and for environments with a different access
point availability. This means that the system does not need to be
trained for each specific deployment which will be further supported
by our emulation and validation results.

\subsection{Emulation Results}
\label{paper6:sec:emulation_results}
The purpose of this emulation is to identify a good detector for the
ComPoScan system and to find the parameters for the identified
detector. These should be used in our validation of the system. An
important goal for a detector is that it is good at detecting movement,
but also that it allows us to make different trade-offs to either favor
communication or positioning.

Several types of emulations have been run to evaluate the detectors on
the collected data. Emulations for Naive Bayes using all features, the
HMM with gap Euclidean distance, the HMM with average Euclidean
distance, and the HMM with signal strength variance have all been run
in an extension to the Loceva toolkit~\cite{King2007e} implemented by
the authors. For the HMM implementation we used the Jahmm
library\footnote{\url{http://www.run.montefiore.ulg.ac.be/~francois/software/jahmm/}}. The
emulations for AdaBoost using Decision Stumps and AdaBoost using Naive
Bayes were run in the Weka toolkit \cite{Witten2005}. Initially,
emulations were run to find good values for the window size used in
the feature calculations and for the history size used by the HMM
detectors. The results were that higher values of window size and
history size made the detectors better at detecting still but worse at
detecting movement. For our data, this will improve a detector's
overall accuracy because our data contains more still than moving
data. However, for the ComPoScan system the overall accuracy is less
important than a detector's ability to detect movement. Therefore, we
focused on finding a window size and a history size that would make
the detectors good at detecting movement without sacrificing too much
on the overall accuracy. Another reason for keeping the window size
low is that this minimizes the start-up time before the system can
start making predictions and for the history size it minimizes the
computational requirements for the Viterbi algorithm used by
HMMs. Based on these criteria, we selected a window size of ten and
also a history size of ten.

To compare the detectors eightfold cross-validation was applied where data was
split into folds depending on the scenario collection round. This makes sure
that test and training data have not been collected at the some point in
time. For the emulation output, movement was chosen as a positive output and
still as a negative output. This means that we can count the number of true
positives (TP), false positives (FP), true negatives (TN), and false negatives
(FN). From these counts we can calculate the true positive rate $tp = 100\%
\times TP / (TP + FN)$ and the false positive rate $fp = 100\% \times FP / (FP
+ TN)$. By varying the parameters of the HMMs and setting different selection
thresholds for Naive Bayes and the two AdaBoost methods a curve of pairs of
$tp$ and $fp$ can be plotted. Such a graph is known as a Receiver Operating
Characteristic (ROC) curve~\cite{Witten2005}. The curve shows which different
trade-offs can be made in terms of $tp$ and $fp$. The more a graph of a
detector stretches towards the upper left corner the better the detector
performs.

\begin{figure}[h]
	\centering
		\includegraphics[width=0.65\textwidth]{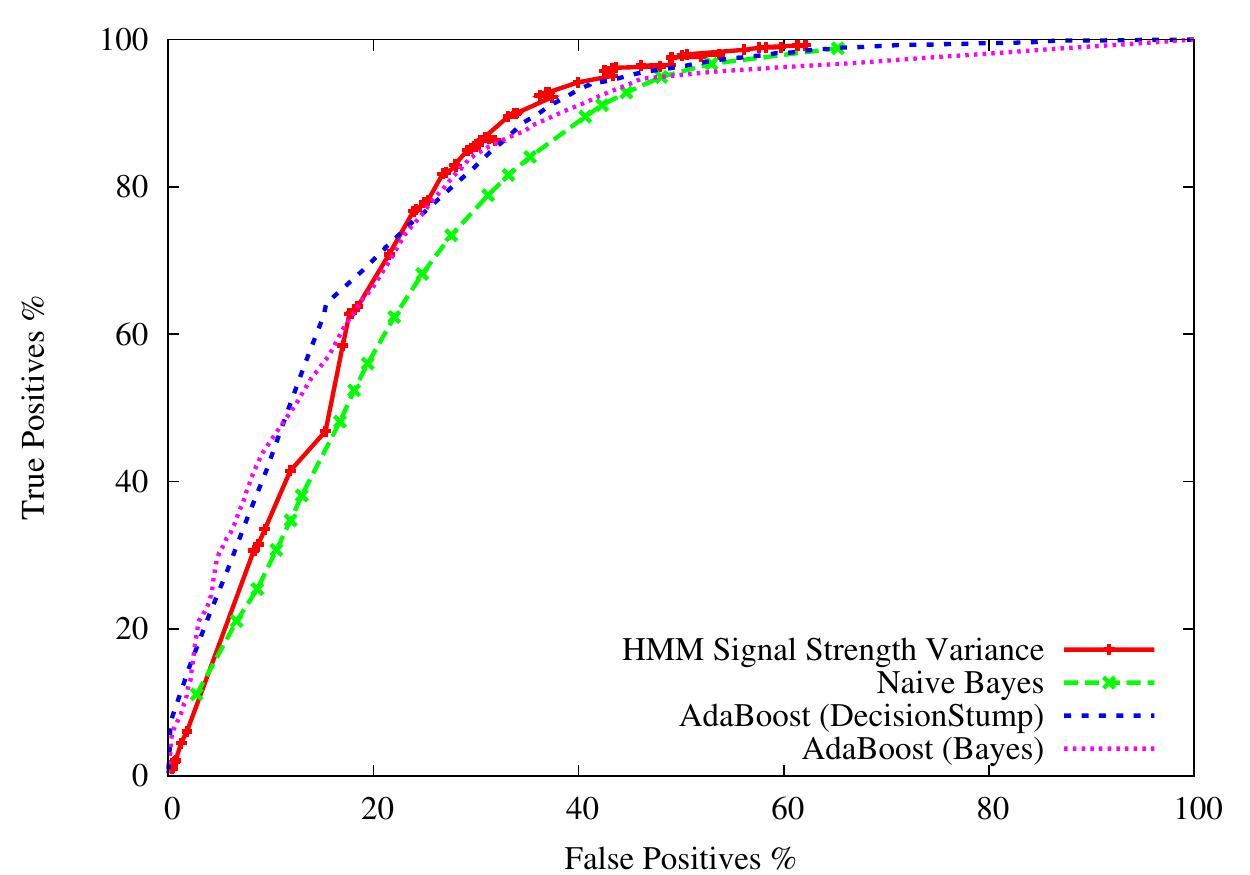}
	\caption{Emulation results as ROC.}
	\label{paper6:fig:ROC}	
\end{figure}

\begin{figure}[h]
	\centering
		\includegraphics[width=0.65\textwidth]{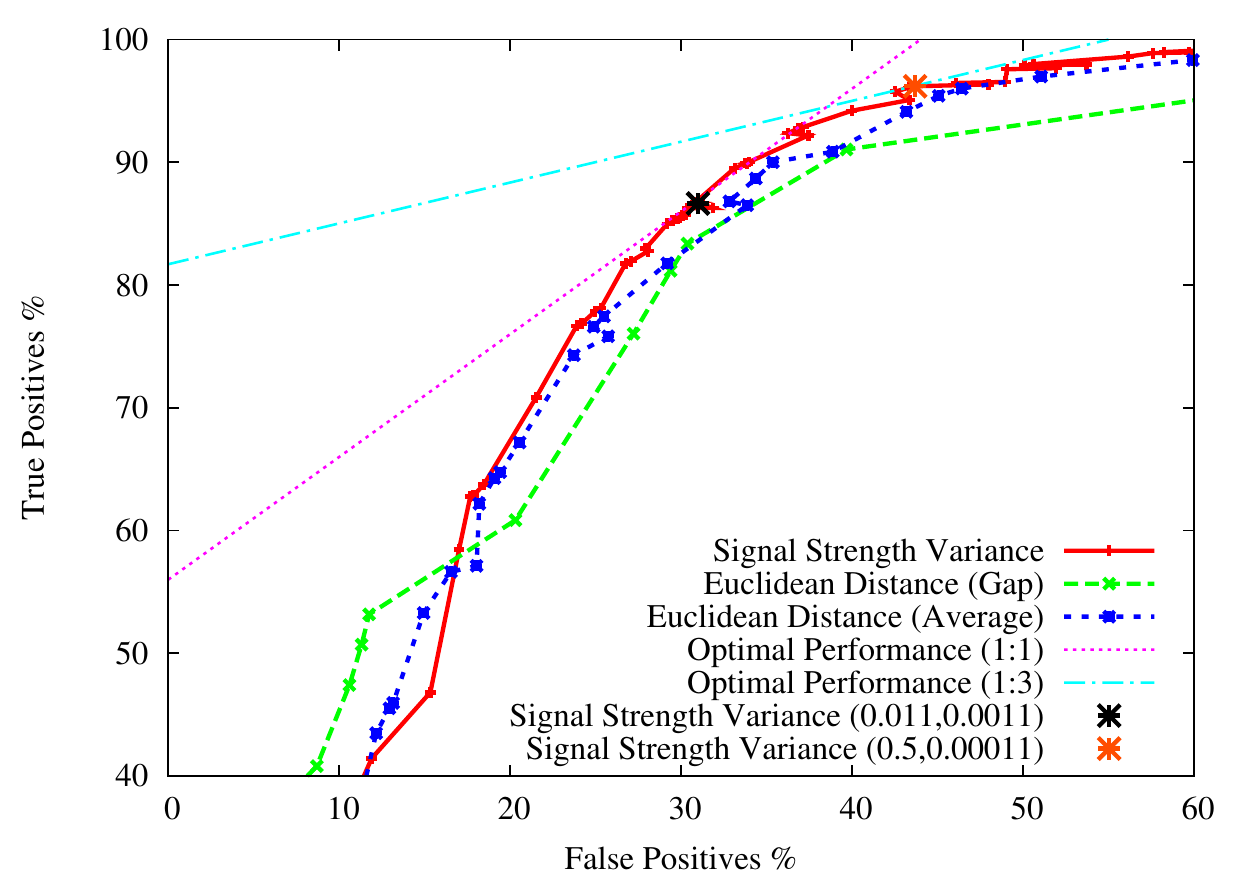}
	\caption{HMM emulation results as ROC.}
	\label{paper6:fig:ROCHMM}
\end{figure}

The ROC curves for Naive Bayes, the two AdaBoost detectors, and HMM
with signal strength variance are shown in Figure~\ref{paper6:fig:ROC}. The
curves show that there are only small differences among the different
detectors. The AdaBoost detector performs best when the $tp$ is below
60, while the HMM with signal strength variance performs best above
80. The Naive Bayes detector performs worst regardless of the
interval. For ComPoScan we are interested in a detector that is good
at detecting movement and therefore we are interested in a detector
that can maximize the $tp$ without increasing the $fp$ too
much. Therefore, the HMM with signal strength variance is the best
choice. We also have emulated the HMM with the other features and the
results are shown, focused on the $tp$ and $fp$ intervals of interest,
in Figure~\ref{paper6:fig:ROCHMM}. Again, there are only small differences
between the detectors. The HMM with gap Euclidean distance performs
best with $tp$ below 60, but in the interval above 80, which we are
interested in, the HMM with signal strength variance performs best
closely followed by the HMM with average Euclidean distance.

To validate our system, we need to fix the parameters for the
prototype. For the HMM with signal strength variance the parameters to
fix are the two transition probabilities of the HMM. To solve this,
in Figure~\ref{paper6:fig:ROCHMM} two lines are plotted which marks the
optimal performance when treating the value of errors in different
ratios. The red line treats false negatives and false positives
equally, and the blue line treats them in the ratio one false negative
to three false positives. On each of the lines we choose one set of
parameters as illustrated in Figure~\ref{paper6:fig:ROCHMM}.

For the parameter set with a probability of 0.011 to change from moving to
being still and 0.0011 to change from still to moving, we ran further
emulations in order to evaluate how the chosen detector addresses the design
requirements listed in Table~\ref{paper6:tab:designrequirements}. The emulations were
run as cross validation with a different number of folds depending on how many
categories they should be split in. Table~\ref{paper6:tab:emulationclients} to
Table~\ref{paper6:tab:emulationaps} list some of the emulation results where each
table entry is named by the test data. So, the first entry in
Table~\ref{paper6:tab:emulationclients} is from an emulation with Centrino
as test data and from the other two cards as training data and so on.

The results for monitor sniffing and active scanning showed that the
detector is working equally well using both monitor sniffing and
active scanning measurements. The results in
Table~\ref{paper6:tab:emulationclients} highlight that there are some
variation across network cards. The TRENDnet network card is best at
detecting movement whereas the Centrino card is worst. If we compare
these results with the distributions for the different network cards
shown in Figure \ref{paper6:fig:plotAnalysisDistClients}, we notice that the
TRENDnet distribution for moving is right shifted compared to the
other distributions. This means that when testing with TRENDnet on
training data from the two other network cards the probability of the
detector predicting still is decreased. So, there are some variations
across network cards, but it is mainly changing the detector's trade-off
between predicting still or moving, not making the detector unable to
detect movement at all.

The results in Table \ref{paper6:tab:emulationenvironment} show that in the
Benjamin building the detector was worse at detecting still compared
to the Hopper building. On the other hand, in the Benjamin building
movement was better detected. Comparing the distributions for the two
environments in Figure~\ref{paper6:fig:plotAnalysisDistEnvironments}, it can
be noticed that for the Benjamin building the still and moving
distributions do have a larger overlap than for the Hopper
building. This can both be attributed to the absence of walls lowering
the variations in signal strength but it can also be attributed to the
difference in access points availability. To analyze this claim,
Table~\ref{paper6:tab:emulationaps} provides results split over the different
access points. For the weakest and least measured access points one
and four, we obtain the best still detection rate because of the lower
signal strength variance as identified from Figure
\ref{paper6:fig:plotAnalysisAPs}. The results also indicate that the access
points one to three from the Hopper building give better still
detection results than the access points four to six in the Benjamin
building.

In all the previous results the feature values were calculated from
measurements for a single access point. However, you can easily extend
the calculation of the feature values to use measurements from several
access points. With active scanning, multiple access points will
normally be measured, but sometimes also monitor sniffing measurement
will be made to several access points that are on the same channel or
on close channels. For signal strength variance, we extend the
calculation to multiple access points by calculating the average value
over the signal strength variance calculated for each single access
point. Denoting the number of access points as $k$, we ran emulations
with a different size of $k$. The results in Table
\ref{paper6:tab:emulationk} indicate that both the detection of moving and
still improves when increasing $k$. In our validation we therefore use
a detector that uses the highest $k$ possible given the current
measurements.

\begin{table}[h]
\center
\caption{Network cards.}
{\small
\begin{tabular}{l|cc}
Card & True positives (\%) & False positives (\%)\\
\hline
Centrino & 76.2 & 12.6 \\
Orinoco & 89.3 & 35.8 \\
TRENDnet & 93.3 & 25.4 \\
\end{tabular}
}
\label{paper6:tab:emulationclients}
\end{table}

\begin{table}[h]
\center
\caption{Physical environments.}
{\small
\begin{tabular}{l|cc}
Environment & True positives (\%)& False positives (\%)\\
\hline
Hopper Building & 83.6 & 20.4 \\
Benjamin Building & 89.9 & 41.7 \\
\end{tabular}
}
\label{paper6:tab:emulationenvironment}
\end{table}

\begin{table}[h]
\center
\caption{Access points (AP).}
{\small
\begin{tabular}{l|cc}
AP & True positives (\%) & False positives (\%)\\
\hline
1 & 78.4 & 10.7 \\
2 & 97.8 & 43.4 \\
3 & 97.1 & 39.3 \\
4 & 95.7 & 32.8 \\
5 & 90.4 & 50.9 \\
6 & 98.6 & 63.2 \\
\end{tabular}
}
\label{paper6:tab:emulationaps}
\end{table}

\begin{table}[h]
\center
\caption{$k$ access points.}
{\small
\begin{tabular}{l|cc}
$k$ & True positives (\%) & False positives (\%)\\
\hline
1 & 82.0 & 29.9 \\
2 & 84.7 & 24.3 \\
3 & 85.7 & 25.3 \\
4 & 85.3 & 23.3 \\
5 & 86.0 & 19.3 \\
6 & 86.7 & 12.2 \\
\end{tabular}
}
\label{paper6:tab:emulationk}
\end{table}

Using emulation, we also have evaluated how ComPoScan impacts position
accuracy. The emulation implements ComPoScan's switching mechanism
between monitor sniffing and active scanning. The emulation in each
step makes a prediction with the chosen detector of being still or
moving. The data used in this prediction depend on what was detected
in the preceding step. So, if moving was predicted in the preceding
step, the detector uses active scanning data, and if still was
predicted it uses monitor sniffing data. Then, in all steps where the
state of the switching mechanism is equal to active scanning, a
position system is allowed to update its current position using the
active scanning measurements. Below, this emulation is compared to the
results from a positioning system that is allowed to update its
position in every step. We refer to this system as traditional
positioning.

For the positioning system, we selected the positioning algorithm that
is probably one of the most studied ones: The Gaussian fit
probabilistic algorithm proposed in~\cite{Haeberlen2004}. For the
positioning system, we applied the fingerprint database as described in
Section~\ref{paper6:sec:localtestenvironment}. To calculate a signal strength
distribution for each access point at each reference point, we randomly
selected twenty measurements out of the 120~previously collected
measurements. As already mentioned, only on the second floor of the Hopper
building a 802.11-based positioning system was deployed. So, all
results presented in the following are collected in this part of the
building. For the emulation, we used the data for the eight scenarios
collected in the Hopper building with the Centrino card collecting
monitor sniffing measurements and the Orinoco card collecting active
scanning measurements.

We compare the emulation results achieved by ComPoScan with the
results obtained by traditional positioning. The average positioning
error with traditional positioning was 3.81~meters and with ComPoScan
it was 3.74~meters.  The results indicate that ComPoScan on average
actually improves the position accuracy with two percent. However, for
two of the eight scenarios ComPoScan decreases the accuracy. From the
emulation we can therefore conclude that ComPoScan does not have a
major impact on the position accuracy and might even improve it in
some cases.


\section{Prototype Implementation}
\label{paper6:sec:prototype}
We wanted to see if our ComPoScan system works in real-world
deployments in the same way as the emulation results suggest. To be
able to deploy the system in real-world environments, we implemented a
prototype that runs on Linux and supports any 802.11 network card that
is able to perform monitor sniffs and active scans.

Based on the emulation results, we selected only the HMM signal strength
variance detector to be implemented. Further, we kept the parameters of this
detector easily configurable to make sure that, during our validation, we can
switch between different configurations to trade communication capabilities
against positioning accuracy.

The implementation is structured into three different parts
that deal with diverse tasks:
\begin{itemize}
\item \emph{Driver:} We wanted the prototype to support at least
  Atheros-based wireless network cards. The reason for this is that we owned a
  NETGEAR WG511T network card that contains the Atheros AR5212 chip-set that
  has not been used for data collection in the emulation. The driver we used
  for this card is modified in the same way as described in
  Section~\ref{paper6:sec:hardwaresoftwaresetup}.
\item \emph{Signal strength measurement system:} Depending on the
  results calculated by the movement detection system, an active scan
  or a monitor sniff is performed to collect signal strength
  measurements. In case that the movement detection system cannot
  calculate any result (e.g., a lack of sufficient signal strength
  samples) an active scan is executed.  To invoke an active scan or a
  monitor sniff, we used the Loclib library~\cite{King2007e}. This
  library collects signal strength measurements from the kernel driver
  and makes them available to user-space applications.
\item \emph{Movement detection system:} After the collection of a
  signal strength measurement, the data is stored with other recent
  measurements in a ten-entries sliding window. For all access points
  that are available in at least eight of the measurements, the signal
  strength variance is calculated. These values are then forwarded to
  the HMM to decide whether the mobile device that provided the signal
  strength measurements is currently moving or not. For the HMM
  implementation we used the same Jahmm library. We trained the HMM to
  detect movement by using the traces we collected in the hall of the
  Benjamin building.
\end{itemize}

Furthermore, to be able to evaluate the impact of ComPoScan on
communication capabilities and the positioning accuracy of a mobile
device, we additionally implemented three sub-systems. The so-called
\emph{network measurement system} gauges throughput, delay and packet
loss. For this, we utilized a tool called
iperf\footnote{\url{http://dast.nlanr.net/Projects/Iperf/}}. Iperf is
a client-server application that measures the maximum throughput
achievable over a given link. To measure round trip delay and packet
loss we implemented a client application that sends out ping requests
every 100~ms. The corresponding server application sends back a ping
response every time a ping request packet arrives. The identification
number, contained in each ping request packet, is copied into the
response. In this way, the client is so able to calculate the round
trip delay by subtracting the time when a ping request packet carrying
a certain identification number was send out from the time when the
corresponding ping response packet arrived. In case that no response
arrives, the request or response packet must be lost during
transmission.

The second and third sub-systems are required to calculate position estimates
based on signal strength measurements. For this, we implemented a \emph{signal
  strength normalization} method that makes signal strength data taken from
different network cards directly comparable. The method we selected is
published by Kjærgaard~\cite{KjaergaardLoca2006}. It finds a linear function to
match signal strength measurements of two different cards by using the least
squares analysis. This approach works pretty well and it can easily be applied,
because only at a few different locations signal strength measurements from
the cards that should be normalized are required. However, normalization can be
completely avoided by using the technique of hyperbolic location
fingerprinting proposed by Kjærgaard et~al.~\cite{KjaergaardPercom2008}. On top of
this stands the \emph{positioning system}. As positioning algorithm we selected
one of the probably most studied ones: The Gaussian fit probabilistic
algorithm proposed in~\cite{Haeberlen2004}.

\begin{figure}[!h]
  \centering{
    \includegraphics[width=0.65\textwidth]{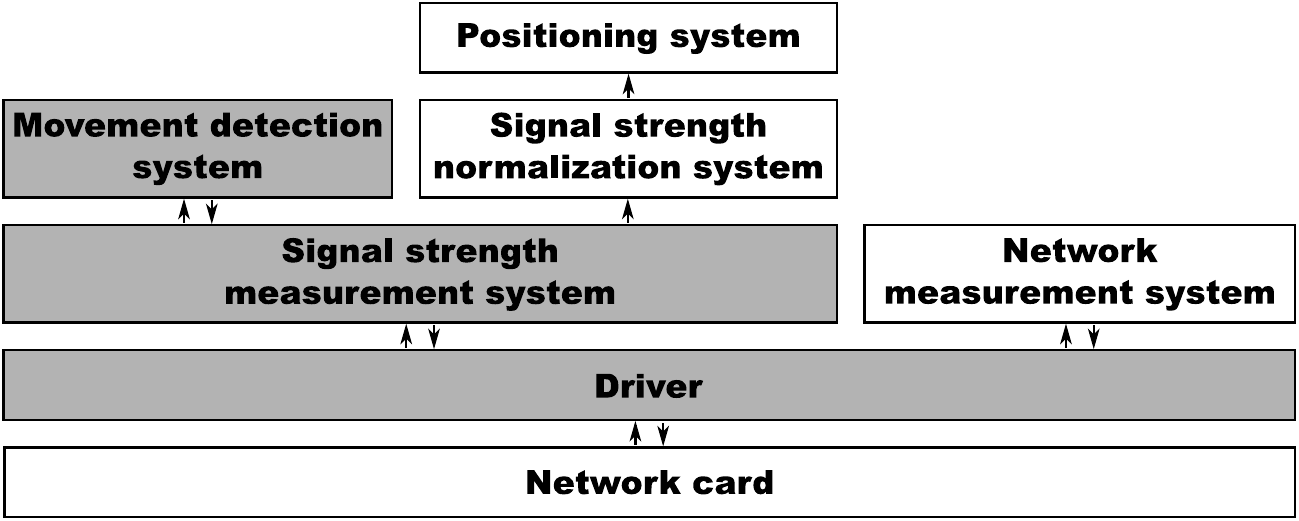}
  }
  \caption{Architecture of the prototype implementation.}
  \label{paper6:fig:architectureprototype}
\end{figure}

In Figure~\ref{paper6:fig:architectureprototype} the architecture of the
ComPoScan prototype implementation is illustrated. The sub-systems
belonging to the ComPoScan system are marked by gray boxes. The three
additional sub-systems required to evaluate the prototype system are
depicted in white boxes. The arrows show how information is
distributed between different parts of the system.

For the evaluation of the prototype we also developed a small tool that writes
a timestamp to a trace file each time its button is pressed. The person who
validates the prototype is supposed to press this button each time walking is
started or stopped.


\section{Real-World Validation}
\label{paper6:sec:validation}
\begin{figure}[h]
  \centering{
    {\includegraphics[width=0.65\textwidth]{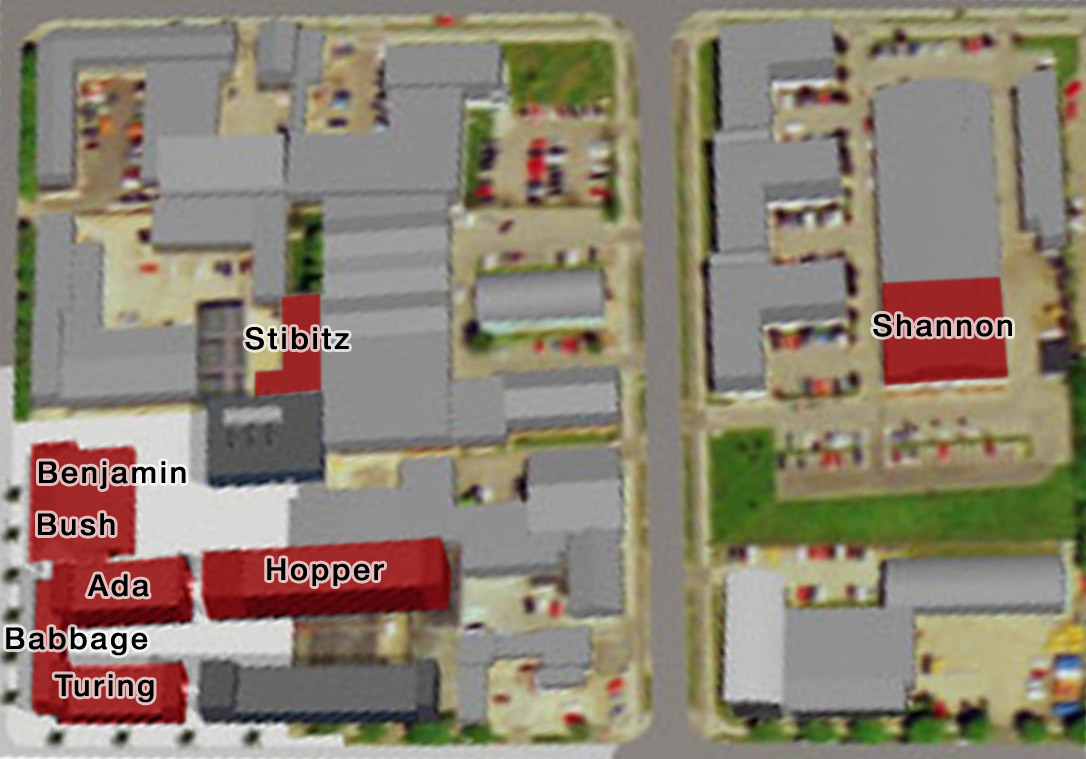}}
  }
  \caption{This map shows the names of the buildings where the ComPoScan
    system has been validated in real-world deployments.}
  \label{paper6:fig:buildings}
\end{figure}

In this section, we present our results obtained from prototype
deployment during a period of more than one week. The system is
deployed in eight buildings of the University of Aarhus, Denmark
(see Figure~\ref{paper6:fig:buildings}). The movement detection accuracy is
illustrated from walks through the Ada, Babbage, Benjamin, Bush,
Hopper, Shannon, Stibitz, and Turing buildings. As the positioning
system is only available in the Hopper building, the position
accuracy measurements are only collected from this place. The
network conditions are also investigated showing the benefit of the
system's configurability by using two parameter sets.

For most real-world experiments we configured the HMM detector in such
a way that we favored stable network conditions over position
accuracy. Or in other words, we wanted to make sure that ComPoScan
only performs active scans if it is quite sure that the person in
question is moving. If it is not stated otherwise, we used the
following parameters: The probability to change from moving to being
still is 0.0011, the probability to change from being still to moving
is 0.011, the window size is ten and the history size is ten.

\subsection{Movement Detection Accuracy}
\label{paper6:sec:movementdetectionaccuracy}

To study the movement detection accuracy of our ComPoScan system, we
recorded a typical route a member of the University of Aarhus would go
from an office of the Hopper building to the Cafeteria located in the
Benjamin building to pick up a cup of coffee. On the way back the walk
contains stops at different locations to chat with colleagues and to
pick up mail and printouts. The path leads additionally through the
Ada, Baddage, Bush and Turing buildings.

We recorded the walk once in the morning when many people move around in the
buildings and once during the night when the building is abandoned. The walk
lasts for more than 25~minutes and contains eleven stops. In both walks, the
person who walked around is standing still in 79~percent of the time.

\begin{figure}[!h]
  \begin{minipage}[b]{1\linewidth}
  \centerline{%
    \subfigure[Real movement\label{paper6:fig:realmovementaccuracy}] 
    {\includegraphics[width=0.99\linewidth]{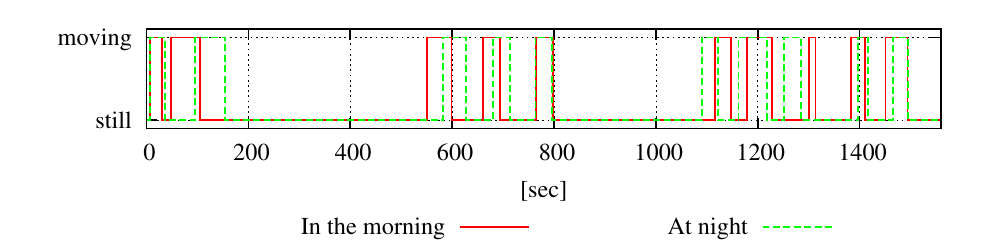}}
  }   
  \centerline{%
    \subfigure[Detected movement\label{paper6:fig:detectedmovementaccuracy}]
    {\includegraphics[width=0.99\linewidth]{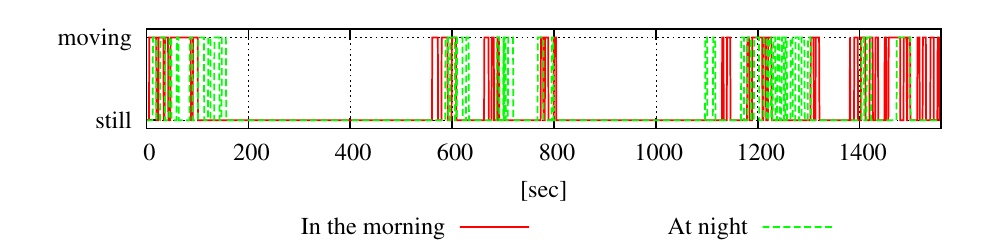}}
  }
  \caption{Movement accuracy.}
  \end{minipage}
\end{figure}

The real movement of the two walks is indicated in
Figure~\ref{paper6:fig:realmovementaccuracy}. Movement as perceived by ComPoScan is
depicted in Figure~\ref{paper6:fig:detectedmovementaccuracy}. As we see from the two
figures, still periods are quite often detected for both walks. The walk
performed in the morning shows correctly detected still periods in 87~percent
of all real still periods. This is in contrast to 94~percent for the walk
during the night. Furthermore, movement is also pretty often
detected. However, the correct detection rates are a bit lower here. For the
walk at night ComPoScan achieves 54~percent and 65~percent for the walk in the
morning. The main reason why ComPoScan detects moving periods better during
the day is the following: During the time we performed our walk many people
arrived at the office to start working or walked to other offices or meeting
rooms to attend meetings. Moving people increase the signal strength variance,
because each time they walk into the path a radio signal is traveling they
attenuate the signal. This means that the signal strength measurements are a
bit more broadened. The scattered measurements add to the signal strength
variance caused by movement which means that movement is easier to detect.

A further analysis of the data shows that there is always a delay between real
movement and movement reported by the HMM detector. Three reasons cause this
delay: First, if we assume that ComPoScan detected the motionless state
correctly then only monitor sniffs are performed. During a monitor sniff only
beacons emitted from access points are examined. These results are stored for
600~milliseconds before forwarded to the HMM detector. This procedure is
chosen to be similar to an active scan. Even though there might be room for
improvements, our further analysis shows that this delay accounts only slightly
to the overall delay. The second reason is the sliding window we utilize to
calculate the signal strength variance. To smooth variations in the signal
strength measurements, we applied a window size of ten entries. It takes some
measurements to propagate an increased signal strength variance through this
window. Third, the history required by the Viterbi algorithm to be able to
calculate the most likely sequence of state changes also adds up to the
overall delay. For instance, the walk during the night shows an average
detection delay of 8.6~seconds with a standard deviation of 13.3~seconds. The
minimum delay we observed is only 156~milliseconds and the maximum detection
delay is 13.4~seconds. The morning walk shows similar delays.

We also configured ComPoScan to use the second parameter set as defined in
Section~\ref{paper6:sec:emulation_results}. Compared to the previously selected
parameter set, this set differently trades movement detection against still
period detection. We expect to see the percentage of
correctly detected movements to go up and the rate of correctly detected still
periods to go down. The comparison of the two parameter sets is based on data
we have collected in the Shannon building. This walk contains five stop
periods which represent about 67~percent of the more than 13~minutes walk. The
results are listed in Table~\ref{paper6:tab:movementaccuracyshannon}.

\begin{table}[h]
\center
\caption{Movement detection accuracy for different configurations in the Shannon building.}
{\small
\begin{tabular}{l|cc}
  & (0.011, 0.0011) & (0.5, 0.00011) \\
\hline
True positives (\%)	& 51.25 & 78.13 \\
True negatives (\%)	& 99.23 & 74.86 \\
\hline
Overall accuracy (\%) & 82.42 & 76.05 \\
\end{tabular}
}
\label{paper6:tab:movementaccuracyshannon}
\end{table}

Even though the overall accuracy drops by six percent, we see that the
second configuration works like expected. The true positives rate
increases from 51.25 to 78.13~percent, which means that the second
configuration correctly detects movement better than the first
configuration. However, this improvement comes with the cost of a drop
of the true negatives rate. The second configuration only recognizes
still periods in 75~percent correctly. Knowing this it is clear that
the overall accuracy is a weak measure for our system. The overall
accuracy depends on the movement pattern and the selected
configuration of our system. To make results comparable the true
positives rate and true negatives rate should be used.

Additionally, the results presented in Table~\ref{paper6:tab:movementaccuracyshannon}
also show that ComPoScan works in different buildings without any further
training. For all our validation walks we only used the data collected in the
hall of the Benjamin building to train the HMM detector. To back up our claim,
that ComPoScan does not require any local training to be functional, we went to
the Stibitz building for another round of validation. The data reveals that
movement is correctly detected in 67.07~percent of all walking periods and
still times are correctly detected in 99.23~percent of all cases. This is
consistent with the other results we have presented.

In summary, we have shown that ComPoScan is able to work well in different
buildings, with different configurations, during different times of a
day, and with a new network card.

\subsection{Positioning Accuracy}
\label{paper6:sec:validationposacc}

As already mentioned, only on the second floor of the Hopper building
a 802.11-based positioning system was deployed. So, all results
presented in this section are collected in this part of the building.
We define \emph{position error} as the Euclidean distance between the
real position of the user and the position estimate computed by the
positioning system. The term average position error refers to a set
of position errors averaged over time.

The positioning system is set up in the same way as described in
Section~\ref{paper6:sec:datacollection}.  The parameters for the normalization system
are gathered by collecting signal strength samples with both the Orinoco and
NETGEAR network card at five randomly selected positions within the operation
area. Based on the least square analysis the parameters are calculated.

During validation, each time the signal strength measurement system
invokes an active scan the measurements are copied and forwarded to
the signal strength normalization system. After normalizing the
measurements, the values are sent to the positioning system. The
positioning system uses only measurements obtained from one active
scan to calculate a position estimate. Before a new position estimate
is calculated the positioning system is reset to discard any knowledge
learned from a previous measurement. We are aware of the fact that
tracking technologies might improve the positioning accuracy, but we
wanted to keep the positioning system as simple as possible to get a
clear insight into the impact of ComPoScan on the positioning
accuracy.

Signal strength measurements collected from a monitor sniff are omitted,
because they usually contain only a sub-set of all available access
points. While it might be possible to increase the positioning accuracy by
using monitor sniffing results, we want to investigate how the position
accuracy drops if movement is not correctly detected and hence signal strength
measurements from active scans are missing.

The Tracer application described in Section~\ref{paper6:sec:datacollection} was used
to collect real position information of the user carrying out the real-world
validation. For this, we stuck labels on the floor to mark prominent places of
the path the user was supposed to walk. Based on the Tracer's trace and known
positions of the labels, we are able to calculate the real position of the
user during the walk. The real position information is later on compared to
the position estimates computed by the positioning system.

\begin{figure}[!h]
  \begin{minipage}[b]{1\linewidth}
  \centerline{%
    \subfigure[Real movement\label{paper6:fig:movementpositioning}] 
    {\includegraphics[width=0.99\linewidth]{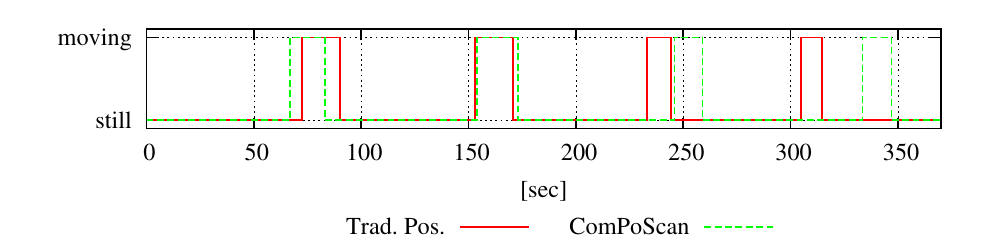}}
  }   
  \centerline{%
    \subfigure[Detected movement\label{paper6:fig:detectedmovementpositioning}]
    {\includegraphics[width=0.99\linewidth]{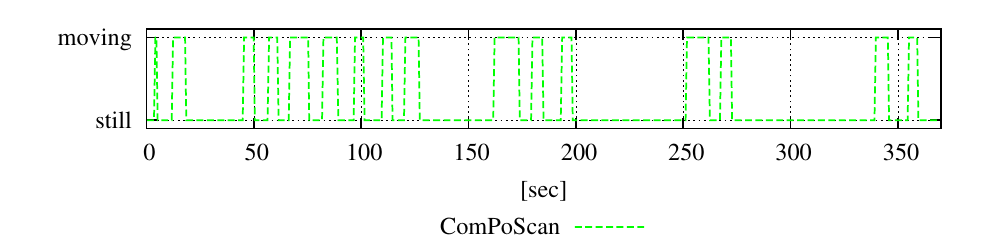}}
  }
  \centerline{%
    \subfigure[Positioning error\label{paper6:fig:positionerror}]
    {\includegraphics[width=0.99\linewidth]{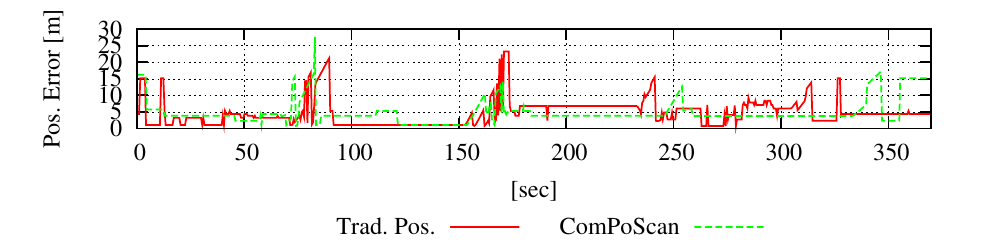}}
  }
  \caption{Positioning accuracy.}
  \end{minipage}
\end{figure}

We selected two walks of 370~seconds each from our data to investigate
the positioning accuracy in detail.  During one walk ComPoScan was
activated, while during the other walk only the positioning system was
running. In the latter setup, the positioning system calculates
position estimates every 600~milliseconds. We refer to this setup as
\emph{traditional positioning}. Real movement and movement detected by
ComPoScan are presented in Figure~\ref{paper6:fig:movementpositioning}
and~\ref{paper6:fig:detectedmovementpositioning}, respectively. The real
movement for the two walks slightly differs, because the person moving
around paused a bit more during the ComPoScan walk.

Figure~\ref{paper6:fig:positionerror} depicts the positioning error for
traditional positioning as well as the ComPoScan system. On average,
ComPoScan achieves an error of 4.68~meters whereas traditional
positioning is slightly worse by obtaining 4.74~meters. From the graph
we see that the curves generated by the two systems look quite
similar. During still periods we observe that the positioning error
produced by ComPoScan is more stable than what can be achieved by
traditional positioning. This is consistent with the average
positioning error only calculated for real still periods. ComPoScan
achieves 4.06~meters in contrast to 4.16~meters produced by
traditional positioning. On the other side, we see from the figure
that the ComPoScan position accuracy is worse than what can be
achieved by traditional positioning during periods of real
movement. On average, the positioning error is 8.22 and 8.12~meters,
respectively. The reason for this is that ComPoScan calculates only a
new position estimate if movement is detected. From the previous
section we know that ComPoScan detects movement always a bit later
than it actually happens. This means if a person already started
walking, ComPoScan is still perceiving the person to be motionless for
a short period of time. The increase in the positioning error is
caused by this delay.

All our traditional positioning data shows an average positioning
error of 4.68~meters. In contrast, the average positioning error over
all ComPoScan data is 4.44~meters. If we compare this to the emulation
results presented in Section~\ref{paper6:sec:emulation_results}, we see a
difference of 0.87~and 0.70~meters, respectively. The difference is
caused by using different network cards to collect the fingerprint
database and while performing signal strength measurement for
positioning. Additionally, the signal strength normalization system
also contributes to the positioning error as shown by Kjærgaard
\cite{KjaergaardLoca2006}.

\begin{figure}[!th]
 \begin{minipage}[b]{1\linewidth}
 \centering{
  \includegraphics[width=0.99\linewidth]{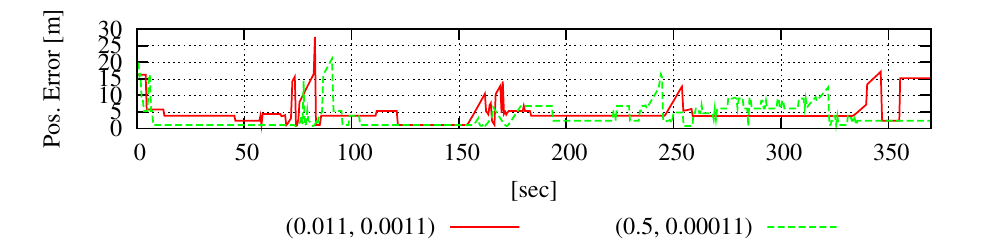}
 }
 \caption{Positioning error for the two different configurations of ComPoScan.}
 \label{paper6:fig:positionaccuracytwoparams}
 \end{minipage}  
\end{figure}

We also configured ComPoScan to use the second parameter set we have chosen in
the emulation section. This parameter set sets the state change probability of
the HMM detector to switch from moving to being still to 0.00011 and the vice
versa probability to 0.5. These parameters are supposed to be more positioning
friendly and hence we expect the positioning error goes down. To be able to
compare the positioning performance with these parameters to the
previously used ones, we walked the path a third
time. Figure~\ref{paper6:fig:positionaccuracytwoparams} shows the position error for
the two configurations. The curve for the previously used parameter set is
exactly what we have seen in Figure~\ref{paper6:fig:positionerror}. If we compare
this curve with the curve produced by the second parameter set, we see that the
previously used parameters are outperformed by the newly applied
parameters. On average, the newly applied parameters achieve a positioning
error of 3.68~meters. This is exactly one meter less than what we obtained
from the previously used parameter set.

To sum up, we have shown that ComPoScan does not harm the positioning
accuracy at all. We even showed that the positioning error goes up a
bit during movement periods and that it slightly drops during still
periods. Further, ComPoScan's configuration parameters can be used to
define its sensibility to compute position updates.

\subsection{Communication Capabilities}

One of the reasons why we came up with the idea of ComPoScan is that
communication is quite weak in terms of throughput, delay, and packet
loss if the 802.11 network card is used for positioning at the same
time. ComPoScan is build in such a way that it trades communication
capabilities against position accuracy. For this it switches to
light-weight monitor sniffing when it detects that the person carrying
a ComPoScan-enabled mobile device is standing still. During monitor
sniffs, untroubled communication with the access point the mobile
device is associated with is possible. In case ComPoScan detects
movement it switches to invasive active scanning to collect enough
data to enable the positioning system to calculate an accurate
position estimate.

The network measurement system as described in Section~\ref{paper6:sec:prototype} is
utilized to generate the results presented in this section. We used an extra
Apple Airport Extreme access point directly connected to a Fujitsu-Siemens
Lifebook T4010 laptop running iperf and the ping server. This setup guaranteed
that the only bottleneck is the wireless channel and not the wired network.

We configured iperf to send a UDP stream for six minutes with a bandwidth of
17.6~MBit/s. This value was determined by a stepwise increase until no gain
could be achieved. The simple UDP transport protocol is selected because the
more sophisticated protocols often bring congestion avoidance strategies that
may interfere with our measurements. As we are the only participant in this
network, the UDP measurements are the upper limit of what is achievable.  We
measured the throughput in the smallest time interval supported by iperf: Every
0.5 seconds. During the time iperf was sending data, we measured the round trip
delay by using our self-developed application. We configured it to send a ping
request every 100~milliseconds.

For the throughput, delay, and packet loss measurements, we walked
around in the Hopper building for six minutes. During this walk, we
stopped five times representing 86~percent of the total time. To be
able to compare our ComPoScan system, we repeated the walk while
performing active scans every 600~milliseconds as a traditional
positioning system would request to do.

\begin{figure}[!h]
  \begin{minipage}[b]{1\linewidth}
  \centerline{%
    \subfigure[Real movement\label{paper6:fig:movementcomm}] 
    {\includegraphics[width=0.99\linewidth]{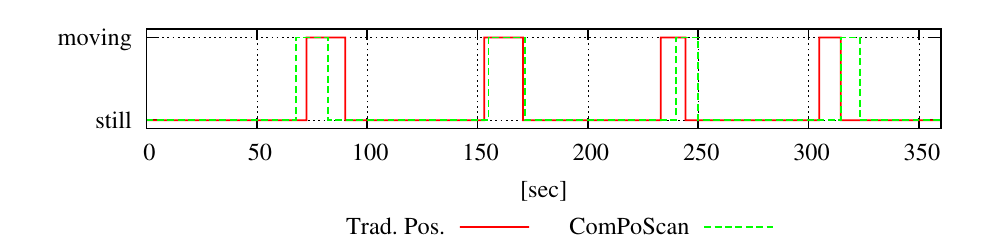}}
  }   
  \centerline{%
    \subfigure[Detected movement\label{paper6:fig:detectedmovementcomm}]
    {\includegraphics[width=0.99\linewidth]{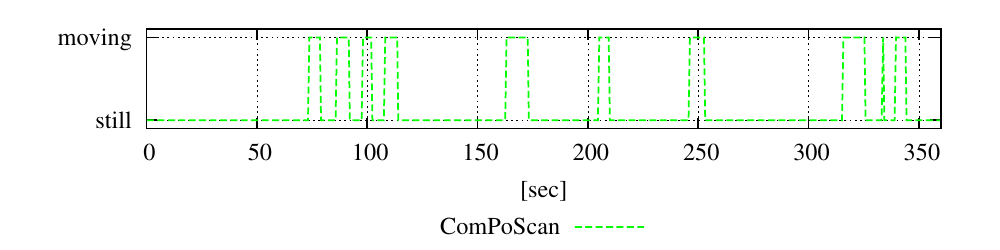}}
  }
  \centerline{%
    \subfigure[Throughput\label{paper6:fig:throughputcomm}]
    {\includegraphics[width=0.99\linewidth]{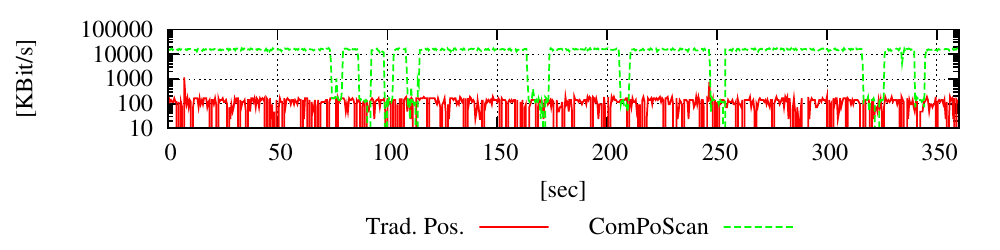}}
  }
  \centerline{%
    \subfigure[Round trip delay\label{paper6:fig:delaycomm}]
    {\includegraphics[width=0.99\linewidth]{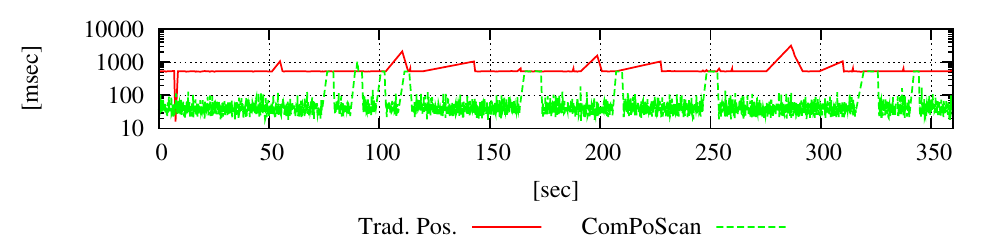}}
  }
  \centerline{%
    \subfigure[Packet loss\label{paper6:fig:packetlosscomm}]
    {\includegraphics[width=0.99\linewidth]{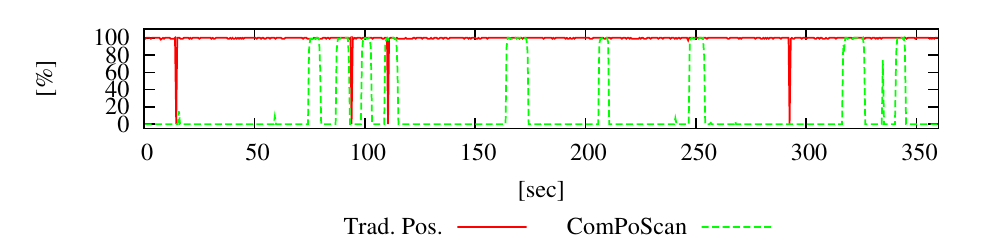}}
  }
  \caption{Communication capabilities.}
  \end{minipage}
\end{figure}

In Figure~\ref{paper6:fig:movementcomm} the real movement for both walks is
presented. Figure~\ref{paper6:fig:detectedmovementcomm} shows how ComPoScan
perceives still and movement periods. The correctly detected rate for
movement here is 58~percent and 84~percent for being still. The
throughput results are depicted in
Figure~\ref{paper6:fig:throughputcomm}. ComPoScan is able to transfer
638.3~MBytes during the six minutes of the experiment whereas a
traditional positioning system reduces the amount of data being
transferred to 5.2~MBytes. This corresponds to an improvement of
factor~122. On average, a throughput of 12.8~MBit/s and 0.1~MBit/s, and
a standard deviation of 5.9~MBit/s and 0.07~MBit/s is achievable,
respectively. During detected still periods, ComPoScan is able to
transfer 14.9~MBit/s on average and 2.9~MBit/s during detected moving
periods. The reason why ComPoScan achieves such a high number during
detected movements is the detector update delay. A detector update is
available every 600~milliseconds right after a monitor sniff or an
active scan returns signal strength measurements.

If we compare the round trip delay, we see a similar picture (see
Figure~\ref{paper6:fig:delaycomm}). For ComPoScan, on average, the delay goes
down to 46.97~milliseconds during detected still periods and up to a
maximum of 1054~milliseconds during detected moving periods. The
average delay during movement is 186.36~milliseconds. This sums up
to a total average delay of 53.46~milliseconds. In contrast,
traditional positioning achieves only an average round trip delay of
566.58~milliseconds and peaks around 3193~milliseconds.

The packet losses are related to the delay. During active scans no
data can be transmitted meaning that the data is dropped after the
different buffers provided by the network stack of the kernel and the
network card driver are filled up. Figure~\ref{paper6:fig:packetlosscomm}
shows the packet loss for ComPoScan and traditional positioning during
the walks. For ComPoScan, the packet loss spikes each time movement is
detected (e.g., around 70, 100, and 200~seconds). During these periods
ComPoScan drops around 72.79~percent of the packets. Only 3.49~percent
of the packets are dropped during detected still periods. On average,
ComPoScan drops 16.07~percent of all ping request and ping response
packets. This is in contrast to traditional positioning where
89.78~percent of the packets are dropped. The reason why we see more
than three percent of packets being dropped by ComPoScan during still
periods is how we count packet loss. We consider the state ComPoScan
reports during the generation of the ping request packet. So, for example,
it might happen that a ping request is sent out when ComPoScan detects
a still period. Directly after the packet left the mobile device,
ComPoScan recognizes movement and starts an active scan. So, the ping
response packet generated by the server application is then dropped,
because it can not be delivered.

We repeated the validation for the communication capabilities by applying the
second parameter set as selected in Section~\ref{paper6:sec:emulation_results}. This
parameter set is supposed to favor position accuracy. The downside of this is
that the communication capabilities might be affected. In the following, we
investigate how severe the impact is.

\begin{figure}[!th]
 \begin{minipage}[b]{1\linewidth}
 \centering{
  \includegraphics[width=0.99\linewidth]{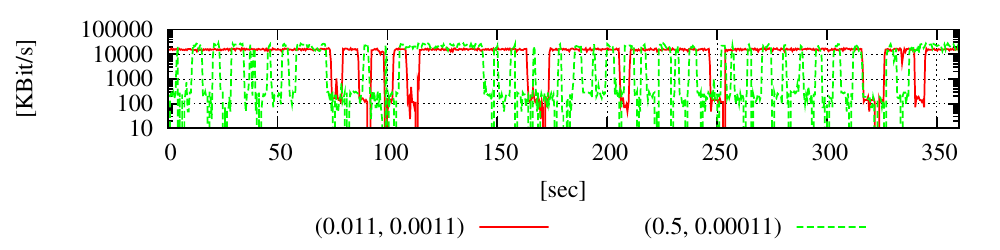}
 }
 \caption{Throughput for the two different configurations of ComPoScan.}
 \label{paper6:fig:throughputcommtwoparams}
 \end{minipage}  
\end{figure}

Figure~\ref{paper6:fig:throughputcommtwoparams} shows the throughput for both
configurations. The spikes in throughput are not always at the same positions
in the graph, because the two different configurations influence the movement
detection system of ComPoScan inducing different detection results. Further,
the person who validated the system walked with slightly different speeds and
stayed still a bit longer at the different places during the two
walks. However, as the walks are both equally long in terms of time and
contain the same number of still periods and still times, the impact of the
two configurations can be compared by looking at the average values over the
total walking time.

For the second configuration, the average throughput is 10.01~MBit/s
with a standard deviation of 10.6~MBit/s. This on average is
2.2~MBit/s less than what we achieved in the previous validation
experiment. The total throughput during the six minutes of the
experiment for the second parameter set is only 482.8~MBytes which
is more than 155~MBytes less compared to the first parameter set. As
shown by the graph, during real still periods ComPoScan using the
second configuration switches quite often back and forth. This is
the reason why we see such a huge standard deviation. The average
delay draws a similar picture: On average, the round trip delay is
81.98~milliseconds compared to 53.46~milliseconds drawn from the
other configuration. The maximum delay for the second configuration
is on the same level as what we observed for the first configuration.
For packet loss, the second parameter set produces a loss rate of
49.62~percent on average. In comparison to the results obtained from
the first parameter set this is an increase of 33.55~percent.

To conclude this section, we have shown that ComPoScan generates communication
conditions that can be used for meaningful data transfer. Further, we
investigated two configurations to show that ComPoScan can be configured to
favor communication capabilities.

\section{Discussion}
\label{paper6:sec:discussion}
In the literature (e.g., \cite{Brunato2005, Smailagic2001}) it is
reported that performing active scans regularly consumes more battery
power than not scanning at all. For instance,
Brunato~et~al. \cite{Brunato2005} state that their HP~iPAQ H5450 PDA
having the 802.11 network card switched off lives for 228~minutes. If
the network card is switched on and associated with an access point
without sending any data, the PDA's battery is depleted in
140~minutes. In case that the PDA scans continuously the lifetime is
only 103~minutes. Inspired by these results, we investigated the
battery lifetime of our IBM~R51 laptop while having the network card
switched off, scanning actively, and performing monitor sniffing. We
followed the recommendations for battery lifetime measurements listed
in~\cite{Brown2006}. Our experiments showed that the six cell
\mbox{li-ion} battery of our laptop provided energy for 192~minutes if
the network card was switched off. In case that active scans were
performed every 0.6~seconds, the laptop lived for 177~minutes. The
lifetime is increased to 184~minutes if monitor sniffing is
performed. These results show that switching on the network card and
selecting different scanning methods matters in terms of battery
lifetime. However, the impact is smaller compared to the results shown
by Brunato. The reason for this is that the battery drain caused by
the different hardware components (e.g., CPU, memory, graphic card) of
a PDA is lower compared to a laptop. The impact of the network card is
smaller for a laptop than for a PDA. However, for
resource-constrained devices such as a PDA, ComPoScan might also be
able to increase their lifetime.

For our movement detection system we heavily rely on monitor sniffing. To be
able to perform a monitor sniff, the 802.11 network card is configured to work
in monitor mode. This mode sets the network card into a listening state that
allows to receive frames sent from a wireless network the network card is not
associated with. Even better, frames from channels close to the channel the
network card is using for communication can be retrieved. Nowadays, most
available network cards support monitor mode. For instance the Intel Centrino
chip-sets (e.g., 2100, 2200, 2915, 3945) and nearly all Atheros chip-sets
(e.g., AR5002G, AR5004X, AR5005, AR5211, AR5212) as well as older chip-sets
such as the Lucent Orinoco chip-sets support monitor mode. Unfortunately, MS
Windows does not support monitor mode by the NDIS driver interface which is
why many drivers for this operating system do not support monitor mode. On the
other hand, Linux and most BSD derivatives provide a wide range of drivers with
build-in support for monitor mode. If the demand for monitor mode grows, we can
expect to see more drivers for MS Windows supporting monitor mode as
well. Therefore, enabling ComPoScan on MS Window is just an implementation
issue.

An easy and simple way to mitigate the impact of scanning on communications is
to reduce the scan frequency. However, this solution comes with the drawback
that the positioning error increases dramatically. For instance, let's imagine
that an active scan is performed only every four seconds. This means that the
positioning system is also only updated every four seconds with new signal
strength measurements. In four seconds a person can walk up to six meters if
we assume a descent walking speed of 1.5~m/s. So, in this scenario, on average,
three meters have to be added to the positioning error of the positioning
system. In indoor environments three meters matter, because they distinguish
between different rooms. From our point of view, such an approach is not a
solution.

Nowadays, modern hard-disks as part of laptops often contain
accelerometers to protect the drive in case it is dropped
accidentally. Many cell phones also contain such accelerometers to
detect automatically if a picture is taken in landscape or portrait
orientation. So, the question is if these accelerometers can be used
to detect movement of a person. Depending on the quality of the
accelerometer and how well it is integrated into a movement detection
system the answer is yes (e.g.,~\cite{Mathie2006,You2006}). Although
movement can be detected by this class of sensors, high-quality
triaxial accelerometers are required and the accelerometers usually
integrated into consumer products are only dual-axis ones which do not
work as well. However, our system provides the advantage that it works
without any additional sensors. This means that all the millions of
802.11-enabled mobile devices already deployed all over the world work
with our system without any hardware modifications. Furthermore, if
accelerometers become omnipresent in mobile devices, our system might
be extended to make use of them and further improve movement
detection.


\section{Conclusions}
\label{paper6:sec:conclusion}
The primary contribution of this paper is the novel ComPoScan system
that can mitigate the effect of scanning on concurrent
communications. ComPoScan is based on movement detection to switch
adaptively between light-weight monitor sniffing and invasive active
scanning. Additionally, we provide an evaluation of the proposed
system both by emulation and by validation in a real-world
deployment. The emulation showed that our movement detection system
works independently of the environment, the network card, the signal
strength measurement technology, and number and placement of access
points. We also showed that ComPoScan does not harm the positioning
accuracy of the positioning system. By validation in a real-world
deployment, we provided evidence for that the real system works as
predicted by the emulation. In addition, we provide results for
ComPoScan's impact on communication where it increased throughput by a
factor of 122, decreased the delay by a factor of ten, and decreased
the percentage of dropped packages by 73~percent. Furthermore, as
mentioned in the discussion, the system is also able to decrease the
power consumption.

In our ongoing work we are trying to address several issues. These
are: First, conceive a system that can make further use of monitor
sniffing measurements for updating the position estimate without
switching into active scanning. Second, switch between monitor
sniffing and active scanning dependent on other metrics (e.g.,
network traffic, user preferences).  Third, evaluate the impact of
using accelerometers to implement movement detection. Fourth,
implement our system on a smaller platform (e.g., PDA) that allows
us to better evaluate the power savings of our system.


\section*{Acknowledgments}
\label{paper6:sec:acknowledgments}
\noindent
The authors acknowledge the financial support granted by the \emph{Deutschen
  Forschungsgemeinschaft} (DFG), the \emph{European Science Foundation} (ESF)
  and the software part of the ISIS Katrinebjerg competency centre.



\markboth{\textit{Bibliography}}{\textit{Bibliography}}
\clearemptydoublepage
\bibliographystyle{abbrv} 
\addcontentsline{toc}{chapter}{Bibliography}
\bibliography{Thesis2008,../../papers/privacy/privacy,../../papers/books,../../papers/standards,../../papers/systems,../../papers/definitions,../../papers/localization/Fingerprinting/Fingerprinting,../../papers/localization/TOA/TOA,../../papers/localization/AOA/AOA,../../papers/localization/VisionSystems/VisionSystems,../../papers/localization/DeadReckoning/DeadReckoning,../../papers/tags/tags,../../papers/mypapers,../../papers/localization/Tracking/Tracking,../../papers/localization/Ad-hoc/ad-hoc,../../papers/localization/LocationMiddleware,../../papers/context/ContextApplications}

\end{document}